# Phonon Dynamics of Topological Quantum Materials

*A Thesis*

*Submitted for the award of the degree*

*of*

**Doctor of Philosophy**

*by*

**Vivek Kumar**

(Reg. No. DI-1702)

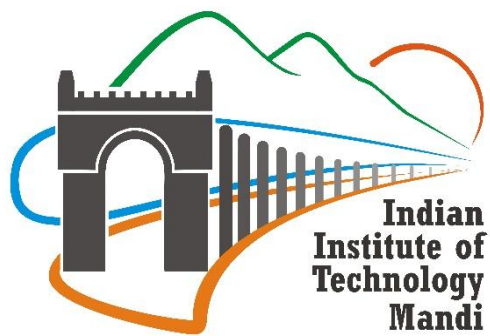

**to**

**School of Physical Sciences**

**Indian Institute of Technology Mandi**

**Kamand, Himachal Pradesh (175005), India**

**March 2025**

# <u>Declaration by the Research Scholar</u>

I hereby declare that the entire work embodied in this Thesis is the result of investigations carried out by **Vivek Kumar** in the **School of Physical Sciences**, Indian Institute of Technology Mandi, under the supervision of **Prof. Pradeep Kumar**, and that it has not been submitted elsewhere for any degree or diploma. In keeping with the general practice, due acknowledgments have been made wherever the work described is based on findings of other investigators.

Place: IIT Mandi                                             Signature:

Date:                                                       Name: Vivek Kumar

Email ID: vivekvke@gmail.com

ORCiD:  https://orcid.org/0000-0001-8836-6776

# <u>Declaration by the Thesis Advisor</u>

I hereby certify that the entire work in this thesis has been carried out by **Vivek Kumar**, under my supervision in the School of Physical Sciences, Indian Institute of Technology Mandi, and that no part of it has been submitted elsewhere for any Degree or Diploma.

Signature:

Name of the Guide: **Prof. Pradeep Kumar**

Date:

Email ID: pkumar@iitmandi.ac.in

ORCiD: https://orcid.org/0000-0002-1079-8592

# Dedicated to Myself

for the joy of discovery

# Acknowledgments

I enrolled in the Integrated PhD in Physics programme at I.I.T. Mandi in 2017, and I have thoroughly enjoyed every moment of my journey. This thesis is devoted to my maternal grandfather, late Sh. Ramkishan (ex. Soldier, Assam Rifles), and my parents for supporting me in pursuing my dream.

I express gratitude to my thesis advisor **Prof. Pradeep Kumar** for his precious guidance and constant support throughout the journey. It would not have been possible without his supervision. He granted me the freedom and creative space to grow, which helped to develop understanding at my own pace. I was fortunate enough to have a guide like him. I am also deeply thankful to my doctoral committee members Prof. C.S. Yadav and Prof. Kaustauv Mukherjee for their positive and critical evaluation of my research work. I would like to specially thank Prof. Samar Agnihotri. He is one of the most influential and amazing personalities I have ever met in my life, whose guidance and insights were very helpful for my academic and personal growth. I would like to thank my school teacher, Mrs. Sarita Ashri, for recognizing and igniting my hidden potential. I also want to thank all the teachers who have ever taught me that have played a significant role in shaping who I am today.

I would like to thank my senior lab mates Dr. Birender Singh and Dr. Deepu Kumar whose support was very crucial in the early stages of my PhD. I thank my lab colleagues Nasaru, Anjali, Sonia, Chaitanya, Atul, Fayaz, Kiran, and Omprakash. Also, I want to thank my friends Adesh, Kaushik, Saurabh, Rasheed, Sonu, Himanshu, Antik, Vivek, Bhumit, Yogesh, Shubhanshu, Azaz, Sunil, Jaskirat, Gurpreet, Joginder, Prakash, and Gautham to make my time enjoyable and memorable. I extend my sincere thanks to the mess worker, medical staff, security guards, cleaning crew, and school office staff for their kind cooperation. I thank all those who have ever helped me in my journey.

It was a journey full of precious lessons that shaped me as a person. Life has been my greatest teacher and taught many lessons and I have embraced it. Since childhood, I found a deep connection with nature and was fascinated by it. The only reason why I pursued physics in the first place, is the pleasure of discovery which still keeps driving me. I think that's the real reward, not money, not fame, or any prestigious title.

At last, I wish to offer a message to all youngsters and society: Learn from nature, it does not discriminate or impose unrealistic expectations. It does not ask a whale to climb the mountains or fly in the sky. Everyone is unique and has infinite potential within them. Follow what drives you, and excites you from within, not to impress or to prove anyone right or wrong but simply to explore and resonate with the universe. Life is not about dwelling on failures or chasing successes, it is about understanding that they are just a consequence of the sequence of events that sum up to an outcome. The journey itself holds more meaning and wonders than the final destination. Embrace change, and keep evolving with time, the whole nature is dynamic and we are part of it. The most important thing for you is you. Mental and physical health should be the top priority. The foundation of any society, nation, or world is an individual. If individuals get the opportunity to reach their true potential, the world will be a better place to live. Be kind and honest to yourself and others around you. Keep exploring and stay curious.

# Abstract

Quantum spin liquid (QSL), a state characterized by exotic low-energy fractionalized excitations and statistics, is still elusive experimentally and may be gauged via indirect experimental signatures. A remnant of the QSL phase may reflect in the spin dynamics as well as quanta of lattice vibrations, i.e., phonons, via the strong coupling of phonons with underlying fractionalized excitations. Inelastic light scattering (Raman) studies on bulk and low-thickness (down to $\sim$ 6-7 layers) $V_{1-x}PS_3$ single crystal evidence the spin fractionalization into Majorana fermions deep into the paramagnetic phase reflected in the emergence of a low-frequency quasielastic response, along with a broad magnetic continuum marked by a crossover temperature $T * \sim 200$ K from a pure paramagnetic state to a fractionalized spin regime qualitatively gauged via dynamic Raman susceptibility. We found further evidence of anomalies in the phonon self-energy parameters, in particular, phonon line broadening and line asymmetry evolution at this crossover temperature, attributed to the decaying of phonons into itinerant Majorana fermions. Inappreciable effect of the long-range antiferromagnetic ordering in flakes as compared to the bulk suggests enhanced quantum fluctuation in the low dimensional regime and as a result prominent QSL state effect.

Topological insulators are characterized by protected gapless surface or edge states but insulating bulk states which is due to the presence of spin-orbit interactions and time-reversal symmetry. An in-depth investigation of a topological nodal line semimetal $PbTaSe_2$ via temperature, polarization-dependent Raman spectroscopy, and temperature-dependent single crystal X-ray diffraction (SC-XRD) measurements. Our analysis shows the signature of electron-phonon coupling as reflected in the Fano asymmetry in the line shape of M1-M4 modes and anomalous temperature variation of line-width of P3-P4 modes. Further polarization-dependent phonon symmetry changes at different temperatures (6K and 300K), discontinuities in bulk phonon dynamics for P2-P5 modes, and disappearance of phonon modes i.e., M1-M5, on decreasing temperatures indicate a thermally induced structural phase transition which is also supported by the SC-XRD results. Hence based on our findings we propose that M1-M4 modes are surface phonon modes, the material undergoes a thermally induced structural phase transition from α to β phase at $T_{\alpha \rightarrow \beta} \sim 150$ K or is in close proximity to the β phase and another transition below $T_{CDW+\beta} \sim 100K$ which is possibly due to the interplay of remanent completely commensurate charge density wave (CCDW) of $1H\text{-}TaSe_2$ and β phase.

Thickness-dependent Raman measurements were performed on $1T\text{-}TaS_2$. This compound is well known for its rich charge density wave phases. Along with that, it has been one of the promising candidates for a quantum spin liquid state as the spins reside on a triangular lattice and it does not show any signature of magnetic ordering. We performed a thickness-dependent Raman measurement in a regime of completely commensurate charge density wave (C-CDW) to a nearly commensurate charge density wave (NC-CDW) with varying temperature (4K-330K) and polarization direction of the incident light. We observed the signature of CDW transition and in addition to that we have also found the signature of a well sought hidden quantum CDW state at low temperature around $T_H \sim 80K$. The emergence of CDW, both normal and hidden one, is marked by the emergence of new phonon modes and distinct renormalized phonon self-energy parameters for the most prominent modes. A transition from metallic to the Mott insulating state is gauged via the Raman response using low-frequency slope $\left( \lim_{\omega \rightarrow 0} \partial \chi^{''} / \partial \omega \right)$, reflected in the renormalized slope below $T_{CDW}$ and $T_H$.

# List of Publications

# Content

# Chapter 1: Introduction

This chapter provides a brief introduction to quantum materials and their significance. It also discusses basic principles and phenomena related to these materials, such as quantum spin liquids, Kitaev model, Mott-insulators, topological-insulators, topological phonons and charge density waves. In the end, an overview of the thesis is provided.

## 1.1 Quantum Materials

Quantum mechanics manifests itself in the physical description of materials at the classical level. In recent years a surge in interest in materials where the quantum effects are active over a wide range of energy and length scales. Such materials consist of superconductors: novel superconductors, heavy fermion superconductors, cuprates; topological quantum materials: topological insulators/superconductors, Dirac/Weyl semimetals, topological heterostructures; strongly correlated materials: Mott-insulators, multiferroics; Weyl and Dirac semimetals, quantum spin liquids, and two-dimensional (2D) materials (Graphene). On the atomic scale, the lattice, charge, spin, and orbital degrees of freedom dynamically couple with each other to produce complex electronic states. Quantum materials inhibit exotic phenomena where strong quantum correlation and long-range entanglement lead to a collective excitation known as quasiparticles, which are very different from fundamental particles and result in phenomena like Mott transition, high-temperature and topological superconductivity, colossal magnetoresistance, and giant magnetoelectric effect. These phenomena are deciding factors for designing next-generation quantum devices that are robust against perturbations such as quantum computers, quantum sensors, quantum communication, imaging devices, etc., which is also considered as the start of the second quantum revolution. **Figure 1.1(a)** shows a summary of concepts in condensed matter and chronological developments. **Figure 1.1(b)** shows different degrees of interactions in strongly correlated condensed matter systems which can be probed using different physical observables.

The topological nature of the quantum wavefunction allows dissipationless electronic currents that are momentum-locked with quantum spin. For example, the presence of vortices in the case of superconductors which are quantized in nature. The quantization of these vortices comes into the picture as the superconducting state has a well-defined phase associated with it, which couples with the magnetic flux. The topological invariant in this case is an integer, the winding number, which conveys how many times the phase winds around the vortex. Another similar quantum mechanical feature is the phenomenon of entanglement, which is non-local in nature. In 1935, Erwin Schrödinger coined the term entanglement and proposed that in such a peculiar situation, the optimal understanding of a whole does not inherently mean optimal comprehension of its components [1]. Such a phenomenon has been seen in the wavefunction of a singlet pair where spins are entangled. Another instance has been experimentally observed where two photons were entangled and teleported successfully over a long macroscopic distance. Such a finding is very crucial for a global quantum communication network [2].

Quantum materials hence, can be thought of as systems where quantum entanglement and topological properties give rise to novel emergent phenomena. In this thesis work quantum materials discussed are $1T\text{-}TaS_2$, $V_{(1-x)}PS_3$, and $PbTaSe_2$. To probe these quantum materials, we have used Raman spectroscopy, single-crystal X-ray diffraction, and density functional perturbation theory. These quantum systems show emergent phenomena like Kitaev spin liquid, Mott insulators, topological phonons, and charge density waves. These phenomena are discussed in brief detail in the sections ahead.

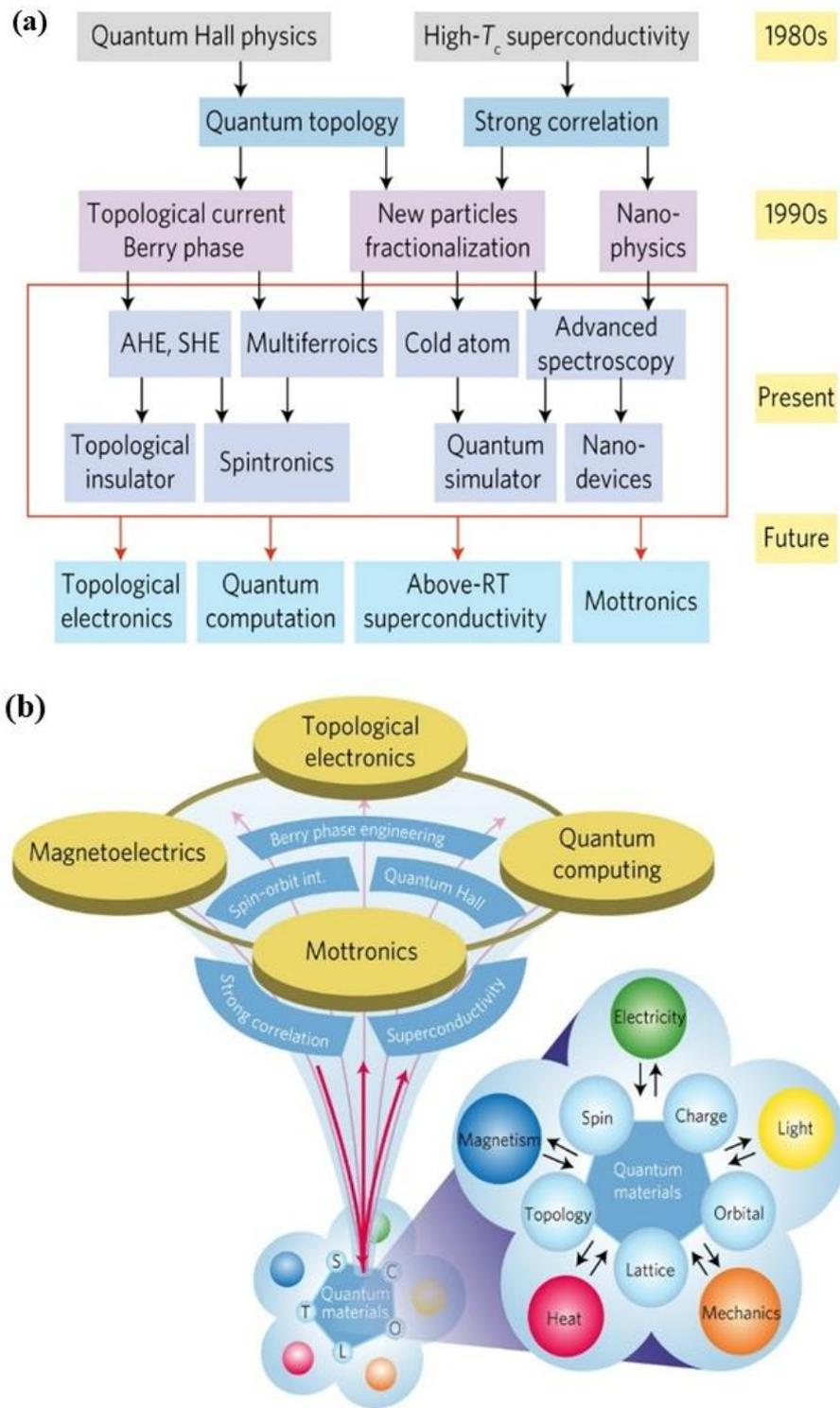

**Figure 1.1:** (a) A brief history of the work done on the quantum materials along with their underlying physics and applications, (b) various possible interactions and corresponding observables in quantum materials that lead to different emergent future applications [3].

## 1.2 Quantum Spin Liquid (QSL)

Magnetism in solids has always been a fascinating phenomenon that has provided significant technological advances over time, having applications in broad fields from data storage, energy generation, magnetic levitation, sensors, medical applications, and electromagnetic devices to next-generation technologies such as quantum computation, quantum communication, energy harvesting, and magnetic refrigeration. The 2D regime of these quantum magnetic materials is going to boost the development of nanoscale engineering.

The underlying physics of typical ferromagnets and antiferromagnets, where the neighboring spins align parallel and antiparallel respectively, was described by the Heisenberg back in 1928 [4]. The origin of exchange interactions can be well understood in the Heisenberg model. In the case of classical spins, the spin quantum number $S \rightarrow \infty$, but we are interested in how a system of quantum mechanical spins behaves. The nearest neighbor Heisenberg interaction Hamiltonian for magnetic solid of an N spin-lattice site where generally $N \rightarrow \infty$ can be written as:

$$H = \frac{1}{2} \sum_{<ij>} J_{ij} \vec{S}_i \cdot \vec{S}_j \qquad \text{- (1.1)}$$

Here $J_{ij}$ is the exchange integral and its origins have quantum mechanical roots which is related to the anti-symmetric nature of the overall wavefunction of the electrons (fermion). The sign of $J_{ij}$ determines the preferred spin orientation in order to minimize the energy i.e. $J_{ij} < 0$ corresponds to Ferromagnetism and $J_{ij} > 0$ corresponds to antiferromagnetism. The spins in this case $\vec{S}_i$ are 3-D vectors with dimensionality D = 3, unlike the Ising model where D = 1. The summation can be taken over a lattice of dimension d (d = 1, 2, or 3). The spin components on the same site follow a commutation relation given as $\left[ S_j^{\alpha}, S_j^{\beta} \right] = i \sum_{\gamma} \varepsilon_{\alpha\beta\gamma} S_j^{\gamma}$ ,

here $\alpha, \beta, \gamma = x, y, z$, whereas the spin components on different sites commute with each other. The spin operator $S_i^2$ has an eigenvalue of $S(S+1)$. Equation 1 can also be written as:

$$H = \frac{1}{2} \sum_{\langle ij \rangle} J_{ij} \left( \frac{S_i^+ S_j^- + S_i^- S_j^+}{2} + S_i^z S_j^z \right) \qquad \text{- (1.2)}$$

Where the spin raising and lowering operators are defined as: $S_j^+ = S_j^x + iS_j^y$ , $S_j^- = S_j^x - iS_j^y$. The limitation here is that the exchange integral $J_{ij}$ is non-zero only for nearest-neighbor lattice sites. For interaction between two spins Heisenberg hamiltonian takes following form:

$$\begin{aligned} H &= J \ \vec{S}_1 \cdot \vec{S}_2 = J \left[ \frac{1}{2} \left( S^{tot} \right)^2 - \frac{3}{4} \right] \\ &= \frac{1}{4} J \ ; \ S^{tot} = 1 \text{ (Triplet) and} \\ &= -\frac{3}{4} J \ ; \ S^{tot} = 0 \text{ (Singlet)} \end{aligned} \qquad \text{-(1.3)}$$

For $J > 0$, the anti-parallel alignment can take advantage of hybridization and reduce their kinetic energy by hopping to the other site where another electron is already present. The hopping is restricted for a parallel spin by the Pauli principle which makes the singlet state favorable than the triplet state.

A more realistic approach in order to describe the behavior of interacting fermions in quantum materials having a strong electronic correlation effect was proposed by John Hubbard in 1963 [5]. This model considers onsite Columb repulsion into the picture and is a modified version of the tight-binding model where non-interacting electron hopping occurs. **Figure 1.2 (a)** shows an arrangement of fermions on a lattice, which is determined by its position in the lattice and magnetic moment, which can be either up or down (blue and red arrow). In this model, the fermions can hop around in the lattice. This model has been successfully employed in order to study a wide variety of complex phenomena are realized via the Hubbard model such as ferro/antiferromagnetism, unconventional superconductivity, charge-density waves (CDWs), quantum spin liquids, etc. The Hubbard Hamiltonian is written as follows:

$$H = -t \sum_{<i, j>\sigma} (c_{i\sigma}^{\dagger} c_{j\sigma} + c_{i\sigma}^{\dagger} c_{j\sigma}) + U \sum_{i} n_{i\uparrow} n_{j\downarrow} - \mu \sum_{i} (n_{i\uparrow} + n_{j\downarrow}) \qquad \text{-(1.4)}$$

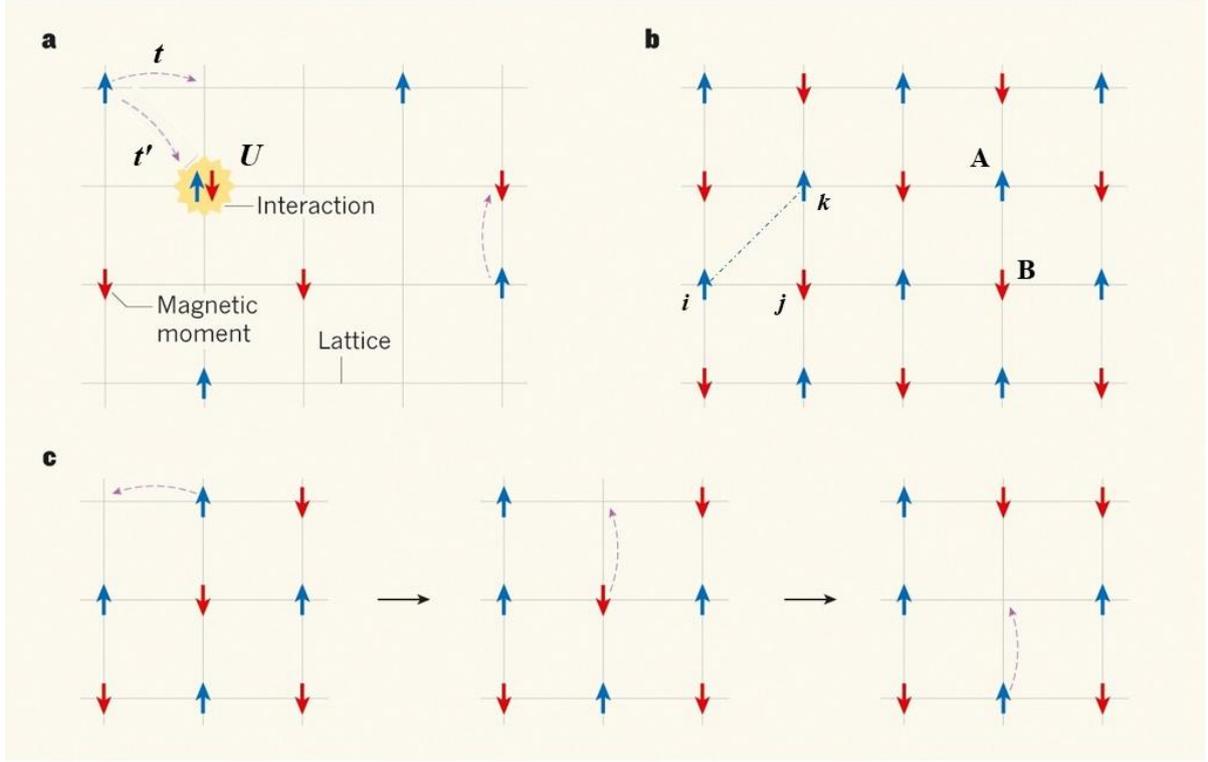

**Figure 1.2:** Key aspects of Hubbard model: (a) Fermions represented by their position in the lattice and magnetic moment (up/blue or down/red arrow). Hopping integral for nearest $t$  and second-nearest $t$ ' is shown by a curved arrow. (b) Low-temperature arrangement of spin for one spin per lattice site, (c) presence of vacancy, where average density is less than one spin per site [6].

Here $n_{i\sigma} = c_{i\sigma}^{\dagger} c_{i\sigma}$ is the number operator which counts the number of electrons with spin ($\sigma$) on the $i^{th}$ site. '$t$' is the hopping interaction integral within the nearest neighboring sites ($<i, j>$) which is also the kinetic energy term and is responsible for the creation of fermion on the $i^{th}$ site ($c_{i\sigma}^{\dagger}$) and the annihilation ($c_{j\sigma}$) at the $j^{th}$ site. The second term '$U$' represents onsite screened coulomb repulsion interaction between electrons and comes into the picture only if the site is doubly occupied. The third term is the chemical potential which dictates the filling of electrons per lattice site. An interesting scenario occurs in the case of half-filling, where there is one electron per site. In this case, the relative strength of $U$ and $t$ i.e., $U/t$ determines the behavior of the material. If the onsite coulomb repulsion $U/t >> 1$, then it is difficult for the electrons to hop to a neighboring site and hence are localized on a particular site and the material is an insulator. Such non-conducting materials which contain an odd number of electrons per site are known as mott-insulators which are predicted to be conductors in accordance with the band theory. If $U/t << 1$ then electrons can hop around and we have a conductor-like behaviour. If $U/t >> 1$ the Hubbard model reduces to the Heisenberg Hamiltonian with $J \sim t^2 / U$ and $\vec{S}_i = \frac{\hbar}{2} c_{\sigma i}^\dagger P_{\sigma\sigma'} c_{\sigma' i}$, here $P_{\sigma\sigma'}$ which represents the three Pauli matrices [7]. If we consider higher-order terms one has to introduce the ring-exchange process, second-nearest neighbor interactions. For example, up to the third order in a strongly coupled Hubbard model, we get the three-spin ring-exchange term which is

$H_3 = -J_3 \sin(\phi_3) \sum_{\Delta_{i,j,k}} S_i \cdot \left( S_j \times S_k \right)$ here $J_3 = -24 t^2 t' / U^2$ is the corresponding three-spin exchange coupling constant described around an elementary triangle $\Delta_{i,j,k}$ on the square lattice as shown in **figure 1.2 (b),** where $t$ and $t'$ are the nearest and second-nearest hopping integrals. A magnetic flux $\Phi_3$ is associated with $\Delta_{i,j,k}$ ($i$, $j$, and $k$ are ordered anti-clockwise) having a scalar spin chirality $X_{i,j,k} = S_i \cdot (S_j \times S_k)$ and giving rise to non-trivial topological effects [8,9].

An interesting case arises if the average density of filling is less than one electron per site. Here, holes play a crucial role as they disturb the magnetic order in the lattice, as shown in **figure 1.2 (c).** Such an exotic nature has been studied by Mazurenko *et al.* [10] in a hole-doped antiferromagnet where they demonstrated that microscopy of cold atoms in optical lattices can be instrumental in order to comprehend the low-temperature Fermi-Hubbard model. Another study showed that doping in the antiferromagnet may give rise to a pseudo gap and high-temperature superconductivity [11].

Heisenberg model for ferromagnets at a temperature above Tc, the thermal fluctuations of order $k_B T$ affect the magnetic ordering which is dictated by the spin-spin exchange interaction constant $J (< 0)$. Here the parallel alignment of spins minimizes the total energy to about $-\left| JS^2 \right|$ per spin, and the ground state is an eigenstate of the Heisenberg hamiltonian. Antiferromagnets can be thought of as consisting of two disjoint sub-lattices (bipartite) A and B, such that the interactions are allowed between sites of different lattices. Having opposite spins the ground state of antiferromagnets is more complicated as compared to that of ferromagnets. Louis Néel proposed that due to negative exchange interaction, the spins at low temperatures will align anti-parallelly [12]. But the Néel state is not an eigenstate of the Heisenberg exchange hamiltonian because the mutual flip of spins will reduce the sublattice order and will eventually disturb the long-range anti-parallel ordering of the spins. In fact, according to Marshall's theorem [13] the absolute ground state for equal-size sublattices A and B is a singlet, i.e. as shown in **figure 1.3 (a, b).** However, this singlet ground state is not uniquely determined and would lead to fluctuations that randomize the spin order.

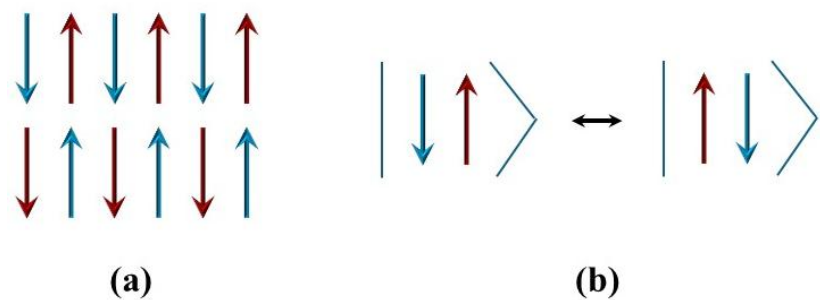

**(a)**　　　　　　　　　**(b)**

**Figure 1.3:** (a) Antiferromagnetic ordering as proposed by Néel, (b) mutual spin flip due to quantum fluctuations which would lead to RVB state or entangled singlet pairs.

In conventional magnetic materials interaction between quantum spins give rise to a phase transition from a disordered spin state at high temperature to an ordered spin state or a spin-solid state at low temperature. Signatures of such a transition are reflected in thermodynamics observables for example, spin entropy reduces to zero as the system attains a unique ground state. The symmetry breaking plays an important role in the emergent states, often it is accompanied by a phase transition which lowers the overall energy of the system. For instance, inversion symmetry is broken in ferroelectric materials below the curie point. In the case of superconductors, U(1) gauge symmetry is compromised below a critical temperature. An interesting scenario appears when the spin entropy is released without breaking any symmetry down to absolute zero, and no local order parameter exists, which is beyond Landau's framework of understanding the phase transitions [14,15]. This novel state forms a highly non-local entangled phase where electron spins behave like a fluid and is known as a quantum spin liquid (QSL).

For a 1D spin chain, the energy per bond of singlet dimers turns out to be $-\frac{3}{4}J$ whereas for the Néel state, it is $-\frac{1}{4}J$. This lower energy of the singlet dimer state reflects the stronger quantum mechanical stabilization owing to entanglement, which is in contrast to the classical alignment in the Néel state. In 1973, Philip W. Anderson proposed that the ground state consists of a superposition of singlet pairs all over the lattice and named it a resonance valence bond state (RVB) which will consume the Néel state [16]. In search of high-temperature superconductors, Anderson proposed that upon doping the parent compound $La_2CuO_4$ which is a Mott-insulator, the singlet pairs are in the RVB state and can be visualized as cooper pairs [17]. Theoretical investigation of high-temperature superconductors still remains elusive while there is a surge in the study of the physics of quantum-disordered spin states [18]. Such states have been proposed for geometrically frustrated magnetic systems having corner-sharing triangles or tetrahedra [19].

Generally, any physical system is considered frustrated if all the components of potential energy are not minimized simultaneously. In the case of magnetic materials having frustrated geometry where interaction of magnetic moments induces an incompatibly in minimizing the systems energies. A simple example of such geometrical frustration is a triangular lattice where the third spin cannot simultaneously align and satisfy $J > 0$ condition and consequently, the spin alignment remains undetermined as also shown in **figure 1.4 (a)** [19]. This problem of geometric frustration was resolved when each spin points to 120° with respect to each other and can lead to long-range magnetic ordering [20,21]. Other instances of geometrically frustrated lattices are kagome and honeycomb structures as shown in **figure 1.4 (c, d)** where the geometrical frustration cannot be overcome by 120°. Orientation as in the case of triangular lattice [20]. Whether the ground state of a kagome lattice is a QSL or not is still under debate but there are theoretical studies that support the prediction [22-25].

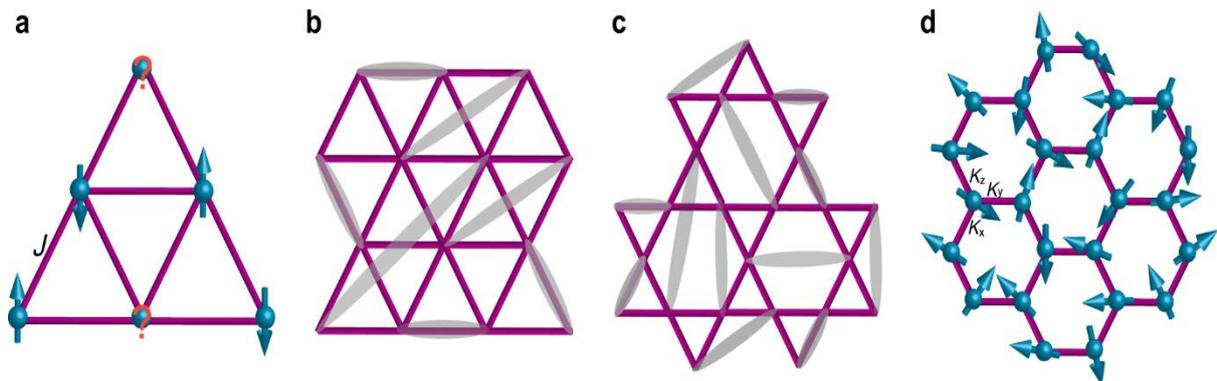

**Figure 1.4:** Instances of geometrically frustrated lattices (a, b) triangular, (c) kagome, and (d) honeycomb. Blue arrows and grey ovals indicate quantum spins and spin-singlet [26].

In general, the fluctuation in a spin liquid can be classical or quantum. In classical limit $S \to \infty$ or $S >> 1/2$ and the non-trivial commutation relation vanishes. Classical fluctuations are dominated by thermal fluctuations and geometric constraints, where the system undergoes a transition among different microstates. The fluctuations tend to decrease or cease totally as $k_B T \to 0$ and the system tends to attain an ordered state. In the case of quantum spins, where $S$ is comparable to ½, the quantum fluctuations do not cease down to zero kelvin known as zero-point energy, owing to the uncertainty principle. In contrast to classical spin liquids, quantum fluctuations in quantum spin liquids are phase coherent and highly entangled and are in a superposition state of non-localized (long-range) spins as shown in **figure 1.4 (b, c)** where light shaded ovals are singlet pairs. This phase exhibits exotic features which arise from its topological character such as fractionalized excitations. These emergent quasiparticles are created in multiple pairs only. Spin systems with high macroscopic ground state degeneracy which induces strong thermal and quantum fluctuations prevent any long-range ordering.

QSLs states are highly probable to be found in magnetic insulators (Mott insulators) which do not show any sort of ordering even at zero kelvin [27-29]. Theory predicts the presence of exotic quasiparticles that emerge from the collective excitation of the interacting spin system called spinons as shown in **figure 1.5.** Spinons carry spin but not charge. They can be bosons or fermions, and the associated excitations may be gapped or gapless. Spin liquid state can be thought of as a deconfined phase where it is related to the lattice gauge theory coupled to the matter field. Spinons are also coupled to an underlying gauge field, which may be U(1) or $Z_2$ in nature. A complete analytical theory is still lacking to comprehend the possibilities these materials may hide.

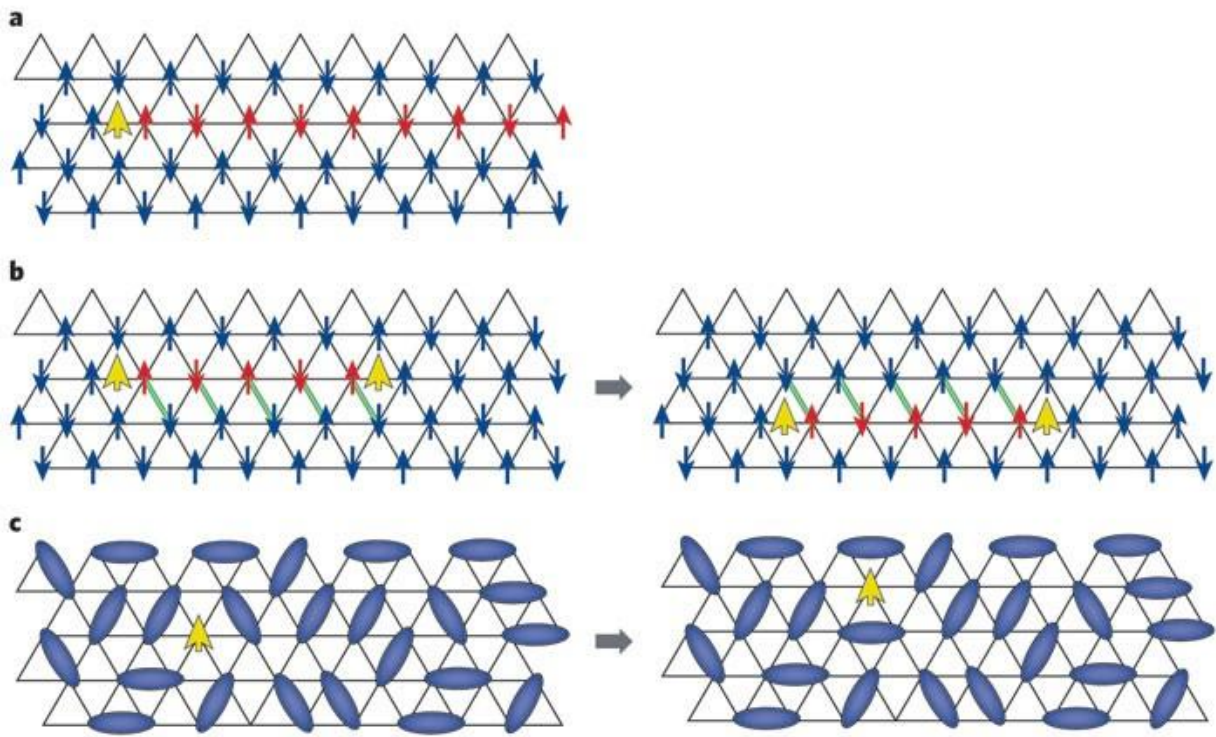

**Figure 1.5:** (a) Formation of spinons (yellow arrow) by flipping the chain of spins (red arrow) on a quasi-1D lattice, (b) bound triplon formation which coherently moves in the lattice by flipping finite no of spins along green bonds, (c) spinons in 2D lattice is an unpaired spin (yellow arrow) [19].

### 1.2.1 Experimental Probes of QSL

It has been a very challenging job to find candidate QSL materials and probe them experimentally. Still, a lot of progress has been made theoretically and experimentally for 1D, 2D and 3D (D- dimension) materials. 1D uniform Heisenberg spin-1/2 chain does not have a gap in the triplet excitation spectrum and is in a disordered state for isotropic case but any sort of anisotropy may result in long-range ordering at $0K$. Very well case studied case for 1D is $X_2Cu(PO_4)_2$ (X=Sr, Ba) [30], $KCuF_3$ [31]. In 2D materials, the reduced dimensionality increases the quantum fluctuations, for example, in the case of $Na_2IrO_3$ [32], $\alpha$-$Li_2IrO_3$ [33]and $\alpha$-$RuCl_3$ [34], all have a honeycomb crystal structure. There have been efforts to search for 3D QSL materials such as hyperkagome $Na_4Ir_3O_8$ [35] and $PbCuTe_2O_6$ [36], hyperhoneycomb $\beta$-$Li_2IrO_3$ [37], stripyhoneycomb $\gamma$-$Li_2IrO_3$ [38], Triangular $\kappa$-$(BEDT-TTF)_2Cu_2(CN)_3$ [39] and $1T$-$TaS_2$ [40].

Investigating for QSL state down to $0K$ is not practically possible; however there can be some experimental techniques that can provide some solid signatures. It is often considered that at a temperature below around two orders of magnitude of the magnetic exchange coupling can be taken as representative of the properties at $0K$ provided there is no other phase transition present. Magnetic susceptibility can provide a crucial hint of magnetic ordering $\chi \approx C/(T - \Theta_{cw})$, here $C$ is the Curie constant and $\Theta_{cw}$ is the Curie-Weiss temperature that is an indication of the strength of exchange interactions. $\Theta_{cw} < 0$ and $\Theta_{cw} > 0$ indicates dominant antiferromagnetic and ferromagnetic interactions respectively. In case of magnetically frustrated materials, by comparing $\Theta_{cw}$ with the temperature ($T_C$) where the order freezes, one can estimate the magnitude of frustration which is defined by the frustration parameter as $f = |\Theta_{cw}|/T_C$. The value of $f > 5-10$, typically indicates strong frustration and the temperature range of $T_C < T < |\Theta_{cw}|$ defines the spin-liquid regime [41].

Macroscopic specific-heat measurements give information about the low energy density of states (DOS) which further can be verified from the theoretical predictions. In the case of magnetic ordering, it shows a λ-type type peak in the specific heat vs temperature curve and reflects the presence of magnetic excitations at low temperatures. But in the case of QSL, such a sharp transition is absent unless a topological phase transition is present [42]. To obtain the residual magnetic entropy, which is the indication of long-range magnetic ordering at low temperatures, one has to carefully subtract the phononic contribution [43,44]. Further microscopic measurements such as muon spin relaxation (µSR) and nuclear magnetic resonance (NMR) which are probes of the dynamic local magnetic environment can provide a signature of magnetic ordering [45,46]. Thermal transport measurement can be useful in determining whether the excitations are localized or itinerant [44,47]. Neutron diffraction can also provide the signature of magnetic ordering in the system [48].

In order to detect the defining feature of QSL, which is the presence of fractionalized excitations, i.e., spinons, which are deconfined from the lattice and have their own dispersions and become itinerant in the crystal. In U(1) gapless QSL spinons form a Fermi sea similar to electrons in metals [49]. In order to detect the fractionalized excitations and to classify the QSL on the basis of the nature of correlation and excitation, crucial signatures are obtained from inelastic and elastic neutron scattering (NS). INS is a spin-1 process where at least one spinon is excited during spin-flip and the spinon pair follows the energy-momentum conservation rule $E_q = E_s(k) + E_s(q-k)$, here $E_s(k)$ is the dispersion of spinons, and $E$ and $q$ are the transferred energy and momentum respectively. As many possible values of k satisfy this condition, hence a broad magnetic continuum both in momentum and energy is observed in the excitation spectra. There are numerous reports for observation of broad continuum, such as for $ZnCu_3(OH)_6Cl_2$ (herbertsmithite) [50], $Ca_{10}Cr_7O_{28}$ [45], $Ba_3NiSb_2O_9$ [51] and $YbMgGaO_4$ [52,53]. This behavior is in contrast to the spin wave excitations in magnetically ordered systems, where a well-defined peak is observed around the ordering wave vector [54].

Electron Spin Resonance (ESR) is used to study the local environment and dynamics of the spin state of unpaired electrons in the system and provides shreds of evidence of fractionalized excitations [55,56]. Raman spectroscopy is a non-destructive technique that provides the signature of fractionalized low-energy excitations that are embedded in the spectral features [57,58]. Last, but not least, terahertz spectroscopy, where electromagnetic radiation of frequency range of terahertz is ideal for investigating low-energy excitation. The temperature-dependent absorption across a wide range of frequencies can provide potential signatures of the QSL state [59,60].

QSLs promise a great deal of application in quantum computation and quantum communication [61] owing to the exotic physics of entangled spins where the excitations are fractionalized in nature. These excitations are robust against perturbations. Another very important feature required for quantum technologies is the presence of Majorana fermions (MFs), which are their own anti-particles and obey non-Abelian statistics in the presence of magnetic fields. Kitaev Spin Liquids are the promising candidates that theoretically inherent all the required characteristics. Next, we will discuss Kitaev spin liquids in detail in the next section.

## 1.2.2 Kitaev Spin Liquid

In 2006 Alexei Kitaev [62], proposed an exactly solvable *S=1/2* spin model on a 2D Honeycomb lattice as shown in **figure 1.4 (d)**. Here, the geometrical frustration factor is absent in fact, the anisotropic interaction between spin-pair on different honeycomb lattice bonds induces a conflict, giving rise to strong frustration and a spin-liquid ground state. The presence of fractionalized excitation, i.e., Majorana fermions, naturally comes out of the solution of this model. These topological excitations are fault-resistant in nature and can be utilized for quantum technologies [63]. The potential signature of Kitaev spin liquid has been evidence for numerous candidates such as $Na_2IrO_3$[64], (α, β and γ) -$Li_2IrO_3$ [33,65,66], α-$RuCl_3$[67], $VPS_3$ [68,69] and $H_3LiIr_2O_6$[70]. All these materials are known to exhibit mott insulating behavior. A clear signature of long-range ordering is also observed in these materials except $H_3LiIr_2O_6$. However, the ordering temperature is around one order of magnitude lesser than the interaction energy scale, which is unlike Curie-Weiss behavior and indicates the presence of magnetic frustration.

The interaction hamiltonian for the Kitaev model on a honeycomb lattice with Ising type of nearest neighbor interactions as shown in **figure 1.6** is given as follows [62] :

$$H = -\sum_{<mn>_\gamma} K_\gamma S_m^\gamma S_n^\gamma \qquad - (1.5)$$

Here $<mn>_\gamma$ is a $\gamma (x-, y- and\ z-type)$ bond and the summation is taken over all the bonds around the hexagonal (honeycomb) lattice. $K_\gamma$ is the nearest neighbor bond-dependent exchange coupling constant, $S = (S_m^x, S_m^y, S_m^z)$ is the spin-1/2 operator at $m^{th}$ site. If we try to align all the spins along a specific direction let say z-axis, $S_m^x S_n^x, S_m^y S_n^y$ and $S_m^z S_n^z$ do not commute and the bond energies along other directions (x and y) are not minimized. Consequently, spins cannot satisfy three different configurations simultaneously and leads to frustration as shown in **figure**

**1.6 (b)**. This geometric magnetic frustration is inherent to latices like honeycomb, kagome [71] and pyrochlore [72]. The ground state of the Kitaev model is highly degenerate even in the classical limit where quantum spin is treated as vectors [73,74]. When quantum fluctuations are in effect the system starts a transition between degenerate states which can be considered as the superposition of all possible classical configurations.

The plaquette (*P*) flux operator *W* defined over each hexagon of the honeycomb lattice commutes with the hamiltonian which is written as:

$$W = \sigma_1^x \sigma_2^x \sigma_3^x \sigma_4^x \sigma_5^x \sigma_6^x \qquad\qquad - (1.6)$$

Here $\sigma_m^\gamma = 2S_m^\gamma$ are the Pauli matrices on-site m, $\hbar = 1$ . The flux operator has a quantized eigenvalues i.e., ±1, which is the $Z_2$ flux through the hexagon, and different flux operators commute with each other which further simplifies the problem as this divides the Hilbert space into sectors of *W* eigenspaces. Majorana fermions are self-adjoint which means they are their own anti-particle. So, any fermion mode gives rise to two Majorana modes for instance, $M_1 = (a + a^\dagger)$ and $M_2 = (a - a^\dagger)$ where $a^\dagger$ and $a$ are creation and annihilation operators. Further, the exact solution of the Kitaev model is achieved by fractionalizing the spin degree of freedom i.e., replacing the spin operator with four Majorana operators such as $\sigma_m^\gamma = i b_m^\gamma c_m$, where $\gamma = x, y, z$, $b_m$ *and* $c_m$ are Majorana operators that satisfy $c_m^2 = 1$, $c_m c_n = -c_n c_m \ (m \neq n)$. The constraint $b_m^x b_m^y b_m^z c_m = 1$ preserves both the *S-1/2* algebra and the local 2D Hilbert space. Hence the revised form of hamiltonian for this model is written as:

$$H = -\frac{1}{4} \sum_{<mn>_\gamma} K_\gamma u_{mn}^\gamma c_m c_n \qquad\qquad - (1.7)$$

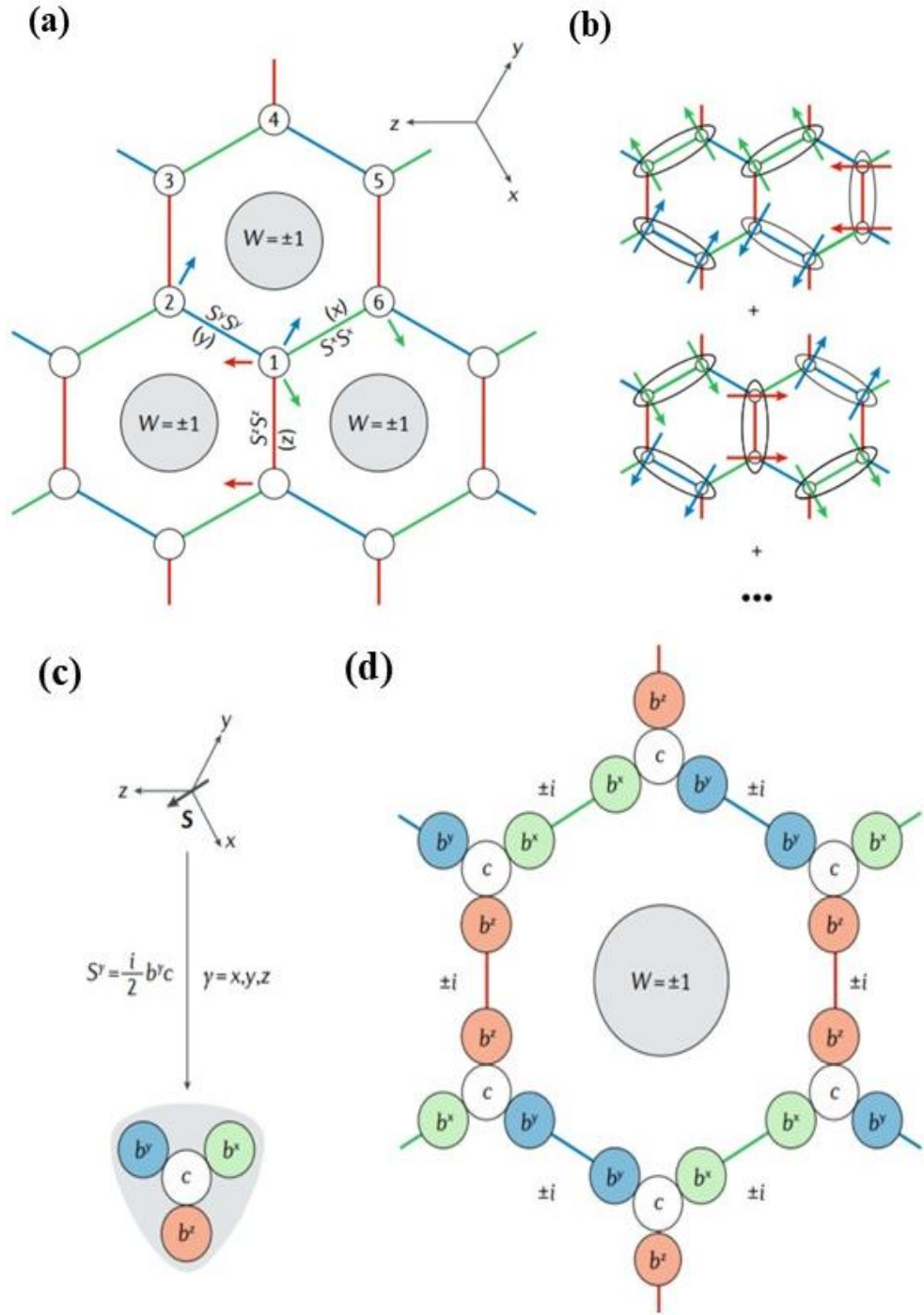

**Figure 1.6:** (a) Spin interaction on honeycomb structure having bond-dependent Ising interactions as shown in green red and blue bonds. $W$ is the Plaquette flux operator having conserved eigenvalues as discussed in the text, (b) Kitaev Quantum spin liquid as superposition of different allowed spin configurations, (c) fractionalization of a single spin into localized $b^\gamma$ and itinerant $c$ Majorana Fermions, (d) modified picture of Kitaev Model on gauge transformation leading to the emergent fractionalized excitation i.e., Majorana fermions [75].

Here $u_{mn}^{\gamma} = b_m^{\gamma} b_n^{\gamma}$ is the nearest neighbor bond operator has an eigenvalue $\pm i$ as also shown in **figure 1.6 (c, d)**. $u_{mn}^{\gamma}$ the operator also commutes with the hamiltonian hence conserved and the Hilbert space is divided into two eigen space of $u_{mn}^{\gamma}$ then it can be replaced by a number and finally, the solution corresponds to free Majorana fermions which correspond to $c$. As there are four MFs assigned to each spin-1/2, another physical subspace is introduced by gauge field transformation where itinerant MFs combine to form a static $Z_2$ gauge field and is associated with $b^{\gamma}$ and restricted eigenvalue of $W_P$ operator as $w_p = \pm 1$. $w_p = 1$ corresponds to the vortex-free ground state and $w_p = -1$ is regarded excited state $Z_2$ vortex, as also shown in **Figure 1.6 (d)**. Hence, $u_{mn}^{\gamma}$, leads to an emergent $Z_2$ gauge field and dictates the phase of the nearest neighbor tunnelling integral of $c$-MFs and are also termed matter fermions.

The type of magnetic anisotropy dictates the ground state phase of the Kitaev magnets. The energy spectrum is gapless for weakly anisotropic coupling constants $K_{\gamma}$, however, the gap appears if one of the couplings is greater than the other two combined. In case of the gapless phase, the Dirac point can acquire a finite gap on the application of an external magnetic field and this topological phase transition is driven by broken time-reversal symmetry perturbations. This gives rise to a chiral spin liquid phase which has non-zero Chern invariants also termed as Majorana chern insulators. For instance, in the presence of external magnetic field $H$, a Majorana gap is induced which is $\Delta \approx H_x H_y H_z / K^2$, where all the exchange couplings are kept constant ($K_x = K_y = K_z = K$) [62].

## 1.3 Topological Phases of Matter

Topology is a well-known branch of mathematics that introduces the mathematical structures that remain unaltered under continuous deformation. It is also known as "rubber-sheet geometry" as objects can be stretched, twisted, and contracted like a rubber without tearing them. For instance, in Euclidian geometry, a square ($\square$) is topologically equivalent to a circle ($\bigcirc$) but not to a symbol of infinity ($\infty$). Such topological properties are also found in quantum systems in nature where it appears due to boundary conditions imposed on the wave function. The boundary conditions can transform a flat surface into a space with twists or holes and is responsible for inducing the topological nature. For instance, in the case of periodic boundary conditions, a line can be transformed into a circle, a hollow cylinder into a torus.

The interlinks of topology, geometry, and properties of quantum systems are well known. For instance, the topological phase which is based on the geometry of single-electron wavefunctions is the integer quantum hall effect (IQHE) where electron motion is constrained in a plane exposed to a strong magnetic field leading to quantized transport properties [76]. As the Hall conductivity was found to be an integer multiple of $e^2/2\hbar$. The phenomenon was explained well in terms of Landau levels which are the eigenstates of the electron moving in the presence of a magnetic field in a free space. However, another deeper perspective was proposed by Thouless, Kohmoto, Nightingale, and den Nijs in 1982 in order to comprehend the physics of electrons in a periodic potential. Their approach considers the topological aspects (TKNN integers or chern numbers) in order to explain the robustness and quantization observed in IQHE [77]. This proposal was further complemented by Haldane a few years later where he showed how IQHE could arise even in the absence of an external magnetic field using a simple crystal model [78]. For the discovery of topological phases of matter, Haldane and Thouless received Nobel prize in 2016.

Quantum Hall effect was proposed for 2D systems but there are instances of 3D Materials having insulating bulk and conducting surface states which is a consequence of the non-trivial topology of the wavefunction and are classified as topological Insulators [79,80]. E-K band diagram representation is shown in **figure 1.7 (a, b)**. Such a topological state can be produced by the presence of spin-orbit coupling [81] instead of the magnetic field which consequently leads to such a novel phase [82]. Some of the instances of materials that are realized to be topological insulators are $Bi_2Se_3$ [83], $Bi_2Te_3$ [84], here the surface contains a Dirac cone of electrons similar to what is observed in the case of graphene. These materials have massless electrons at the surface states where spin-momentum are locked. Three-dimensional topological Insulators are reported to exhibit a quantized magnetoelectric effect [85] and further produce the quantum anomalous Hall effect [86].

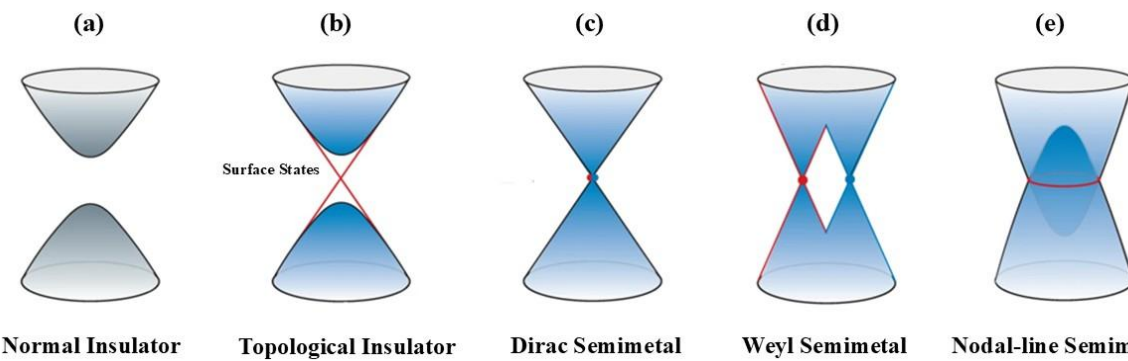

**Figure 1.7:** Band Diagram representation of different quantum states (a) normal Insulator, (b) topological Insulator, (c) Dirac semimetal, (d) Weyl semimetal, and (e) nodal-line semimetal [87].

Recently there has been a shift in focus to topological semimetals and metals as well. The theoretical prediction of Weyl [88,89] and Dirac semimetals [90] and their experimental realization have revived this field of research [91,92]. Topological semimetals are an apt platform where both bulk and surface states participate in various physical processes which is in contrast to topological insulators where the bulk bands are far away from the Fermi energy and can be treated separately. Identification of topological fermi arcs on the surface and chiral magnetic effects in the bulk, topological semimetals promises the presence of exotic topological states [88]. Dirac's work in 1928, where he presented an effect of unification of quantum mechanics and relativity in order to describe the nature of electrons and the solutions predicted three distinct kinds of relativistic particles: the Dirac, Majorana, and Weyl fermions.

When it was applied to a condensed matter system describes a point where four bands touch. In Dirac semimetal, the Fermi surface is a point-like if the Fermi energy passes through the Dirac points. Their schematic band diagram is shown in **figure 1.7 (c)**. In 1929 Hermann Weyl proposed that the massless electron also exhibits chirality (fixed-handedness) [93]. In the case of Weyl semimetal the presence of symmetry breaking (time-reversal or inversion) leads to the splitting of four-fold degenerate Dirac cones into two-fold degenerate Weyl cones as shown in **figure 1.7 (d)**. Each of these cones is marked (blue and red) with an associated topological integer 'charge'. The total number of Weyl points in a material sums up to zero as proposed by Nielsen-Ninomiya in particle physics [94]. The surface electrons in the case of Weyl semimetal exhibit the Fermi arc (line of zero-energy states). Weyl points are deeply related to the topological phase transition between the topological and normal insulator. Weyl fermions are strange quasi-particles that are not elementary or fundamental particles but are emergent phenomena. Such particles do not conserve charge, in condensed matter systems there are always multiple Weyl points but they might occur at different energies such that the total charge cancels out. In the case of nodal-line semimetal the nondegenerate conduction and valence band touch at a closed loop (line) in the momentum space as shown in **figure 1.7 (e)** which is in contrast to Dirac and Weyl semimetals which nearly meet at an isolated point [95]. Nodal line semimetals exhibit $\pi$ Berry phase along the closed loop at band crossing which provides topological robustness, torus-shaped fermi surface, and drum head-like surface states which may lead to high density states at fermi-level and consequently increased conductivity which might provide a path to understand high-temperature superconductivity [96,97]. Owing to these interesting features quantum oscillations have also been observed in nodal-line semimetals [98]. Some of the candidate materials of nodal-line semimetals are $SrIrO_3$ [99], $PbTaSe_2$ [100], and $Ca_3P_2$ [101].

### 1.3.1 Topological Phonons

Lord Rayleigh in 1887 first discussed the occurrence of localized waves at the surface of isotropic elastic media. Generally, the surface properties are assumed to be very different from the bulk properties due to lower dimensionality and broken symmetry of chemical bonds. Such an instance can be seen in the case of phonon dispersion for bulk and surface [102]. Surface lattice dynamics or surface phonons are assumed to be the excitations that are localized to the surface only, however, an interesting scenario arises when there is an interplay of surface dynamics and topology. A new physics may emerge due to the strong coupling of topological electrons and surface lattice vibrations which are not present in the bulk and that potentially can lead to a surface superconducting phase where the formation of cooper pairs is mediated by topological phononic edge state [103,104].

In contrast to topological electrons the topological phonons show several distinct properties such as the equation of phonon motion is governed by classical Newton equations whereas electrons follow the Schrödinger equation, there is no restriction of the Pauli exclusion principle on phonons and can be thermally excited as they follow Bose-Einstein statistics. Further, as compared to spin-1/2 and charged electrons (fermions) phonons are electrically neutral with spin 0 (Bosons) and can only couple indirectly with electric and magnetic fields [105]. However, in a periodic lattice, both can be described by Bloch wave functions and hence the concepts of Berry phase, Berry connection, Berry curvature, and Chern number apply to both quasiparticle excitations.

Topological Phonons have been realized both experimentally and theoretically [105-108]. Theoretically, the concepts of Berry curvature and topological invariants along with applications of the phonon tight-binding model and nonequilibrium Green's function method have been applied to investigate various topological phonon states, such as topological optical phonons, topological acoustic phonons, and higher-order topological phonons, in crystalline materials [109] such as MSi family (M = Fe, Co, Mn, Re, Ru) [110]. Single Weyl points have been reported in TiS, ZrSe, and HfTe [111]. Whereas the materials like CuI and CdTe have been reported to have type-II Weyl points [112,113]. In 2021, Liu *et al.* proposed charge-four Weyl phonons in BiIrSe and $Li_3CuS_2$ [114]. Additionally, in 2022, Liu *et al.* found a new type of topological nexus phonon and a multitude of symmetry-protected nodal line phonons in the phonon spectrum, exhibiting a quantized Berry phase of $\pi$ in Silicon [115]. Topological phonons and electronic structure has been reported for $Li_2BaX$ (X = Si, Ge, Sn, and Pb) semimetals [116]. SnIP a quasi-one-dimensional van der Waals material with a double helix crystal structure, has been reported to exhibit topological nodal rings/lines in both the bulk phonon modes and their corresponding surface states, which provides an exotic platform to study degenerate lines in surface phonons states and related transport properties [117].

Experimentally, it has been very challenging to probe the presence of topological phonons due to due to low energy phonon bands (typically at 0.1-10 meV). However, inelastic X-ray scattering has been used to measure the phonon dispersion of FeSi [118] and $MoB_2$ [119]. Jin *et al.* observed topologically degenerate phonon points in the phonon spectra of MnSi and CoSi via inelastic neutron scattering and also found a distinct relation between the chern numbers of a band-crossing node and the scattering intensity modulation in momentum space around the node [120]. In order to obtain a global map of phonon bands, high-resolution electron energy loss spectroscopy (HREELS) turned out to be very useful. Evidence of nodal-ring and Dirac phonons has been found in the phonon spectra of graphene in the entire 2D Brillouin zone using HREELS. Observations of truly chiral phonons in α-HgS [121] and Te [122] were reported suing Raman spectroscopic technique. In 2018, Zhu *et al.* detected chiral phonon mode in a monolayer of $WSe_2$, probed using the circular dichroism (CD) in the transient infrared spectroscopy [123].

In such materials, under the breaking of time-reversal symmetry, there are one-way edge states and there is no backscattering even in the presence of impurities. Such a feature is crucial in making phono diodes where unidirectional phonon conduction is present [124-126]. These non-trivial phonon states can be characterized by nodal point phonons, nodal ring phonons, and nodal straight-line phonons [111,127,128]. Also one can use the robustness of topological phonons for frequency filtering [129], heat transfer [130], and infrared photoelectronic [131]. Quantum phenomena like quantum anomalous Hall or quantum valley Hall-like effects can be observed in topological phononic materials due to the presence of time-reversal symmetry-breaking [107]. Different topological phononic states has been proposed in acoustic systems, metamaterials, and Maxwell frames [125,132,133]. Hence, hunting for topological surface phonons would be a very challenging job for condensed matter physicists but it will open new realms in designing the future quantum materials.

## 1.4 Charge Density Waves

Some metals undergo a transition such as ferromagnetic as in the case of Iron and nickel, superconducting for lead and aluminum where electron forms cooper pairs when they are cooled. Another kind of transition occurs in correlated condensed matter systems where a conduction electron charge density undergoes a modulation in some of the periodic metals at low temperatures. It was first proposed by Sir Rudolf Ernest Peierls in 1930 [134] and then by Herbert Fröhlich in 1954 [135]. As temperature is lowered below a certain transition temperature $T_C$, there is a presence of fermi surface instability which induces periodic spatial modulation of charge density. The modulation of electron density also perturbs the underlying ionic core potential which shifts them to new equilibrium positions and hence a structural transition. Fröhlich realized that and came up with a comprehensive theory of 1-D superconductivity and tried to explain the phenomenon with a model where electrons and the underlying lattice go under modulation collectively [135]. In 1941, in order to explain the phenomenon of superconductivity John Bardeen argued that there is a presence of lattice distortion in the superconducting state which induces energy gaps which consequently would lead to an increase in diamagnetic property [136]. Later this argument was rejected as it could not be realized in the case of three-dimensional Fermi surface of trivial superconductors. The phenomenon of superconductivity in conventional superconductors and its origin are discussed in the BCS theory [137] but there is no mention of Charge Density Waves (CDWs) however it becomes important in the case of unconventional superconductors like the cuprates [138]. The origin of charge density wave transition has been discussed to be rooted in concepts like nesting Peierls' instability, Fermi surface, and Kohn anomaly which are discussed ahead.

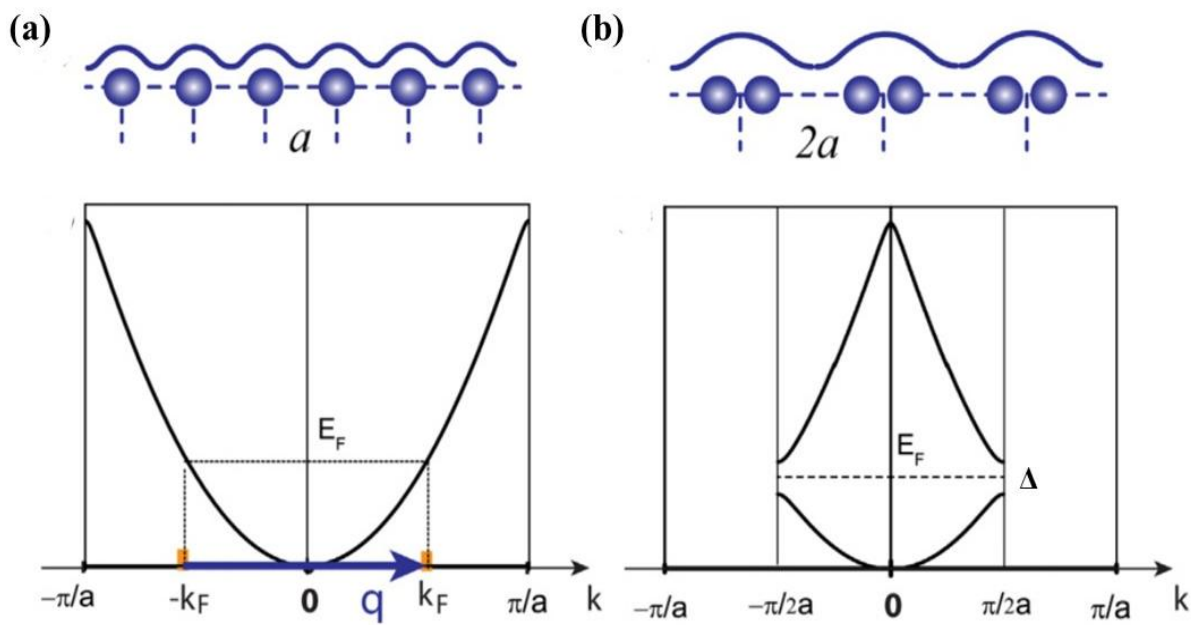

**Figure 1.8:** Representation of Peierls' transition in a one-dimensional atomic chain and corresponding band structure (a) before and (b) after the lattice distortion [139].

The picture of CDW transition in a 1D periodic atomic chain with a lattice constant $a$ and containing one electron per atomic site was proposed by Peierls' where he describes the inherent instability of such a system [140,141]. The modulation of electron density, lattice, and corresponding free electron band structure above and below the transition temperature $T_C$ is shown in **figure 1.8**. This 1D electronic system due to electron-phonon coupling (EPC) undergoes a metal-insulator transition on lowering temperature as a gap arises at the Fermi wavevector $q = \pm 2K_F$, here $K_F = \pi/2a$ which is accompanied by a perturbation in lattice periodicity to **2a** (two atoms per unit cell). This reconstruction of the unit cell affects the band structure in the new Brillouin Zone. The electronic energy is lowered $\Delta E_{elc.} < 0$ due to the opening of an energy gap at $\pm k_F$, while the lattice distortion increases the energy of the underlying lattice $\Delta E_{latt.} > 0$. The equilibrium lattice distortion is provided by minimizing the total energy $E_{tot.} = \Delta E_{elc.} + \Delta E_{latt.}$.

The underlying mechanism for the Peierls' transition could be explained by assuming a free electron gas model using the Lindhard response function. Lindhard susceptibility $\chi(q) = \chi' + i\chi''$ describes the response of charge density to the underlying lattice potential, here $\chi'$ is the real part and $\chi''$ is the imaginary part of susceptibility. The real part of susceptibility $\chi'$ depends on the entire band structure and is given as $\chi'(q) = (1/q) \cdot \ln \left| (q - 2K_F / q + 2K_F) \right|$. $\chi''$ only concerns with the Fermi surface, which means that Fermi surface nesting via a wavevector $q$ will show a peak in $\chi''$ that is not necessarily reflected in $\chi'$ at the same $q$ [142,143]. This is because $\chi'$ can have contributions away from the Fermi level, which are neglected in $\chi''$. However, for the 1D case it happens at the same $q$ i.e., $q = \pm 2K_F$. $\chi'$ is plotted for 1D, 2D, and 3D in **figure** 1.9 (a), the divergent nature at $q = 2K_F$ is a reflection of instability that leads to dimerization of the 1D chain [144]. There is a lack of experimental evidence of Fermi surface nesting-driven CDW transitions in real materials. The divergence in $\chi'$ alone at a particular $q$ does not guarantee the CDW transition in the Peierls' picture as one has to take care of the presence of electron-phonon coupling or the contribution from the lattice.

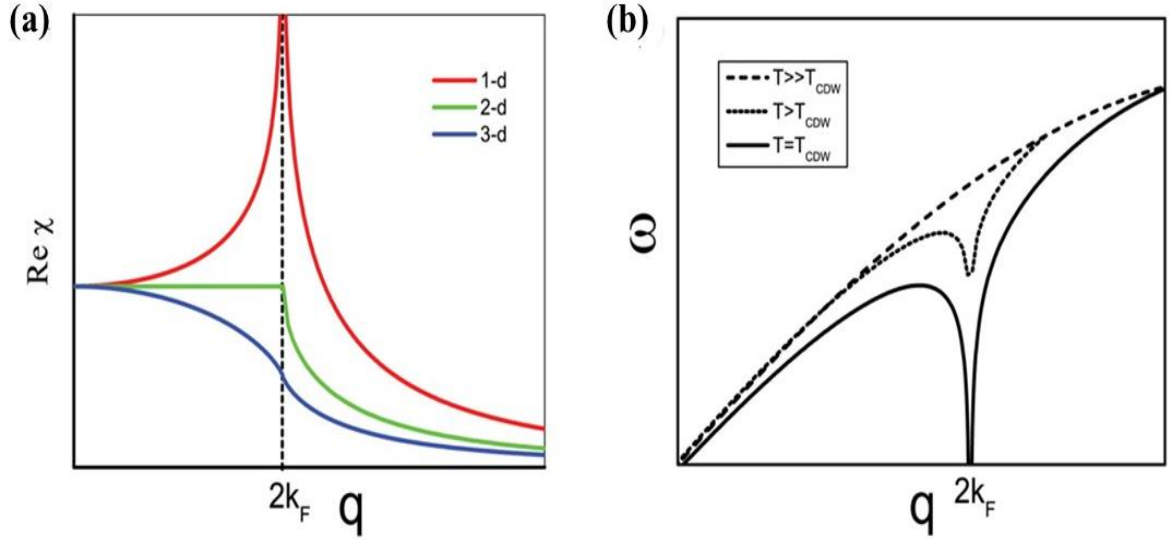

**Figure 1.9:** (a) Real part of Lindhard function for 1D, 2D, and 3D free electron gas model, (b) phonon dispersion at different temperatures for 1D atomic chain [144].

As discussed, the modulation of charge also affects the lattice in strongly correlated systems (electron-phonon), hence the changes in the Fermi surface are also reflected in the phonon dispersion. This is what was pointed out by Kohn in 1959, that the zero energy electronic excitations at $q = 2K_F$ will screen any lattice vibration of a similar wavevector and the phonon mode will undergo renormalization with a lowered energy (phonon softening) and is known as Kohn anomaly [145,146]. As shown in **figure 1.9 (b)** the phonon renormalization around $q = 2K_F$ shows temperature dependence as below $T_C$ ($T_{CDW}$) phonon energy decreases drastically and becomes imaginary, indicating the formation of a new lattice. This frozen phonon leads to a second-order phase transition into the CDW state [141]. However, above $T_C$ the dip in the phonon dispersion is not accompanied by lattice reconstruction. For a weak coupling, phonon frequency can be written as [147]:

$$\omega_q^2 = (\omega_q^0)^2 - |\lambda|^2 \, \Theta(q,0) \qquad \text{- (1.8)}$$

Here $\lambda$ is the electron-phonon coupling constant and $\Theta(q,0)$ is bare charge density susceptibility at the $q = 2K_F$.

The modulation in the electronic charge density can be written as:

$$\eta(r) = \eta_0 + \eta_A \cos(q \cdot r + \varphi) \qquad \text{- (1.9)}$$

here $q = 2K_F$, $\eta_A$ is the amplitude of the modulation and $\varphi$ is a phase that determines the position of the charge density wave relative to the underlying lattice. Different types of CDW states can be classified on the basis of a comparison of CDW periodicity and original lattice periodicity. If the periodicity of CDW is proportional to *a,* then it is a commensurate charge density wave (CCDW); otherwise incommensurate (ICDW) and may be quantified as:

$$\frac{2\pi}{q} = \frac{m}{n} a \ , \ (m, n \ \in \ \mathbb{R}) \qquad \text{- (1.10)}$$

Peierls' description of CDW was for 1D systems, but in order to explain the phenomenon in higher dimensionality, the concept of Fermi surface nesting is required [148]. Fermi surface nesting causes the interaction between the lattice vibrations and electrons at the Fermi surface and is a key to stabilizing CDWs [139]. Fermi surface nesting is observed in 1D metal where the partially filled bands are dispersive along one direction upon translation by $\vec{q}_{CDW}$ which creates a CDW state having spatial periodicity $2\pi / |\vec{q}_{CDW}|$. Hence the Peierls', transition model boils down to Fermi surface nesting, singularity in Lindhard response function, Kohn anomaly in the phonon dispersion, structural distortion, and a metal-insulator transition. According to Ginzburg–Landau theory of second-order phase transition, the translation symmetry of the lattice is spontaneously broken and leads to excitations. An analogy of the ground state free energy can be made with a mexican hat as shown in **figure 1.10 (a)** and an order parameter

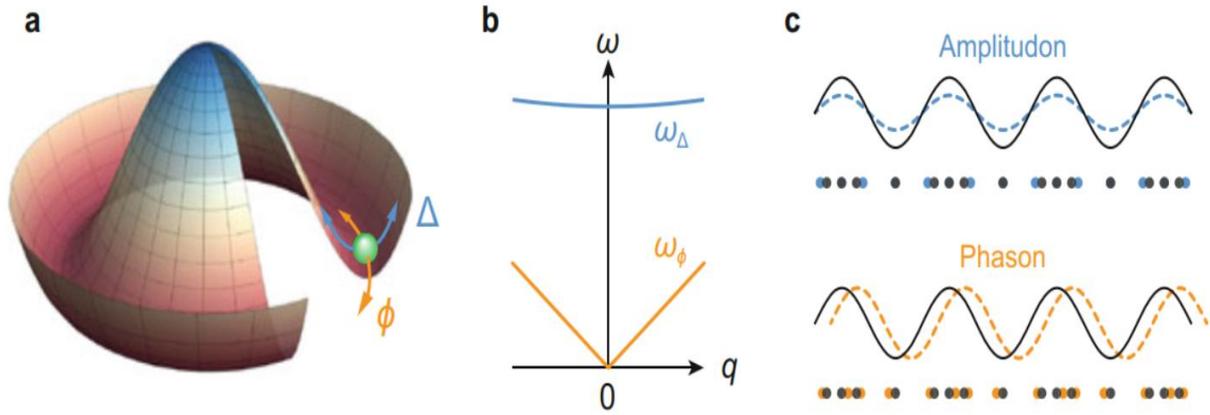

**Figure 1.10 (a)** Mexican hat potential of a complex CDW order parameter $\Delta e^{i\phi}$, **(b)** dispersion of amplitudons and phasons in the vicinity of $\Gamma$-point, **(c)** schematic representation of excitations (amplitudons and phasons) where dashed curve is the change in the electron density whereas circles shows variation in the atomic positions [143].

can be defined as $\Omega = \Delta e^{i\phi}$ for $\Delta > 0$ and $\phi \in [-\pi, \pi)$. $\Delta$ is the amplitude of ionic displacement relative to the original state or deviation of the charge density in the transition and $\phi$ represents the phase of the charge density wave relative to the original lattice. The elementary excitations in the ground state of CDW are: amplitudons and phasons which could be thought of as two orthogonal deviations from the minimum energy of the mexican hat potential (**figure 1.10 (a)**). The dispersion relation of amlitudons and phasons in the long-wavelength limit is shown in **figure 1.10 (b)**. In the presence of these excitations, the order parameter can be rewritten as $\Omega(x,t) = [\Delta_o + \delta(x,t)]e^{i[\phi_o + \theta(x,t)]}$, where $x$ is along the direction of the CDW wavevector, $\Delta_o$ and $\phi_o$ are the amplitude and phase in the equilibrium state and $\delta(x,t)$ and $\theta(x,t)$ represents the fluctuations as also shown in **figure 1.10 (c)**. The atomic movement in the case of phason is not a uniform displacement that follows the electron density [143].

There are various experimental probes for CDW such as electronic transport, neutron scattering, scanning tunneling microscopy, Raman spectroscopy, X-ray scattering,

Temperature-dependent resistivity and Magnetic susceptibility, etc. [139]. Some of the well-studied CDW systems in their low and three-dimensional regime are 1T-TaS$_2$ [149], NbSe$_2$ [150], 2H-TaSe$_2$ [151], etc. These materials have been proposed to be apt for applications in nanodevices, memory devices, voltage oscillators, quantum computing, electrodes in supercapacitors, photodetectors, etc. [152,153].

## 1.5 Raman spectroscopic Investigation of Quantum materials

Atoms and molecules in matter are in continuous motion even at absolute zero temperature, owing to quantum mechanical restrictions. Phonons (quasi-particles) are the quanta of lattice vibrations, just like photons are for light. Phonons emerge as the consequence of the breaking of continuous translational and rotational symmetry in solids. Inside the material, any quantum phenomenon emerges as a collective behavior of the system, and in this process, the underlying lattice does not remain untouched. So, the lattice degrees of freedom can couple with electronic and spin degrees of freedom and can provide crucial information about the presence of other quasiparticles such as magnons, excitons, Majorana fermions, etc. It is crucial in quantum materials to identify low-energy excitations (magnetic or electronic) and the ground state properties. Light is an electromagnetic wave and consists of an oscillating, self-sustaining electric field $\vec{E}(\omega,t)$ and magnetic field $\vec{B}(\omega,t)$. It interacts with the electronic spin via two processes, which are direct and indirect coupling. In direct coupling, there is magnetic dipole interaction between the oscillating magnetic field of light and the electronic spin of the magnetic ion. In indirect coupling there is an electric dipole interaction that happens between the oscillating electric field of light and the electrons of magnetic ions, mediated by spin-orbit coupling [154,155]. Instead of two channels of interaction, the strength of indirect coupling is found to be much greater than the direct coupling.

Inelastic light scattering, Raman measurement can simultaneously provide signatures of scattering from lattice, spin, and charge degrees of freedom via spin-phonon and electron-phonon couplings as shown in **figure 1.11**. Raman spectroscopy is a very user-friendly and non-destructive technique based on inelastic light scattering to investigate the exotic features of quantum materials, which we have discussed in sections above. Unlike other spectroscopic techniques, such as INS, even a micron-sized sample will serve the purpose. Temperature and polarization-dependent Raman scattering can provide crucial information about material properties such as the presence of structural/magnetic transitions, exotic quasiparticles, etc. This information is embedded in the Raman spectra as background continuum, peak line shape, position, full-width half maxima (FWHM), and spectral weight or area under the curve of typical Raman spectra. A detailed discussion regarding the theory and instrumentation is included in Chapter 2 of this thesis.  In this section, we will discuss the utility of the Raman Spectroscopic technique to probe exotic states, which are quantum spin liquids, topological phonons, and charge density waves.

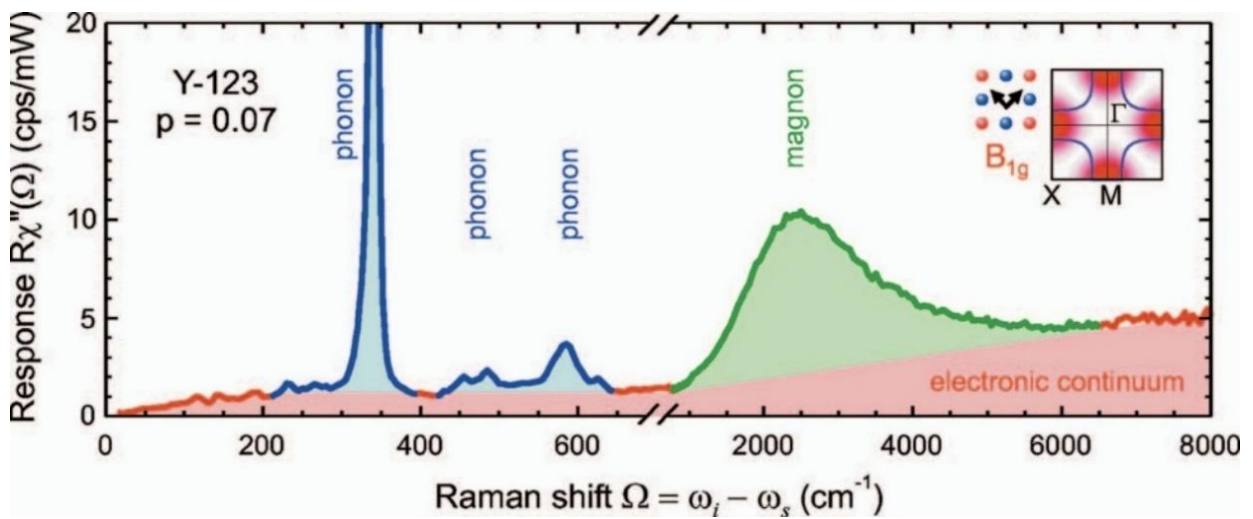

**Figure 1.11:** Raman scattering response unveiling different kinds of quasi-particle excitations over a wide range of energy from an underdoped cuprate [156].

### 1.5.1 Magnetic Excitations

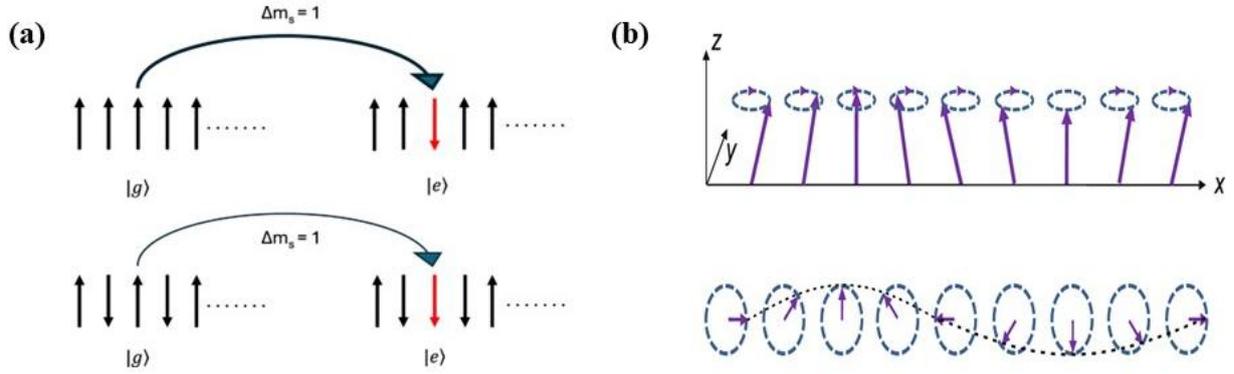

**Figure 1.12:** Raman scattering as a probe of magnetic excitations in (a) Ising type 1D Ferro and antiferromagnetic ordering and (b) spin precession around the quantization axis (z) and the x-y plane projection of the spins representing a propagating spin-wave in a 1D ferromagnetic chain [157].

Raman scattering from magnetic insulators is comprised of signals from phonons and magnetic excitations. The magnetic excitations in long-range ferro and antiferromagnets are magnon or also known as spin waves. Magnons are bosons with integer spin and hence obey Bose statistics. Magnons are the excitation that happens via spin flip as also shown in **figure 1.12 (a)** for a perfectly aligned 1D chain of Ferro and antiferromagnets having Ising-type interaction, where $|g\rangle$ and $|e\rangle$ are the ground state and excited states, respectively. In the process of spin-flip, the total change in spin is $\Delta m_s = 1$ and it costs energy of $\Delta E = 8|J|S^2$. At any finite temperature, instead of perfect alignment the spins precess about a quantization axis which is the direction of the internal field as shown in **figure 1.12 (b)** along with the x-y projection where the dotted pattern signifies a propagating spin wave or magnon. The dispersion relation of this spin wave is given as: $\hbar\omega = 4JS\left[1 - \cos\left(qa\right)\right]$, where $a$ is the lattice constant and $q$ is the wavevector [158].

The Stokes/anti-Stokes Raman light scattering from one magnon involves two photons and one magnon, while for two magnons pair of magnons is created/annihilated. In terms of energy and momentum conservation a one-magnon process in Stokes scattering where an incident photon $(\omega_i, \boldsymbol{k}_i)$ scatters $(\omega_S, \boldsymbol{k}_S)$ after interacting with the spin and generates a magnon $(\omega_i, \boldsymbol{k}_i)$ in the system, which can be written as: $\hbar\omega_i = \hbar\omega_S + \hbar\omega_M$ and $\boldsymbol{k}_i = \boldsymbol{k}_S + \boldsymbol{q}$. Here a single flip occurs via a transition from $S^z = S$ ground state to $S^z = S - 1$ via electrical transition with $L = 1$ as the intermediate state due to the presence of spin-orbit coupling $\lambda \boldsymbol{L} \cdot \boldsymbol{S}$, here $\lambda$ is the spin-orbit interaction term, and the process is schematically shown in **figure 1.13 (a).**

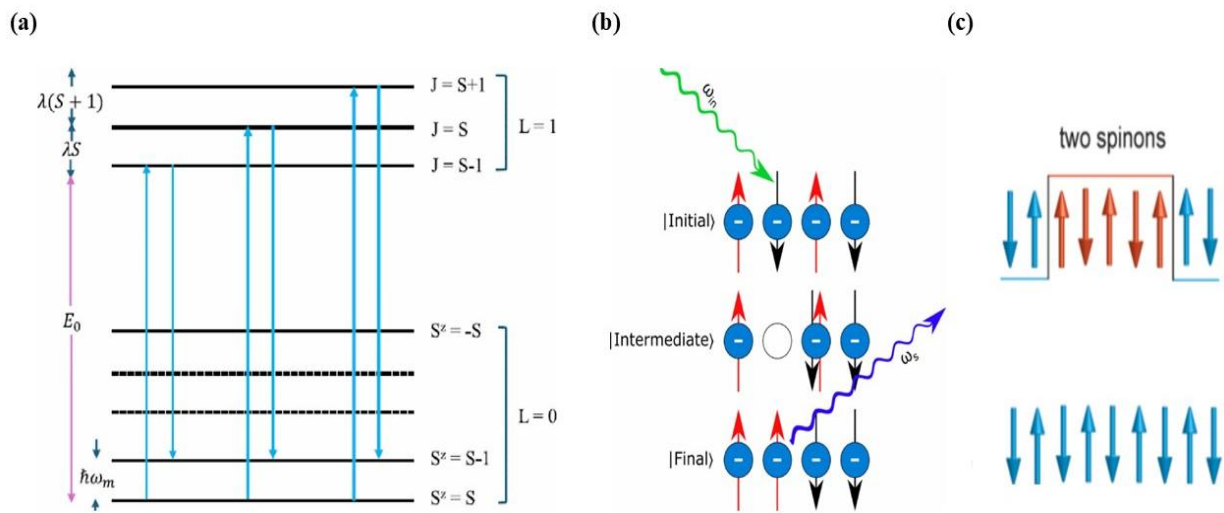

**Figure 1.13:** A schematic representation of (a) principle of one-magnon scattering for a ferromagnet as discussed in the text, (b) schematic representation of two-magnon scattering for an antiferromagnetic spin chain [157] and (c) spinons with S=1/2 where spin flips are surrounded by domain walls in a spin chain.

In antiferromagnetic systems, two-magnon scattering arises from the exchange scattering mechanism which is based on double spin-flip due to Coulomb interaction where the total z-component of spins is conserved [159]. Hence, this is a creation or annihilation of an even number of magnons which leads to a larger scattering cross-section as compared to one magnon case. In the case of two magnon as shown in **figure 1.13 (b)** Stokes Raman scattering the incident photon generates two magnon for which the energy and momentum conservation can be written as: $\hbar\omega_i = \hbar\omega_S + 2\hbar\omega_M$, $\boldsymbol{k}_i - \boldsymbol{k}_S = \boldsymbol{q}_1 + \boldsymbol{q}_2$. Generally, the laser used in the Raman scattering which will generate the magnon has a wavelength in visible range for instance for $532nm$ the transferred momentum is $k \sim 10^{-3} \overset{o}{A}{}^{-1}$, so one can only probe excitation near the zone center as Brillouin zone boundary is at $\pi \overset{o}{A}{}^{-1}$ for a lattice constant of $1\overset{o}{A}$. The presence of spin waves is reflected in the temperature-dependent Raman spectra as an appearance of new broad or sharp peaks (Magnon) in the spin-solid phase, also it will reflect in the phonon energies and lifetime (FWHM) and line shapes.

Raman scattering has been smoking gun evidence for probing the fractionalized excitation such as spinons (Majorana fermions) as shown in **figure 1.13 (c)** which follow Fermi-Dirac statistics and are the key feature of quantum spin liquids. As discussed in section 1.2.2, in the case of Kitaev Spin Liquids the elementary spin-1/2 fractionalizes into two itinerant MFs and localized $Z_2$ flux [160,161]. Raman scattered light couples with itinerant MFs either via the creation and annihilation of a pair of fermions or the creation of one and annihilation of another fermion. These features are reflected in the multiparticle continuum background signal which is due to itinerant MFs, phonon anomalies in frequency, linewidth, and quasi-elastic scattering which comes into the picture due to $Z_2$ flux and line shapes such as Fano asymmetry. A careful subtraction of the bosonic signal from phonons is needed to obtain the scattering response from the other quasi-particles. We have discussed this in detail in the thesis chapter 3.

## 1.5.2 Topological Phases

Exotic topological phases have been well investigated by polarization, laser wavelength, and temperature-dependent Raman spectroscopy due to the presence of strong electron-phonon interactions [162]. Resonance effects may also provide a signal in the Raman spectra such as the Fano line shape which arises due to the coupling of surface phonons and the electronic continuum [163]; phonon linewidths which is an indication of the lifetime of the phonons etc. At any finite temperature, the phase-space symmetry is broken owing to electron-phonon coupling, which limits the lifetime of surface electrons. Polarization-dependent Raman scattering is an excellent method to investigate surface and bulk phonons, as both bulk and surface electronic excitations are sensitive to the polarization direction of the phonons. The presence of surface phonons along with a thermally induced structural transition has been proposed in a topological semi-metal $PbTaSe_2$ via Raman measurements [108] as also shown in **figure 1.14**. This situation enhances Raman scattering cross section due to electron-hole pair creation that can provide information about unoccupied electronic states and electron-phonon coupling [164,165].

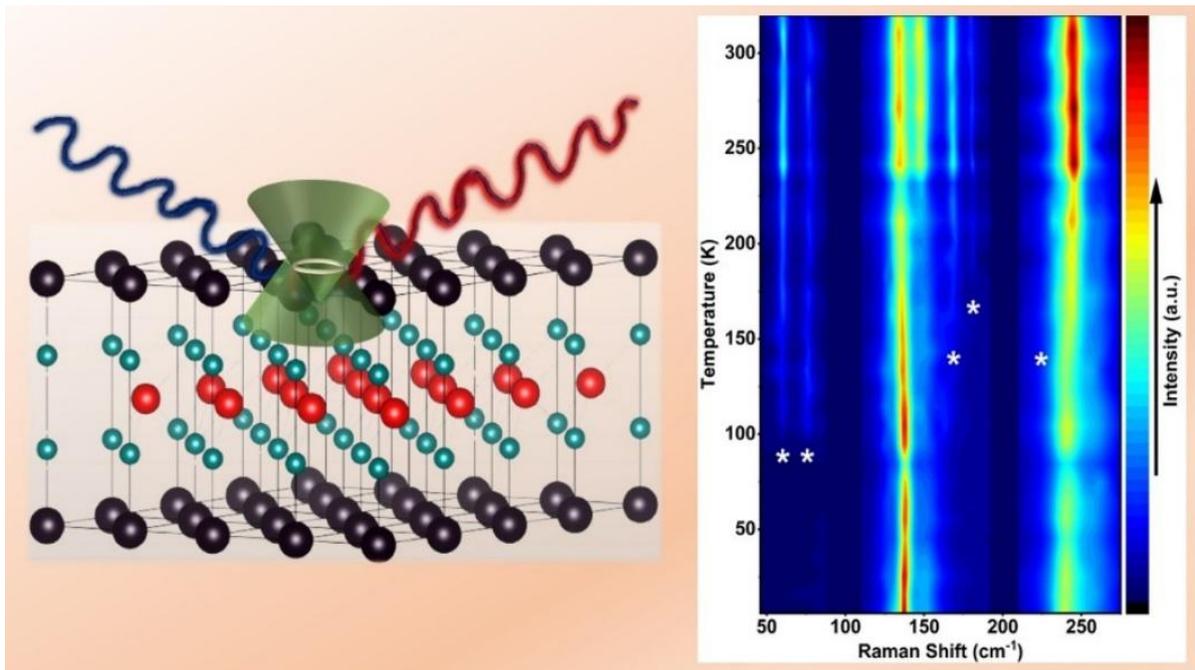

**Figure 1.14**: Inelastic light scattering unveils the interplay of surface lattice dynamics and topology for nodal-line semi-metal $PbTaSe_2$ [108]. '*' indicates the surface phonons.

### 1.5.3  Structural Transitions

Raman Spectroscopy is sensitive to lattice distortions. Group theoretical analysis of a crystal structure predicts the number of Raman and infrared active phonon modes at the gamma point which is subjected to vary for different structure symmetry. On lowering temperature some material undergoes a structural change where symmetry increases, which may result in the appearance or disappearance of new phonon modes in the Raman spectra, and such is the case of charge density waves. These new modes are Phasons and amplitudons, which are also discussed in detail in section 1.4 [166,167].

## 1.6 Thesis Work Overview and Significance

In this section, we will discuss the findings of this research thesis and their significance. Quantum computing has been a far-sought dream to be realized after theoretical predictions. All modern computers work on the classical 'bit' (binary) i.e. 0's and 1's, which obeys logical operations via Boolean algebra. These binary terms 0 and 1 can be realized as two states of on and off of a current or charge accumulation via suitable electronics. On the other hand instead of classical bits quantum computers use 'qubit' that can be expressed as a superposition of two quantum states $|0\rangle$ and $|1\rangle$ as $|\psi\rangle = a|0\rangle + b|1\rangle$ , here $a$ and $b$ are the complex numbers. For $N$ qubits there will be $2^N$ such complex variables and hence can store more information as compared to the classical bit simultaneously. Nature has its own natural qubit as electron spin which can be written as a superposition of up and down states as $a|\uparrow\rangle + b|\downarrow\rangle$. A 500-qubit system will have variables more than total number of atoms in the universe. In solid materials, there are $\sim 10^{23}$ electrons which means $\sim 2^{10^{23}}$ complex variables. A PC with 16GB RAM, the performance can be mimicked by a 37-qubit system for instance a $1.5\ nm^3$ sample of $KNiF_3$ (antiferromagnet). However, the challenge is to control a huge number of qubits and maintain the quantum coherency of the states due to the environmental disturbance surrounding the qubits. Strategies to overcome the problem being pursued are using spin in a quantum dot [168] and superconducting qubits [169]. Another promising way that come to the rescue is using topological quantum computing, where one can utilize materials sustaining Majorana Fermions which obey non-abelian statistics and are robust to environmental perturbations [3]. At present many multi-national companies such as Intel, IBM, Google AI, etc. are working towards achieving quantum computation and communication, and hence an in-depth scientific search for such quantum materials is crucial for the future quantum revolution.

The quantum systems investigated in this work are $V_{1-x}PS_3$, $PbTaSe_2$, and $1T\text{-}TaS_2$, these are discussed in **chapters 3, 4, and 5,** respectively in the thesis. $V_{1-x}PS_3$ and $1T\text{-}TaS_2$ are investigated in the lower thickness regime and bulk, while $PbTaSe_2$ is in bulk only.

The thesis contains a total of six chapters:

**Chapter 1** presents a detailed introduction and the importance of the investigation done on the quantum materials having properties like quantum spin liquids, topological surface phonons, Mott-insulator and charge density waves.

**Chapter 2** briefs about the history and fundamentals of the Raman scattering phenomenon and also includes details experimental and computational techniques used in the research done in this thesis.

**Chapter 3** presents an in-depth Raman investigation of a non-stoichiometric single crystal $V_{0.85}PS_3$. Here, computational methods and experimental measurements are performed at extreme temperature and pressure conditions. This chapter is divided into two parts.

**Part A** provides a detailed investigation on bulk $V_{0.85}PS_3$ via Raman spectroscopy as a function of temperature (4K-330K) and spectral range of 5-760 $cm^{-1,}$ incident light polarization dependence of spectral response. The findings are also supported by resistivity and magnetization measurements.

**Part B** gives an in-depth temperature-dependent (4K-330K) Raman study in a spectral range of 5-760 cm$^{-1}$ on various low thickness flakes down to ~8-9 layers of $V_{0.85}PS_3$. A density functional perturbation theory (DFPT) based $\Gamma$-point phonon frequencies are also calculated.

**Chapter 4** presents a comprehensive temperature (6K-320K) and incident light polarization dependent inelastic light scattering (Raman) study in a spectral range of 10-450 cm$^{-1}$ on a topological semimetal $PbTaSe_2$. The investigation is also supported by temperature-dependent single-crystal XRD experiment and zone-centered DFPT-based phonon calculation.

**Chapter 5** provides a temperature and incident light polarization direction-dependent inelastic light scattering (Raman) study on various low thicknesses of $1T-TaS_2$. DFPT based computational methods are also invoked to calculate the dispersion curve and zone-centered phonon frequencies.

**Chapter 6** gives summary of the research work done in this thesis and also provides further future possibilities and investigations in order to explore the underlying physics and applications associated with the quantum systems probed in the thesis.

# Chapter 2: Methodology - Theory, Instrumentation, and Computation

This chapter includes the historical and theoretical framework of the principle of Raman scattering. In addition to that the details of experiments and related instrumentation of temperature and polarization-dependent Raman and temperature-dependent single-crystal X-ray diffraction instrumentation is mentioned. In the end, a short description of on density functional theory used for phonon calculations.

## 2.1 Raman effect

Electromagnetic radiation in the classical description consists of self-sustaining oscillating electric and magnetic fields. Quanta of electromagnetic waves are known as photons, which have a zero rest mass but carry energy and momentum, so have a relativistic mass. Light can be produced by oscillating charges or fields for e.g., heating a substance or bombarding it with a stream of electrons. The emitted light is referred to as primary radiation. If light is induced by exposing a substance to strong radiation, then such an emission is a secondary radiation. Scattering of light by atoms and molecules is another kind of secondary radiation that gives color to the sky, the blue color of the oceans, and the opalescence of large masses of ice. Light can interact in multiple ways with matter, where it may be absorbed, emitted, and scattered depending on various quantum principles. For example, the phenomenon of fluorescence and phosphorescence is where light is absorbed and emitted subsequently on interaction with a substance, whereas light gets elastically scattered in Rayleigh scattering, Mie/Tyndall scattering, where the wavelength of light remains unchanged. Light gets inelastically scattered in the case of the Compton effect and Raman scattering.

In 1928 Sir C.V. Raman and K.S. Krishnan experimentally discovered additional spectral signatures in the spectrum of liquid benzene and coined it 'A New Type of Secondary

Radiation', which opened a new pathway into molecular spectroscopy [170]. Raman realized that these results were related to the Kramers-Heisenberg phenomenon and was later awarded the Nobel Prize in 1930 for the discovery. This phenomenon of inelastic light scattering was first proposed by Adolf Smekal in 1923 [171].

In the process of light-matter interaction, the oscillating electromagnetic field perturbs the charge distribution and modifies the state of the system via the exchange of energy and momentum. Predominantly, ~99.9% of light scatters elastically, i.e., without losing any energy, and is known as Rayleigh scattering. The extent of inelastically scattered light is much less ~ 0.1%, which could be higher or lower in energy as compared to the incident wave. The inelastic process has its roots in the induced electronic polarizability of the system by the quasi-particle excitations. The part of the inelastically scattered radiation having energy more than the incident is known as anti-Stokes, and the part having lesser energy is known as Stokes scattering. In the case of crystals, the transfer of energy creates a quanta of lattice vibration known as a phonon. Raman scattering is an excellent tool to investigate various quasi-particle excitations associated with lattice, spin, or charge degree of freedom in solids. In the Stokes scattering, the incident light creates and in anti-Stokes it annihilates such quasi-particle excitations in the material. **Figure 2.1** shows a schematic representation of these scattering processes. Here, incident radiation of energy $\omega_i$ modifies the material state to a virtual state, and in the process of returning to the ground state, it emits the three radiation of energy $\omega_{AS} > \omega_i$, $\omega_S < \omega_i$ and $\omega_R = \omega_i$. In the Raman scattering process, energy and momentum remain conserved. So, if we consider quasi-particle excitation from lattice vibrations, a phonon with energy $\omega$ and wave vector $\boldsymbol{q}$ is either created or annihilated during the inelastic scattering process. The Stokes (redshift) and anti-Stokes (blueshift) process can also be expressed as

$\hbar\omega_S = \hbar\omega_i - \hbar\omega$; $\boldsymbol{k}_S = \boldsymbol{k}_i - \boldsymbol{q}$ and $\hbar\omega_{AS} = \hbar\omega_i + \hbar\omega$; $\boldsymbol{k}_{AS} = \boldsymbol{k}_i + \boldsymbol{q}$, respectively as also shown in a vectorial representation of both processes in **figure 2.2**.

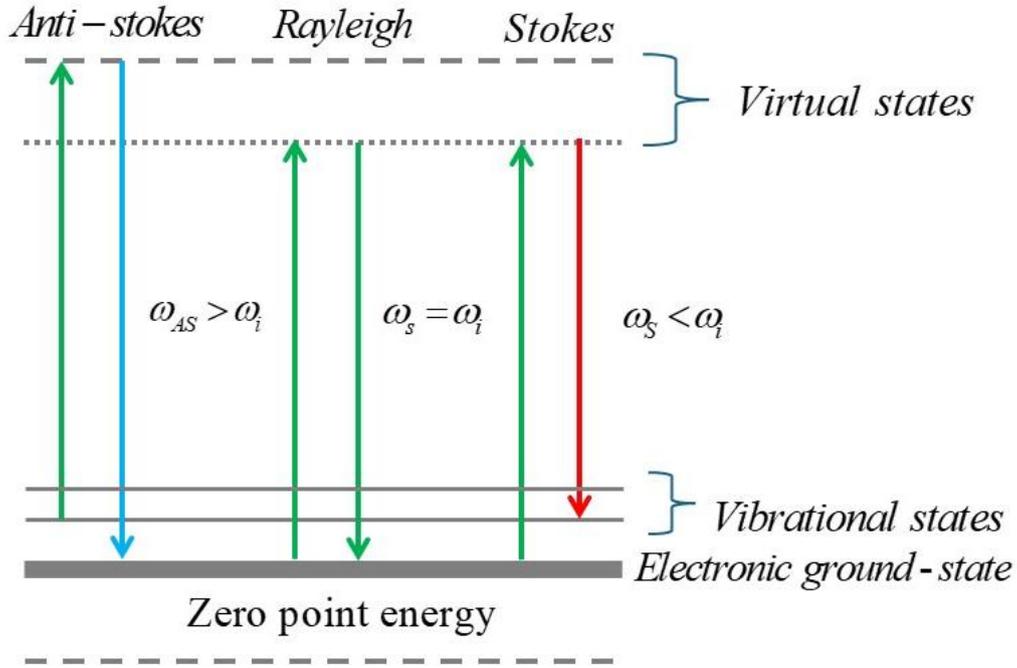

**Figure 2.1:** Pictorial representation of the elastic scattering i.e., Rayleigh, and inelastic scattering i.e., Stokes and anti-Stokes.

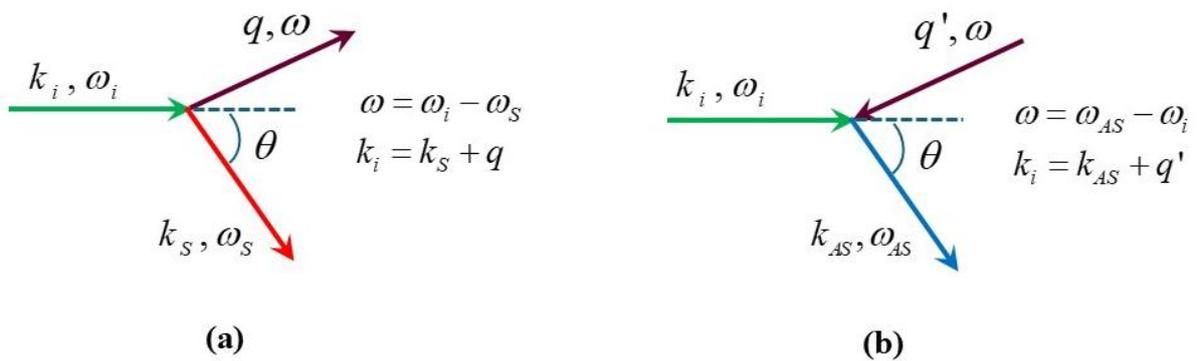

**Figure 2.2:** A vectorial representation of **(a)** Stokes, and **(b)** anti-Stokes scattering components showing conservation of wavevector and energy in the inelastic scattering process.

In the vectorial representation, one can express the crystal momentum generated from the incident radiation in the material, i.e. phonons $q$ as $q^2 = k_i^2 + k_S^2 - 2k_i k_S \cos\theta$. Phonons have a very less energy as compared to the incident light and hence the transfer of momentum and energy is very small. For instance, wavevector transfer for an incident radiation in the visible region is $|q| \sim 2k_i \sin(\theta/2) \sim \dfrac{4\pi n}{\lambda_i}\sin(\theta/2)$, here $\theta, n$ and $\lambda_i$ are scattering angle, the refractive index of the medium, and the incident light wavelength respectively. For a material of $n = 1.3$, and for a typical value of $\lambda_i \sim 500\ nm$, in backscattering configuration i.e., $\theta = 180^o$, the value of $|q| \sim 10^5\ cm^{-1}$ whereas the Brillouin zone boundary typically corresponds to $|k|_{max} = \dfrac{\pi}{a} \sim 10^8\ cm^{-1}$, here $a$ is the lattice constant. Hence in one-phonon or first-order Raman scattering one can only scan the phonon branches (optical / acoustic) in a small region around the center of the Brillouin zone i.e., $\Gamma - point$ which is further limited by the crystal symmetry. A schematic representation of the same is also shown in **figure 2.3**. However, the excitations from the whole Brillouin zone reflect in the process of two-phonon or second-order Raman scattering excitations and there is no momentum transfer due to the creation of a phonon with $\pm q$ wavevector.

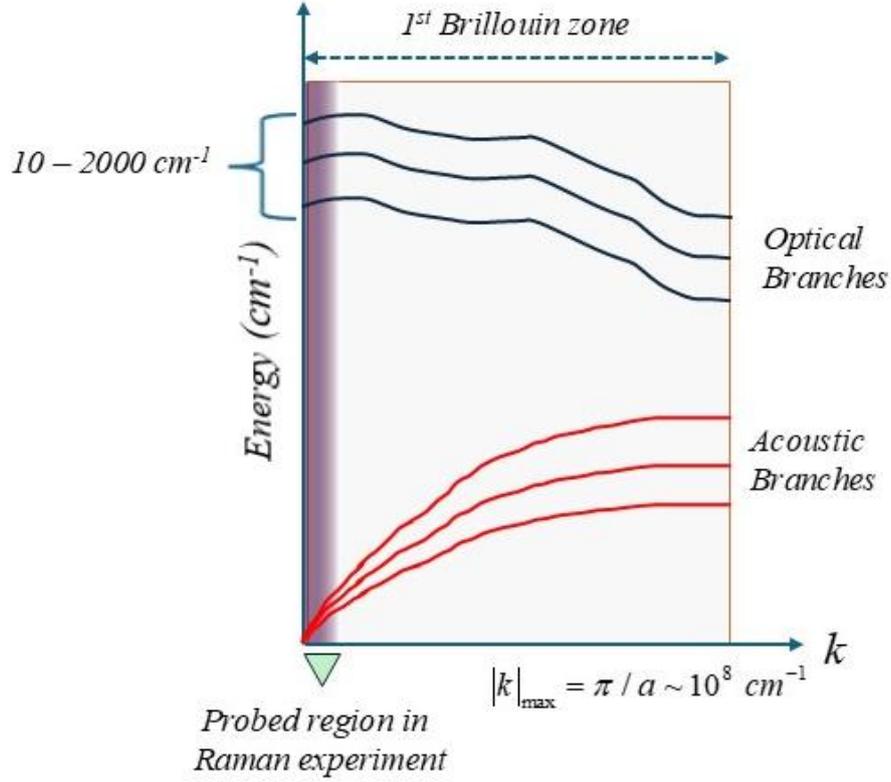

**Figure 2.3:** Schematic representation of Raman scattering process by a visible light where phonons only around the Brillouin zone center can be probed.

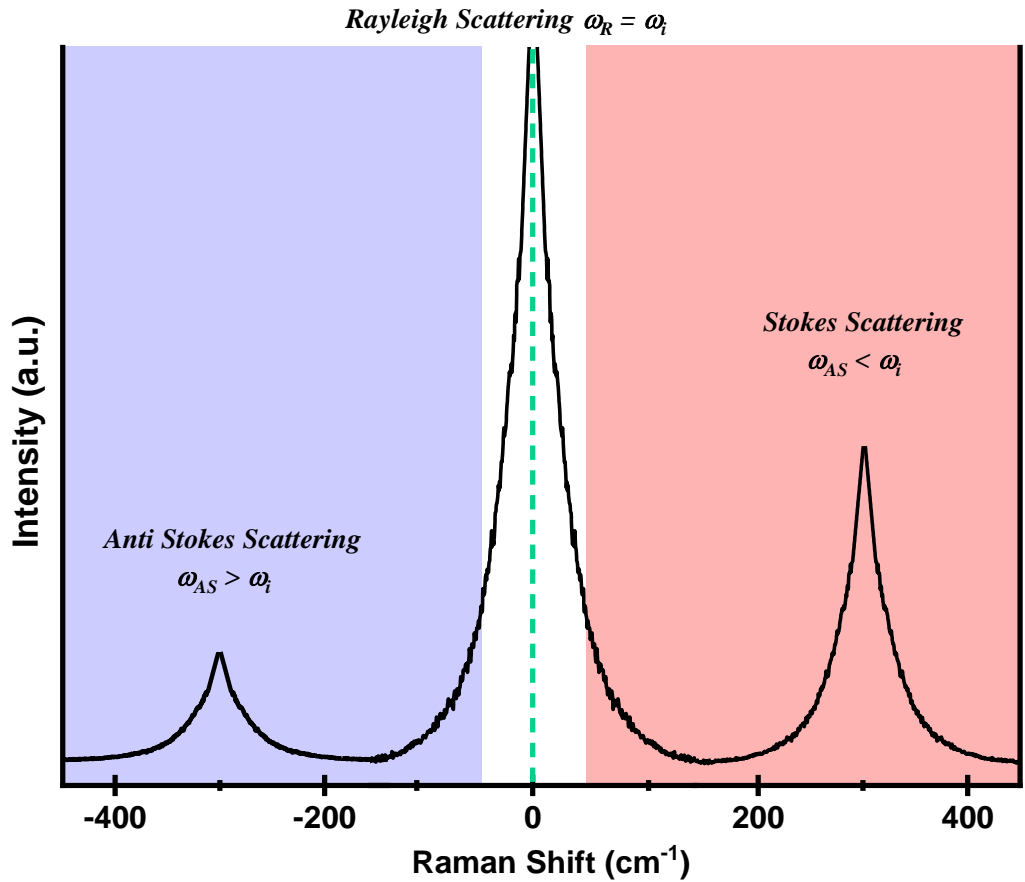

**Figure 2.4:** A typical Raman spectrum showing the elastic and inelastic part of the scattered radiation.

The difference in the energy of Stokes and anti-Stokes process is generally quantified by Raman shift which signifies the energy of the created or annihilated quasi-particles in the system. A typical Raman spectrum containing Stokes, anti-Stokes, and Rayleigh region is shown in **figure 2.2**. Phonons are bosonic excitations and the probability of their creation or annihilation is dictated by the Bose-Einstein distribution function i.e., $n(\omega,T) = [e^{\hbar\omega/k_B T} - 1]^{-1}$. The intensity of Stokes and anti-Stokes scattering is closely related to the $n(\omega,T)$. As in Stokes scattering phonon is created so the number of phonons is increased by 1 i.e., from $n$ to $n+1$;

hence, the corresponding intensity is proportional to $n+1$, whereas for anti-Stokes it is proportional to $n$. As evident from the Raman spectra in **figure 2.4,** the intensity of anti-Stokes in less than Stokes, and the ratio can be expressed as $I_S/I_{AS} = n+1/n = e^{\hbar\omega/k_B T}$ and it can be utilized to estimate the local temperature of the material when exposed to laser radiation [172].

## 2.1.1 Raman Effect: Classical Consideration

Oscillating induced or permanent dipoles generate electromagnetic radiation. The macroscopic or classical theory of the Raman effect is based upon induced polarizability of the molecules, which is a measure of how easily the electron cloud of a molecule can be distorted by the incident electric field of the incident light. The process of Raman scattering can be explained as follows: The oscillating electric field strength ($\vec{E}_i$) of the radiation (laser) which interacts with the material can be given as: $\vec{E}_i = \vec{E}_o\left(\vec{k}_i, \omega_i\right)\cos\left(\vec{k}_i \cdot \vec{r} - \omega_i t\right)$. The induced dipole moment due to the incident radiation can be expressed as a power series as follows:

.
$$\vec{\mu}_{ind} = \alpha \cdot \vec{E} + (1/2) \cdot \beta \cdot (\vec{E})^2 + (1/6) \cdot \gamma \cdot (\vec{E})^3 .... \qquad \text{-(2.1)}$$

Here $\alpha, \beta \text{ and } \gamma$ are polarizability, hyperpolarizability, 2nd hyperpolarizability tensors respectively. The contribution from $\beta \text{ and } \gamma$ is negligible as compared to $\alpha$ hence can be neglected. Hence equation 2.1 can be rewritten as:

$$\vec{\mu}_{ind} = \alpha \cdot \vec{E} \qquad \text{-(2.2)}$$

Polarizability tensor $\alpha$ is a measure of system susceptibility to polarization. It depends on the shape and dimensions of the bond, which changes during vibrations and hence is a function of symmetry and magnitude of atomic displacement. Hence, $\alpha$ depends on the normal coordinate $u$ of molecular vibrations and can be expressed as a Taylor series around the equilibrium position $u_0$ as:

$$\alpha = \alpha_o + \sum_k \left(\frac{\partial \alpha}{\partial u_k}\right)_{u_0} \cdot u_k + \frac{1}{2} \sum_{k,l} \left(\frac{\partial^2 \alpha}{\partial u_k \partial u_l}\right)_{u_0} \cdot u_k \cdot u_l + \dots \qquad - (2.3)$$

Here $\alpha_0$ is the polarizability in the equilibrium configuration; $u_k$ *and* $u_l$ are normal coordinates corresponding to the $k^{th}$ *and* $l^{th}$ normal vibrations have a frequency $\omega_k$ and $\omega_l$ respectively. If we consider that different vibrations are totally independent, only the first two terms survive in equation 2.3. and reduces to

$$\alpha = \alpha_o + \sum_m \left(\frac{\partial \alpha}{\partial u_i}\right)_{u_0} \cdot u_m \qquad - (2.4)$$

The lattice vibrations (normal coordinates) to first approximation oscillate as a harmonic oscillator and can be written as:

$$u_i = u_o(\vec{q}, \omega) \cos(\vec{q} \cdot \vec{r} - \omega t) \qquad - (2.5)$$

Here $\omega$ is the vibrational frequency for atoms. Using equations 2.4 and 2.5 the expression for $\vec{\mu}_{ind}$ reduces to:

$$\vec{\mu}_{ind} = \alpha_o \cdot \vec{E}_i + \frac{1}{2} \sum_m \left(\frac{\partial \alpha}{\partial u_m}\right)_{u_0} \cdot u_m \cdot \vec{E}_o\left(\vec{k}_i, \omega_i\right) \cdot \left[\cos\left((\vec{k}_i + \vec{q}) \cdot \vec{r} - (\omega_i + \omega)t\right) + \cos\left((\vec{k}_i - \vec{q}) \cdot \vec{r} - (\omega_i - \omega)t\right)\right]$$

$$- (2.6)$$

The first term in equation 2.6 corresponds to the Rayleigh scattering, while the second term contains the inelastically scattered radiation terms i.e., $\cos\left((\vec{k}_i + \vec{q}) \cdot \vec{r} - (\omega_i + \omega)t\right)$ and $\cos\left((\vec{k}_i - \vec{q}) \cdot \vec{r} - (\omega_i - \omega)t\right)$ depicts anti-Stokes and Stokes Raman scattering terms respectively.

It is evident from the expression that for a normal mode to be Raman active $\frac{\partial \alpha}{\partial u_i} \neq 0$ the

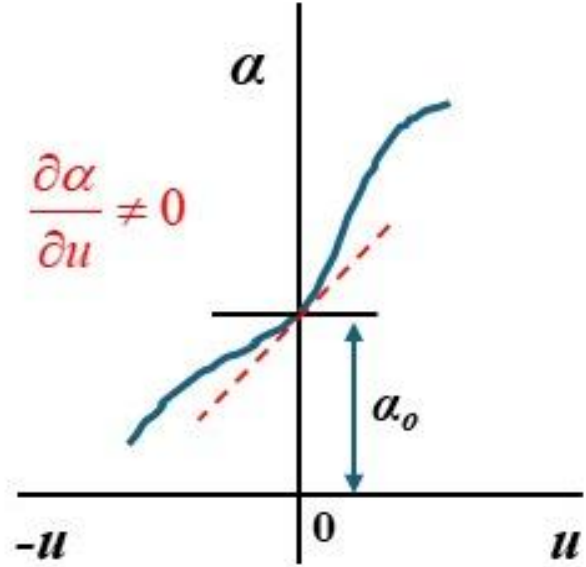

**Figure 2.5:** Variation of $\alpha$ with normalized coordinate $u$ for a Raman active phonon mode.

condition must be followed as also shown in **figure 2.5**. The pre-factor ( $R_m = \left( \dfrac{\partial \alpha}{\partial u_m} \right)_{u_0} \cdot u_m$ ) on

the right side in equation 2.6 is referred as the Raman tensor and it dictates the amplitude of

the scattered light for a given vibrational mode *m*. Using this one may express the intensity *I*

of the scattered light which arises from incident light when scattered by the *m*<sup>th</sup> vibrational

mode as:

$$I \propto \left| \hat{e}_S \cdot R_m \cdot \hat{e}_i \right|^2 \hspace{3cm} \text{- (2.7)}$$

Here $\hat{e}_S$ *and* $\hat{e}_i$ are the unit vectors showing the polarization direction of scattered and incident light. The component of the Raman tensor is dependent on the crystal and vibrational mode symmetry and can be derived from group theory. Hence, the polarized excitation and detection can provide information about crystal symmetry.

## 2.1.2 Raman Effect: Quantum Description

Above we discussed the classical treatment of the Raman scattering process where a classical wave nature of the incident radiation is considered which induces dynamic fluctuations in the polarizability by redistributing charge density of the material and provided selection rules for a normal mode to be Raman active. Classically there should not be any inelastic scattering at zero kelvin but we know that quantum fluctuations do not vanish and zero-point energy exists which causes atoms to vibrate even down to absolute zero. The classical description has its limitations and for a real picture, we should consider the quantization of electromagnetic radiation which consists of quanta of energy i.e., photons that interact with discrete or continuum energy levels of the material. The scattering phenomenon which is not apparent from the classical treatment consists of the resonance effect, phonon statistics, the interaction of electrons with other quasiparticles, the effect of laser energy, multi-phonon Raman scattering, etc.

In quantum mechanical treatment for a first-order Raman scattering process involves electronic excitation by an incident photon to a higher-energy state, which could be a real/virtual eigenstate from there, it is mediated by the creation or annihilation of phonons owing to coupling between electronic and lattice vibrational energy levels.

Such a process can be easily visualized via the band structure diagram as shown in **figure 2.6** and the Feynman diagram as shown in **figure 2.7**. This can be divided into three-step process i.e. (A) electron-photon interaction; generation of electron-hole pair due to absorption of the incident photon which excites an electron from the ground $|g\rangle$ state to a (virtual or real) state $|m\rangle$ with energy $E_m$, (B) electron-lattice interaction, creation or annihilation of phonon which brings the system to a state $|n\rangle$ (virtual or real) with energy $E_n$, (C) electron-photon interaction, finally recombination of electron-hole pair via emitting a scattered photon coming back to the ground state, which is taken as zero of energy. This photon-in and out process is limited by the conservation of momentum and energy, which is also reflected in **figure 2.6**. The scattering intensity in terms of the strength of electron-radiation ($\hat{H}_{ER}$) and electron-lattice ($\hat{H}_{EL}$) coupling can be calculated via perturbation theory as depicted in the Feynman diagram in **figure 2.7**. The resulting expression of the Stokes scattering intensity can be written as [173]

$$I(\omega_S, \omega_i) \propto \left| \sum_{m,n} \frac{\langle g|H_{ER}(\omega_S)|n\rangle \langle n|H_{EL}|m\rangle \langle m|H_{ER}(\omega_i)|g\rangle}{[\hbar\omega_i - E_m + i\Gamma_m][\hbar\omega_S - E_n + i\Gamma_n]} \right|^2 \delta(\hbar\omega_i - \hbar\omega_S - \hbar\omega) \qquad \text{- (2.8)}$$

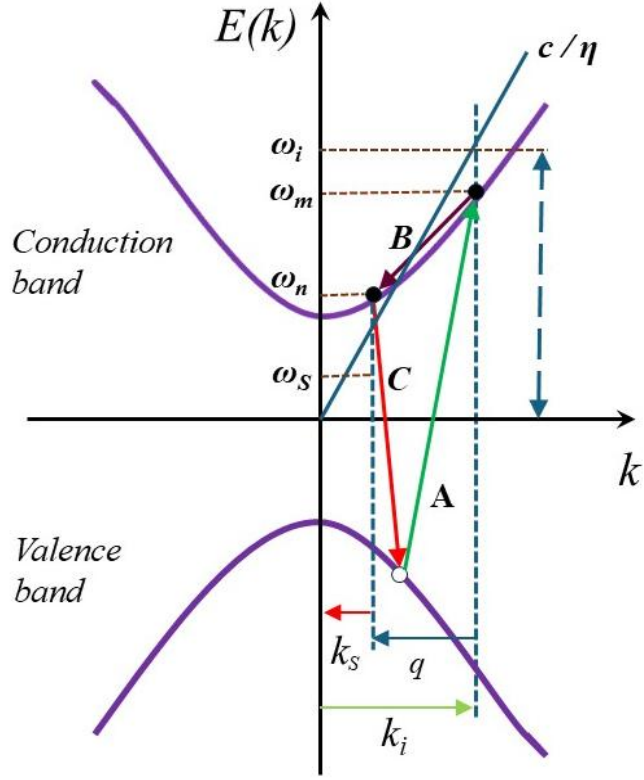

**Figure 2.6:** Pictorial representation of Stokes scattering process via band structure representation.

Here $\Gamma_{m,n}$ is the broadening factor introduced to incorporate the finite lifetime of the emission/absorption process. $\langle m|H_{ER}(\omega_i)|g\rangle$, $\langle g|H_{ER}(\omega_S)|n\rangle$ are the matrix elements representing the absorption and emission of photons and $\langle n|H_{EL}|m\rangle$ represents electron-phonon coupling.

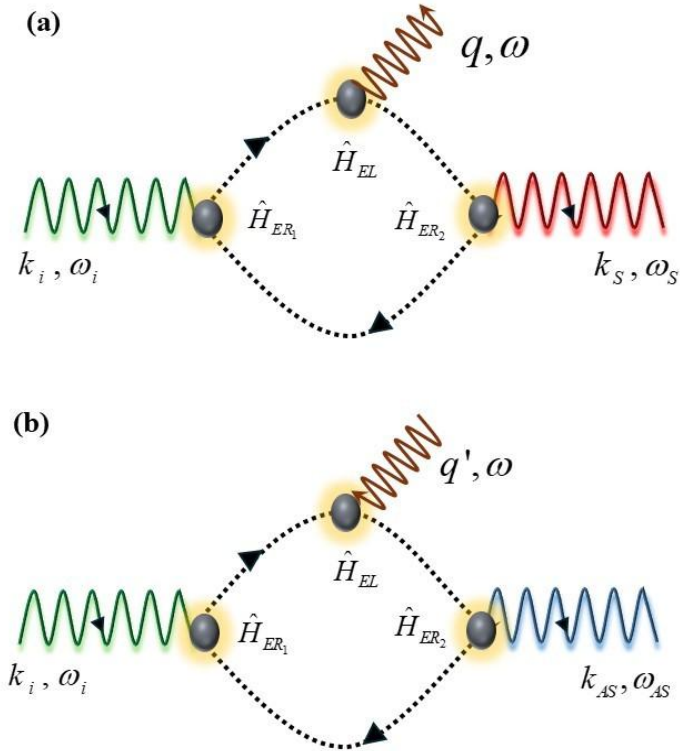

**Figure 2.7:** Feynman diagram for the (a) Stokes and (b) anti-Stokes Raman scattering process, respectively. $\hat{H}_{ER}$ and $\hat{H}_{EL}$ represents electron-radiation and electron-lattice interaction terms.

## 2.2 Instrumentation

### 2.2.1 Micro-Raman Spectrometer

For work done in this thesis, the HORIBA LabRAM HR Evolution spectrometer is used to perform Raman experiments. This setup is designed to work for UV-VIS-NIR (**U**ltra **V**iolet-**VIS**ible-**N**ear **I**nfra **R**ed) which means a range of 200-2200 nm laser source. It is a micro-Raman instrument where an optical microscope is integrated with the Raman spectrometer. An optical microscope with different power magnifies the sample area and focuses the laser beam on the sample surface. This enables to collection Raman spectra from microscopic samples or microscopic regions of large samples, which is crucial to determine the homogeneity and quality of the sample. Here, the spectra are collected in a backscattering configuration. The main components of Raman instruments consist of (a) Laser, (b) optical parts, (c) spectrograph, (d) grating, and (e) detector. A schematic representation of the components of the spectrometer and optical path for LabRAM HR Evolution is shown in **figure 2.8**.

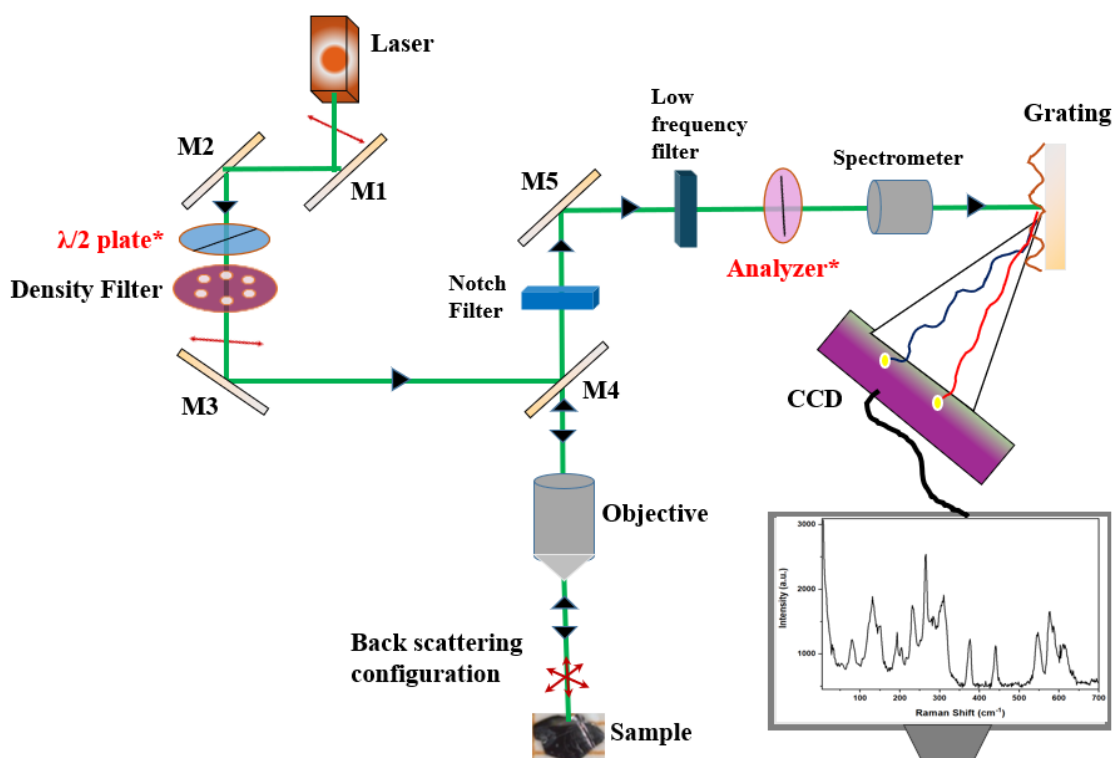

**Figure 2.8:** Schematic representation of components of the Micro-Raman spectrometer. '*' symbol indicates the additional components placed in carrying out polarization-dependent experiments.

## (a) Laser

Early Raman spectrometers employed mercury arc lamps where a 435.8 nm line was used as the excitation source. In the late 1960s mercury lamp was replaced by LASER, which is a stable and very intense source of light**.** A highly monochromatic and polarized laser (light amplification by stimulated emission of radiation) is used as an excitation source in the experiments conducted in this thesis work. The most commonly used lasers are 325nm (ultra-violet), 532 nm (visible, Green), 632 nm (visible, Red) and 785 nm  (near-infrared). The choice of laser plays an important role as the Raman scattering efficiency is $\propto \lambda^{-4}$ where $\lambda$ corresponds to laser wavelength. For instance, Raman scattering with 532nm is a factor of ~ 4.7 times more efficient than at 785nm and a 325 nm Raman scattering is a factor of ~14 times more efficient than at 633nm. The disadvantage of UV lasers is that due to the high energy sample is prone to local burning and also results in a higher fluorescence signal. Hence one has to optimize according to the response from the sample.

## (b) Optical parts

The laser light has to pass through different optical components before and after interacting with the sample in order to finally extract the inelastic scattered part. Such optical components are as follows: a combination of mirrors to direct the beam path, an objective lens with different magnifications to focus the laser on the sample, and a notch or bandpass filter that attenuates a specific range of wavelength and allows others outside the blocked region with an aim to eliminate the Rayleigh signal. The Rayleigh signal is further filtered out with an additional low pass filter which transmits only the longer wavelengths of inelastically scattered light and directs it to the monochromator. In this current work, a 100x and 50 x long working distance objective lens are used. The laser power is controlled by an intensity filter placed in the path of incoming laser light which can be controlled to 100% , 50% , 25% , 10% , 5% , 3.5% , 1% , 0.1% and 0.01%.

## (c) Spectrograph

In order to efficiently detect even a fraction of light undergoing the Raman scattering, an effective spectrograph setup is crucial. A schematic diagram of a typical spectrograph is shown in **figure 2.9**. It contains an entrance slit from which the scattered light enters the spectrograph which is collimated by a mirror towards a diffraction grating where the light gets dispersed and finally is directed to the detecter i.e. Charge Coupled Device (CCD) via a focusing mirror. The longer the focal length (e.g., the distance between the dispersing grating and detector) of the spectrometer, the higher the spectral resolution.

## (d) Diffraction Grating

In order to split the inelastically scattered light in its constituent wavelength a diffraction grating in employed as shown in **figure 2.9**. The resolution ($R$) of a diffraction grating is significantly dependent and is proportional to the number of grooves or lines per millimeter *(N)* of the grating. $R = \dfrac{\lambda}{\Delta\lambda} = mN$ here $\lambda, \Delta\lambda$ *and* m is the wavelength of the incident light, the smallest resolvable wavelength difference, and the order of diffraction, respectively. In our instrumental setup, we have 600 and 1800 grooves per millimeter grating, which are used to perform experiments.

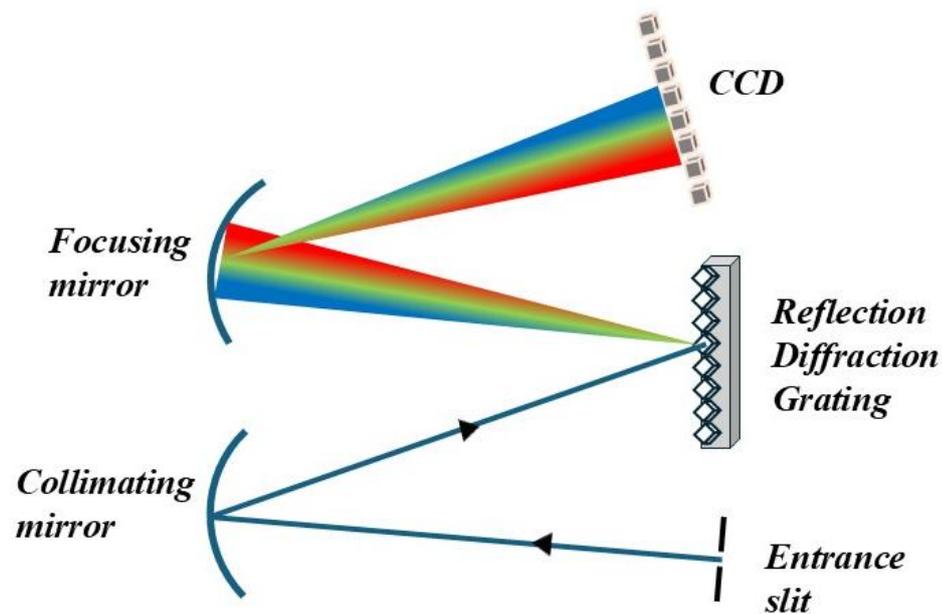

**Figure 2.9:** Schematic representation of spectrograph optic path.

### (e) Detector

In a typical Raman spectrometer, the Raman scattered light is dispersed using the diffraction grating, and this dispersed light is then projected onto the long axis of the CCD array. It is a silicon-based multichannel array detector for a wide range of UV, visible, and near-infra light. CCD is highly sensitive to the energy of the incident photon. The resolution and energy range of detection is determined by the pixel (Silicon p-n junction) density and size of the multichannel array. It works on the principle of the photo-electric effect, where each element of the CCD interacts with light to build up charge proportional to the intensity of the light, and at the end of the measurement, the electronics read out the individual element charge density in the form of voltage. The performance of a CCD is highly affected by the thermal fluctuations in the absence of any incident light, also known as dark current. Hence, CCD requires a degree of cooling for proper operation and high-grade spectroscopy. Generally, it is done via liquid nitrogen and/or using a Peltier cooling phenomenon. In our case Raman spectrometer is equipped with a Peltier-cooled ($70°C$) CCD detector having 1024 x 256 pixels density and the dark current noise level is less than 0.002 electron/pixel/sec.

## 2.2.1.1 Polarization-dependent Raman Study

The polarization direction of the incident and scattered light provides crucial information about the material properties. The in and out photons are related via the Raman tensor, which dictates how the polarization of the incoming photon will change in the Raman scattering process. As this scattering is mediated by lattice vibrations hence the polarization of the scattered photon is determined by the symmetry of the vibrational modes and hence there are different Raman tensors for different normal modes of vibrations. The polarization-dependent study plays a crucial role in determining the symmetry of the vibration modes, molecular structure, anisotropy, molecular orientation, crystallinity, and amorphicity of the material, etc. Within the semiclassical approach, the expression for intensity of Raman scattering is discussed in section 2.1.1 and is given by equation 2.7.

## Controlling Incident light Polarization direction and Polarization analyzer:

The laser light used in the Raman scattering experiment is linearly polarized which excites the sample. The polarization direction of this incident light can be precisely controlled by placing retarders (lambda plates) as shown in **figure 2.8**. We have used a $\lambda/2$ plate in the path of the incident light where a rotation of angle $\theta$ of the $\lambda/2$ plate rotates the polarization direction of the incident light by $2\theta$. Such a rotation is attributed to the birefringence property which introduces the fast and slow axis of the retarder. A schematic diagram is shown in **figure 2.10** for the same.

The scattered light polarization direction can be analyzed by using an analyzer after the scattered light. The experiment can be performed in different configurations (i) Incident light rotated; analyzer kept fixed, (ii) incident kept fixed, analyzer rotated and (iii) sample is rotated while incident light polarization direction and analyzer is kept fixed.

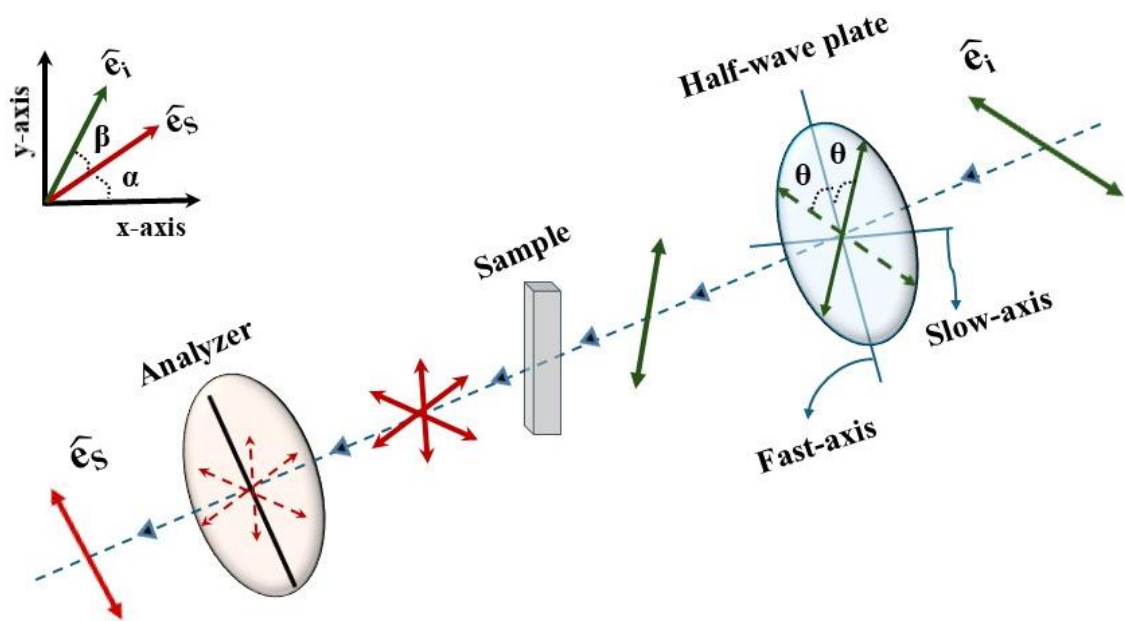

**Figure 2.10:** Schematic representation of working of half-wave $(\frac{\lambda}{2})$ plate and analyzer.

## 2.2.1.2 Temperature-dependent Raman Study

Temperature affects the properties of a material substantially. Temperature-dependent Raman scattering measurements can be utilized to analyze properties of materials such as structural, electronic, magnetic, etc. It can be used to detect structural, magnetic, and metal-insulator phase transitions, the presence of exotic quasi-particles such as Majorana fermions, magnons, excitons, etc.. On varying temperatures, one can control the effect of thermal fluctuations and at low temperatures, it allows other collective excitations to emerge which couple to the lattice dynamics and may reflect in the Raman spectra in a variety of ways for example dynamic change in self-energy parameters phonons (frequency or line-width) or in the intensity. The analysis of temperature-dependent measurements can also give information about materials' thermal stability, the effect of strain, defects, etc. such an analysis is crucial for the characterization of future quantum technologies. To perform a controlled temperature-dependent measurement we have used MONTANA Cryostation which works on closed cycle refrigeration (CCR) technique where a continuous helium flow controls the temperature over a range of 3K to 350K providing a peak-to-peak temperature and vibration stability of < 10 mK and 5nm, respectively.

## Closed-cycle cryostat (CCR) Setup

The main parts of the cryostat are: (i) Compressor, (ii) Control Unit, (iii) Cryostat and Sample chamber, and (iv) user interface computer. A schematic diagram is also shown in **figure 2.11**.

**(i) Compressor:** It controls the flow of helium which uses an onboard microcontroller to control the compressor capsule and cold-head drive motor to achieve the required temperature. It contains pressor sensors that monitor any leakage or improper connection of cold-head.

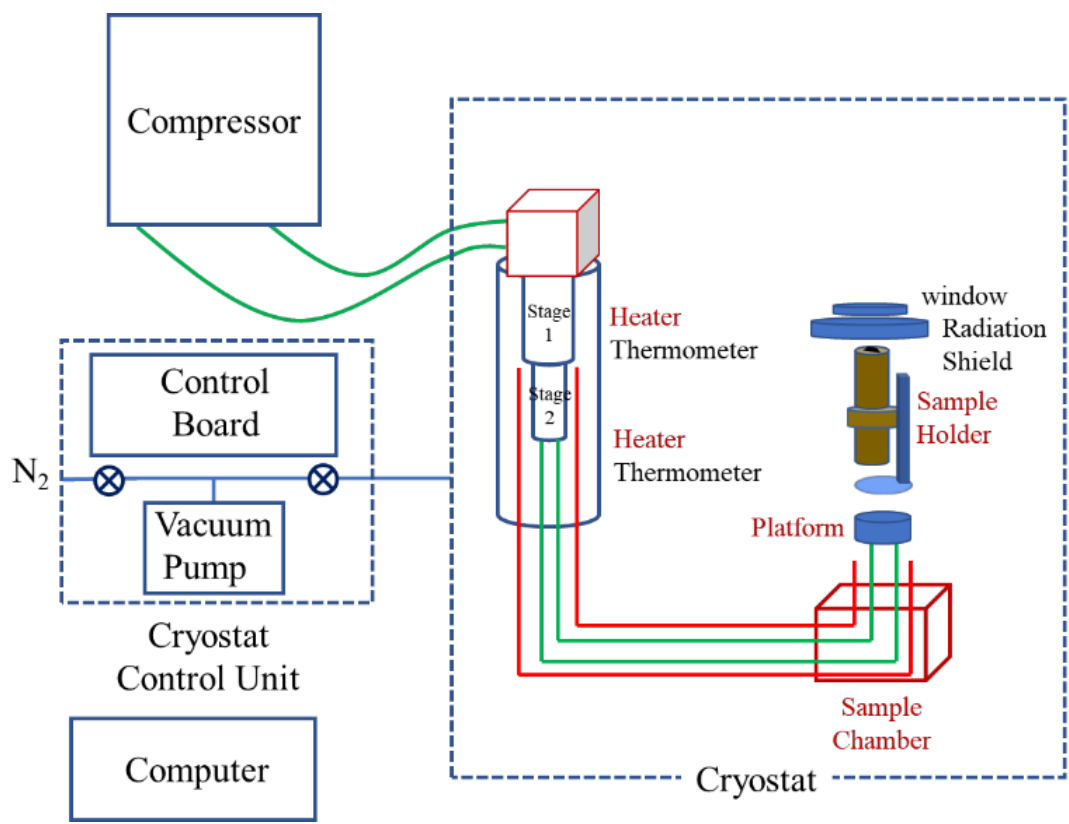

**Figure 2.11:** Schematic representation of main components of closed-cycle cryostat (CCR) setup. (ref. CCR manual)

**(ii) Control Unit:** This unit contains the electronics that control the function of the cryostat such as the system power supplies, the vacuum pump, and the high-frequency filter for the cryo-cooler motor.

**(iii) Cryostat:** The cryostat consists of the cooling tower and the sample chamber which are connected by a base plate-related assembly. It is connected to the control unit and to the compressor for helium supply and return.

**(iv) User Interface computer:** A user interface PC is required to run and control the configuration parameters of the cryostat which is connected via USB to the control unit.

In order to mount the sample in the cryostat chamber we have used a thermal greece GE Varnish. A nitrogen gas supply is needed to connect to the control unit before cooling down.

Once the cooling is initiated via the user interface, the control unit supplies that nitrogen gas to the sample chamber at very low pressure (~10 psi) in order to remove impurities. After that nitrogen flow stops, the vacuum pump starts, and the liquid helium supply begins to flow to achieve the required temperature. The target temperature is maintained with a high precision of ~ 0.1 mK and one can take the measurement.

## 2.2.2 Single Crystal X-Ray Diffraction (SC-XRD) Measurements

Single-crystal X-ray diffraction is a non-destructive technique that provides the measure of interatomic distances with very high resolution, unit cell dimensions, bond lengths, bond angles, details of site ordering, and structural phase transitions.

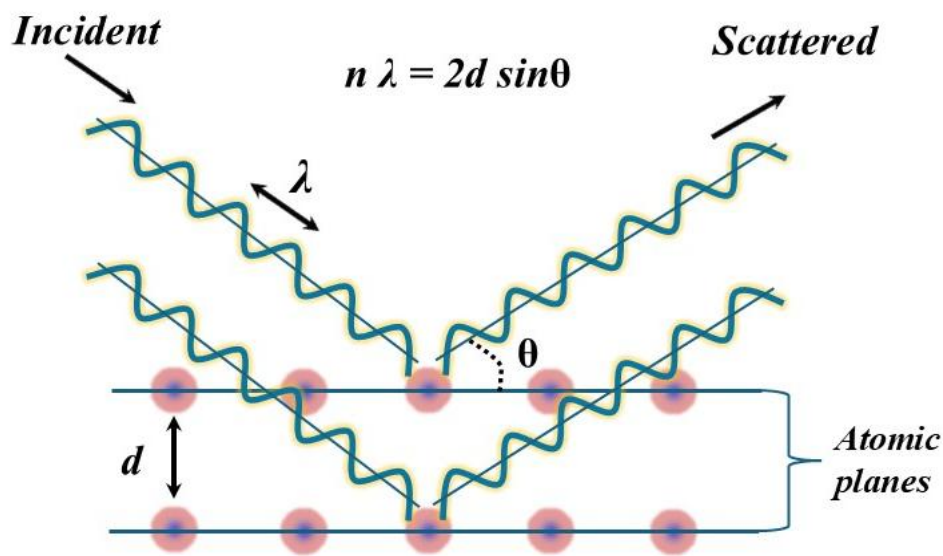

**Figure 2.12:** Schematic diagram of XRD diffraction and Braggs condition from crystal planes.

The fundamental principle of SC-XRD was laid by Max von Laue in 1912 where he discovered that the crystalline solids act as a 3D diffraction grating when X-rays are of comparable wavelengths with the lattice spacing of the crystal planes. The interaction of incident X-ray with the material produces a constructive interference with the diffracted ray when the brags law conditions are satisfied which is $n\lambda = 2d\sin\theta$, here $n, \lambda, d \; and \; \theta$ are integer, x-ray wavelength, interplanar spacing, and diffraction angle, which is measured by the goniometer.

A schematic diagram of the diffraction process is shown in **figure 2.12** A typical X-ray diffractometer consists of three basic components i.e., X-ray source (tube), sample holder, and detector.

A typical view of the diffractometer is shown in **figure 2.13** where the key components marked in numeric are: (1) X-ray tube; to generate x-rays, (2) 4-circle Kappa goniometer, (3) X-ray shutter, (4) collimator; for refining the x-ray beam, (5) beamstop; to protect detector from damage and (6) beryllium window; to collect scattered X-rays to the CCD detector. The X-rays are generated by a micro-focus sealed tube, which is mounted on the goniometer and powered by the high-voltage X-ray generator. In this thesis, in order to reveal the possibility of structural transition, we carried out a temperature-dependent SC-XRD measurement using a SuperNova X-ray diffractometer system with a copper (Cu, $\lambda \sim 1.54$ A˙) source used to radiate the sample at ambient pressure. Liquid nitrogen is utilized to vary temperature from 87K till room temperature.

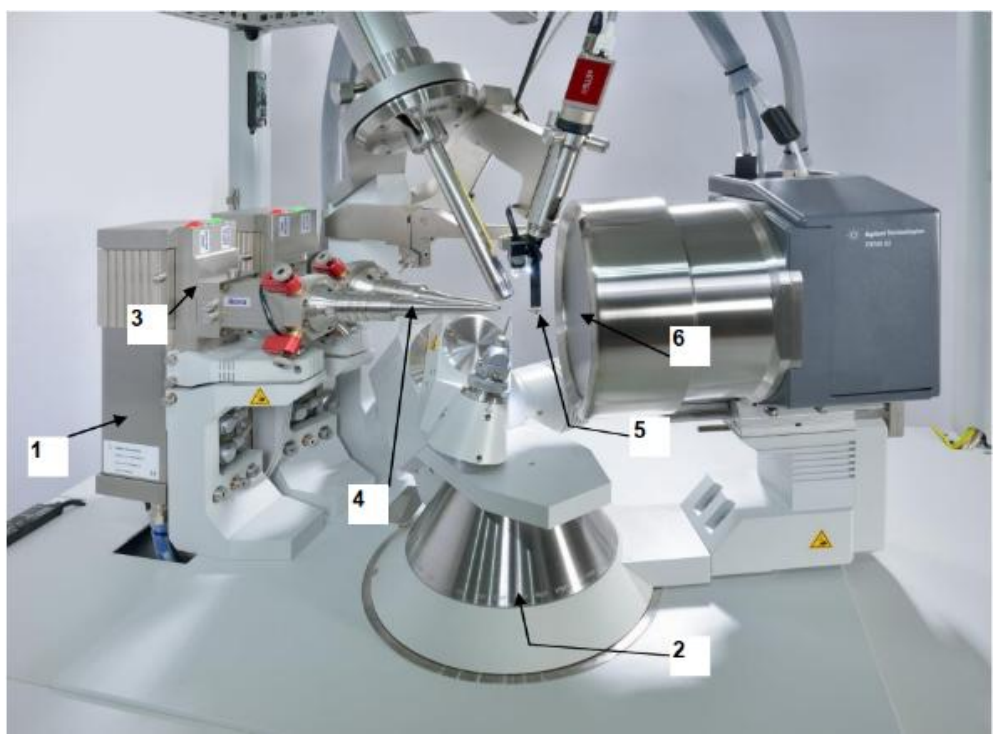

**Figure 2.13:** A typical diffractometer with key components. (ref. X-Ray manual)

## 2.3 First Principle Calculations

First-principles or ab initio calculation is a method based on quantum mechanics which is extensively used in material science, where one can control the properties of matter at the level of individual atoms and molecules. Density Functional Theory (DFT) is one of the successful ab initio quantum mechanical modeling methods that provides solutions to the fundamental equations that describe complex quantum many-body systems, which is practically useful. We have performed the first principle calculations to find phonon dispersion within the framework of Density Functional Perturbation Theory (DFPT), using Quantum Espresso, which is an integrated suite of open-source computer codes based on DFT, plane waves, and pseudopotentials that is meant for electronic structure calculation and material modeling [174].

Solving a many-electron problem using the Schrödinger equation is a difficult task. It can be simplified if the electrons are categorized into valence and inner core electrons. Core electrons strongly bound and partially screen the nucleus and do not play a significant role in determining the properties of the material, hence one can approximate the problem to an ionic core that interacts with valence electrons. For a many-electron system, the Schrödinger equation takes the following form:

$$i\hbar \frac{\partial \Phi(\boldsymbol{r},\boldsymbol{R};t)}{\partial t} = \left( -\sum_{\mu} \frac{\hbar^2}{2M}\nabla^2_{\boldsymbol{R}_{\mu}} - \sum_{i} \frac{\hbar^2}{2m}\nabla^2_{\boldsymbol{r}_i} + V(\boldsymbol{r},\boldsymbol{R}) \right) \Phi(\boldsymbol{r},\boldsymbol{R};t) \qquad \text{- (2.9 a)}$$

$$V(\boldsymbol{r},\boldsymbol{R}) = -\sum_{i,\mu} \frac{Z_I e^2}{\left|\boldsymbol{r}_i - \boldsymbol{R}_{\mu}\right|} + \frac{1}{2}\sum_{i \neq j} \frac{e^2}{\left|\boldsymbol{r}_i - \boldsymbol{r}_j\right|} + \frac{1}{2}\sum_{\mu \neq \nu} \frac{Z_{\mu} Z_{\nu} e^2}{\left|\boldsymbol{R}_{\mu} - \boldsymbol{R}_{\nu}\right|} \qquad \text{- (2.9 b)}$$

Here $\Phi(\boldsymbol{r}_1,\boldsymbol{r}_2...\boldsymbol{r}_N,\boldsymbol{R}_1,\boldsymbol{R}_2...\boldsymbol{R}_N;t)$ is many-body wave function ($\boldsymbol{r}$, m); ($\boldsymbol{R}$, M) are notations for electrons and ions (nuclei). The first and second term on the right side of equation 2.9 a, represents kinetic energy term for nuclei and electrons respectively. The potential energy (coulombic interaction) $V(\boldsymbol{r},\boldsymbol{R})$ is a sum of electron- nuclei, electron-electron, nuclei - nuclei interactions as shown in equation 2.9 b. It is a very difficult task to determine the solution of this complex equation so one has to make some approximations. For example, the Born-Oppenheimer approximation, where electrons are visualized as moving in a constant potential created by a static nucleus can be considered as fixed sites as the nucleus is much heavier than the electrons ($M/m \sim 1836$). Hence, one can neglect the contribution from the motion of nuclei, and the potential energy term due to nuclei-nuclei coulombic interaction can be treated as a constant. Still, the solution is way too complicated as it contains *3N* variables in the many-body wavefunction. This is where the density functional formalism comes into the picture as it provides an effective way, under certain approximations, to determine and understand material properties[175,176]. In this formalism, the problem of many-body interaction is reduced to a self-consistent one-particle problem. Instead of dealing with many-electron wavefunctions, the electron density, $\eta(\boldsymbol{r})$ is considered as the main variable, and the electronic ground state is a functional of $\eta(\boldsymbol{r}) = \sum_{n=1}^{N} |\Psi_n(\boldsymbol{r})|^2$ .The ground state energy for a given electron density under the

Kohn-Sham approach can be given as:

$$E[\eta(\boldsymbol{r})] = T_S[\eta(\boldsymbol{r})] + E_H[\eta(\boldsymbol{r})] + E_{xc}[\eta(\boldsymbol{r})] + \int \eta(\boldsymbol{r})V(\boldsymbol{r})d\boldsymbol{r} \qquad \text{-(2.10 a)}$$

$$T_S[\eta(\boldsymbol{r})] = -\frac{\hbar^2}{2m}\sum_i \int \Psi_i^*(\boldsymbol{r})\nabla^2\Psi_i(\boldsymbol{r})d\boldsymbol{r} \qquad \text{- (2.10 b)}$$

$$E_H[\eta(\boldsymbol{r})] = -\frac{e^2}{2}\int \frac{\eta(\boldsymbol{r})\eta(\boldsymbol{r}')}{|\boldsymbol{r}-\boldsymbol{r}'|}d\boldsymbol{r}\,d\boldsymbol{r}' \qquad \text{- (2.10 c)}$$

Here $V(\boldsymbol{r}) = -\sum_{\mu}\frac{Z_{\mu}e^2}{|\boldsymbol{r}-\boldsymbol{R}_{\mu}|}$ , is an external static (nuclei/ions) potential acting on non-interacting

electrons ($\eta(\boldsymbol{r})$) and $T_S[\eta(\boldsymbol{r})]$ is the kinetic energy of the non-interacting electrons. In real systems, electron-electron interaction is present which is taken care by the Hartree energy term $E_H[\eta(\boldsymbol{r})]$ that comes in the picture due to electrostatic interactions between electrons. $E_{xc}[\eta(\boldsymbol{r})]$ is the exchange-correlation energy, which includes all the remaining many-particle interaction energy terms. The total energy functional and $\eta(\boldsymbol{r})$ for a real interacting system can be calculated by varying $\eta(\boldsymbol{r})$ for a non-interacting system and using the variational principle one gets the Kohn-Sham equations by minimizing energy with respect to $\Psi_i$ and is given as:

$$H_{SCF}\Psi_i(\boldsymbol{r}) = \left(-\frac{\hbar^2}{2m}\nabla^2 + V(\boldsymbol{r}) + V_H(\boldsymbol{r}) + V_{xc}(\boldsymbol{r})\right)\Psi_i(\boldsymbol{r}) = \varepsilon_i\Psi_i(\boldsymbol{r})$$

$$H_{SCF}\Psi_i(\boldsymbol{r}) = \left(-\frac{\hbar^2}{2m}\nabla^2 + V_{SCF}\right)\Psi_i(\boldsymbol{r}) = \varepsilon_i\Psi_i(\boldsymbol{r}) \qquad - (2.11)$$

$$V_{SCF} = V(\boldsymbol{r}) + V_H(\boldsymbol{r}) + V_{xc}(\boldsymbol{r})$$

For a given set of arrangements of atoms, the ground state energy obtained from Kohn-Sham equations is solved self-consistently where the effective potential $V_{SCF}(\boldsymbol{r})$ is a functional of $\eta(\boldsymbol{r})$. Here $V_{SCF}(\boldsymbol{r})$ is the effective potential, $V_H(\boldsymbol{r}) = \frac{\delta E_H[\eta(\boldsymbol{r})]}{\delta\eta(\boldsymbol{r})} = e^2\int\frac{\eta(\boldsymbol{r}')}{|\boldsymbol{r}-\boldsymbol{r}'|}d\boldsymbol{r}'$ and $V_{xc}(\boldsymbol{r}) = \frac{\delta E_{xc}[\eta(\boldsymbol{r})]}{\delta\eta(\boldsymbol{r})}$ are Hartree and exchange-correlation potentials. Still to get the exact solution one needs information about the $\frac{\delta E_{xc}[\eta(\boldsymbol{r})]}{\delta\eta(\boldsymbol{r})}$ but the issue is, $E_{xc}$ is not fully known and hence an approximate functional dependence has to be introduced to describe the exchange term. Some of the approximation methods that assume the exchange-correlation energy are- Local Density Functional (LDA), where the functional depends only on the electron density which is assumed to be uniform all over the space. It is a good approximation to first order but tends to give some errors. To correct it one has to use the inhomogeneity of electron density that comes under another approach which is Generalized Gradient Approximation (GGA) where $\nabla\eta(\boldsymbol{r})$ is also taken into account and produces more accurate results [177,178].

## 2.3.1 Phonon Calculations

Phonons are quanta of lattice vibrations that have a discrete energy spectrum and their quasi-particle formulation has been well understood in the quantum field theory framework [179]. Phonons play an important role in describing the mechanical, acoustic, electrical, optical, and thermodynamic properties of materials. Under the harmonic approximation, the phonon energies can be determined by the eigenvalues of the dynamical matrix $\boldsymbol{D}$ evaluated as:

$$\det\left|\boldsymbol{D} - \omega^2 \boldsymbol{I}\right| = 0 \qquad\qquad - (2.12)$$

The Dynamical matrix in matrix element form is expressed as the reduced Fourier transform of the harmonic force constant matrix $\xi_{mm'}^{\gamma_A \gamma}$ given as:

$$D_{m'm}^{\gamma'\gamma}(\boldsymbol{q}) = \frac{1}{\sqrt{M_{\gamma'} M_\gamma}} \sum_A \xi_{m'm}^{\gamma'_A \gamma} e^{i\boldsymbol{q}.A} \qquad\qquad - (2.13)$$

Here $M_{\gamma()}$ are the atomic masses , $r_m$ ($m = 1,2,3$ ; represents cartesian components) is the position of the atom $\gamma$ and $\gamma'$ in a unit cell. $\gamma'_A$ indicates the atom $\gamma'$ in the unit cell with the lattice vector $\boldsymbol{A}$ at $r_{\gamma'_R} = r_{\gamma'} + A$ and $\boldsymbol{q}$ represents the wavevector of the phonon. The force constant matrix is also known as the Hesse matrix and can be written in terms of second-derivatives Born-Oppenheimer energy $E$ matrix as follows:

$$\xi_{m'm}^{\gamma'\gamma} = \frac{\partial^2 E}{\partial R_{\gamma'm'} \partial R_{\gamma m}} = E_{\gamma'm'\gamma m}^{(2)} \qquad\qquad - (2.14)$$

To find the phonon frequencies for a $\boldsymbol{q}$ value one has to solve the dynamical matrix and for that interatomic force constant (IFCs) has to be evaluated which can be done using methods such as **(a)** Direct method i.e. Frozen Phonons and **(b)** Modern linear response method i.e., Density functional perturbation theory (DFPT). They are sometimes used in parallel in order to test the effect of anharmonicity in the results of the direct method. Another alternative method where we do not need to find IFC is molecular dynamics, which can also be invoked to calculate phonon frequencies via the fourier transform of the velocity-velocity autocorrection function.

**(a) Frozen Phonons:** This direct method of phonon frequency calculations is based on the ground state calculations for an ideal periodic crystal where the ions are finitely displaced with respect to their equilibrium position. In order to calculate forces caused by the small displacement of an atom which perturbs other atoms in the lattice and then calculates forces between every atom using the Hellman-Feynmann theorem to get the force constant matrix [180]. This method is computationally cheaper than the linear response method but the drawback is that in order to determine the phonon spectrum completely one needs a very large supercell with a size larger than the effective range of the lattice interactions. So, the application of this traditional method is limited to phonon calculation at zone-center. The force acting on an atom due to displacement can be written in terms of derivatives of energy as $F = -\dfrac{\partial E}{\partial R}$ and the corresponding force constant matrix elements take the following form:

$$\xi_{m'm}^{\gamma'\gamma} = -\frac{\partial F_{\gamma m}}{\partial R_{\gamma'm'}} \approx -\frac{F_{\gamma m}(R_{\gamma'm'} + \Delta R_{\gamma'm'}) - F_{\gamma m}(R_{\gamma'm'})}{\Delta R_{\gamma'm'}} \qquad \text{- (2.15)}$$

Here $F_{\gamma m}$ is the $m^{th}$ cartesian component of the force acting on an atom $\gamma$ when another atom $\gamma'$ in the crystal is displaced by $\Delta R_{\gamma'm'}$ from equilibrium position $R_{\gamma'}$ in a $m'$ direction.

## (b) Density Functional Perturbation Theory (DFPT)

In order to overcome the limitations of the Direct method there is an alternative method to calculate IFC directly within perturbative schemes which works directly in the $k$-space and provides the dynamical matrix at any $q$ without considering supercells. DFPT for phonons requires first-order changes in the density, wavefunction, external and effective potential, and ion-ion energy.

The derivative of Kohn-Sham orbitals with respect to atomic positions is given as:

$$H_{SCF} \frac{\partial \Psi_i(\boldsymbol{r})}{\partial R_k} = \left( \frac{\partial V_{SCF}}{\partial R_k} - \frac{\partial \varepsilon_i}{\partial R_k} \right) \Psi_i(\boldsymbol{r}) \qquad \text{- (2.16)}$$

This equation is known as Sternheimer equation and its solution gives first order change in electronic quantities to small atomic position displacements in a self-consistent way [181].

The dynamical matrix can be calculated at any $\boldsymbol{q}$ point and then the real space IFCs are obtained after fourier transformation. The whole process of calculating phonon dispersion can be divided into a sequence of four steps as (i) self-consistent calculation (SCF) of total energy, (ii) calculation of phonon frequencies using DFPT, (iii) calculation of interatomic force constants (IFCs) in real space and (iv) calculation of phonon dispersion.

# Chapter 3

# Part-A: Fluctuating Fractionalized Spins in Quasi Two-dimensional Magnetic $V_{0.85}PS_3$

## 3A.1 Introduction

The family of transition metal phosphorous tri-chalcogenides ($TM$PX$_3$, $TM$ = V, Mn, Fe, Co, Ni or Zn and X = S, Se) with a strong in-plane covalent bonds and weak van der Waals gap between the layers of magnetic atoms have appeared as an intriguing candidate for exploring the quasi-2D magnetism, where the inter-planar direct- and super-exchange magnetic interactions are substantially quenched [68,182,183]. The underlying magnetic ground state ($|GS\rangle$) in these materials is affected by the trigonal distortion ($TM$X$_6$) due to a change in the local symmetry from $O_h$ to $D_{3d}$, which consequently lifts the degeneracy of the $d$-orbitals. Magnetic studies on these materials reveal the emergence of quite different magnetic $|GS\rangle$ on varying $TM$ atoms; for example, FePS$_3$ shows an Ising-type transition at $T_N \sim 123$ K, NiPS$_3$ /MnPS$_3$ undergoes an $XY$/ Heisenberg-type transition at $T_N$ = 155 K / 78 K, respectively [184].

We note that in this $TM$PX$_3$ family, vanadium-based systems remain largely unexplored beyond their basic properties. The recent reports on $V_{0.9}PS_3$ revealed an insulator-to-metal phase transition at a pressure of ~ 120 Kbar without any structural transition and suggested a Kondo-type effect within the metallic phase. It was also advocated that this system lies in the close proximity of the quantum spin liquid (QSL) state, a topologically active phase [68]. The insulator-to-metal transition is also associated with antiferromagnetic (AFM) to paramagnetic transition and opens the possibility of a highly entangled spin liquid phase due to honeycomb lattice via potential Kitaev interactions. Interestingly, recently a new kind of Kondo behavior has been proposed and is attributed to gauge fluctuations from bond defects in spin-liquids [185]. These experimental observations and theoretical predictions suggest a key route to the observation of fascinating QSL states in these quasi 2D quantum magnetic materials. An important characteristic of these systems is that despite all being isostructural, a magnetic lattice is a 2D honeycomb structure formed by TM ions, that have different spin dimensionality. For example, $|GS\rangle$ dynamics of MnPS$_3$ member of this family is described by isotropic ($J_\perp = J_\parallel$) Heisenberg Hamiltonian [$H = -\sum_{ij} J_\perp (S_{ix} \cdot S_{ix} + S_{iy} \cdot S_{iy}) + J_\parallel S_{iz} \cdot S_{iz}$]; FePS$_3$ using Ising ($J_\perp = 0$) model, whereas NiPS$_3$ is understood using anisotropic ($J_\perp > J_\parallel$) model. This complexity is further enhanced in the $|GS\rangle$ of $V$-based system, V$_{0.85}$PS$_3$, also reflected in the magnetic measurements with $T_N \sim 60$K, where the magnetic susceptibility exhibits intriguingly different behavior compared with other members of this family, suggesting the presence of an additional exotic competing interaction in the Hamiltonian, such as Kitaev type, to completely understand the underlying magnetic $|GS\rangle$. We note that a similar behavior of $\chi_{mol}$ below $T_N$ was also reported for Cu$_2$Te$_2$O$_5$Br$_2$ [186,187] which could be understood by quantum critical transition between AFM state to QSL state.

Signatures of the exotic $|GS\rangle$ in this system may emerge from the presence of strong quantum fluctuations due to entanglement of the underlying spins within the honeycomb lattice. Interest in the field of QSL was renewed with the seminal work of Kitaev in 2006 [62] and subsequently certain conditions were laid down for the realization of a QSL state such as [64] presence of hexagonal honeycomb lattice, Mott-insulator, edge sharing octahedra in the structure and spin-orbit coupling. A large number of systems have been proposed, however so far there is not a single system which perfectly display QSL state as a true $|GS\rangle$. Although numerous proposed systems do show strong signature of a QSL state or a proximate QSL state [188-190]. However, in all those cases an ideal QSL state is preempted by the long-range magnetic ordering at low temperature. Despite this, the signature of a QSL state may be captured as fluctuation in the short-range ordering regime much before setting up of the long-range ordering. Recently, it was shown that in a magnetic system, CrSi/GeTe$_3$, with $S = 3/2$ and very weak spin-orbit coupling does reflect the signature of a QSL state understood using $S = 3/2$ XXZ-Kitaev model given as $H_{XXZ-K} = \sum\limits_{<i,j>} \frac{J_\perp}{2}(S_i^+ S_j^- + S_i^- S_j^+) + J_\parallel\, S_i^z S_j^z) + K \sum\limits_{<i,j>_\alpha} S_i^\alpha S_j^\alpha$ [191,192]. Also, signature of fractionalized excitations was reported in a magnetic $S = 1/2$ Kagome AFM systems attributed to a remnant QSL state [193]. In the case of a putative QSL candidate α-RuCl$_3$, it was shown that the low temperature zigzag AFM state is stabilized by quantum fluctuations with spin liquid state as a proximate metastable state [194].

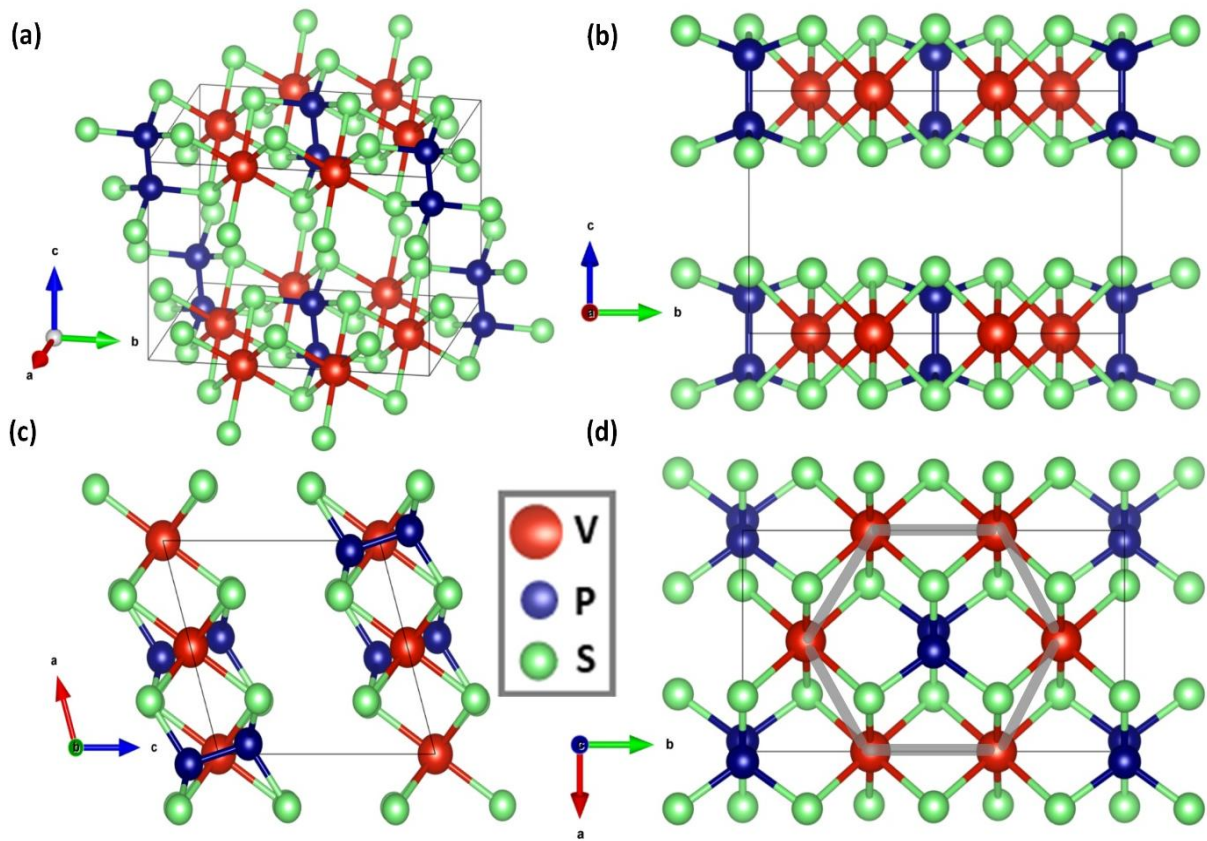

**Figure 3.1:** Crystal structure of VPS$_3$ plotted using (VESTA) as viewed along **(a)** arbitrary direction. **(b)** *a*-axis**, (c)** *b*-axis and **(d)** *c*-axis, thick grey solid lines connecting vanadium atoms represent 2D honeycomb lattice formation. Red, blue and green color spheres represent Vanadium, Phosphorus and Sulphur, respectively.

Generally, for a quantum spin liquid system frustration, dimerization, interchain/interlayer coupling, vacancy/defects leading to bond disorder, and spin-phonon coupling have an impact on the dynamics of the low-energy excitations [186,195,196]. In 2D quantum systems, spin liquid $|GS\rangle$ or remnant of spin liquid phase is expected to be a consequence of exotic topologies such as hexagonal honeycomb structure as both the triangular and square lattice have AFM like $|GS\rangle$ [186]. For $V_{1-x}PS_3$, we do have $p$-$p$ dimer formation in the $b$-$c$ plane **(see figure 3.1)**, vacancy, hexagonal honeycomb lattice, hinting that it may have a remnant of QSL state as strong quantum fluctuations. Inherent coordination flexibility and electronic configuration of V ion make it possible to realize exotic exchange with topologies.

Motivated by these suggestions and possibilities of a QSL state in these quasi-2D magnetic systems, we carried out an in-depth Raman scattering studies on a single crystal of $V_{0.85}PS_3$ to understand the underlying exotic properties. Evidence of a QSL state or its remnant may be uncovered via observation of the quantum fluctuations of the associated spin degrees of freedom and their coupling with the lattice degrees of freedom through spin-phonon coupling. Inelastic light (Raman) scattering is an excellent technique to probe such dynamic quantum fluctuations reflected via the emergence of the quasi-elastic response at low energy and a broad continuum in the Raman response $\chi^{"}(\omega,T)$ [156,197-200], smoking gun evidence of a QSL state. We note that Raman signature of QSL phase has been reported for different class of materials. Sandilands *et al.* [57] reported Raman studies on the α-RuCl₃. Their measurements revealed unusual magnetic scattering typified by a broad continuum which survived till very high temperature (~ 100K) as compared to the magnetic ordering temperature (~ 14K) along with the phonon anomalies suggestive of frustrated magnetic interactions. They suggested that their observations may be understood using the combined Heisenberg-Kitaev model and advocated that α-RuCl₃ may be close to a QSL ground state. Similar broad signature was also reported in other reports on α-RuCl$_3$ and γ-Li$_2$IrO$_3$ [57,58,201]. Another class of putative QSL candidates are spin-*1/2* frustrated Kagome compound called herbertsmithite e.g., ZnCu$_3$(OH)$_6$Cl$_2$. In these systems as well a similar broad continuum and phonon anomalies in the Raman as well as other optical measurements are identified with the possible fractionalized excitations associated with the QSL phase [58,202-205]. Another class of promising QSL candidates are the BEDT-TTF molecule-based organic systems with highly frustrated triangular lattices [206-208]. We note that in these system Raman as well as other optical techniques have been used to uncover the underlying QSL phase via observing a broad continuum and distinct in-gap excitations at low temperature and low energy regime.

Here for V$_{0.85}$PS$_3$, we observed a strong low energy quasi-elastic response with lowering temperature and a broad continuum; quite startling, it starts emerging much above the long-range magnetic ordering temperature. These characteristic features clearly suggest the presence of strong underlying quantum fluctuations. Surprisingly, the corresponding estimated dynamic Raman susceptibility, $\chi^{dyn}$, the amplitude is not quenched below $T_N$ similar to the observed magnetic susceptibility; as expected for a conventional magnetically ordered system, signaling that it emerges from a proximate QSL state or its remnant. Our observations evince the signature of a remnant QSL state as a fluctuating part suggests that this system lies in proximity of a QSL ground state. This also suggests that the low-temperature ordered phase may be proximate to the quantum phase transition into a spin liquid $|GS\rangle$. The anomalies observed in the $\chi^{dyn}$ maps parallel with the anomalies seen in the self-energy parameters of the phonon modes i.e., peak frequencies and line-widths. Here, we show the experimental evidence supporting the existence of a remnant QSL phase in V$_{0.85}$PS$_3$ using Raman spectroscopy.

## 3A.2 Experimental Details

Experimental techniques used are Raman Scattering, powder X-ray diffraction (pXRD), resistivity, and magnetization measurements.

### 3A.2.1 Powder X-ray Diffraction

The grown crystals' structural characterization and phase purity were investigated by powder X-ray diffraction (pXRD) at room temperature. Measurements were performed on a polycrystalline sample by grinding as-grown single crystals. The pXRD measurement was carried out at an STOE powder diffractometer in transmission geometry with Cu-Kα1 radiation (the wavelength (λ) is 1.540560 Å) from a curved Ge (111) single crystal monochromator and detected by an MYTHEN 1K 12.5∘ linear position sensitive detector manufactured by DECTRIS.

### 3A.2.2 Resistivity and Magnetization Measurements

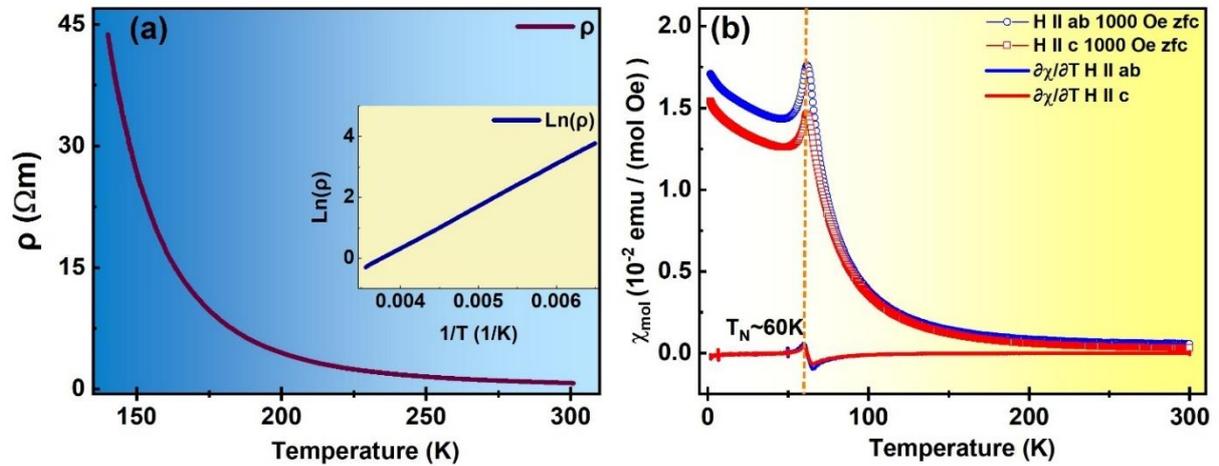

**Figure 3.2: (a)** Resistivity as a function of temperature and inset shows ln(ρ) vs 1/T, **(b)** Molar magnetic susceptibility $\chi_{mol}$ in Z.F.C. mode with H ∥ ab plane and H ∥ c axis along with respective derivatives.

Resistivity was measured as a function of temperature as shown in **figure 3.2(a)**. Inset shows the log of resistivity as a function of temperature, Arrhenius fitting gives the bandgap of ~ 180 meV. Magnetic susceptibility measurement was performed in Z.F.C. mode at $H_{ex}$ = 1000 Oe shown in **figure. 3.2 (b)** and shows $T_N \sim 60$ K respectively, which is very close to ($T_N \sim 62K$) what was reported by Coak *et al.*, [68].

### 3A.2.3 Raman Scattering

The inelastically scattered light was collected via a micro-Raman spectrometer (LabRAM HR Evolution) in a backscattering configuration. The sample was irradiated by a linearly polarized TE-cooled, 532 nm (2.33 eV) laser and focused via 50x LWD objective with N.A. of 0.8. Laser power was kept low (< 0.5 mW) to prevent the local heating effect. The scattered light was detected by a Peltier-cooled CCD after getting dispersed by 600 groves/mm grating. The sample is kept under a high vacuum chamber with a pressure $\sim$ 90 μTorr. The sample temperature is regulated over a range of 4K - 330 K with ± 0.1 K accuracy using a closed-cycle He-flow cryostat (Montana). Further to unveil the symmetry of phonon modes, we performed polarization-dependent measurement where we varied the polarization direction of the incident light using retarders half-wave plate while the analyzer is kept fixed. A schematic diagram of the experimental setup is shown in **figure 2.8** and the working of the half-wave plate along with the plane projection of the polarization direction of the incident and scattered light is shown in **figure 2.10** of Chapter 2.

### 3A.3 Results and Discussion

### 3A.3.1 Structural Analysis

The measured pXRD pattern, shown in **figure 3.3**, was indexed in the monoclinic crystal system, in agreement with the literature[209]. The crystal structure model proposed by Klingen *et al.* for FePS$_3$ [210] was sufficient to describe pXRD pattern. X-ray data were modeled by the Rietveld refinement in the monoclinic space group C2/m using the FULLPROF software. No additional reflections were observed, the fit corresponds to a single phase demonstrating the purity of V$_x$PS$_3$ crystals. As reported by Ouvrard *et al.* [209] and according to our EDX measurements, the $V_xPS_3$ compound is non-stoichiometric. The refinement also confirms this. It was possible to achieve the lowest value of the R-factor with a vanadium occupancy ratio of 0.85. The vanadium content corresponds then to $V_{0.85}PS_3$. The non-stoichiometric $V_xPS_3$ in the *TMPS_3* family of materials is known, which is explained by mixed valency and partial vacancy substitution [68,209,211]. The calculated peak positions and intensities are in agreement with the experimental data. Crystal structures projecting different crystallographic axes of $V_{0.85}PS_3$ are shown in **figure 3.4**. A mix of $V^{3+}$ for $V^{2+}$ is charge compensated by vacancies on the transition. The graphical representation was prepared using VESTA [212]. The obtained lattice parameters and reliability factors are summarized in **table 3.1.**

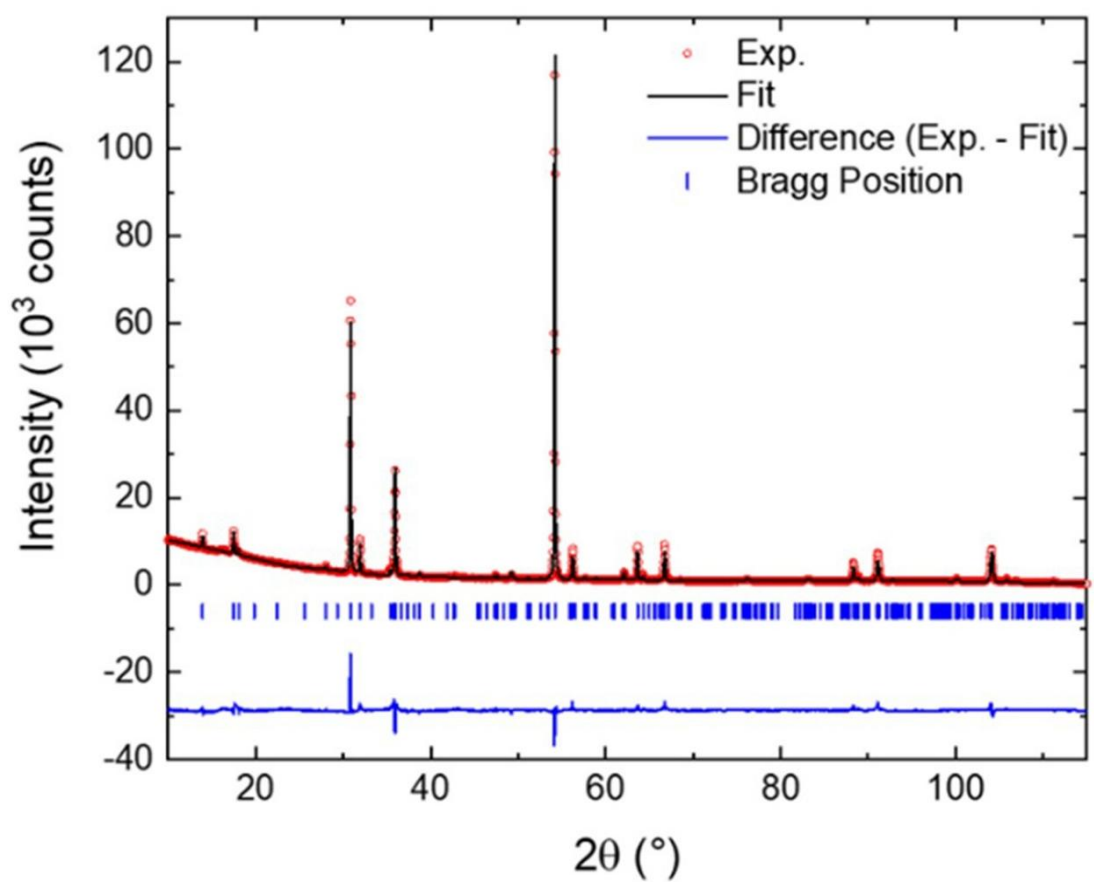

**Figure 3.3:** Rietveld analysis of $V_{0.85}PS_3$. Calculated pattern based on the Rietveld analysis (black line), the difference between measured and calculated pattern (blue line), and the calculated Bragg positions for a trigonal unit cell.

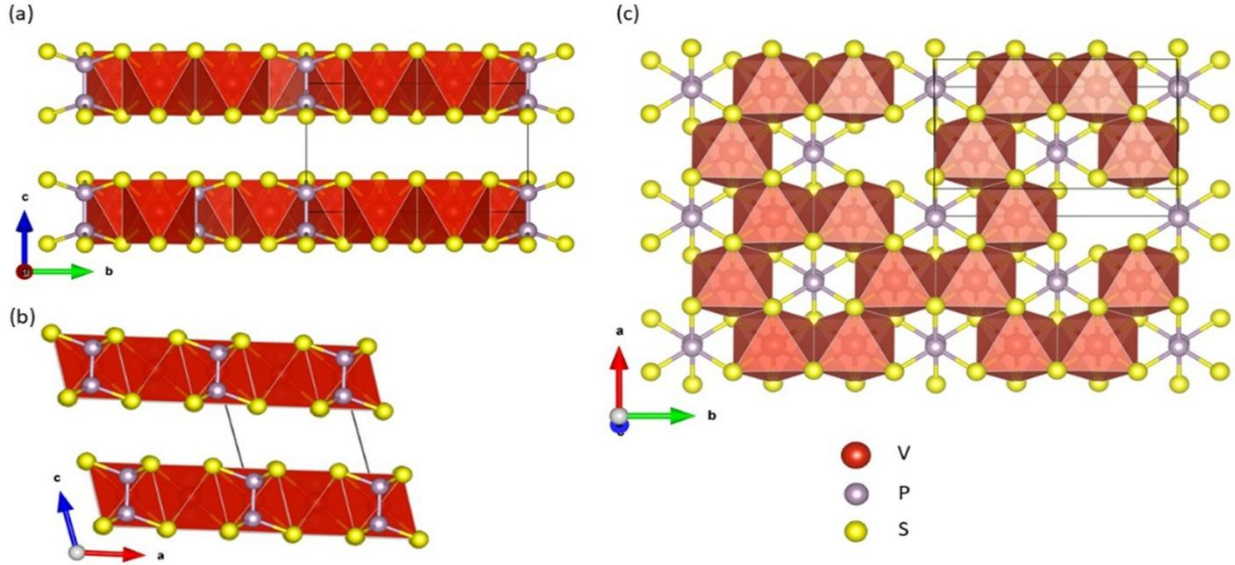

**Figure 3.4:** Schematic of the crystal structure of $V_{0.85}PS_3$.

|  | pXRD |
|---|---|
| *Composition* | $V_{0.85}PS_3$ |
| *Space group* | C2/m (No. 12) |
| *Wavelength (Å)* | 1.540560 |
| *2θ range (°)* | 10 – 115 |
| *Step Size (°)* | 0.015 |
| *Temperature (K)* | 293 |
| *a* (Å) | 5.8589(3) |
| *b (Å* | 10.1497(4) |
| *c* (Å) | 6.6450(7) |
| *β (°)* | 106.906(5) |
| *V* (Å$^3$) | 378.08(4) |
| *Goodness-Of-Fit* | 32.9 |
| *Bragg R-factor* | 17.8 |
| *RF-factor* | 30.7 |

**Table 3.1:** Structural parameters and residual factors of Rietveld refinement.

## 3A.3.2 Resistivity and Magnetization Measurements

Magnetic susceptibility increases slowly as the temperature is decreased from 300K and shows a sharp decrease around 60K. In the paramagnetic phase, it shows no deviation, but once the spin-solid phase is reached the magnetic susceptibility starts to show anisotropic behavior, which suggests that it is comparatively easier to magnetize the sample in *ab* plane than along *c*-axis. Interestingly below ~ 50K, susceptibility again starts increasing, which is very unlikely in the case of conventional antiferromagnets where net magnetization goes to zero sharply below Néel temperature. For Fe, Mn, and Ni, $\chi_{mol}$ *vs T* showed the behavior of a typical antiferromagnetic (AFM) system i.e., $\chi_{mol}^{\parallel}$ parallel to the magnetization axis shows a sharp drop below $T_N$, whereas $\chi_{mol}^{\perp}$ perpendicular to the magnetic axis remains nearly constant with a slight increase owing to the presence of spin waves [182,184,213]. Surprisingly in the case of $V_{1-x}PS_3$ both $\chi_{mol}^{\parallel}$ and $\chi_{mol}^{\perp}$ shows similar (increases with decrease in temperature) behavior below $T_N$, unlike a typical AFM system. Such a complex magnetic behavior could be a reflection of a quantum spin disordered state. Increase in $\chi_{mol}^{\parallel}$ and $\chi_{mol}^{\perp}$ below $T_N$ reflect the dominance of only short-range ordering. This can be a signal of a proximate QSL state or maybe a remnant of the QSL state existing as fluctuation even within the spin solid phase. We note that a similar behavior of $\chi_{mol}$ below $T_N$ was also reported for $Cu_2Te_2O_5Br_2$ [186,187] which could be understood by the quantum critical transition between the AFM state to QSL state.

We also extracted the net effective moments by fitting the inverse magnetic susceptibility data using modified Curie-Weiss law in the temperature range of 200-300K with linear extrapolation below 200K and the extracted value of effective magnetic moment is $\mu_{eff}^{exp.} = 3.61\mu_B$ and $\mu_{eff}^{exp.} = 3.68\mu_B$ for in-plane (H parallel to *ab* plane) and out-of-plane (H parallel to *c* axis) measurements, respectively. The theoretical value of effective magnetic moment for $V^{3+}$ (S=1) is $\mu_{S=1} = 2.83\mu_B$ ; and for $V^{2+}$ (S=3/2) is $\mu_{S=\frac{3}{2}} = 3.87\mu_B$ and for average spin $\mu_{S=1.32} = 3.50\mu_B$. We found the experimental value of $\mu_{eff}^{exp.} = 3.61\mu_B$ which is within ~ 4% error of theoretically estimated value using average spin.

### 3A.3.3 Temperature-dependent Phonon Analysis

In the stoichiometric structure of VPS₃, the factor group analysis predicts a total of 30 non-degenerate modes, $\Gamma = 8A_g + 7B_g + 6A_u + 9B_u$, within the irreducible representation at the gamma point, out of which 15, $\Gamma_{Raman} = 8A_g + 7B_g$, are Raman active with symmetric $A_g$ and antisymmetric $B_g$ lattice vibrations; and 15, $\Gamma_{infrared} = 6A_u + 9B_u$, are of infrared in nature, details are summarized in **table 3.2**. **Figure 3.5** shows evolution of the Raman spectra with temperature. We found 15 modes which is consistent with group theory prediction as well. We have fitted the Raman spectra at different temperatures with Lorentzian function and extracted corresponding phonon self-energy parameters.

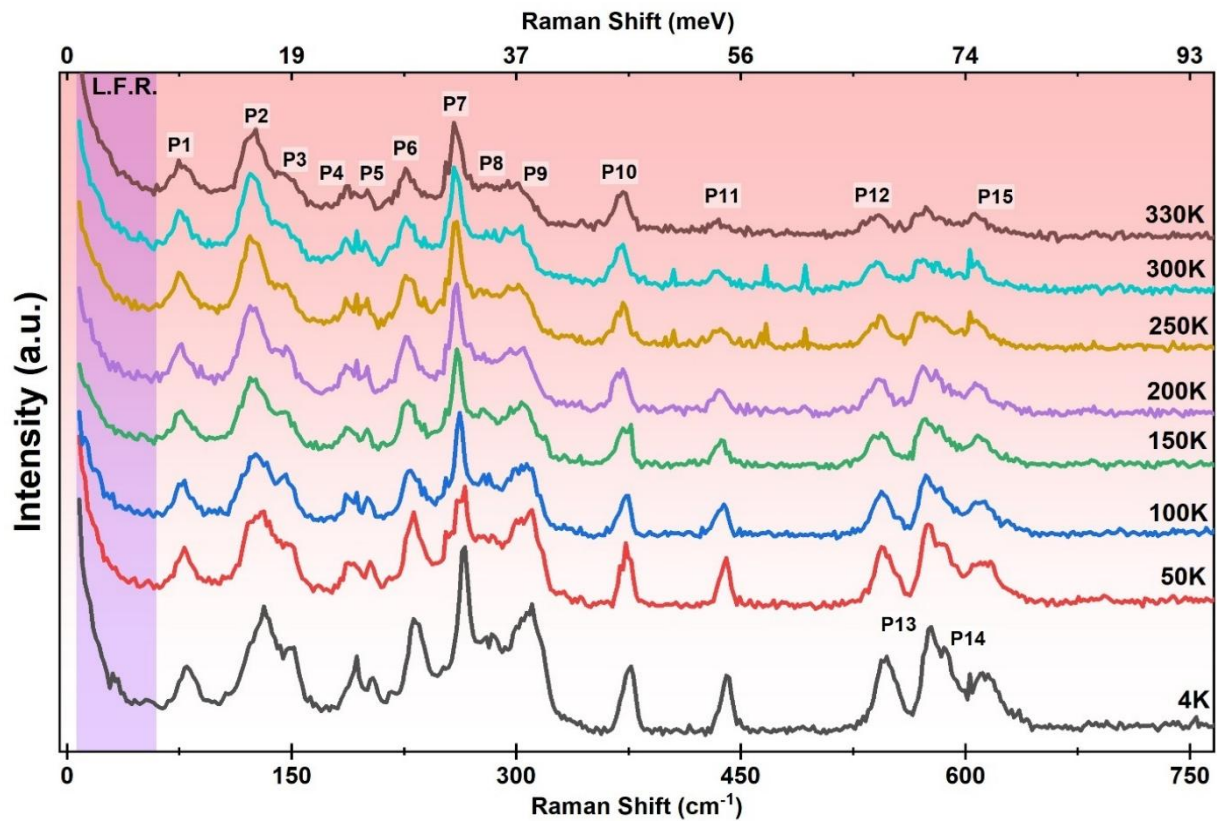

**Figure 3.5:** Raw Spectra of bulk $V_{0.85}PS_3$ along with peak labels at different temperatures within the spectral range of 5-760 cm⁻¹. Low-Frequency Region (LFR) is shaded in purple.

| Atoms | Wyckoff site | Γ-point mode decomposition | Raman Tensor |
|:---:|:---:|:---:|:---:|
| V | 4g | $A_g + A_u + 2B_g + 2B_u$ | $R_{A_g} = \begin{pmatrix} a & 0 & d \\ 0 & b & 0 \\ d & 0 & c \end{pmatrix}$ |
| P | 4i | $2A_g + A_u + B_g + 2B_u$ | |
| S | 4i | $2A_g + A_u + B_g + 2B_u$ | $R_{B_g} = \begin{pmatrix} 0 & e & 0 \\ e & 0 & f \\ 0 & f & 0 \end{pmatrix}$ |
| S | 8j | $3A_g + 3A_u + 3B_g + 3B_u$ | |
| | $\Gamma_{Raman} = 8A_g + 7B_g$ | $\Gamma_{Infrared} = 6A_u + 9B_u$ | |

**Table 3.2:** Wyckoff positions of different atoms in conventional unit cell and irreducible representations of the phonon modes of monoclinic (C2/m [C$_{2h}$] V$_{1-x}$PS$_3$) at the gamma point. R$_{A_g}$ and R$_{B_g}$ are the Raman tensors for A$_g$ and B$_g$ phonon modes.

The effect of the thermal part of anharmonicity can be visualized in a temperature-dependent variation of phonon frequencies and linewidth in a three-phonon process using the following functional forms [214]:

$$\omega(T) = \omega_o + A\left(1 + \frac{2}{e^x - 1}\right) \qquad \text{-- (3.1)}$$

$$\Gamma(T) = \Gamma_o + C\left(1 + \frac{2}{e^x - 1}\right) \qquad \text{-- (3.2)}$$

respectively, here $\omega_o$ and $\Gamma_o$ are frequency and line width at absolute zero; $x = \dfrac{\hbar \omega_o}{2k_B T}$, $A$ and $C$ are the self-energy constants [215]. Three phonon contributions is fitted and shown by a thick red curve in **figure 3.6** in a temperature range of 60K to 200K for temperature-dependent frequency and linewidth of P1 - P7, P9-P15 modes; and the estimated deviation from cubic anharmonic model below 60K is indicated by an extrapolated dashed red curve. This extrapolated curve below 60K is based on the constant parameter obtained by fitting in the range of 60-200K. We obtained a negative value of 'A' for all the modes which implies phonon display blueshift with decreasing temperature, which is considered as normal behavior when fitted using cubic anharmonic model, derived parameters are summed up in **table 3.3**. We clearly spot the deviation from the three-phonon process as phonon modes blueshifts below the transition temperature, which can be attributed to the interaction of magnetic and lattice degree of freedom. Interestingly, above 200K we observe temperature-independent behavior for most of the phonon modes. We observed that beyond which it loses intensity and temperature dependence considerably. In fact, it peaks around P6-P7 modes.

**(a)**

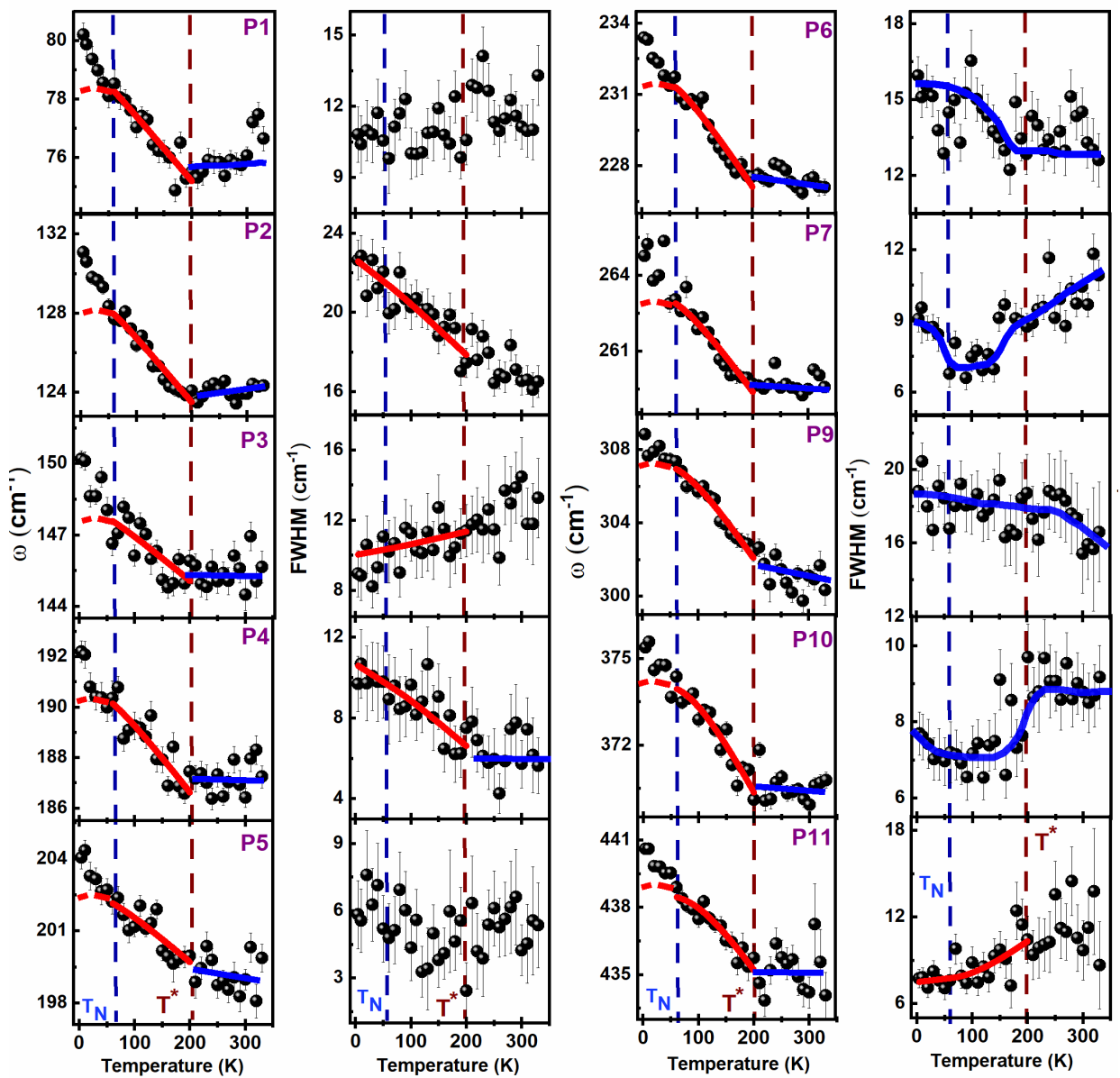

**(b)**

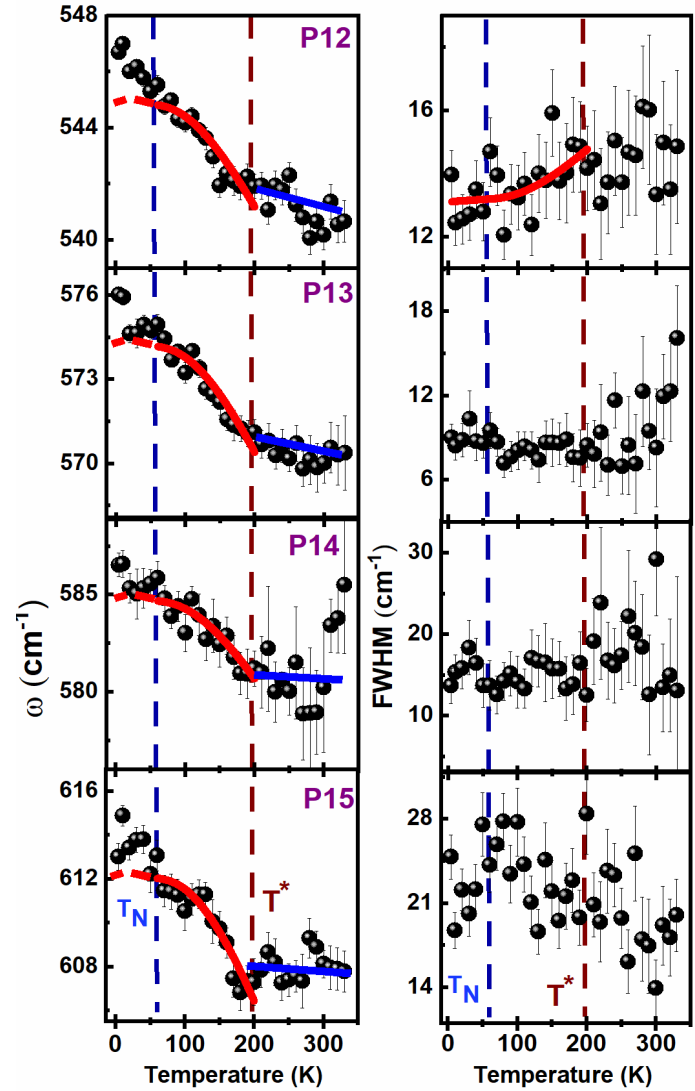

**Figure 3.6:** (a,b) Temperature-dependent evolution of features (frequency and FWHM) of the phonon modes. Some of phonons were fitted with the anharmonic model as mentioned in the text indicated by solid red lines and solid blue lines are guides to the eye. Dotted vertical lines represent $T_N$ (~ 60K) and cross-over temperature $T^*$ (~ 200K).

P1-P9 phonon modes reside on that part of spectra where there is a significant effect of the background continuum. In **figure 3.7** we have shown a variation of phonon self-energy parameter 'A' with increasing phonon frequencies. A slow but gradual linear increase in 'A' is observed for P1-P9 and it increases drastically for P10-P15.

For variation in FWHM, the self-energy constant 'C' is expected to be positive as the phonon population decreases with a decrease in temperature which increases phonon lifetime and we observe it for modes the P3, P11, and P12, whereas P2, P4, and P6 shows opposite behavior. Interestingly P13 and P14 below 200K remain almost constant till 4K which is anomalous behavior.

| Modes (Symmetry) | $\omega_o$ | A | $\Gamma_o$ | C |
|---|---|---|---|---|
| P1  (Bg) | 79.7 ± 0.4 | - 00.6 ± 0.1 | | |
| P2  (Bg) | 130.4 ± 0.4 | - 01.6 ± 0.1 | 23.3 ± 0.7 | - 1.2 ± 0.2 |
| P3  (Bg) | 149.1 ± 0.7 | - 01.1 ± 0.2 | 09.8 ± 1.1 | 0.4 ± 1.4 |
| P4  (Bg) | 192.5 ± 0.7 | - 01.9 ± 0.3 | 11.7 ± 1.1 | - 1.7 ± 0.5 |
| P5  (Bg) | 203.9 ± 0.5 | - 01.5 ± 0.2 | | |
| P6  (Ag) | 234.7 ± 0.4 | - 03.1 ± 0.2 | | |
| P7  (Ag) | 266.2 ± 0.4 | - 03.0 ± 0.2 | | |
| P9  (Ag) | 312.5 ± 0.5 | - 05.3 ± 0.4 | | |
| P10 (Ag) | 379.5 ± 0.6 | - 05.4 ± 0.5 | | |
| P11 (Bg) | 445.1 ± 0.8 | - 06.6 ± 0.7 | 02.5 ± 2.3 | 5.2 ± 2.0 |
| P12 (Ag) | 556.0 ± 1.5 | - 11.7 ± 1.4 | 08.0 ± 2.8 | 5.1 ± 2.6 |
| P13 (Bg) | 588.0 ± 1.6 | - 13.9 ± 1.5 | | |
| P14 (Ag) | 600.1 ± 2.5 | - 15.4 ± 2.3 | | |
| P15 (Ag) | 637.1 ± 3.1 | - 25.0 ± 2.9 | | |

**Table 3.3:** Anharmonic fitting parameters of the phonon modes and symmetry assignment.

P2 and P4 show significant increase in FWHM below 200K. Linewidth of the mode P10 shows quite interesting behavior, below 200K it shows a sharp drop and remains nearly constant till ~60K, and below 60K it starts increasing. Similarly, the linewidth of mode P7 shows a drop around 200K and then remains constant till ~ 60K, and at lower temperatures, it starts increasing. Different extent of variation of FWHM for different phonon modes suggests that underlying magnetic degree of freedom are interacting in different fashion with different energy phonon modes.

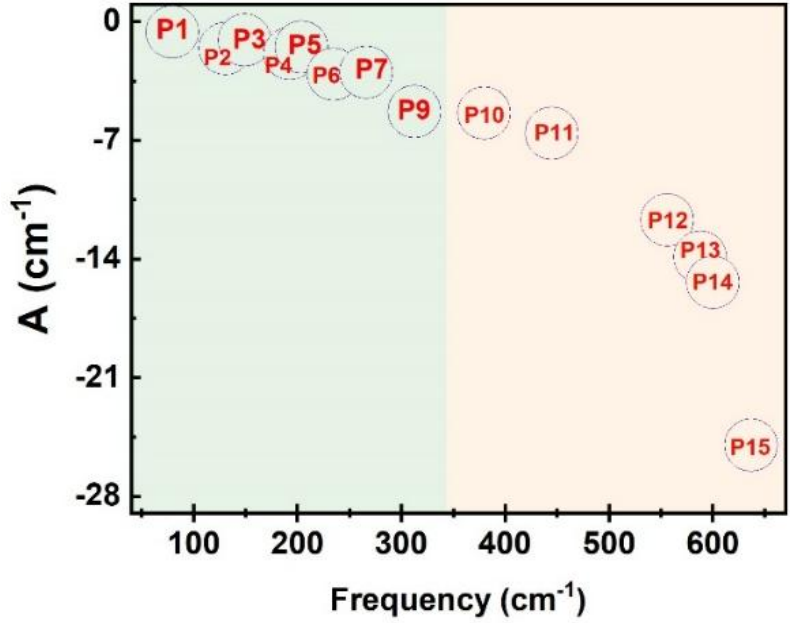

**Figure 3.7**: Variation of the phonon self-energy parameter 'A' with phonon frequencies.

### 3A.3.4 Spin-Phonon Coupling

To further understand the coupling of the lattice degree of freedom with the underlying magnetic degree of freedom, we invoked spin-phonon coupling and tried to understand the phonon's behavior at low temperatures. As the anharmonic effect is more prominent at higher temperatures and at low temperatures, below $T_N \sim 60K$ in the spin-solid phase due to magnetic interaction another decay channel becomes active, which could be responsible for phonon renormalization. Interestingly, the change in the phonon modes frequency is quite large for some of the modes below $T_N$, for e.g., the P1 mode undergoes 3% change in frequency, P2 and P3 $\sim 2\%$.

We fitted the frequency variation with temperature below $T_N$ for different modes (as shown in **figure 3.8**) using the following relation of spin-phonon coupling [216],

$$\Delta\omega \approx \omega_{sp} - \omega_o^{ph}(T) = -\lambda * S^2 * \phi(T) \qquad \text{--(3.3)}$$

where $S$ is spin on the magnetic ion and $\phi(T)$ is the order parameter given as

$\phi(T) = 1 - \left(\dfrac{T}{T_N^*}\right)^\gamma$ ; $\gamma$ is a critical exponent. We have kept $T_N^*$ as a variable and the obtained

value is close to the value obtained from magnetic measurements. Due to the presence of mixed valency, the value of 'S' is taken as a weighted average of spin on $V^{2+}$ and $V^{3+}$: $S = \frac{0.30*1+0.55*1.5}{0.30+0.55} = 1.32$; derived parameters are tabulated in **table 3.4**.

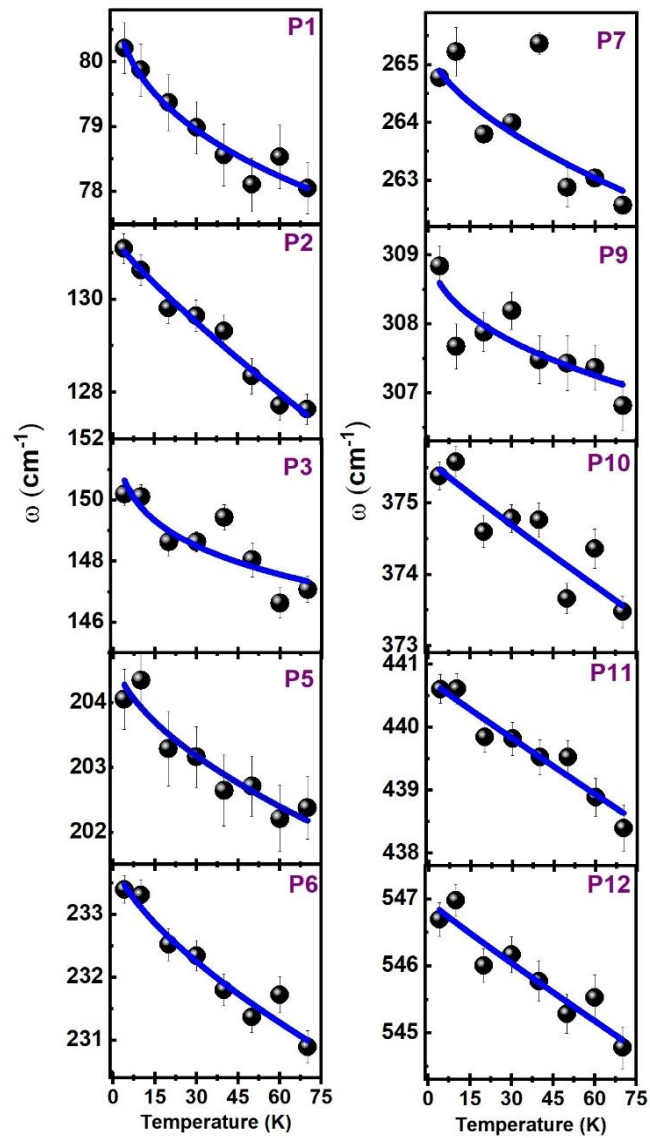

**Figure 3.8:** Spin-phonon coupling fitted with function (solid-blue curve) mentioned in text in spin-solid phase.

| | Spin-Phonon Coupling | | |
|---|---|---|---|
| | $\lambda$ | $\Upsilon$ | $T_N^*$ |
| **P1:** | -2.0 ± 0.9 | 0.3 ± 0.3 | 69.8 ± 08.8 |
| **P2:** | -2.1 ± 0.2 | 0.9 ± 0.2 | 67.0 ± 03.5 |
| **P3:** | -1.8 ± 0.5 | 1.0 ± 0.7 | 65.8 ± 09.1 |
| **P5:** | -1.4 ± 0.6 | 0.5 ± 0.4 | 60.6 ± 07.4 |
| **P6:** | -1.7 ± 0.4 | 0.6 ± 0.3 | 75.8 ± 09.3 |
| **P7:** | -1.3 ± 0.3 | 1.9 ± 1.8 | 68.7 ± 10.6 |
| **P9:** | -1.3 ± 1.2 | 0.4 ± 0.7 | 93.9 ± 49.7 |
| **P10:** | -1.2 ± 0.3 | 0.9 ± 0.7 | 76.8 ± 17.0 |
| **P11:** | -1.3 ± 0.1 | 1.1 ± 0.4 | 74.1 ± 06.3 |
| **P12:** | -1.2 ± 0.2 | 0.9 ± 0.5 | 75.3 ± 10.6 |

**Table 3.4**: Spin-Phonon coupling fitting parameters for some of the prominent phonon modes.

### 3A.3.5 Polarization Dependent Study

To understand the symmetry of the phonon modes. We employed angle-resolved polarized Raman scattering experiment. We performed our experiment by varying the incident light polarization direction $20^o$ from $0^o$ to $360^o$ using a half wave retarder or $\lambda/2$ plate, whereas the analyzer has been kept fixed. The intensity of the inelastically scattered light can be written as:

$$I_{Raman} \propto \left| \hat{e}_s^t.R.\hat{e}_i \right|^2 \qquad \text{--- (3.4)}$$

where '$t$' symbolizes transpose, $'\hat{e}_i'$ and $'\hat{e}_s'$ are the unit vectors of incident and scattered light polarization direction. '$R$' represents the Raman tensor of the respective phonon mode [217-219] .In the matrix form, incident and scattered light polarization direction unit vector can be decomposed as: $\hat{e}_i = \left[ \cos(\alpha+\beta) \ \sin(\alpha+\beta) \ 0 \right]$ ; $\hat{e}_s = \left[ \cos(\alpha) \ \sin(\alpha) \ 0 \right]$, where '$\beta$' is the relative angle between $'\hat{e}_i'$ and $'\hat{e}_s'$ and '$\alpha$' is an angle of scattered light from the x-axis, when polarization unit vectors are projected in x ($a$-axis) - y ($b$-axis) plane as shown in **figure 2.10** of Chapter 2.2.1.1. The angle dependency of intensities of $A_g$ and $B_g$ modes using Raman tensor mentioned in **table 3.2** can be written as:

$$I_{A_g} = \left| a \cos(\alpha) \cos(\alpha+\beta) + b \sin(\alpha) \sin(\alpha+\beta) \right|^2 \qquad \text{-- (3.5)}$$

$$I_{B_g} = \left| e \cos(\alpha) \sin(\alpha+\beta) + e \sin(\alpha) \cos(\alpha+\beta) \right|^2 \qquad \text{--(3.6)}$$

which are expected to have a phase difference of $\pi/2$ in intensity variation. Here $\alpha$ is an arbitrary angle from the $a$-axis and is kept constant. Therefore, without any loss of generality it is taken as zero, giving rise to the expression for the Raman intensity as $I_{A_g} = \left| a \cos(\beta) \right|^2$ and $I_{B_g} = \left| e \sin(\beta) \right|^2$. Polarization dependence of the phonon mode's intensity and evolution of the raw spectrum is shown in **figures 3.9 and 3.10,** respectively.

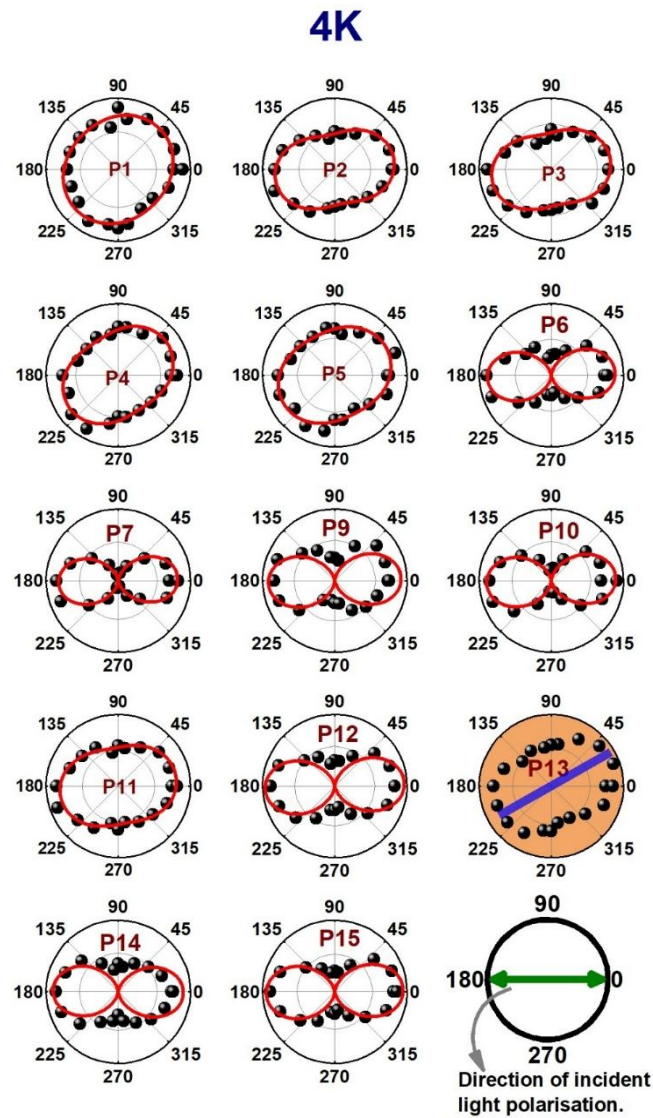

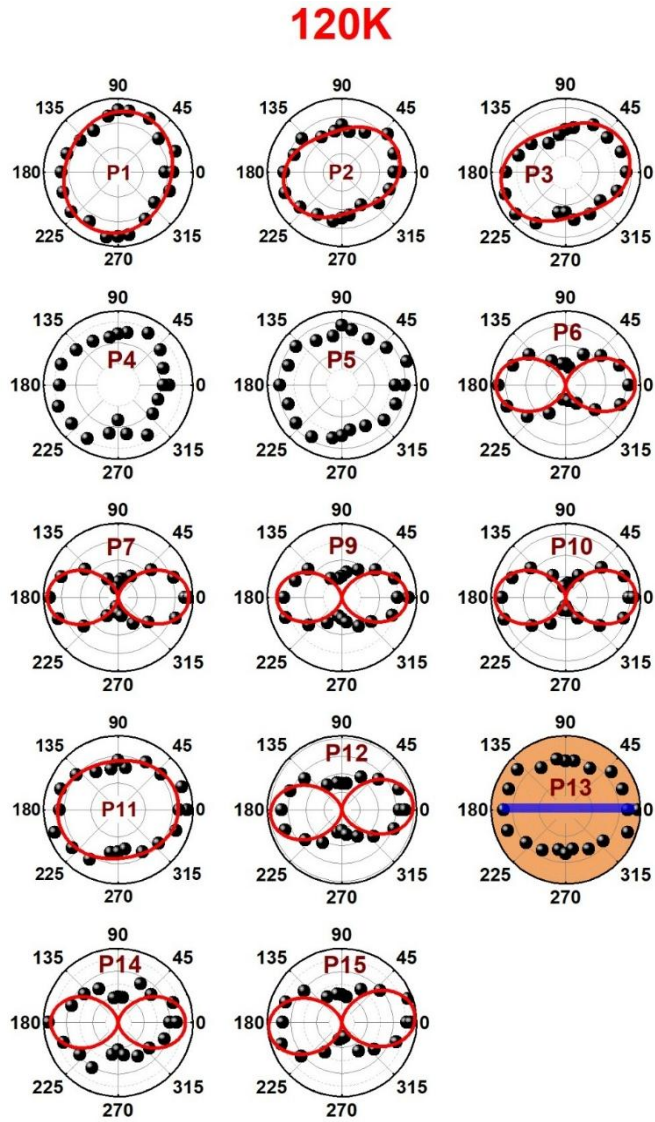

**Figure 3.9:** (Top and bottom panel) Polarization dependent intensity of phonon modes at 4K and 120K.

The high energy modes P6-P10, P12, P14-P15 shows typical of an $A_g$ symmetry mode intensity consistent with group theoretical predictions i.e., two maxima at zero and $\pi$, the red lines are the fitted curve using $I_{A_g} = \left| a \cos(\beta) \right|^2$ and the fitting is quite good. However, the modes P1-P5, P11 and P13 have elongated dumble shape with maxima around $\pi/4$ and $3\pi/4$ suggesting $B_g$ symmetry. The tentative symmetry assignment for the observed phonon modes is given in **table 3.5**.

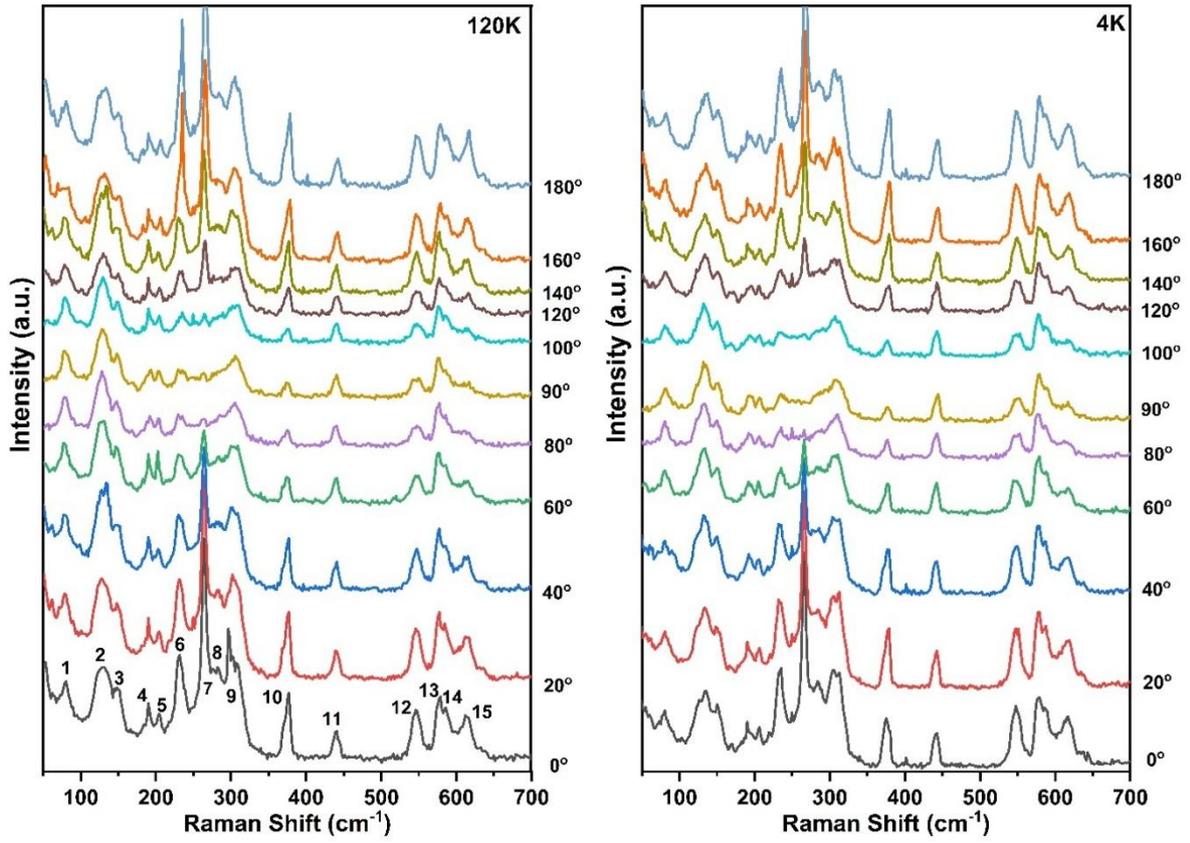

**Figure 3.10:** Polarization dependent evolution of the spectrum in the low temperature magnetic phase (4K) and at high temperature non-magnetic phase (120K).

| Phonon Mode's Symmetry assignment | | | |
|---|---|---|---|
| **Mode** | **Symmetry** | **Mode** | **Symmetry** |
| **P1** | $B_g$ | **P9** | $A_g$ |
| **P2** | $B_g$ | **P10** | $A_g$ |
| **P3** | $B_g$ | **P11** | $B_g$ |
| **P4** | $B_g$ | **P12** | $A_g$ |
| **P5** | $B_g$ | **P13** | $B_g$ |
| **P6** | $A_g$ | **P14** | $A_g$ |
| **P7** | $A_g$ | **P15** | $A_g$ |

**Table 3.5**: Phonon modes label and their symmetry.

We did polarization dependent measurements below and above long-range magnetic ordering i.e. at 4K and 120K. Quite interestingly and fascinating observation is that for some of the modes the major axis, the axis along which we see maxima in the intensity as a function of angle, is rotated in the low-temperature magnetic phase (4K) as compared to the non-magnetic phase (120K). For e.g., P13 modes major axis at 120K (above $T_N$) is at ~ 0 and is shifted to ~ $\pi/4$ at 4K (below $T_N$). Modes P4 and P5 are quasi-isotropic at 120K and the major axis at 4K is at ~ $\pi/4$ . For mode P1 major axis at 120K is at ~ $\pi/2$ and is shifted to ~ $\pi/4$ at 4K. Such a rotation of the major axis suggests the possibility of the tunability of the scattered light in this quasi-2D magnet via symmetry control. As one enters the spin-solid phase both time reversal and spin rotational symmetries are broken and these broken symmetries do have an impact on the underlying spin degrees of freedom possibly resulting in magneto-optical Raman effect, reflected via rotating the scattered light by ~ 30-45 degrees. Our results highlight the possibility of controlling the quantum pathways of the inelastically scattered light in these 2D magnetic systems and this control may provide a way in the future to manipulate these pathways for use in quantum technology.

### 3A.3.6 Temperature Evolution of the Broad Magnetic Continuum

In this part, we focus on the magnetic continuum observed, which is also the central theme of this chapter. Magnetic Raman scattering gives rise to a broad continuum originating from underlying dynamical spin fluctuations and may be used to gauge the fractionalization of quantum spins expected for proximate spin liquid candidates [57,58,220-225]. To investigate the possible emergence of the fermionic excitations, we carried out a detailed analysis of the temperature evolution of the observed broad magnetic continuum in the Raman spectra. The fractional spin excitations play a key role in dictating the temperature evolution of the background continuum of the Raman spectra because their occupation is determined by the Fermi distribution function.

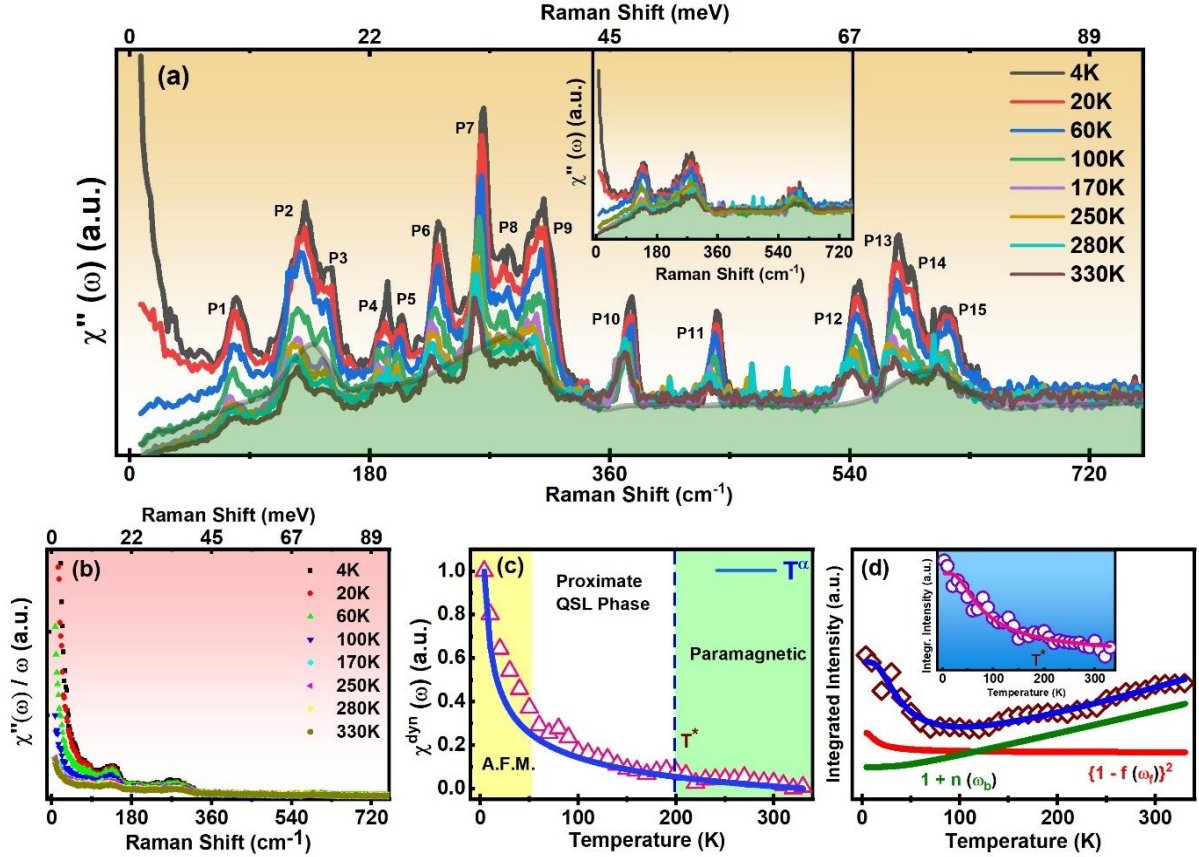

**Figure 3.11: (a)** Temperature evolution of the Raman response $\chi''(\omega,T)$ [measured raw Raman intensity/$1+n(\omega,T)$]. Inset shows the phonons subtracted Raman response. Labels P1-P15 represent phonon modes. **(b)** Temperature dependence of the phonons subtracted Raman conductivity $\chi''(\omega,T)/\omega$. **(c)** Temperature dependence of $\chi^{dyn}(T)$ obtained from Kramer - Kronig relation. The solid blue line is a power-law fit $\chi^{dyn} \sim T^\alpha$. Background-colored shading reflects different magnetic phases. **(d)** The main panel shows integrated raw spectra intensity in the energy range 1 to ~ 95 meV, the blue solid line is fitting by a combined bosonic and fermionic function, $[a + b\{1+ n (\omega_b,T)\}] + c \{1- f (\omega_f,T)\}^2]$. Solid green and red line show temperature dependence of bosonic and fermionic function. Inset shows the magnetic contribution to the Bose- corrected integrated intensity and $T^*$ (~ 200K) represents the temperature where spin fractionalization starts building up. The pink solid line represents fitting by the two-fermion scattering function $a+b\left[1-f(\omega,T)\right]^2$; where $f(\omega,T)=1/[1+e^{\hbar\omega_f/k_BT}]$ is the Fermi distribution function.

First, we focus on the temperature evolution of the integrated intensity of the background continuum to determine the fractionalized excitations energy scale [224], where the raw Raman intensity $I(\omega)$ is integrated over a range of 1.0 meV - 95 meV as: $I = \int_{\omega_{min}}^{\omega_{max}} I(\omega,T)\, d\omega$. **Figure 3.11(d)** shows the temperature evolution of the integrated intensity of the background continuum. As can be seen from **figure 3.11(d)**, the integrated intensity of the background continuum shows a non-monotonic temperature dependence, at high temperature the intensity variation is mainly dominated by a conventional one-particle scattering corresponding to thermal Bose factor given as: $I(\omega,T) \propto [1+n(\omega,T)]$; where $[1+n(\omega,T)] = 1/[1-e^{-\hbar\omega_b/k_BT}]$. However, at low temperatures, a significant deviation from the conventional bosonic excitations is observed below ~150 K, and the intensity shows monotonic increase with decreasing temperature down to the lowest recorded temperature (4K). To understand the temperature dependence of the integrated intensity of the background continuum, we fitted with a function having a contribution from both Bosonic and two-fermion (related to the creation and annihilation of the pair of fermions, its functional form is given as: $a+b[1-f(\omega,T)]^2$, where $f(\omega,T) = 1/[1+e^{\hbar\omega_f/k_BT}]$ is Fermi distribution function with zero chemical potential [224]), excitations. The fitting outcome reveals that the temperature dependence of the integrated intensity below ~ 150 K is mainly dominated by the fermionic excitations and the corresponding fractionalized energy scale for the fermions is $\omega_f = 10.3$ meV (~ 85 cm$^{-1}$). The temperature evolution of the integrated intensity of Bose- corrected spectra is shown in **figure 3.11(d)** inset, clearly indicating the significant enhancement of magnetic contribution below ~ 150-200 K, as fitting well with the two-fermion function, $a+b[1-f(\omega,T)]^2$. Our analysis suggests the signature of fractionalized fermionic excitations in the $V_{0.85}PS_3$, a key signature of the proximate QSL phase, in line with the other 2D honeycomb putative QSL candidates, as well as $Li_2IrO_3$ [57,58]. In a conventional long-range ordered magnetic system, the background continuum develops into relatively sharp modes (magnons with spin -*1*) below $T_N$ [226]. We note that in other members of this family, for e.g. for FePS$_3$, additional low frequency (below $\sim 150$ cm$^{-1}$) sharp modes were observed below $T_N$ attributed to the magnetic excitations [227]. On the other hand, we did not observe any such sharp peaks below $T_N$, or development of the broad underlying continuum into sharp modes; additionally, we observed Fano line asymmetry and phonon anomalies in line with the theoretical suggestion for phonons in putative QSL candidates. Also, we observed that the quasi-elastic response increases with decreasing temperature, on the other hand in the case of conventional magnetic systems quasi-elastic response increases with increasing temperature due to thermal fluctuations.

Now, we will discuss the Raman response, $\chi^{"}(\omega, T)$, which shows the underlying dynamic collective excitations at a given temperature, where $\chi^{"}(\omega, T)$ is calculated by dividing raw Raman intensity with the Bose factor, $I(\omega, T) \propto \left[ 1 + n(\omega, T) \right] \chi^{"}(\omega, T)$. The Raman response, $\chi^{"}(\omega, T)$, is proportional to stokes Raman intensity given as: $I(\omega, T) = \int_0^\infty dt\, \mathrm{e}^{i\omega t} \left\langle R(t)\, R(0) \right\rangle \propto \left[ 1 + n(\omega, T) \right] \chi^{"}(\omega, T)$; where *R(t)* is the Raman operator and $\left[ 1 + n(\omega, T) \right]$ is the Bose thermal factor. **Figure 3.11(a)** shows the temperature evolution of the $\chi^{"}(\omega, T)$. We note that $\chi^{"}(\omega, T)$ is composed of the phononic excitations superimposed on a broad continuum extending up to $\sim 95$ meV. The detailed analysis of the self-energy parameters of all the observed phonon modes i.e., peak frequency and linewidth is given in the section 3A.3.3. An in-depth analysis of this broad continuum may provide further information about the underlying nature of the dynamical spin fluctuations via the dynamic Raman susceptibility ($\chi^{dyn}$). Interestingly the Raman response shows a significant increase in the spectral weight on lowering the temperature [see **figure 3.11(a)** and its inset], and quite surprisingly it continues to increase upon entering into the spin solid phase, unlike the conventional systems where it is expected to quench below $T_N$. This characteristic scattering feature is typical of the scattering from underlying quantum spin fluctuations. For further probing the evolution of this broad magnetic continuum and underlying quantum spin fluctuations we quantitatively evaluated $\chi^{dyn}$, as shown in **figure 3.11(c)**. $\chi^{dyn}$ at a given temperature is evaluated by integrating phonon-subtracted Raman conductivity, $\chi''(\omega)/\omega$, shown in **figure 3.11(b)**, and using Kramers - Kronig relation as:

$$\chi^{dyn} = \lim_{\omega \to 0} \chi\ (k=0,\omega) \equiv \frac{2}{\pi}\int_0^\Omega \frac{\chi''(\omega)}{\omega}d\omega \qquad \text{--(3.7)}$$

where $\Omega$ is the upper cutoff value of integrated frequency chosen as ~ 95 meV, where Raman conductivity shows no change with further increase in the frequency. With lowering temperature $\chi^{dyn}$ shows nearly temperature independent behavior down to ~200 K as expected in a pure paramagnetic phase, on further lowering the temperature it increases continuously till 4K. The relative change in the temperature range of 330K-200K increases only by ~19%, in 200K-60K increases by ~ 73%, and 126% increase in the 60K-4K range. In the quantum spin liquid phase, the Raman operator couples to the dispersing fractionalized quasi-particle excitation and reflects the two-Majorana fermion density of states [222]. Therefore, an increase in the $\chi^{dyn}$ below ~ 200 K reflects the enhancement of Majorana fermion density of states and marks the cross-over from a paramagnetic to the proximate spin liquid state. Remarkably, the temperature dependence of the phonon modes also showed anomalies around ~ 200 K, reflecting the strong coupling of fractionalized excitations with the lattice degrees of freedom (discussed in later section). For conventional antiferromagnets as the system attains ordered phase dynamical fluctuations should be quenched to zero, contrary here we observed a significant increase in dynamic Raman susceptibility hinting at strong enhancement of dynamic quantum fluctuations [186,225,228,229]. The diverging nature of $\chi^{dyn}(T)$ as T → 0K, clearly suggests the dominating nature of quantum fluctuations associated with the underlying collective excitations down to the lowest temperature. This is also consistent with recent theoretical understanding, where it was advocated, that dynamic correlations may have unique temperature dependence in systems with quantum spin liquid signatures and the fractionalization of the quantum spins contributes to dynamic spin fluctuations even in the high-temperature paramagnetic phase [230]. Therefore, naturally, the signature of spin fractionalization is expected to be visible in the dynamical measurable properties such as dynamic Raman susceptibility as observed here. It was also shown that in the low-temperature regime dynamical structure factor, related to spin-spin correlation function, shows a quasi-elastic response with lowering temperature and was suggested as evidence for fractionalization of spins.

Next, we focus on the very low-frequency region (LFR) i.e., 1-9 meV, where we observed the emergence of a strong quasi-elastic response at low temperature, see **figure 3.12 (a and b)**. We evaluated the dynamic Raman susceptibility, $\chi_{LFR}^{dyn}$ for this low energy range (see **figure 3.12 (c)**). The observed $\chi_{LFR}^{dyn}$ remains nearly constant till ~ 200 K and shows a monotonic increase with further decreasing the temperature down to 4K. We fitted both $\chi^{dyn}$ and $\chi_{LFR}^{dyn}$ using a power law as $\chi^{dyn} \sim T^{\alpha}$ ($\alpha = -0.34$ for $\chi^{dyn}$, see solid blue line in **figure 3.11 (c)**; and $\alpha = -0.67$ for $\chi_{LFR}^{dyn}$, see solid line in **figure 3.12 (c)**). Here we observed power law behavior of $\chi^{dyn}$ and $\chi_{LFR}^{dyn}$ much above $T_N$ unlike the conventional pure paramagnetic phase where it is expected to show saturation. The power law dependence of $\chi^{dyn}$ and $\chi_{LFR}^{dyn}$ even well above the long-range magnetic ordering temperature reflects the slowly decaying correlation inherent to the quantum spin liquid phase and triggers the fractionalization of spins into itinerant fermions around T* ~ 200 K [42,231]. We note that this anomalous temperature evolution of the background continuum along with phonon anomalies, discussed later, cannot be captured by the conventional long-range ordered magnetic scattering, rather, it reflects the presence of fractionalized excitations which are intimately linked with the quantum spin liquid phase, in line with the theoretical suggestions for a QSL state. It may be mentioned that the underlying continuum may have its origin in the two-magnon excitations, though a detailed magnetic field-dependent Raman measurement is required to further shed light on the nature of this continuum.

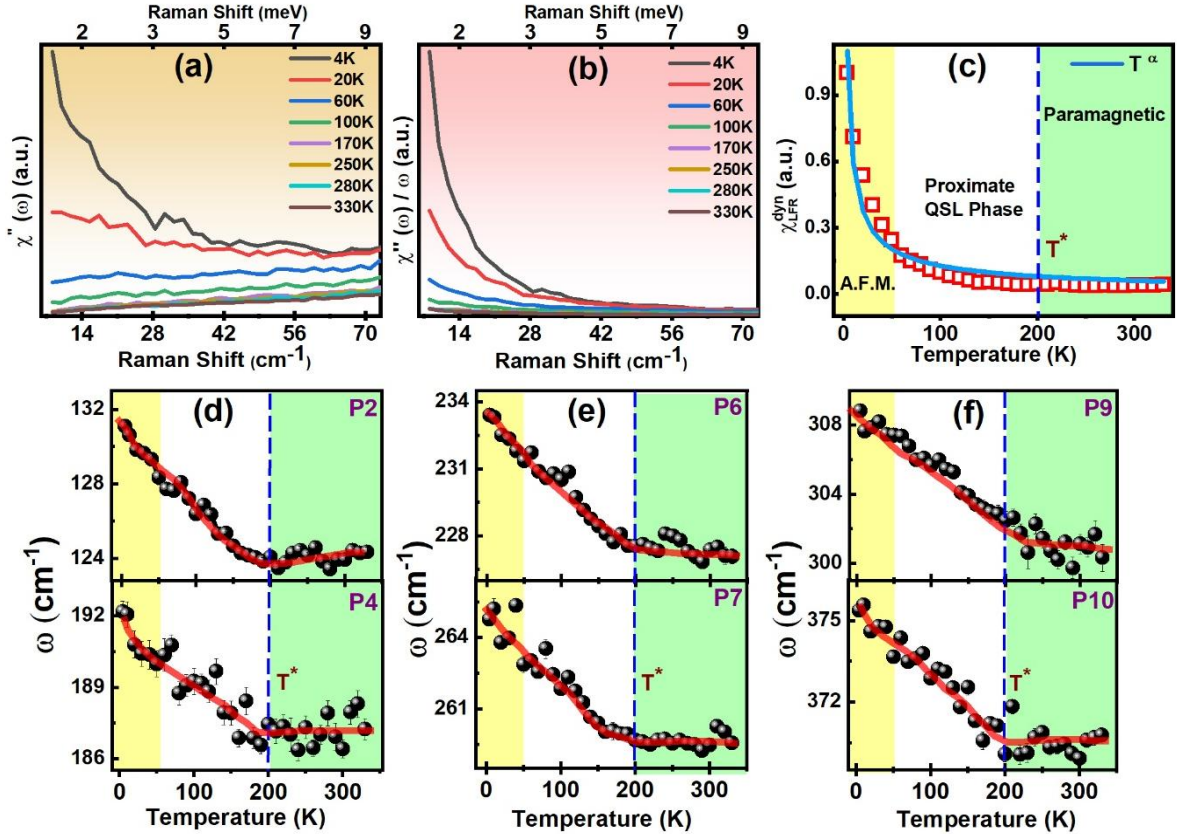

**Figure 3.12: (a), (b)** Shows temperature evolution of the Bose-corrected spectra i.e., Raman response, $\chi''(\omega,T)$, and Raman conductivity, $\chi''(\omega,T)/\omega$, in the low-frequency region LFR (1-9.0 meV), respectively. **(c)** Temperature dependence of $\chi_{LFR}^{dyn}(T)$ obtained from Kramer's Kronig relation. The solid blue line is a power-law fit $\chi_{LFR}^{dyn} \sim T^{\alpha}$. Background-colored shading reflects different magnetic phases. **(d, e, and f)** Mode's frequency evolution as a function of temperature for the modes P2, P4, P6, P7, P9 and P10. Red solid lines are a guide to the eye. $T^*(\sim 200K)$ represents the temperature where spin fractionalization starts building up.

We wish to note that in a recent report for the case of a putative QSL candidate RuCl$_3$ [194], QSL phase is predicted with long-range ordering at T$_N$ ~ 7K (Zigzag AFM state). It is shown that in the high-temperature paramagnetic phase, the quasi-elastic intensity of magnetic excitation has a broad continuum and the low-temperature AFM state is quite fragile with competition from FM correlation and QSL phase; in fact, the AFM state is advocated to be stabilized by Quantum fluctuations leaving QSL and FM states as proximate to the $|GS\rangle$ and at slightly higher temperature Kitaev QSL state becomes prominent. Furthermore, it was advocated that FM and QSL states proximate to AFM $|GS\rangle$ are essential to understanding the anomalous scattering continuum. Based on our observations and phonon anomalies (discussed in the next section) this broad magnetic continuum is attributed to the fractionalized excitations in this material with quasi 2D magnetic honeycomb lattice.

### 3A.3.7 Mode's Asymmetry and Anomalous Phonons

Interaction of the underlying magnetic continuum with the lattice degrees of freedom may reflect via the asymmetric nature of the phonon line shape, known as Fano-asymmetry, and may provide crucial information about the nature of underlying magnetic excitations responsible for the magnetic continuum. This asymmetry basically describes the interaction of a continuum with a discrete state (Raman active phonon modes here) and this effect has its origin in the spin-dependent electron polarizability involving both spin-photon/phonon coupling [232-234]. For the Kitaev spin liquid candidates, recently it was advocated that spin-phonon coupling renormalizes phonon propagators and generates Fano line shape resulting in observable effect of the Majorana fermions and the $Z_2$ gauge fluxes, a common denominator for a Kitaev quantum spin liquid $|GS\rangle$ [235,236]. The evolution of phonon modes asymmetry in putative QSL candidate materials seems ubiquitous [57,237-239] suggesting the intimate link between the QSL phase and phonon asymmetric line shape. Additionally, it was also shown that the lifetime of the phonons decreases with decreasing temperature i.e., an increase in linewidth with decreasing temperature, attributed to the decay of phonons into itinerant Majorana fermions [240]. This is opposite to the conventional behavior where phonon linewidth decreases with decreasing temperature owing to reduced phonon-phonon interactions.

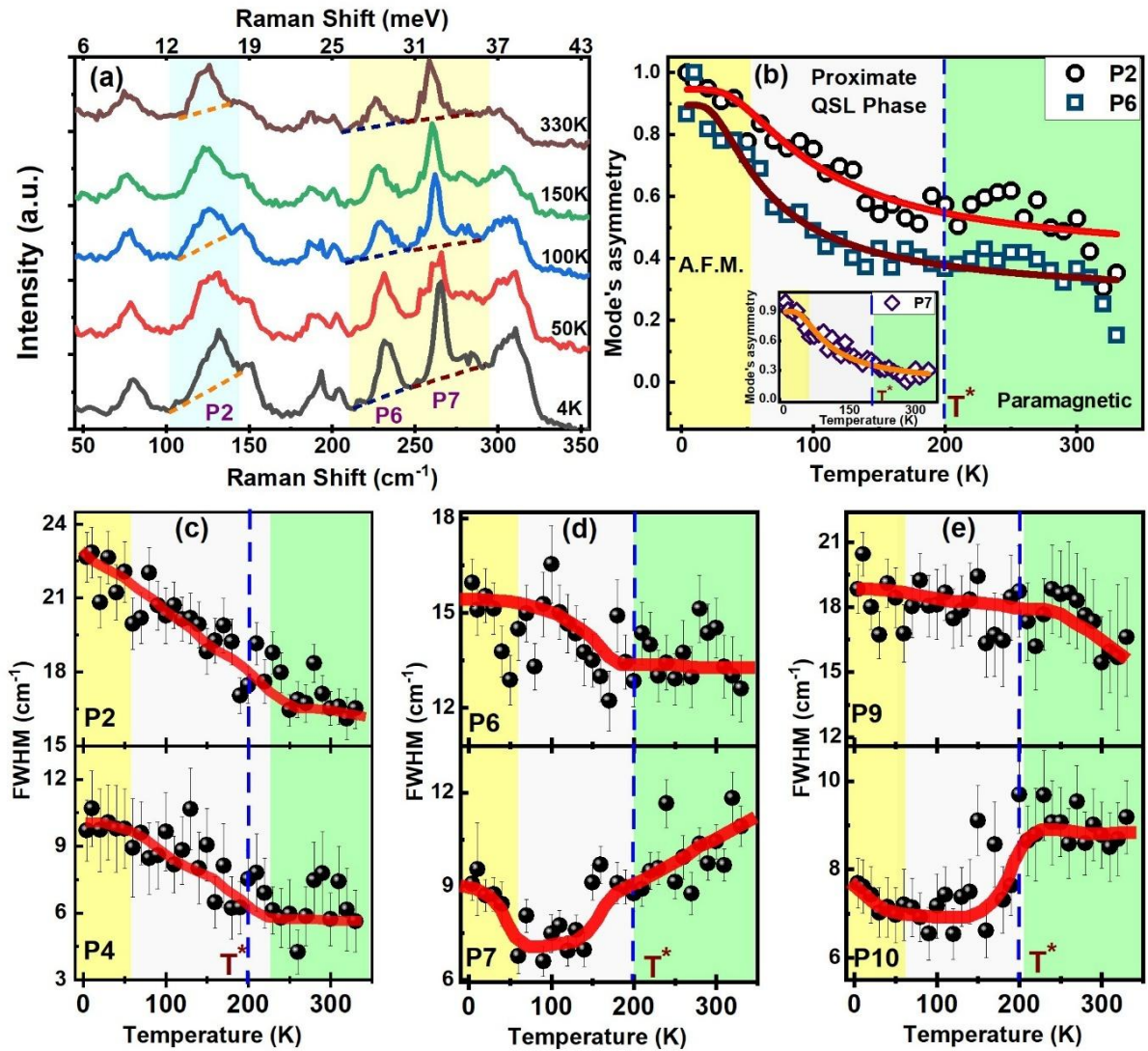

**Figure 3.13: (a)** Shows the raw Raman spectrum in the frequency range of ~ 45 – 360 cm⁻¹ showing the evolution of the asymmetry for the modes P2, P6, and P7. **(b)** Shows the evolution of the phonons modes P2, P6 and P7 (as inset) normalized asymmetry gauged via slope. Background color shading reflects different magnetic phases. **(c, d and e)** Shows the linewidth for the modes P2, P4, P6, P7, P9 and P10. Red solid lines are a guide to the eye. $T^*$ (~ 200K) represents the temperature where spin fractionalization starts building up.

The temperature-dependent Raman spectra in a frequency range where couple of modes (P2, P6, and P7) show strong asymmetric line evolution is shown in **figure 3.13 (a)**. It's very clear that these phonon modes gain asymmetric line shape with decreasing temperature. As expected, these modes are superposed on the broad underlying magnetic continuum in the region where the spectral weight of the continuum is dominating.

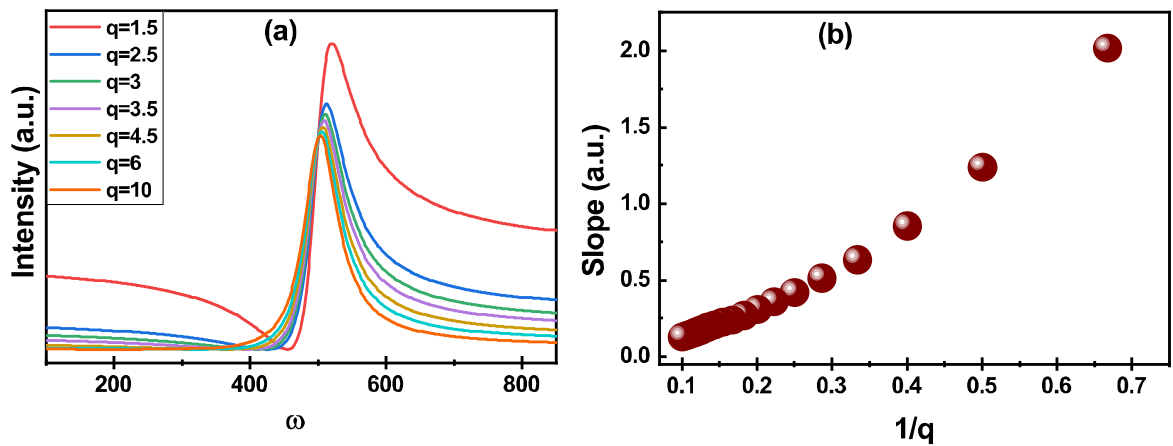

**Figure 3.14**: (a) Evolution of a mode for different $q$ parameter, evolution shows high asymmetry at low $q$ value and approaches Lorentzian functional form at high $q$ value. (b) Plot of slope v/s asymmetry parameter ($1/q$).

The asymmetric nature of these three phonon modes is gauged via the slope method. Here we have adopted a slope method because Fano function fitting resulted in a large error. The slope ($\Delta Y / \Delta X = \Delta Inten. / \Delta \omega$) for a mode is evaluated by keeping the x-axis range (frequency here) the same and the corresponding rise in the intensity as a function of temperature. The Fano function is defined as $F(\omega) = I_0 (q + \varepsilon)^2 / (1 + \varepsilon^2)$; where $\varepsilon = (\omega - \omega_0)/\Gamma$ and $1/q$ represents the asymmetry. The asymmetry parameter ($1/q$) characterizes the coupling strength of a phonon to the underlying continuum: a stronger coupling ($1/q \to \infty$) causes the peak to be more asymmetric and in the weak-coupling limit ($1/q \to 0$) the Fano line shape is reduced to a Lorentzian line shape. We generated this function by keeping $\omega_0$, $\Gamma$ and $I_0$ fixed for different values of $q$, please see **figure 3.14(a)**, and it's very clear from this figure that as $1/q$ increases asymmetry is increased. Then, we estimated the slope as defined above for this generated function and found the one-to-one correspondence between slope and asymmetry parameter $1/q$, please see **figure 3.14(b)**, the higher the slope more the asymmetry, and the lesser the slope lesser the asymmetry.

**Figure 3.13(b)** shows the normalized slope for these three modes **(P2, P6 and P7)**. Interestingly, the asymmetry gauged via slopes shows strong temperature dependence [see **figure 13.13(b)**]; it has a high value in the long-range-ordered phase at low temperature, above $T_N$ (∼60 K) it continuously decreases till ∼200 K, and thereafter it remains nearly constant up to 330 K, clearly suggesting the presence of active magnetic degrees of freedom far above $T_N$. A pronounced feature of this mode asymmetry is that it conjointly varies with dynamic Raman susceptibility on varying temperatures [discussed above; see **figure. 3.11 (c) and 3.12 (c)**], implying that its asymmetric line shape is also an indicator of spin fractionalization or emergence of spin liquid phase, and the increased value below 200 K ($T^*$- defined as a crossover temperature) may be translated to a growth of finite spin fractionalization. We also tried to fit the temperature evolution of the slopes with the two-fermion scattering form $a + b[1 - f(\omega, T)]^2$, (see **figure 3.13 (b)**) and the extracted fermionic energy scale is also found to be similar to that estimated from intensity fitting of the continuum background (see **figure 3.11 (d)**). It is quite interesting that the evolution of the mode's asymmetry for these phonon modes maps parallel to the thermal damping of the fermionic excitations. In materials with Kitaev QSL phase as $|GS\rangle$, spins are fractionalized into the Majorana fermions, as a result of this the underlying continuum emerging from spin fractionalization couples strongly with lattice degrees of freedom as evidenced here. The temperature evolution of these modes, i.e., phonon mode's line asymmetry, in line with the theoretical prediction suggests the fractionalization of spins.

For a normal phonon mode behavior, as the temperature is lowered then phonon peak energy is increased and linewidth decreases attributed to the anharmonic phonon-phonon interactions. Interestingly, a large number of modes showed a change in phonon frequency at the cross-over temperature $T^*$ ($\sim 200$K) signaling the effect of spin fractionalization on the phonon modes, see **figure 3.12 (d, e, and f)**. Startlingly, some of the modes showed anomalous evolution of the linewidth below $T^*$ i.e., linewidth increases with decreasing temperature. **Figure 3.13(c, d, and e)** shows the temperature dependence of the linewidth of the phonon modes which show anomalous behavior. All these modes show clear divergence from the normal behavior starting at the crossover temperature $T^*$, implying an additional decay channel, similar to the temperature scale associated with the phonon mode's line asymmetry and magnetic continuum reflected via dynamic Raman susceptibility. This is also consistent with the theoretical predictions; hence our observation clearly evidenced the emergence of fractionalised excitations in this quasi-2D magnetic system starting from the crossover temperature reflected via phonon anomalies and broad underlying magnetic continuum.

## 3A.4 Conclusion

In conclusion, we have performed in-depth inelastic light scattering (Raman) studies on $V_{1-x}PS_3$ single crystals. Where we focused on the background continuum showing distinct temperature dependence via dynamic Raman susceptibility and the phonons anomalies. Our results on background continuum and phonons self-energy parameters evince anomaly at a similar temperature range suggesting the cross-over from a normal paramagnetic phase to a state where spin fractionalization begins and marked the onset of proximate quantum spin liquid phase. Our studies evinced the signature of spin fractionalization in this quasi-2D magnetic honeycomb lattice system. In addition to the observation of a broad magnetic continuum and its anomalous temperature evolution, our results on the evolution of the mode's line asymmetry and phonon anomalies, in particular for the phonon modes lying on the underlying magnetic continuum, opens the possibility to experimentally identifying the theoretically predicted effects of fractionalized excitations of QSL phase in putative spin liquid candidates.

# Part-B: Dynamics of Phonons and Magnetic Continuum in Thin Flakes of $V_{1-x}PS_3$

## 3B.1 Introduction

Interestingly, understanding the nature of the QSL state in the putative 3D QSL candidates, e.g., $\alpha$-RuCl$_3$, Li$_2$IrO$_3$, Na$_2$IrO$_3$, as a function of layer dependence has not been explored much. It was advocated some time back that the signature of a QSL state may be observed in a single layer of 1T-TaS$_2$ [241] and subsequently, it was also reported for a few layers of other 2D systems [242,243]. The quantum spin liquid phase is expected to be most prominent in a single layer and is expected to diminish with increasing layers subject to the relative strength of interlayer coupling ($J_\perp$) with respect to the Kitaev coupling ($K$). As theoretically advocated only weak interlayer coupling in layered Kitaev material exhibits a Z$_2$-fractionalized spin liquid phase and it goes into the dimer paramagnet phase for large interlayer coupling [244,245]. Also, a gapped/gapless QSL phase has been suggested depending on the even/odd number of layers provided weak interlayer coupling [246]. Therefore, it is very pertinent to probe these potential QSL candidates as a function of layer thickness to shed light on the role of dimensionality in the nature of the QSL phase.

In this chapter, we have taken up such a study to probe the dynamics of underlying excitations as well as phonons on a few layers of $V_{1-x}PS_3$ system, a putative QSL candidate, using inelastic light scattering (Raman) measurements as a function of temperature and density functional theory-based calculation of the phonons. Our measurements as a function of layers revealed the signature of quantum spin liquid, which is more prominent in the thinnest layer ($\sim$ 8-9 layers), probably due to the lowered dimensionality where quantum effects become more prominent.

Raman scattering is an excellent probe to simultaneously determine the dynamics of lattice, spin, and charge degrees of freedom. It can provide signatures of one/two magnon excitations in ferro/antiferromagnets and fractionalized excitations, e.g., Majorana fermions (MFs) in Kitaev QSL candidates. Fractionalized excitations are one of the key features of the QSL ground state. In the Kitaev model, elementary spin S=1/2 fractionalizes into two classes of MFs i.e., itinerant ones and localized $Z_2$ fluxes. Itinerant MFs are efficiently detected via inelastic light scattering while the detection of localized $Z_2$ fluxes still remains a challenge [160]. QSL signatures manifest in the quasi-elastic response, background continuum, phonon-anomalies, Fano asymmetry, etc. [58,70,247]. $V_{0.85}PS_3$ provides an excellent opportunity to explore the material in the low dimensional regime especially down to a few layers. Hence, we probe four mechanically exfoliated $V_{0.85}PS_3$ flakes F1, F2, F3, and F4 of thickness 5 nm (~ 8-9 layers), 10.0 nm, 16.2 nm, and 21.7 nm, respectively; using inelastic light scattering i.e., Raman spectroscopy measurements. We even tried samples with lower thickness, but could not get the exfoliated sample thickness lower than ~5nm.

## 3B.2 Experimental and Computational Details

The thickness of a mono-layer (ML) is $\sim 3.25 \overset{o}{A}$ , bi-layer (BL) is $\sim 9.60 \overset{o}{A}$ and vdw-gap is $\sim 3.11 \overset{o}{A}$, which can be estimated using a VESTA-generated structure as shown in **figure 3.15 (a, b)**. Here, the magnetic vanadium atom forms a honeycomb structure and is octahedrally coordinated by S atoms. Temperature-dependent Raman spectroscopic measurements are performed under the configuration as mentioned in chapter 3A.2.3, on four different mechanically exfoliated flakes of $V_{0.85}PS_3$ i.e., 5 nm (F1), 10.0 nm (F2), 16.2 nm (F3) and 21.7 nm (F4) and are placed on a $SiO_2$ substrate. The AFM images for four flakes are shown in **figure 3.15 (c)**.

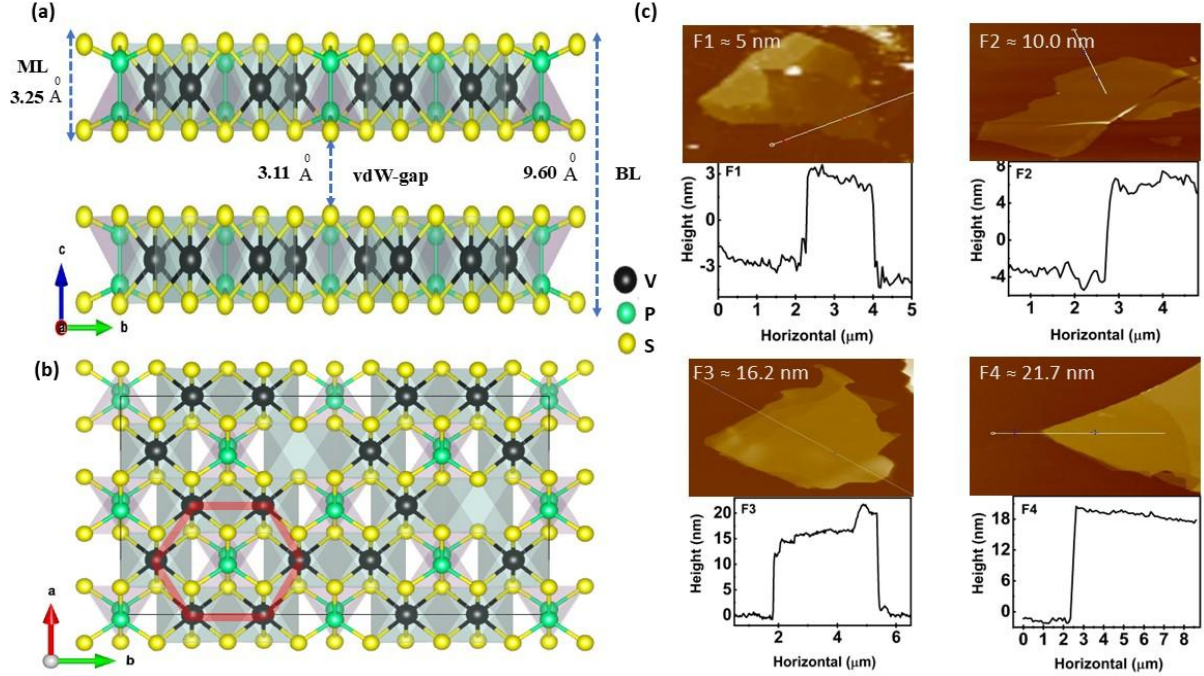

**Figure 3.15:** (a) Showing layered structure of $V_{0.85}PS_3$ with monolayer and bilayer thickness. (b) Shows the crystal structure of $V_{0.85}PS_3$ along a- and c-axis respectively. Red solid line indicates the honeycomb structure formed by magnetic ions. (c) Atomic Force microscope images of F1, F2, F3 and F4 flakes having thicknesses of 5 nm, 10.0 nm, 16.2 nm and 21.7 nm, respectively.

Structural optimization and Zone-centered phonon frequencies were calculated utilizing a plane-wave approach as implemented in the QUANTUM ESPRESSO package [174]. For simplicity the phonon calculations are done only for $VPS_3$ bulk. The linear response method within Density Functional Perturbation Theory (DFPT) is used to get the dynamical matrix. Semi-Local Unified Pseudopotentials Format (SL-UPF) with the Perdew-Burke-Ernzerhof (PBE) exchange-correlation functional is used. The kinetic energy and charge-density cutoff are 100 Ry and 800 Ry, respectively. The Monkhorst-pack scheme with 12 x 12 x 12 k-point dense mesh is used for the numerical integration of the Brillouin zone. A pictorial representation of atomic displacement of Raman active phonon modes at Γ-point is shown in **figure 3.16** and obtained phonon frequencies, and optical activity at Γ-point are listed in **table 3.6**.

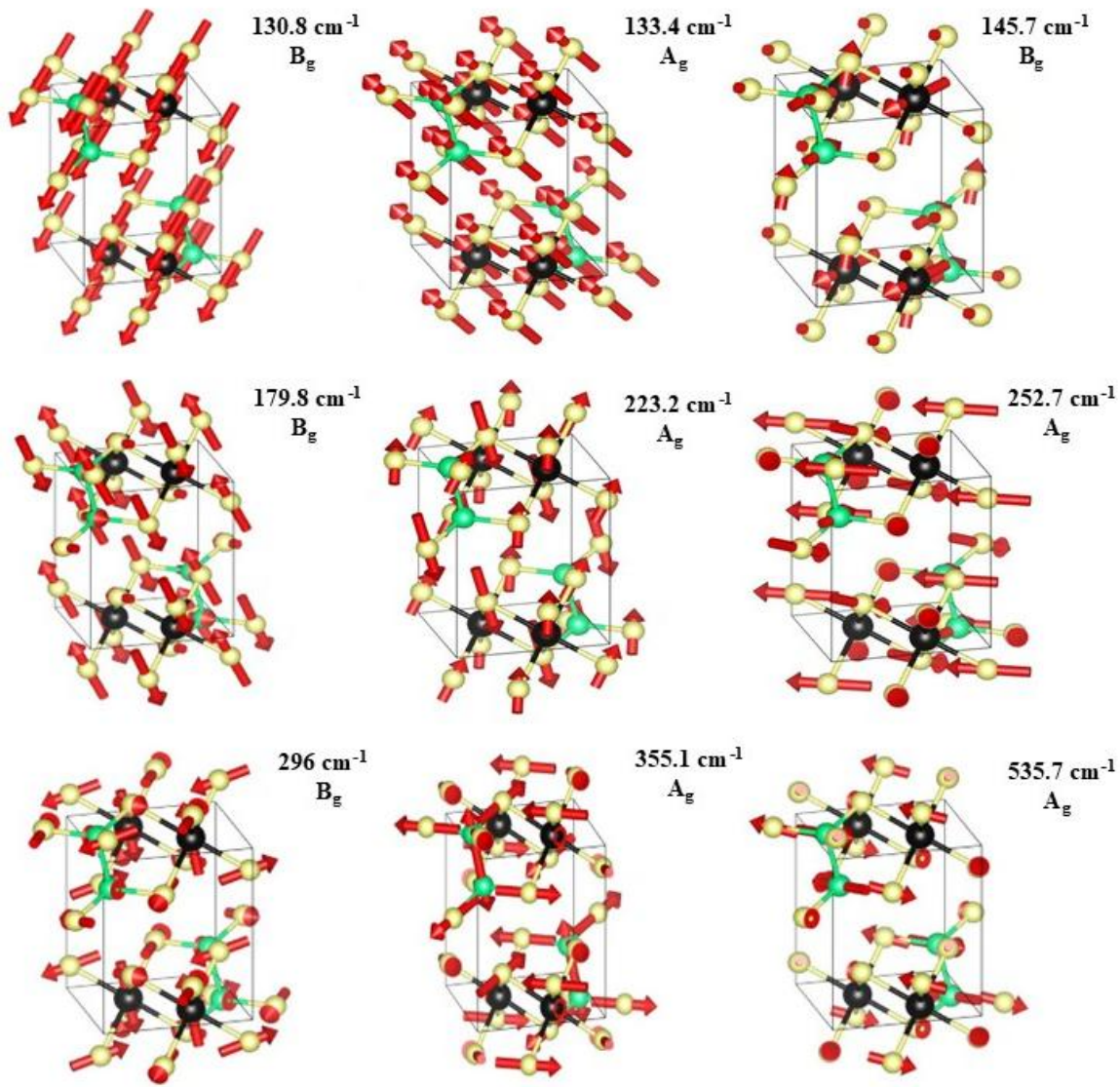

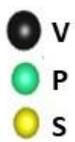

V
P
S

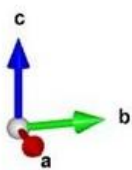

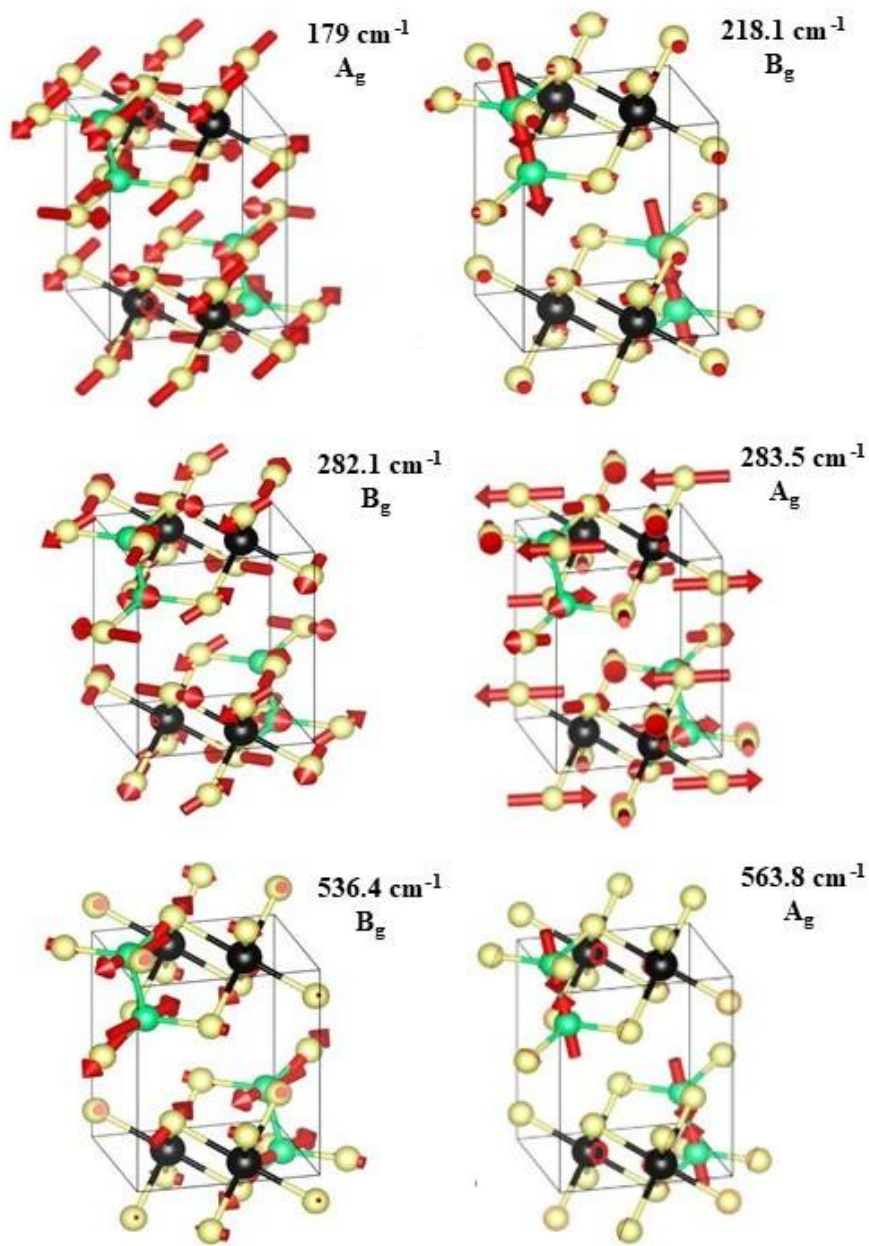

**Figure 3.16:** Pictorial representation of the atomic displacements of the Raman active phonons at Γ-point calculated using DFT-based calculation.

| Mode # | Frequency (cm⁻¹) | Symmetry | Activity | Mode # | Frequency (cm⁻¹) | Symmetry | Activity |
|---|---|---|---|---|---|---|---|
| 1 | -45.8 | $A_u$ | I | 16 | 238 | $A_u$ | I |
| 2 | 11.3 | $B_u$ | I | 17 | 252.7 | $A_g$ | R |
| 3 | 25.4 | $B_u$ | I | 18 | 253.2 | $A_u$ | I |
| 4 | 130.8 | $B_g$ | R | 19 | 253.6 | $B_u$ | I |
| 5 | 133.4 | $A_g$ | R | 20 | 282.1 | $B_g$ | R |
| 6 | 145.7 | $B_g$ | R | 21 | 283.5 | $A_g$ | R |
| 7 | 150.8 | $A_u$ | I | 22 | 290.2 | $B_u$ | I |
| 8 | 163.3 | $B_u$ | I | 23 | 296 | $B_g$ | R |
| 9 | 176.7 | $B_u$ | I | 24 | 355.1 | $A_g$ | R |
| 10 | 179 | $A_g$ | R | 25 | 424.1 | $B_u$ | I |
| 11 | 179.8 | $B_g$ | R | 26 | 535.7 | $A_g$ | R |
| 12 | 218.1 | $B_g$ | R | 27 | 536.4 | $B_g$ | R |
| 13 | 223.2 | $A_g$ | R | 28 | 552.4 | $B_u$ | I |
| 14 | 224.2 | $A_u$ | I | 29 | 554.5 | $A_u$ | I |
| 15 | 228.7 | $B_u$ | I | 30 | 563.8 | $A_g$ | R |

**Table 3.6:** DFT calculation of phonon frequency and mode symmetry for bulk VPS₃.

## 3B.3 Results and Discussion

### 3B.3.1 Temperature Evolution of the Raman Spectra and Phonon Dynamics

Lattice dynamics plays a crucial role in determining the underlying physics of the material and is affected by anharmonicity, electron-phonon coupling, spin-phonon coupling, etc. It can detect the presence of structural, magnetic transition, and other exotic quasi-particle excitations. We obtained a strong Raman signal for all the flakes which is in contrast to what has been recently reported by C. Liu *et al.* [248]. We observed a total of 14 Raman active modes for F1 flake labelled as P1-P14 in **figure 3.17(a)** in which one of the bulk modes at ~ 545 cm⁻ [1] is overshadowed by high-intensity silicon peak, which is expected due to reduced thickness as shown in **figure 3.17 (a, b)**. Other than the phonon modes we observed a very low-frequency quasi-elastic region (shaded in light pink) and underlying background magnetic continuum (shaded in light dark green) in **figure 3.17(a)**. This background magnetic continuum is dependent on the thickness of the material as evident from **figure 3.17(b)**.

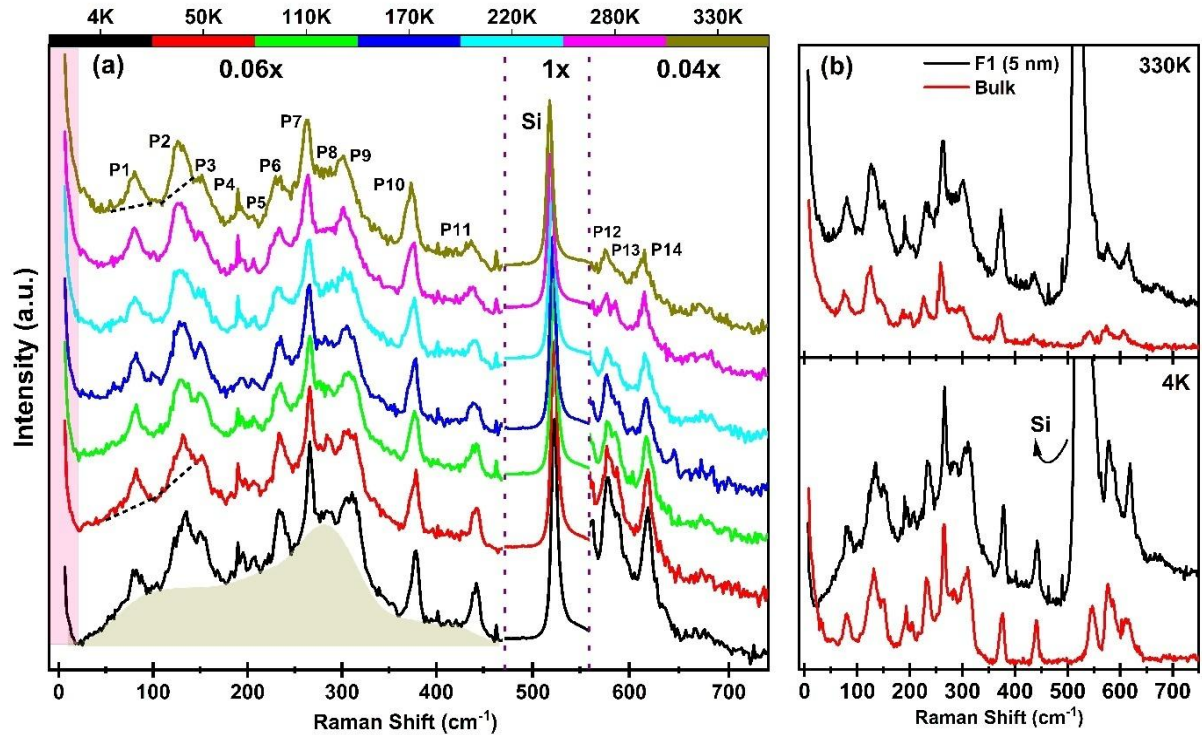

**Figure 3.17: (a)** Temperature evolution of the raw spectra for F1 flake (5 nm). Dashed thick black line shows Fano asymmetry for the modes P1 and P2 mode. **(b)** Comparison of the raw Raman spectra for F1 and bulk at 4K and 330K.

Temperature-dependent variation of phonon mode frequency in magnetic materials can be understood as a combination of lattice anharmonicity due to phonon-phonon interaction and spin-phonon coupling that comes into the picture due to the correlation of spins with the lattice, which renormalizes the phonon dynamics induced by modulation of spin exchange integral [249] in the long-range ordered phase. We will discuss first the effect of anharmonicity and then the spin-phonon coupling term. The paramagnetic phase is dominatingly governed by the anharmonic terms. Temperature dependence of the phonon frequencies and linewidth having the contribution from a three-phonon process can be described using the relation mentioned in equations 3.1 and 3.2.

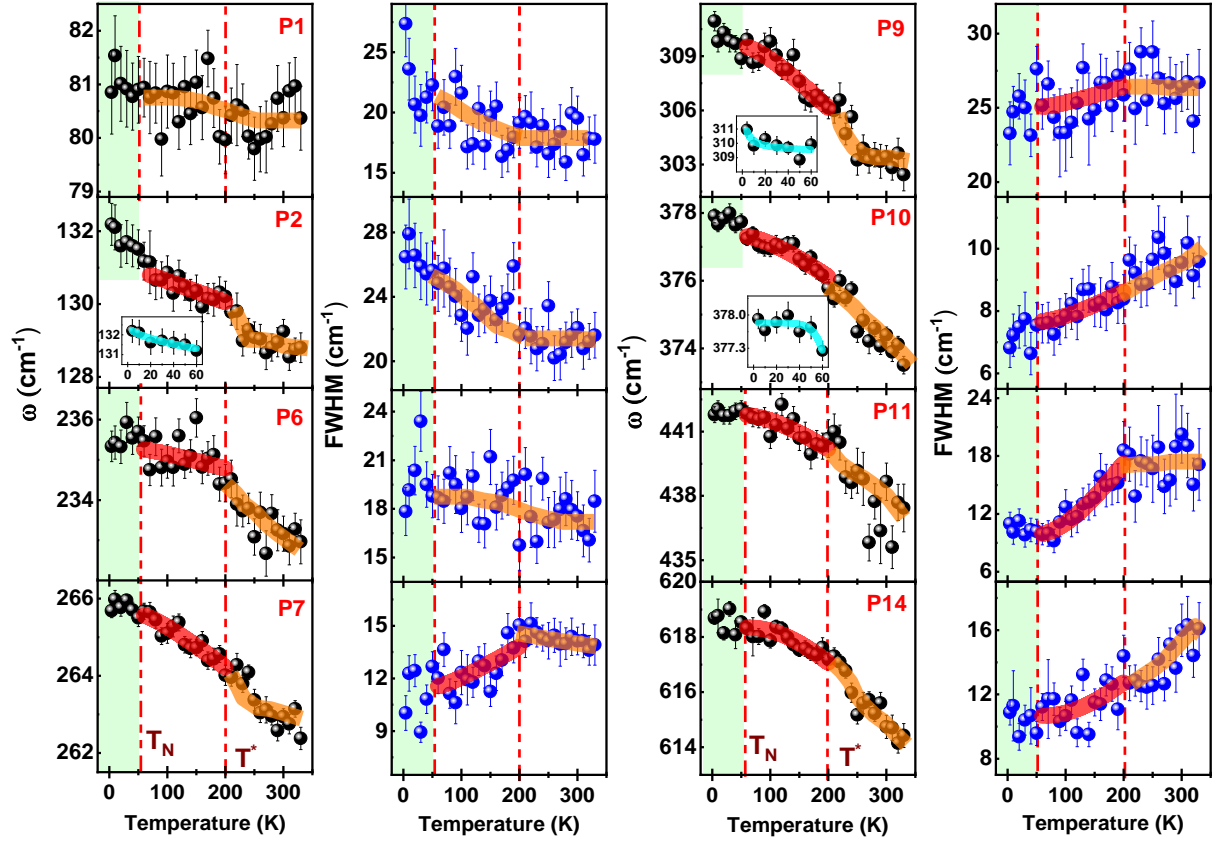

**Figure 3.18:** Temperature variation of phonon frequency and FWHM for modes P1, P2, P6, P7, P9, P10, P11 and P14 for F1 flake (5 nm). Red solid line is the anharmonic fit, dashed red line indicates $T_N \sim 60K$ and $T^* \sim 200K$. Inset (thick cyan solid line) shows fit to spin-phonon coupling function as mentioned in the text. Orange thick solid line is guide to eye. Green shaded region shows a spin-solid phase.

Conventionally, as per the anharmonic model, on decreasing temperature, the phonon frequencies blueshifts and phonon-phonon interaction reduces which is reflected in the decreased Full-width at half-maxima (FWHM or $\Gamma$). Phonon linewidth $(\Gamma)$ is a function of lifetime $(\tau)$ i.e., $\Gamma \propto 1/\tau$, and is affected by different decay channels. The temperature evolution of frequency and linewidth of the modes P1, P2, P6, P7, P9, P10, P11, and P14 for F1 flake is shown in **figure 3.18**. The temperature evolution of frequency and linewidth of the prominent modes for other flakes is shown in **figure 3.19 (a, b)**. We observed anomalous behavior in frequency and linewidth of most of the mode's above 200K which deviate from anharmonic nature so have fitted them i.e., P2, P6, P7, P9, P10, P11, and P14 modes using **equation 3.1 and 3.2** in a temperature range of 200K to 70K as below 60K it is reported to show antiferromagnetic ordering. The fitted derived parameters, frequency at 4K, and the symmetry of these modes are tabulated in **table 3.7**.

In the temperature range of 330K-60K, on decreasing temperature frequency of the mode P1 increases slightly; P2, P7, and P9 modes frequency remains nearly constant till 200K and then increases till 60K; P6 mode slowly blueshifts till 200K and then increase slightly till 60K. Modes P10, P11, and P14 blueshift monotonously with decreasing temperature till 60K with slight discontinuity at 200K.

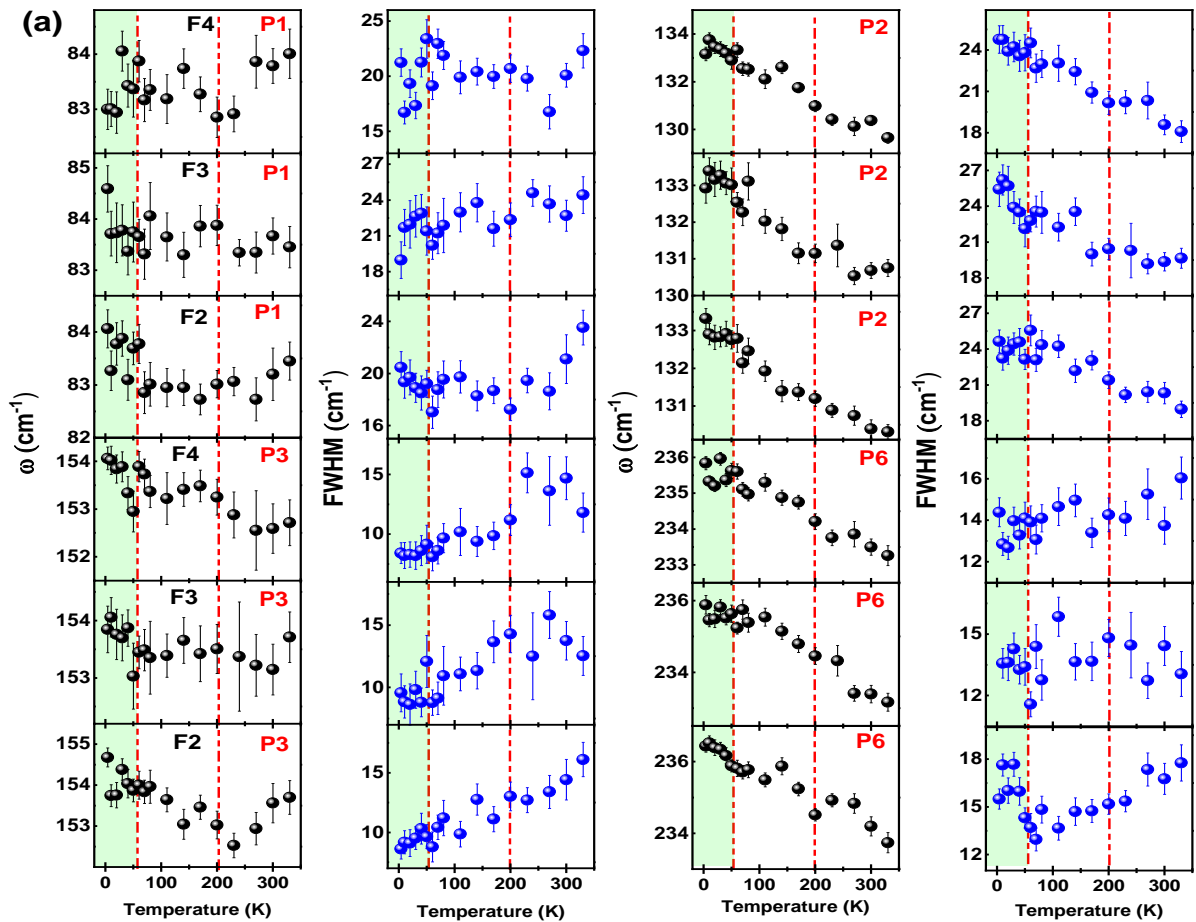

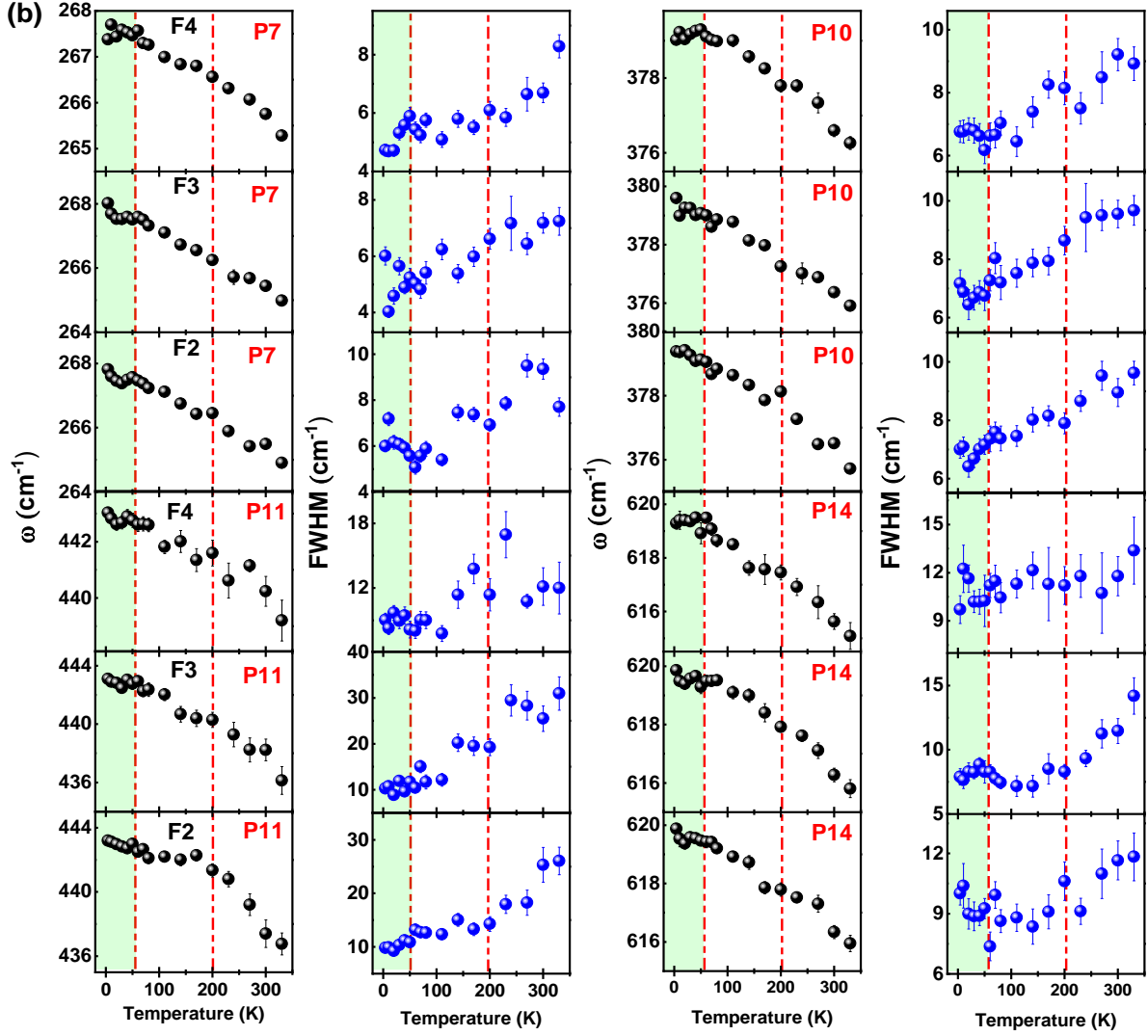

**Figure 3.19:** Temperature dependent frequency and linewidth variation of (a) P1-P3, P6; (b) P7, P10, P11 and P15 modes for F2, F3 and F4 flakes. Dashed red line indicates $T_N \sim 60K$ and $T^* \sim 200K$. Green shaded region shows a quasi-spin-solid phase.

Interestingly, below 60K i.e., the spin-solid phase (green shaded region); P7 and P11 show a slight redshift while P2, P9, and P10 show a blueshift, and P1, P6, and P14 show nearly temperature-independent behavior till 4K. In the temperature range of 330K-60K, the linewidth of the modes P10 and P14 shows a conventional trend i.e., decreases with decreasing temperatures. P7, P9, and P11 show nearly constant behavior till 200K and then decrease monotonously till 60K. Mode P6 shows a minimal increase with decreasing temperature till 60K. However, surprisingly for the modes P1 and P2 the line width remains nearly constant till ~200K and then increases with decreasing temperatures till 4K. The anomalous behavior in the linewidth of the modes P1 and P2 below ~ 200K may be captured within the picture of scattering of phonons from fractionalized fermionic excitations in the QSL candidates [250].

| Modes # | Symmetry | ω$_{exp}$ | ω$_o$ | A | Γ$_o$ | C |
|---|---|---|---|---|---|---|
| P1 | B$_g$ | 80.9 ± 0.8 | - | - | - | - |
| P2 | B$_g$ | 132.2 ± 0.6 | 131.3 ± 0.2 | -0.3 ± 0.1 | ± | ± |
| P3 | B$_g$ | 152.3 ± 0.6 | - | - | - | - |
| P4 | B$_g$ | 193.4 ± 0.7 | - | - | - | - |
| P5 | B$_g$ | 206.9 ± 0.7 | - | - | - | - |
| P6 | A$_g$ | 235.3 ± 0.4 | 235.6 ± 0.4 | -0.3 ± 0.2 | - | - |
| P7 | A$_g$ | 265.7 ± 0.2 | 266.8 ± 0.2 | -1.2 ± 0.2 | 9.4 ± 1.2 | 1.9 ± 0.7 |
| P8 | - | 284.8 ± 1.3 | - | - | - | - |
| P9 | A$_g$ | 310.9 ± 0.5 | 313.4 ± 0.7 | -3.7 ± 0.5 | 23.5 ± 1.5 | 1.5 ± 1.0 |
| P10 | A$_g$ | 377.9 ± 0.2 | 379.1 ± 0.3 | -1.8 ± 0.2 | 6.2 ± 0.5 | 1.4 ± 0.4 |
| P11 | B$_g$ | 441.8 ± 0.4 | 444.9 ± 0.8 | -3.2 ± 0.7 | -4.9 ± 1.6 | 14.7 ± 1.5 |
| P12 | B$_g$ | 578.2 ± 0.3 | - | - | - | - |
| P13 | A$_g$ | 587.9 ± 0.6 | - | - | - | - |
| P14 | A$_g$ | 618.7 ± 0.2 | 623.8 ± 1.2 | -5.4 ± 1.1 | 1.8 ± 5.0 | 8.8 ± 4.6 |

**Table 3.7:** Experimentally observed frequency at 4K for F1 flake and anharmonic fit derived parameters as mentioned in the text.

Incidentally, these modes also show Fano asymmetry discussed in detail in section 3B.3.4. The anomalous behavior of the linewidth below ~ 60K i.e., green-shaded region, can be seen as they deviate from the conventional trend.

Such an anomalous behavior of frequency and linewidth of the phonons below ~ 60K could be an indicator of a spin-solid phase where multiple competing interactions are active. A similar nature is also observed for the other flake (F2-F4) modes shown in **figure 3.19 (a, b)**. In the model put forward by Swetlana et al [250], spins are fractionalized into Majorana fermions and $Z_2$ fluxes, and at low temperatures, below the flux phase transition, the spin-phonon coupling is effectively reflected in coupling between phonons and the Majorana fermions. In QSL, the scattering of phonons involves two channels namely particle-particle and the particle-hole channels. With decreasing temperature, the increase in these modes linewidth reflects increased particle-particle scattering i.e. reduced lifetime, leading to more broadening. The increased particle-particle scattering with decreasing temperature is also expected to be reflected in the asymmetric line shape of these modes at lower temperatures, which is indeed observed in our measurements (see section 3B.3.4 for more details).

## 3B.3.2 Spin-Solid Phase and Effect of Dimensionality

Here, we will discuss the dynamic coupling between the phonons with spin excitations and the effect of dimensionality in the spin-solid phase i.e., below 60K. Spin excitations contribute to the phonon self-energy, $\Lambda = \Lambda' + i\Lambda''$, Here $\Lambda'$ is the real part of the self-energy and determines the renormalization of phonon frequency whereas the imaginary part $\Lambda''$ dictates the lifetime or the line width of the phonon mode. The spin-phonon coupling occurs in the spin-solid phase and affects the phonon dynamics below ~ 60K. It is also reflected in the reduced lifetime and broader linewidth in the magnetic systems with significant spin-phonon coupling. Lattice vibrations may lead to the modulation in the exchange interaction and can be expressed as:

$J_{ij} = J_0 + \dfrac{\partial J_{ij}}{\partial q} q_{ij} + \dfrac{1}{2} \dfrac{\partial^2 J_{ij}}{\partial q^2} q^2_{ij}$ . Here, $J_0$ is a constant signifying the equilibrium exchange coupling, $q_i$ is the $i$th atom displacement. The first term gives rise to the bare spin Hamiltonian i.e., $H_{spin} = \sum_{i,i \neq j} J_0 \vec{S}_i . \vec{S}_j$ and the last two terms signify $H_{spin-phonon}$. The renormalization of phonon energy occurs due to alteration in the exchange integral and is proportional to the spin-spin correlation which is given as [251,252]:

$$\Delta \omega = -\frac{1}{2 \mu_\alpha \omega_\alpha} \sum_{i,i \neq j} \frac{\partial^2 J_{ij}}{\partial q_\alpha^2} \vec{S}_i . \vec{S}_j \qquad \text{--(3.8)}$$

Here $\omega_\alpha$ is the mode frequency, $\mu_\alpha$ is the reduced mass associated with a mode, $J_{ij}$ is the exchange interaction between $i^{th}$ and $j^{th}$ atomic spin, $q_\alpha$ is the atomic displacement associated with a phonon mode, $\vec{S}_i . \vec{S}_j$ is the spin-spin correlation function. One can easily see that only those phonons which modify the underlying exchange interaction, contribute to the renormalization process. One can write equation 3.8 in an equivalent form as in equation 3.3. We have followed the line of argument as in Chapter 3A.3.4. $\Delta \omega$ can be positive or negative depending on the spin-phonon coupling coefficient which is dictated by the kind of interaction i.e., symmetry and amplitude of vibrations of a particular phonon mode. The fitting is shown (solid cyan line) for modes P2, P9, and P10 in **figure 3.18** (inset) for F1 flake and derived parameters are tabulated in Table 3.8.

| Mode | $\beta$ | $\Upsilon$ | $T_N^*$ |
|------|---------|------------|---------|
| P2 | -0.7 | 0.5 | 59.9 |
| P9 | -1.3 | 2.0 | 60.0 |
| P10 | -0.3 | 7.7 | 60.0 |

**Table 3.8:** Spin-Phonon coupling fitting parameters for some of the prominent phonon modes for F1 flake.

As discussed in previous section, the anomalous behavior of the frequency and linewidth of the phonon modes below ~ 60K, suggests presence of additional competing magnetic interactions other than just the anti-ferromagnetic ones. We observed a slight blueshift for the modes P2, P9, P10 and redshift for the modes P7 and P11 while P1, P6 and P14 remain nearly constant for F1 flake as shown in **figure 3.18**. Similar nature of modes can be observed for other flakes as shown in **figure 3.19**. Quite interestingly, we did not find significant effect of spin-phonon coupling due to long range magnetic ordering in the low thickness samples as compared to the bulk [69] reflecting a strong quantum fluctuation effect. For example, the maximum phonon modes hardening/softening below $T_N$ for bulk is of the order of ∼ 3.5 cm$^{-1}$, on the other hand for the lowest thickness (F1) it reduces to ∼ 1 cm$^{-1}$. The change in phonon frequency below $T_N$ for a few prominent modes P1, P2, P6, and P11 for Bulk and F1 (5nm) flakes is shown in **figure 3.20**. We note that the effect of interlayer coupling in the Kitaev materials has also been proposed in recent theoretical reports [244-246].

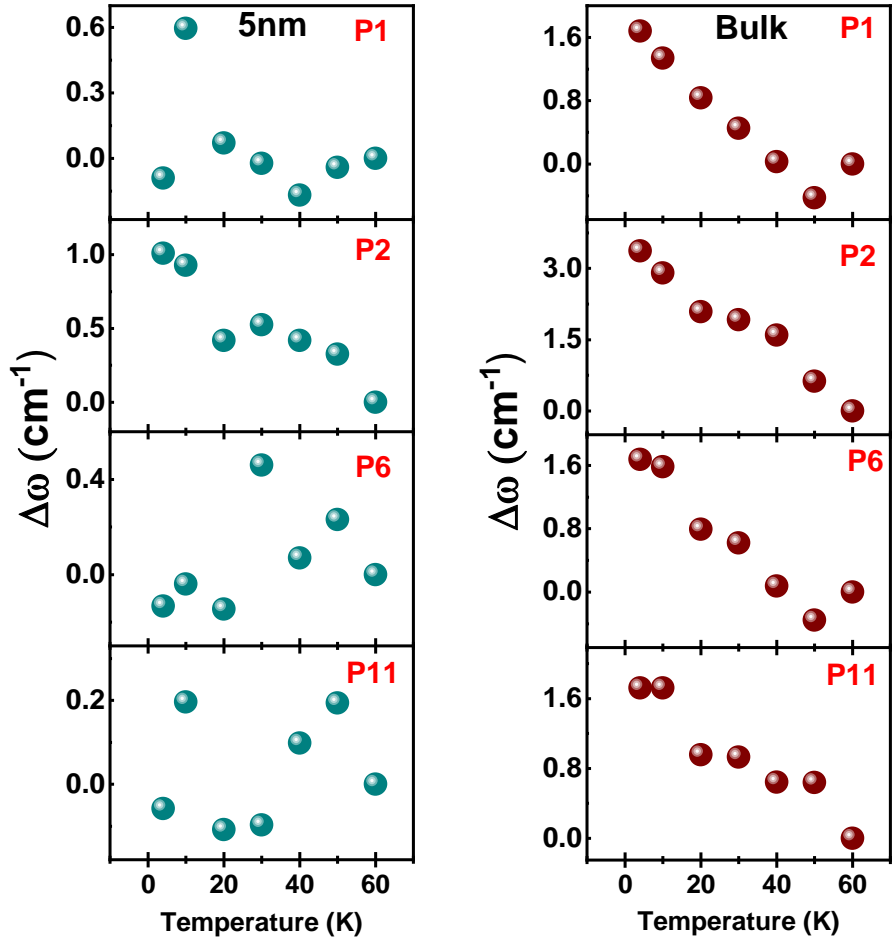

**Figure 3.20:** Variation of $\Delta\omega$ ($= \omega_T - \omega_{60K}$) for F1 (5nm) and bulk in spin-solid phase for modes P1, P2, P6 and P11.

Further, the effect of dimensionality can be seen in this spin-solid phase as a clear change in slope of both frequency and FWHM with increasing thickness is observed for the phonon modes as shown in **figure 3.19 (a, b)**. For example, in the case of modes P7, P10, and P14, we can clearly see that there is a signature of phonon softening with an increase in thickness. This observation further underlies that in addition to enhanced quantum spin fluctuation due to inlayer Kitaev interactions, the effect of the inter-layer exchange interactions also dictates the true ground state in the low-dimensional system. Indeed, probing thin layers of samples can be a good direction to search for the quantum spin liquid candidates.

### 3B.3.3 Signatures of Quantum Spin Liquid

The signature of underlying electronic and magnetic Raman scattering is embedded in the Raman spectral background. Mobile charge carriers in the metallic systems yield a continuum that is weakly energy-dependent, fractionalized excitations in QSL systems follow Fermi-Dirac statistics and one/two-magnon often gives sharp/broad Raman peak. Raman spectrum can probe different dynamics within the system simultaneously such as magnon, orbiton, charge dynamics, fractionalized excitations [156,239,252-255]. Two magnon spectral signatures generally result in a broad mode which gains intensity considerably in the spin-solid phase. Electronic continuum comes from the dynamics of mobile charges which are absent in the case of insulators. However, this is not the case for magnetic insulators and the Raman background continuum may arise from the spin dynamics which is different from the one for the mobile charge carriers. In the case of long-range ordered magnets, magnon give rise to sharp modes [227,256], which is in contrast to our case.

In particular, Kitaev materials where spins are arranged on a quasi-2D honeycomb lattice, due to the presence of bond-dependent anisotropic magnetic interactions various degenerate states of different spin alignments are equally probable throughout the material, which results in magnetic frustration and no particular ordering is observed down to even zero kelvin. The characteristic feature of Raman scattering from these systems is the broad background known as continuum scattering or multi-particle scattering, Fano asymmetry in the phonon modes line-shape, strong quasi-elastic scattering as observed in the case of α-RuCl$_3$ [57] and β-Li$_2$IrO$_3$ [58]. In the framework of Fleury-Loudon-Elliott theory [229], the magnetic Raman scattering intensity in 3D Kitaev systems is given by the density of states of a weighted two-Majorana spinon and is written as: $I(\omega) = \pi \sum_{m,n;k} \delta(\omega - \varepsilon_{m,k} - \varepsilon_{n,k}) \left| C_{m,n;k} \right|^2$, where $\varepsilon_{m,k}$ is a Majorana spinon band dispersion with $m, n$ as band indices and $C_{m,n;k}$ is the matrix element creating two Majorana excitations [222]. The itinerant Majorana fermions contribute in two ways to the inelastic light scattering process. One is the creation or annihilation of a pair of fermions and the second is the creation of one fermion and annihilation of the other fermion. These two processes have a distinct scattering response and the spectral weight of the former one is proportional to $\left[ \left(1 - f(E_1)\right)\left(1 - f(E_2)\right) \delta(\hbar\omega - E_1 - E_2) \right]$, and for the latter, it is given as $\left[ f(E_1)\left(1 - f(E_2)\right) \delta(\hbar\omega + E_1 - E_2) \right]$ [224]. Here, $f(E)$ is the Fermi-Dirac distribution function $f(\omega, T) = 1/[1 + e^{\hbar\omega/k_B T}]$, $\omega$ is the Raman shift and $\hbar$ is Plank's constant, $E_1$ and $E_2$ are the energies of the two fermions. The analysis of the broad continuum can provide information regarding the nature of the underlying dynamical spin fluctuations via the qualitative analysis of the dynamic Raman susceptibility ($\chi^{dyn}$).

As quantum effects are expected to enhance in the low dimensional regime so here, we have focused on the V$_{1-x}$PS$_3$ system down to a few layers. First, we focus on the temperature evolution of the integrated raw intensity of the background continuum (phonon subtracted) as shown in **figure 3.21 (a)** for flake F1, where the raw Raman intensity *I(ω)* is integrated over a range of 7 cm$^{-1}$ – 460 cm$^{-1}$ given as: $I = \int_{\omega_{\min}}^{\omega_{\max}} I(\omega, T) \, d\omega$ ; as shown in the inset of **figure 3.21(a)**. Remarkably temperature dependence of the spectral weight does not follow the typical bosonic nature which is expected for both magnons and phonons in insulating magnets. Such a nature of the spectral weight is very compelling and reflects on the presence of an anomalous contribution which is non-bosonic in nature. To extract that vital information we calculated the Raman response which is the imaginary part of the scattering response, $\chi''(\omega, T)$, and qualitatively analyzed the dynamics of underlying non-bosonic collective excitations, which at a given temperature is written as, $I(\omega, T) \propto \left[1 + n(\omega, T)\right] \chi''(\omega, T)$ or $I(\omega, T) = \int_0^\infty dt\, e^{i\omega t} \left\langle R(t)\, R(0) \right\rangle \propto \left[1 + n(\omega, T)\right] \chi''(\omega, T)$; where $R(t)$ is the Raman operator and $\left[1 + n(\omega, T)\right] = 1/[1 - e^{-\hbar\omega/k_B T}]$ is the Bose thermal factor. It shows the remaining, presumably the dominating magnetic spectral contribution which persists much above the spin-solid phase. **Figure 3.21 (b)** shows the temperature evolution of the background continuum Raman response or Bose-corrected spectra i.e., $\chi''_{cont.}(\omega, T)$.

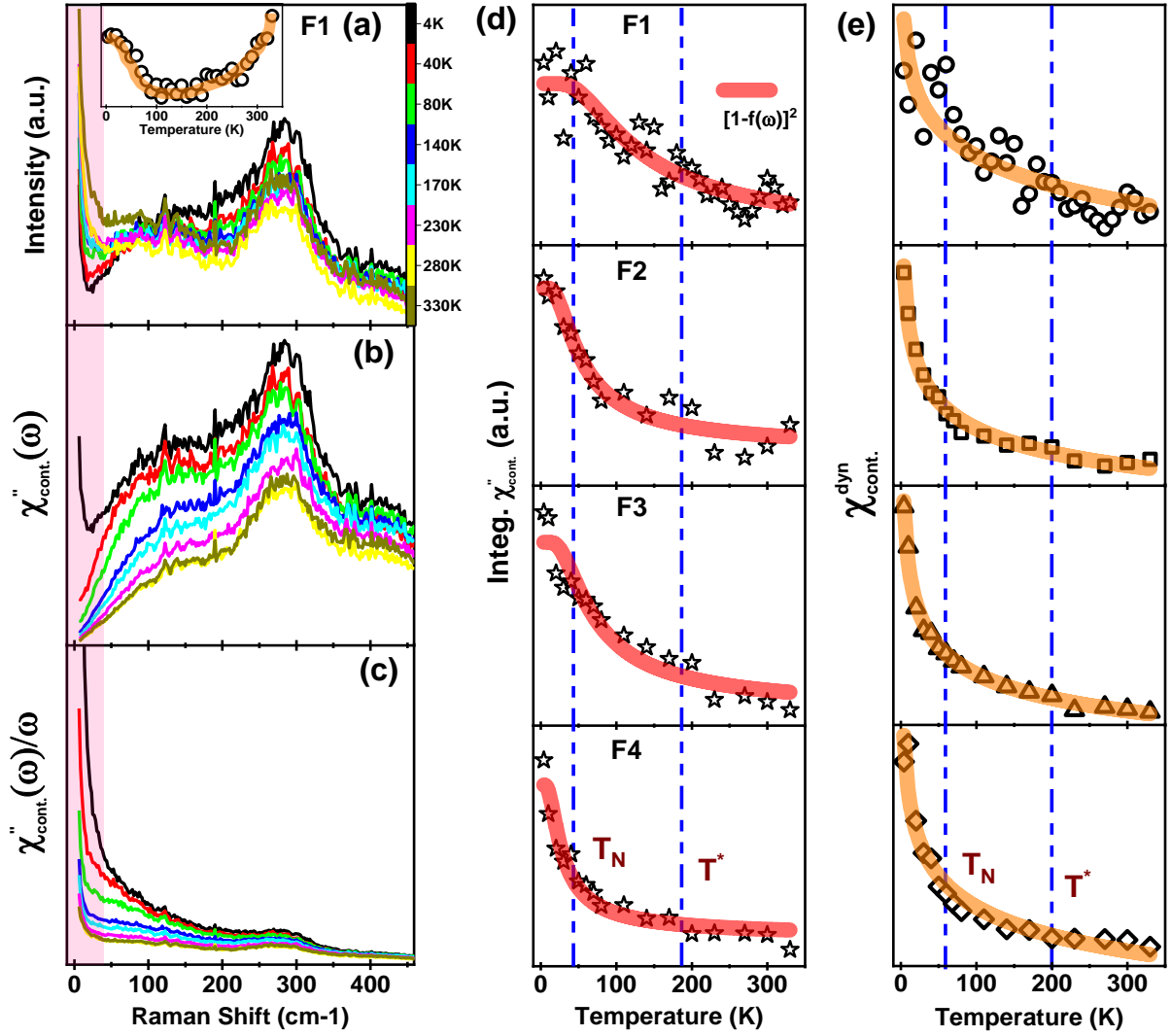

**Figure 3.21:** **(a)** Temperature evolution of the phonon subtracted background continuum raw spectra, inset shows the integrated intensity (orange solid line is guide to the eye). **(b)** Raman response, $\chi^{"}_{cont.}(\omega)$, **(c)** Raman conductivity, $\chi^{"}_{cont.}/\omega$, for a spectral range of (7 cm$^{-1}$ – 460 cm$^{-1}$); **(d)** Integrated Raman response, $\chi^{"}_{cont.}(\omega)$, for all the four flakes i.e. F1, F2, F3 and F4 and red solid line shows the two-fermionic fit ; and **(e)** $\chi^{dyn}_{cont.}(\omega)$ obtained from the Kramers-Kronig relation and the orange solid line is a guide to the eye. The shaded light pink region indicates a quasi-elastic part. The dashed blue line indicates T$_N$ ~ 60K and T$^*$ ~ 200K.

We further notice that the asymptotic two-fermion-scattering of form $\left(1 - f\left(E\right)\right)^2$, with $f\left(E\right)$ being Fermi distribution function and $E$ is the energy for the fermions as discussed above, shows a good fit to the integrated Raman response as shown in **figure 3.21 (d).** It provides compelling and reasonable evidence of identification of the nature of fundamental excitations and can be interpreted that the spins thermally fractionalize into itinerant MFs. The extracted value of energies for all the flakes (F1-F4) for this background continuum is mentioned in **table 3.9**. Interestingly the value of the energy ($E$) scale increases with decreasing thickness which suggests that the energy range of the underlying quantum fluctuations/fractionalized excitations increases with decreasing thickness owing to enhanced quantum effects in a low dimensional regime.

For further probing the evolution of this broad magnetic continuum and the underlying quantum spin fluctuations we quantitatively evaluated $\chi^{dyn}$. $\chi^{dyn}$ is in the dynamic limit of $\chi^{static} = \lim_{k \to 0} \chi(k, \omega = 0)$ and can be calculated at a given temperature by integrating phonon-subtracted Raman conductivity, $\dfrac{\chi^*_{cont.}(\omega)}{\omega}$ , shown in **figure 3.21 (c)**, and using Kramers - Kronig relation as given by equation 3.7 where $\Omega$ is the upper cutoff value of integrated frequency chosen as ~ 460 cm$^{-1}$, where Raman conductivity shows no change with a further increase in the frequency.

| Two fermion scattering $a + b[1 - f(\omega, T)]^2$ | $\omega$ (cm$^{-1}$) |
|---|---|
| **Fano asymmetry (F1)** | |
| **P1** | $85.0 \pm 13.43$ |
| **P2** | $125 \pm 35.6$ |
| **(7 cm$^{-1}$ - 460 cm$^{-1}$)** | |
| **F1** | $138.14 \pm 22.5$ |
| **F2** | $56.63 \pm 7.63$ |
| **F3** | $73.84 \pm 12.67$ |
| **F4** | $29.5 \pm 4.96$ |
| **QES (7 cm$^{-1}$ - 50 cm$^{-1}$)** | |
| **F1** | $46.74 \pm 4.57$ |
| **F2** | $20.69 \pm 1.76$ |
| **F3** | $25.79 \pm 2.22$ |
| **F4** | $31.96 \pm 3.99$ |

**Table 3.9:** Two fermion scattering fitting parameters for Fano asymmetry of P1 and P2 mode for F1 flake, and for different frequency regions i.e., background continuum and QES, for F1-F4 flakes as mentioned in the text.

For all the flakes on lowering temperature $\chi^{dyn}_{cont.}$ shows nearly temperature-independent behavior down to ~200 K as shown in the **figure 3.21 (e)** which is expected in a pure paramagnetic phase where there is no correlation between the spins. On further lowering the temperature it increases continuously till 4K. The quantum spin liquid phase is highly correlated topological phase with no long-range ordering and in this phase the Raman operator couples to the dispersing fractionalized quasi-particle excitation and reflects the two-Majorana fermion density of states [222].

We have also observed phonon anomalies at temperature ~ 200K, which can be considered as a cross-over temperature where the system goes from a paramagnetic phase to a proximate quantum spin-liquid state. For conventional antiferromagnets as the system attains ordered phase dynamical fluctuations should be quenched to zero, on the contrary here we observed a significant increase in dynamic Raman susceptibility which points towards strong enhancement of dynamic quantum fluctuations [186,225,228,229]. The diverging nature of $\chi_{cont.}^{dyn}(T)$ as T → 0K, clearly suggests the dominating nature of quantum fluctuations associated with the underlying collective excitations on lowering temperature. Such a high value of cross-over '$T^*$' temperature is consistent with earlier studies where it was advocated that in systems with quantum spin liquid signatures and the fractionalization of the quantum spins contribute to dynamic spin fluctuations even in the high-temperature paramagnetic phase [230].

Next, we focus on the quasi-elastic region of the Raman spectra in the low energy range of (7 cm$^{-1}$ – 50 cm$^{-1}$) as shown in **figure 3.21 (a)**, where the signatures of a QSL can be uncovered in the presence of underlying dynamic quantum spin fluctuations [57,58,240]. It arises from diffusive fluctuations of a four-spin time correlation function or fluctuations of the magnetic energy density [257]. The temperature dependence of quasi-elastic scattering intensity can provide information about the evolution of low-energy excitations. We observed that the quasi-elastic response increases with decreasing temperature, on the other hand in the case of a conventional magnetic system quasi-elastic response increases with increasing temperature due to thermal fluctuations. We have also evaluated the Raman response spectral weight shown in **figure 3.22 (a)**. Interestingly, it also follows two-fermion scattering very well (thick red curve) and the obtained value of energies are mentioned in table 3.9. Further, $\chi_{QES}^{dyn}$ is also calculated as shown in **figure 3.22 (b).** Interestingly, the qualitative nature of $\chi_{QES}^{dyn}$ also reflects a similar nature as observed for the background continuum where it remains nearly constant till $T^*$ and increases on further lowering temperatures till 4K. The signatures of spin fractionalization are very evident from the analysis of the background continuum and the quasi-elastic part of the spectra.

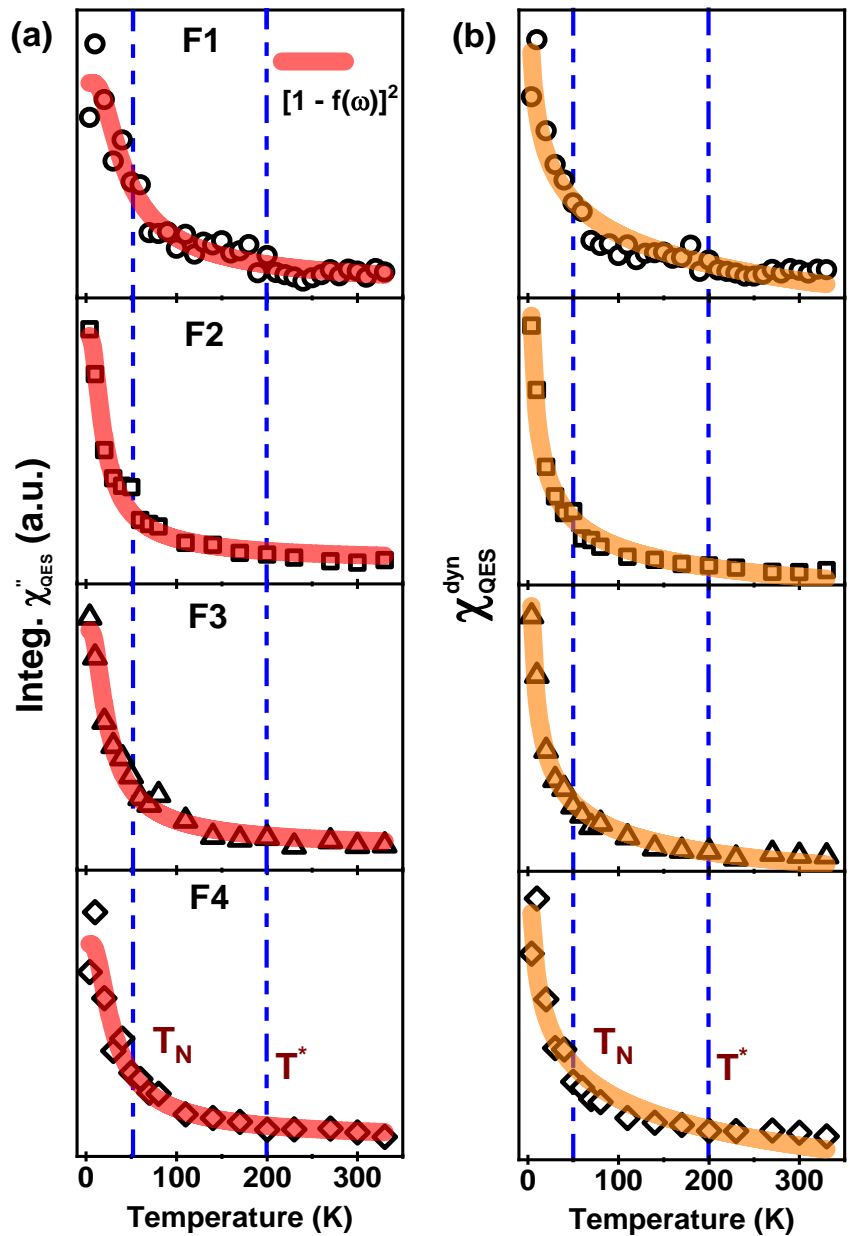

**Figure 3.22: (a)** Integrated Raman response $\chi''_{QES}(\omega)$ and red solid line shows the two-fermionic fit and **(b)** $\chi^{dyn}_{cont.}(\omega)$ obtained from the Kramers-Kronig relation for the quasi-elastic region ($7\ cm^{-1} - 50\ cm^{-1}$) and the orange solid line is a guide to the eye. The dashed blue line indicates $T_N \sim 60K$ and $T^* \sim 200K$.

### 3B.3.4 Mode's Asymmetry and Anomalous Phonons

Fano asymmetry has its origins in the interaction of discrete states (phonons here) with a continuum of excitations which manifests itself via the spin-dependent electron polarizability involving both spin-photon/phonon coupling [232-234]. In a Kitaev honeycomb system, spins are thermally fractionalized into the itinerant Majorana spinons. As a result, the continuum from the spin fractionalization strongly couples to lattice vibrations and may lead to the Fano-asymmetry [57,235,250,258].

We have used the slope method to qualitatively analyze the nature of the phonon modes asymmetry and have shown earlier that it is an equally effective way to estimate the asymmetry as mentioned in Chapter 3A.3.7. Interestingly, Fano asymmetry is observed in the modes where temperature dependence of the spectral weight of the magnetic continuum is considerably large (low-frequency region below 150 cm$^{-1}$). We observed the temperature dependence of line asymmetry for modes P1 and P2 and analyzed it using the slope method for all the flakes. The line width of these two modes also increases on lowering temperatures from $T^*$ (~200K) where it diverges from the expected behavior arising from anharmonicity (phonon-phonon interaction). This clearly indicates an additional relaxation channel. For F1 it is shown in **figure 3.23** and for F2, F3, and F4 it is shown in **figure 3.24**. Quite strikingly P1 and P2 mode's temperature-dependent asymmetry closely matches with the two-fermion scattering function. Further, a more interesting thing observed is that the obtained value of energies (**table 3.9**) is in close proximity to the position of P1 and P2 modes at 4K i.e., ~81 cm$^{-1}$ and 132 cm$^{-1}$ respectively. Such an intriguing nature suggests that the underlying continuum arises from the spin fractionalization which strongly couples to the lattice vibrations that mediate the Kitaev interactions and is indicator of the thermal fractionalization of spins into the MFs as also advocated theoretically for QSL candidates.

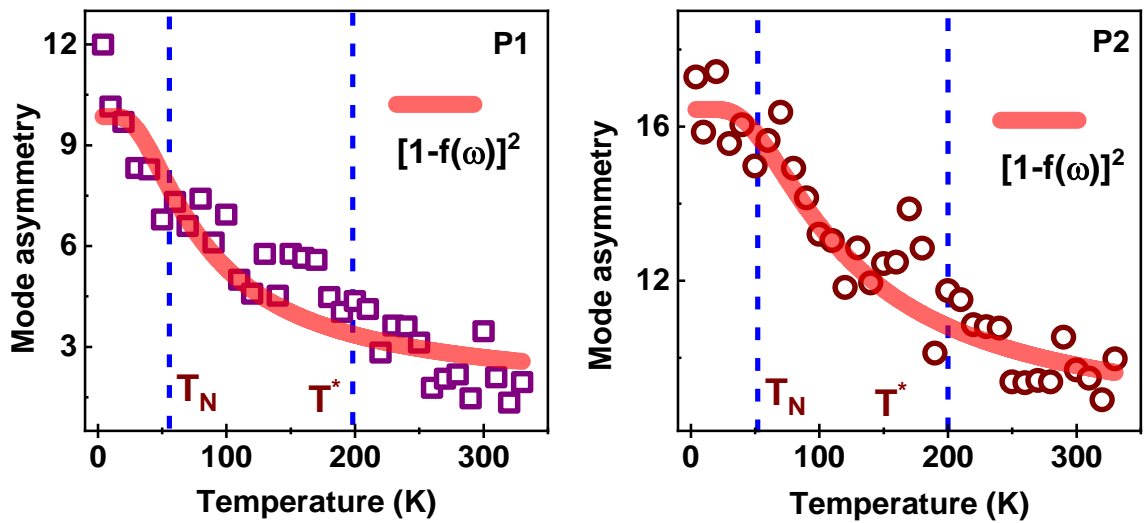

**Figure 3.23:** Mean slope for modes P1 and P2 mode for F1 flake, representing the asymmetric nature of the line shape. The red solid line shows the two-fermionic fit. The dashed blue line indicates $T_N \sim 60K$ and $T^* \sim 200K$.

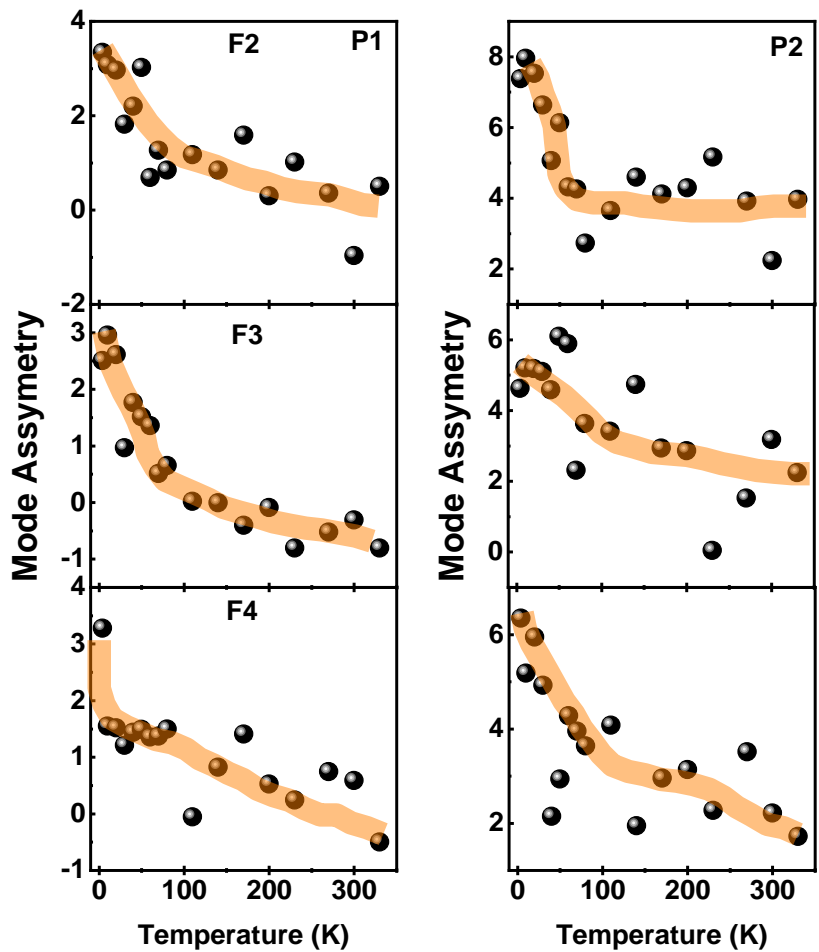

**Figure 3.24:** Modes line shape asymmetry of P1, P2, P6, and P7 modes for F2, F3, and F4 flakes.

## 3B.4 Conclusion

In a nutshell, we have carried out a comprehensive thickness-dependent inelastic light scattering study on a non-stoichiometric single crystal of $V_{0.85}PS_3$. In the temperature evolution of the Raman spectra, we observed signatures of a broad background magnetic continuum, quasi-elastic scattering, phonon anomalies, and Fano-asymmetry. An anomalous increase in line width of some of the phonon modes, especially P1 and P2, indicates the presence of an additional decay channel for the optical phonons below a crossover temperature $T^* \sim 200K$, where spin-fractionalization begins deep into the paramagnetic phase. The low energy modes P1 and P2 also show Fano asymmetry which suggests that the underlying magnetic continuum strongly couples with the lattice vibration and indicates the fractionalization of spins.

For low-thickness samples, the effect of long-range ordering coupling with the lattice is feeble as reflected in the small relative change ($\sim 1$ cm$^{-1}$) in frequency of the phonons as compared to the bulk counterpart ($\sim 3.5$ cm$^{-1}$) in this phase below $\sim T_N$. The suppression of the long-range ordering effect in low-thickness layers is attributed to underlying dominating quantum spin fluctuations hint towards the dimensional tunability of the underlying magnetic interactions. The presence of a possible quantum spin liquid phase is also confirmed by the analysis of a broad background continuum and quasi-elastic region which shows signatures of the presence of non-bosonic excitations for all the flakes. These quantum effects are found to be enhanced by decreasing thickness. Our results in this putative quasi-2D QSL candidate further open up new realm in this field of research and set a paradigm for future quantum materials. Still, we would like to keep it an open quest to investigate this system both theoretically and experimentally especially for the monolayer/bilayer in order to shed further light on the nature of underlying physics which will further excite future research in quantum spin liquid candidates.

# Chapter 4

# Surface Phonons and Possible Structural Phase Transition in a Topological Semimetal PbTaSe₂

## 4.1 Introduction

The discovery of topological insulators (TIs) has prompted a deep interest of condensed-matter physicists in finding new methods to generate exotic quantum phases in novel quantum materials. Other than their rich theoretical realm, topological semimetals got promising spintronic applications in magnetic nano-devices owing to the presence of spin-momentum-locked surface states and quantum computing [259]. The presence of Majorana fermions, symmetry-protected robust metallic edge, and surface modes with insulating bulk ground states have been important features of topological materials [260]. After the recent emergence of Weyl semimetals, an imperative shift in focus toward Topological semimetals has been observed because of such promising properties and applications [260-267].

In the league of such materials, PbTaSe₂ has attracted the interest of the condensed matter community. R. Eppinga *et al.*, in the 1980s reported it as one of the prepared intercalated compounds of the post-transition elements i.e., In, Sn, Pb and Bi with TX₂ (T= Ta and Nb; X = S and Se) [268]. It is a spin-orbit coupled non-centrosymmetric superconducting topological semimetal and has been reported to possess a nonzero $Z_2$ topology with fully spin-polarized Dirac surface state and topological nodal-line fermions [261,269-272]. D. Multer *et al.*, reported evidence of robust nature of topological surface states at Fermi energy against dilute magnetic impurities [273]. M. N. Ali *et al.* found the existence of a gapped graphene-like, but heavier Dirac cone at K point in the electronic structure which is extracted from the hexagonal Pb layer and generates 3D massive Dirac fermions. Strong spin-orbit coupling which is inherent to heavy Pb atom, introduces a gap of 0.8 eV which can contribute to large Rashba splitting. The absence of inversion symmetry is a crucial component in lifting spin degeneracy of electronic bands and the formation of topological nodal lines near the Fermi level [274]. The topological character of the electronic structure has been identified in experimental and first principal DFT calculations. Transport and magnetization measurements indicated a metallic behavior until it reaches $T_c \sim 3.79$ K below which it exhibits a type-II Bardeen-Cooper-Schrieffer (BCS) superconductor-like character at ambient pressures [261,274-276]. Interestingly, the structure of stochiometric $PbTaSe_2$ has been found to be extremely sensitive to pressure. It undergoes a series of transitions from $P\bar{6}m2$ ($\alpha$-phase) to $P6_3mc$ ($\beta$-phase) at sub-GPa to $P6/mmm$ ($\gamma$-phase) at ~7.5 GPa and finally to $Pmmm$ ($\delta$-phase) at ~ 44 GPa successively [277-280]. In ambient surrounding conditions ,XRD pattern revealed that $PbTaSe_2$ crystalizes into a hexagonal crystal structure with space group $P\bar{6}m2$ ($\alpha$-phase, #187, $D_{3h}$) and extracted lattice constants are $a = b = 3.45 \overset{0}{\text{A}}$, $c = 9.35 \overset{0}{\text{A}}$ and $\alpha = \beta = 90^o$ and $\gamma = 120^o$. In this phase, an alternate hexagonal stacking of $TaSe_2$ and Pb layers is present, the position of the Ta atom is such that it breaks inversion symmetry as shown in **figure 4.1(a, b).** Pb, Ta, and Se atoms occupy 1a, 1d, and 2g Wyckoff positions, respectively [261,279].

Charge density waves (CDWs) are periodic spatial modulations of the electronic charge density in certain materials. CDWs originate from the coupling of electronic structures with the lattice distortions which was initially postulated by Peierls and later by Fröhlich in order to characterize metal-insulator transition [281]. Transition metal dichalcogenides (TMDs) are known to sustain CDWs and their applicability in optoelectronics and energy storage devices. Bulk $2H-TaSe_2$ has shown the presence of CDW instability below ~122 K where it undergoes from metallic to incommensurate CDW (ICDW) followed by a completely-commensurate CDW (CCDW) below ~90K [282-286]. The conundrum of competition between superconductivity and CDWs is well known, for example in the case of compressed topological Kagome metal X (Cs, Rb) $V_3Sb_5$, $Cu_xTiSe_2$ and TMDs like $NbSe_2$ and $1T-TaSe_{2-x}Te_x$ [287-292].

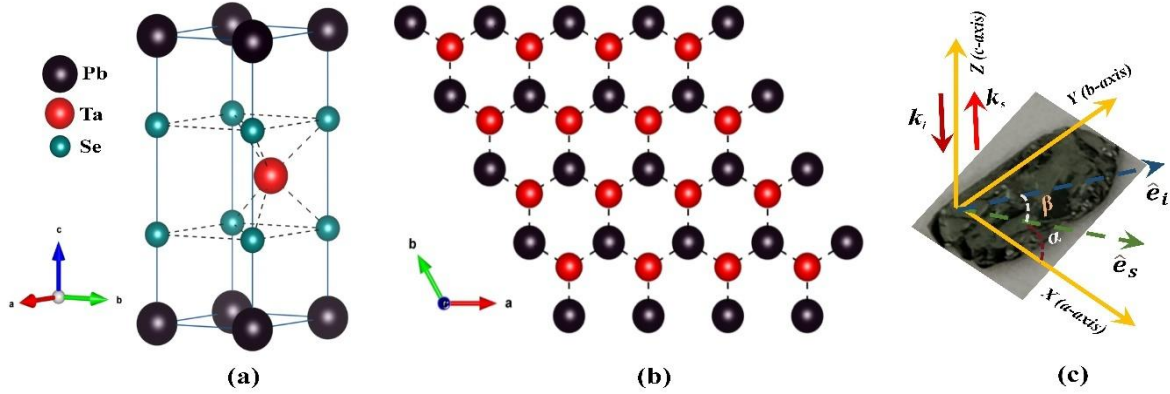

**Figure 4.1:** Shows the structure of the PbTaSe$_2$ in **(a)** standard orientation **(b)** along the c-axis and **(c)** Plane projection of polarization direction of incident and scattered light.

Phonons play a vital role when it comes to determining the lattice and electronic transport behavior of materials. Raman spectroscopy is the phenomenon of light-matter interactions which can provide information on underlying electron-photon, electron-phonon, electron-electron interactions, etc. [293,294]. The presence of strong electron-phonon coupling in topological semi-metals like TaAs, NbAs, and WTe$_2$ has also been investigated using Raman spectroscopy [293,295-297]. It is sensitive to the crystal symmetry of the material as different vibrational modes, including surface phonons, will exhibit specific Raman activity based on the selection criteria associated with the crystal symmetry. Surface phonons are localized near the boundary or surface of a material. Their behavior is influenced by the disruption of the crystal lattice at the surface. Surface phonons can substantially affect thermal conductivity, electronic transport, and other material properties, particularly in nanoscale systems where the surface-to-volume ratio is high. Raman spectroscopy has proved to be an excellent non-destructive probe to investigate the presence of CDWs [284,298], superconductivity [299,300], and surface phonons [301,302].

It has been theoretically predicted from the phonon spectrum that superconductivity arises as a consequence of the introduction of a Pb atom in the TaSe$_2$ lattice [270]. Still, how the underlying CDW phase, superconductivity, and topological nature interplay with the evolution of temperature in PbTaSe$_2$ has not been pondered upon much. Hence, motivated by these facts we investigated this stoichiometric compound via the Raman spectroscopic technique supported by DFT calculation and SC-XRD measurements.

## 4.2 Experimental and Computational Details

### 4.2.1 Single Crystal X-Ray Diffraction (SC-XRD) Measurements

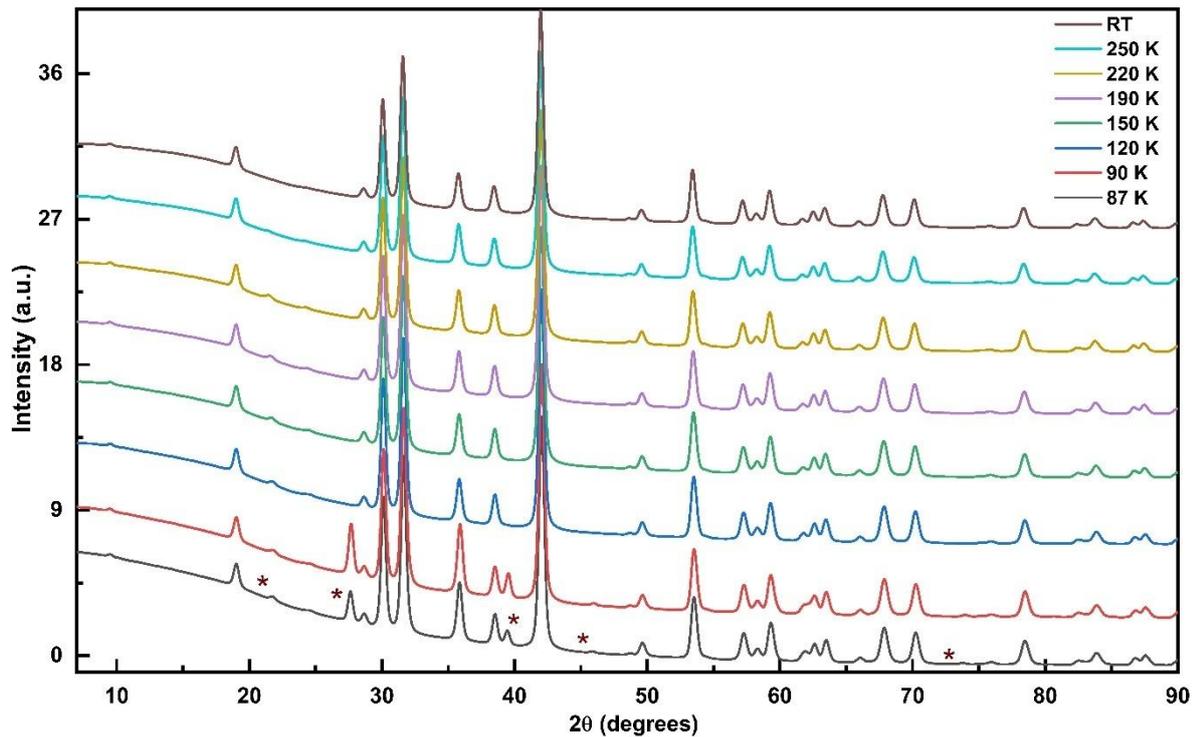

**Figure 4.2:** Raw SC-XRD spectra at different temperatures. '*' indicates peaks which changes with varying temperatures.

To reveal the possibility of temperature-dependent (TD) structural transition, we carried out a TD SC-XRD measurement using SuperNova X-ray diffractometer system with a copper (Cu, λ ∼ 1.54 A˚) source to radiate the sample at ambient pressure. Liquid nitrogen is utilized to vary temperature from 87K till room temperature.

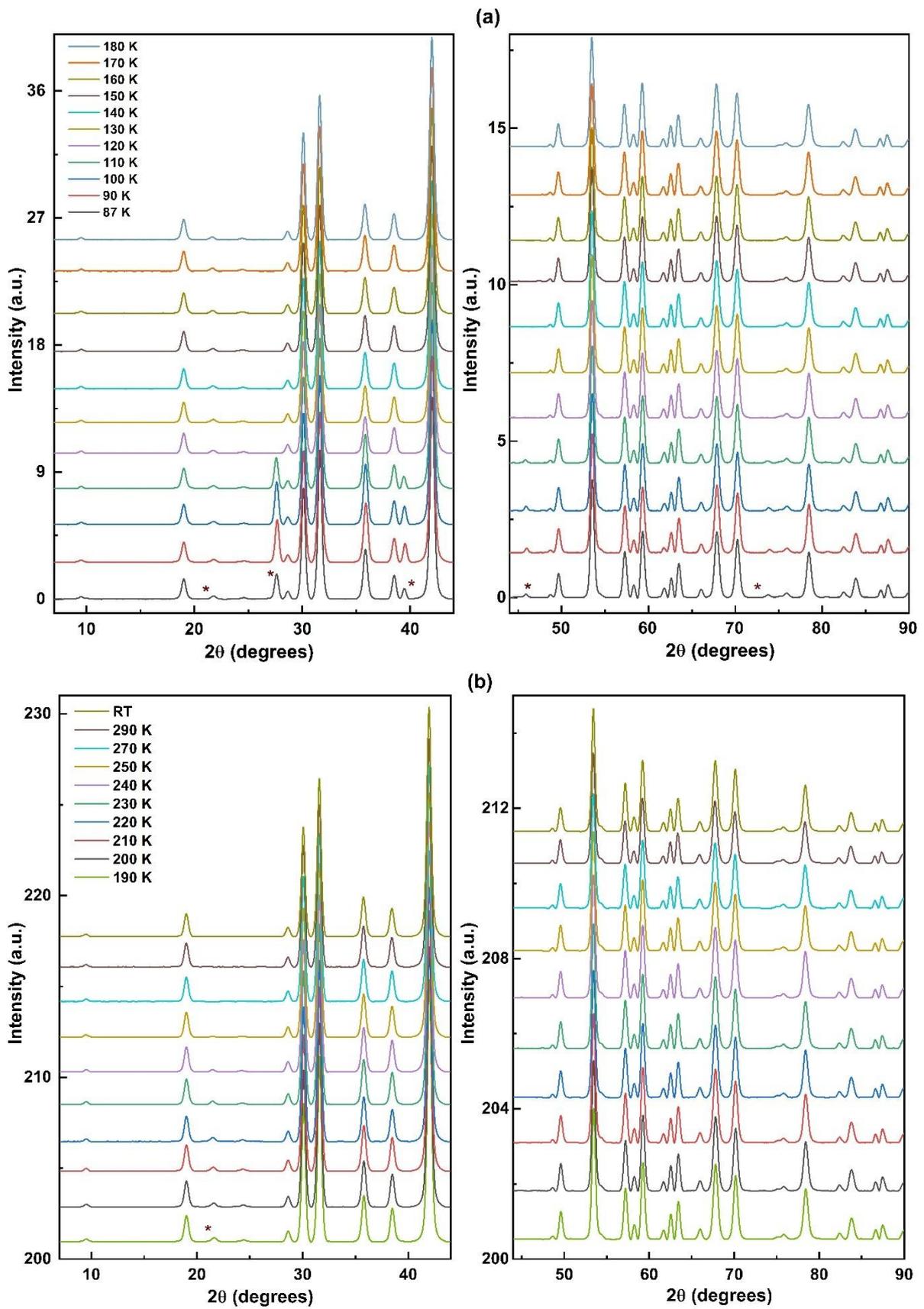

**Figure 4.3:** (a), (b) Shows background subtracted SC-XRD spectra at different temperatures. '*' indicates peaks which changes with varying temperatures.

A clear signature of structural transitions is observed as the appearance of new peaks on decreasing temperatures which are indicated with the '*' symbol in the raw SC-XRD spectra as shown in **figure 4.2** and **figure 4.3** (background is removed). Appearance of new peaks at ~ $27^o$, $39^o$, $45^o$ and $72^o$ occurs below 120K whereas peak at ~$22^o$, show broadening above ~200K. The XRD spectra obtained at room temperature are consistent with the previous report by Sankar *et al.* [276].

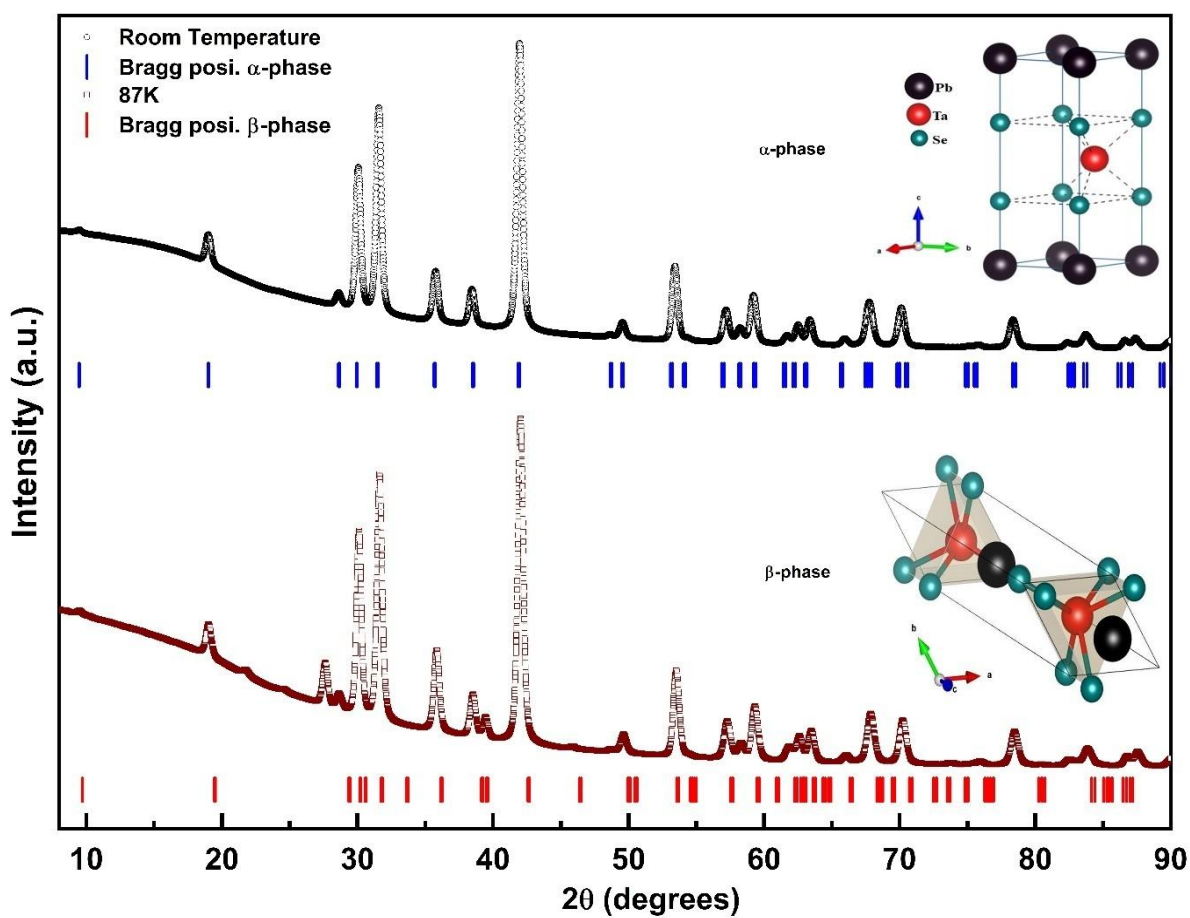

**Figure 4.4:** Raw SC-XRD spectra at 87K and room temperatures along with Bragg positions generated using VESTA for α and β-phase.

We also generated XRD patterns from VESTA using structural parameters reported by Udhara S. Kaluarachchi *et al.* [277] and have plotted the generated Braggs positions for α and β-phase. For comparison purposes we have plotted it for 87K and room temperature XRD spectra, as shown in **figure 4.4**. Our observed data for room temperatures do match with the α-phase. In the case of 87K XRD spectra, we found a deviation from the generated Braggs position for the β-phase. Still, we observe that some new peaks $\sim 45^o$ and $\sim 72^o$ which were not present in the α-phase do match with the generated Braggs positions of the β-phase. So, we suspect that it is in proximity to the β-phase if not exactly.

## 4.2.2 Raman Spectroscopic Measurements

Raman spectroscopic experiment (temperature and polarization dependent) was done with the setup mentioned in Chapter 2**.** A 633nm (1.96 eV) linearly polarized solid-state LASER is used to illuminate the sample. The sample temperature is modulated over a range of 6K-320K. We have also performed a polarization-dependent study to probe the symmetry of the phonon modes. Here we have done the measurement in a configuration where the incident light polarization direction is rotated keeping the analyzer fixed at 14K and room temperature ($\sim$ 300K). A pictorial representation of the plane projection of this configuration is shown in **figure 4.1 (c)**.

## 4.2.3 Phonon Calculation

Structural optimization and Zone-centered phonon frequencies were calculated utilizing a plane-wave approach as implemented in QUANTUM ESPRESSO [303]. The linear response method within the Density-Functional Perturbation Theory (DFPT) is used to get a dynamical matrix. Projector-Augmented Wave (PAW) pseudopotentials and Perdew-Burke-Ernzerhof (PBE) is used as an exchange-correlation functional. The kinetic energy and charge-density cutoff is taken at 40 Ry and 320 Ry, respectively. The Monkhorst-pack scheme with 30 x 30 x 30 k-point dense mesh is used for the numerical integration of the Brillouin Zone (BZ). Electron-phonon interaction is calculated by interpolation over the BZ as reported by M. Wierzbowska *et al.* [304]. Obtained phonon frequencies, optical activity, and corresponding electron-phonon interaction energies are tabulated in **table 4.1**. Frequencies are found close to the experimentally observed modes and calculated mode symmetry is in agreement with the group theory prediction for α-phase as well.

| Modes # (Symmetry) | $\omega_{DFT}$ (cm$^{-1}$) | Optical Activity | e-ph (GHz) |
|---|---|---|---|
| 1 ($E'$) | -6.9 | I+R | 0.05 |
| 2 ($E'$) | -6.9 | I+R | 0.05 |
| 3 ($A_2''$) | 4.9 | I | 0.05 |
| 4 ($E'$) | 23.2 | I+R | 0.17 |
| 5 ($E'$) | 23.2 | I+R | 0.17 |
| 6 ($A_2''$) | 63.9 | I | 0.51 |
| 7 ($E''$) | 150.3 | R | 1.40 |
| 8 ($E''$) | 150.3 | R | 1.40 |
| 9 ($E'$) | 221.1 | I+R | 4.84 |
| 10 ($E'$) | 221.1 | I+R | 4.84 |
| 11 ($A_1'$) | 228.3 | R | 4.80 |
| 12 ($A_2''$) | 272.9 | I | 2.66 |

**Table 4.1:** DFT Calculation for α-phase (P$\bar{6}$m2).

## 4.3 Results and Observations

### 4.3.1 Group Theory

$2H$-TaSe$_2$ at high temperatures possess a hexagonal P6$_3$/mmc ($D_{6h}^4$) symmetry and the irreducible representation of Raman active modes is given as $\Gamma_{Raman} = A_{1g} + 2E_{2g} + E_{1g}$ [286]. PbTaSe$_2$ in P$\bar{6}$m2 (α-phase, #187, D3h) has an effective of 4 atoms per unit cell as also clear from **figure 4.1(a).** Group theoretical prediction of phonon modes at Γ-point is given as $\Gamma = A_1' + 3A_2'' + 3E' + E''$ ($\Gamma_{acoustic} = A_2'' + E'$ and $\Gamma_{optical} = A_1' + 2A_2'' + 2E' + E''$), out of which Raman active modes are $\Gamma_{Raman} = A_1' + 3E' + E''$ and $\Gamma_{Infrared} = 3A_2'' + 3E'$ are infrared active modes [305,306]. $A_1'$ is symmetric with respect to (w.r.t.) principle-axis and reflection in the horizontal plane of symmetry; $A_2''$ is symmetric w.r.t principle-axis and anti-symmetric reflection in the horizontal plane of symmetry. $E'$ and $E''$ both are doubly degenerate modes where the former is symmetric and the later is anti-symmetric to the reflection in the horizontal plane of symmetry. **table 4.2** summarizes $\Gamma$-point phonon mode decomposition for respective Wyckoff sites along with Raman tensors for $\alpha$-phase.

| Atoms | Wyckoff site | $\Gamma$-point mode decomposition | Raman Tensors |
|-------|-------------|-----------------------------------|---------------|
| Pb | 1a | $A_2'' \, (I.R) \; + E' \, (I.R. + R)$ | $E'(x) = \begin{pmatrix} d & 0 & 0 \\ 0 & -d & 0 \\ 0 & 0 & 0 \end{pmatrix} ; E'(y) = \begin{pmatrix} 0 & -d & 0 \\ -d & 0 & 0 \\ 0 & 0 & 0 \end{pmatrix}$ |
| Ta | 1d | $A_2'' \, (I.R) \; + E' \, (I.R. + R)$ | $E'' = \begin{pmatrix} 0 & 0 & 0 \\ 0 & 0 & c \\ 0 & c & 0 \end{pmatrix} ; E'' = \begin{pmatrix} 0 & 0 & -c \\ 0 & 0 & 0 \\ -c & 0 & 0 \end{pmatrix}$ |
| Se | 2g | $A_1'(R) \; + A_2'' \, (I.R.) \; + E'(I.R. + R) + E''(R)$ | $A_1' = \begin{pmatrix} a & 0 & 0 \\ 0 & a & 0 \\ 0 & 0 & b \end{pmatrix}$ |
| | | $\Gamma_{\text{Raman}} = A_1' \; + 3E' + E'' \qquad \Gamma_{\text{Infrared}} = 3A_2'' + 3E'$ | |

**Table 4.2** Wyckoff positions of different atoms in conventional unit cell and irreducible representations of the phonon modes of hexagonal (#187; $P\bar{6}m2$ [$D_{3h}$] PbTaSe$_2$) at the gamma point, along with Raman Tensors of the Raman active phonon modes.

### 4.3.2 Temperature Evolution of the Phonon Modes

**Figure 4.5 (a)** shows the temperature evolution of the raw spectra, we observe a broad spectral background (light brown shaded region) with phonons on top. $I_B$ quantifies the background spectral weight which increases monotonously with increasing temperatures as shown in the inset. $I_B$ shows a bosonic behavior and is in line with the temperature-dependent quasi-elastic behavior, which increases with increasing temperature. Glamazda *et al.* have also observed such a background and attributed it to the spectral response from electrons by Pb defects, high-temperature phonons, and phonon anomalies of (Pb-Pb) mode, which further reflects on the charge dynamics of this topological material [306]. Interestingly they observed very broad modes in pure sample i.e., for PbTaSe₂ which is in contrast to our case as we have observed comparatively much sharp modes. This indicates that the effect of Pb defects is not that prominent on the phonon spectrum but may lead to a broad background. It is well understood that the low-energy continuum in the electronic Raman response may reflect the quasi-elastic scattering of incoherent charge carriers [156]. So, we attribute the root of this low-energy bosonic background in the quasi-elastic scattering which may originate from the underlying incoherent dynamics of the carriers and becomes dominant at higher temperatures due to thermal effect as also found to increase with increasing the temperature in our observation.

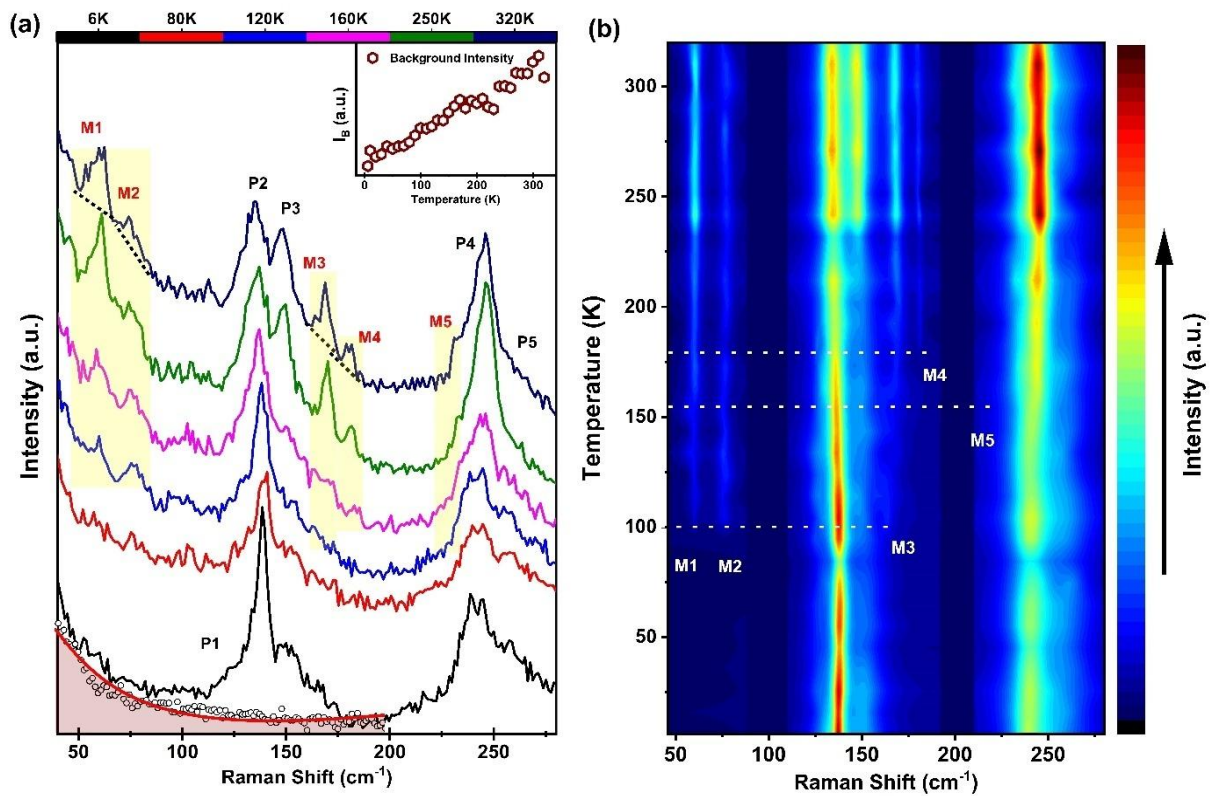

**Figure 4.5:** (a) Temperature evolution of the Raman Spectra. Prominent modes labeled as P1-P5. The shaded yellow region shows the emergence of new modes on increasing temperatures i.e., M1-M5 and the dotted black line shows the asymmetry in the line shape of M1-M4 modes. Circular data points indicate the background electronic contribution in the spectrum. Inset shows the temperature evolution of the raw background spectral weight (I_B) as shown in a light brown shaded region. The red solid line indicates the fit by a collision-dominated model as mentioned in the text. (b) Contour map of Raman intensity as a function of temperature showing the evolution of M1-M5 modes.

Such a background may also arise from the diffusive hoping of conduction electrons and can be qualitatively analyzed by a collision-dominated model [307,308]. We have fitted this potential electronic contribution to the spectrum with this model (solid red line) which is obtained by subtracting the phonons as shown in **figure 4.5 (a)**. In this case, the Stokes scattering intensity is given by $I(\omega) = \left[1 + n(\omega)\right] \times \left\{(\omega \cdot \Gamma \cdot B)/(\omega^2 + \Gamma^2)\right\}$; here $\left[1 + n(\omega)\right] = 1/\left[1 - e^{-\hbar\omega/k_B T}\right]$ is Bose-Einstein thermal factor, $\Gamma$ determines the width of the Lorentzian scattering profile and is associated with the carrier scattering rate, $B$ represents scattering amplitude. We have used $\Gamma = \Gamma_0(T) + \alpha \cdot \omega^2$ and obtained a better fit to the electronic scattering contribution; here $\alpha$ indicates the electron correlation effects [309]. The temperature variation of the derived parameters are shown in **figure 4.6**.

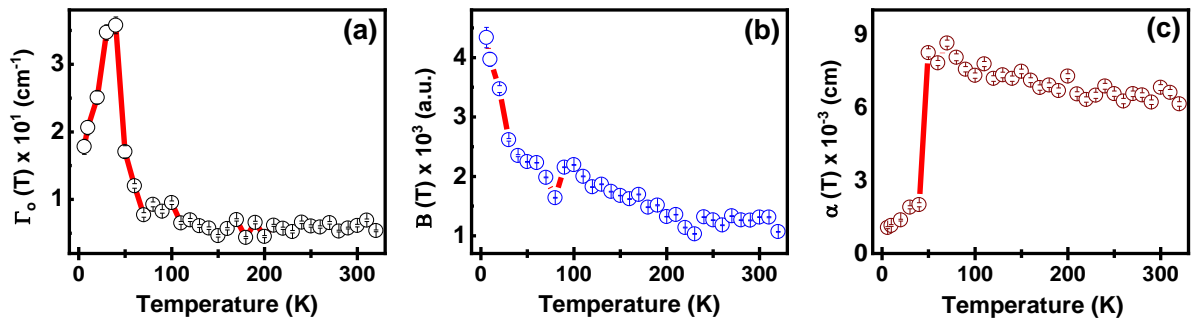

**Figure 4.6:** Temperature variation of fitted parameters for Collision-dominated model.

We observed five prominent modes at the lowest recorded temperature 6K labeled as P1-P5 for convenience. P2-P5 modes are consistent throughout the temperature range of 6K-320K. P2-P3 mode shows a $E'(x)$ or $A'_1$ kind of symmetry at 14K (refer to section 4.3.4 for further discussion). Interestingly in addition to these, some weak modes appear in certain temperature ranges only and are labeled as M1-M5, see **figure 4.5 (a, b)** which shows temperature evolution of the raw Raman spectra and color contour map of Raman intensity as a function of temperature. M1-M4 modes exhibit asymmetry in the line shape which is indicated by a dotted black line. We have fitted the Raman spectra at different temperatures with the Lorentzian function as shown in **figure 4.7** and extracted the corresponding variation of phonon features (frequency, linewidth, and intensity) with temperature. Experimentally observed mode frequencies are mentioned in **table 4.3.**

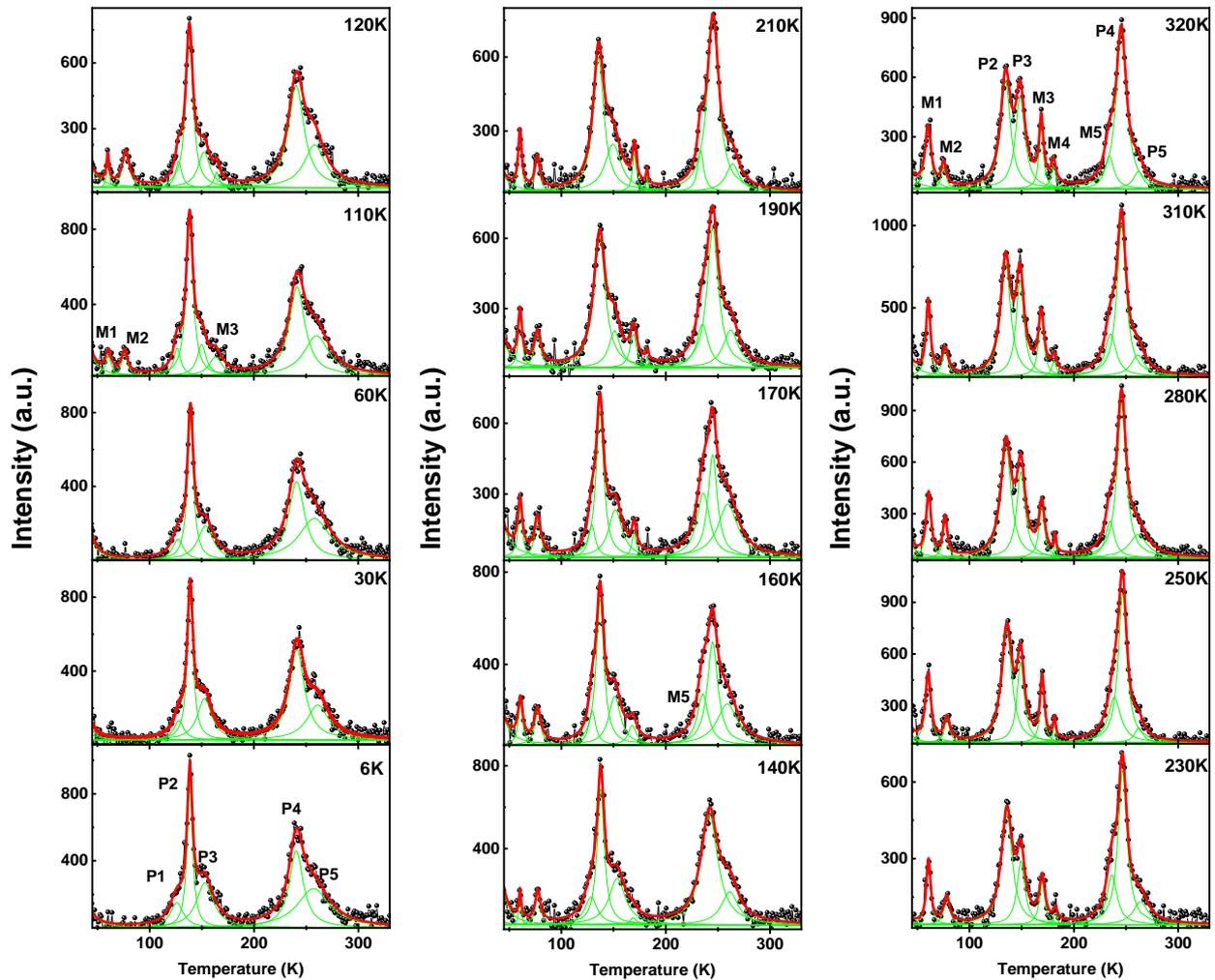

**Figure 4.7:** Lorentzian fitted raw spectra at different temperatures.

A non-linear temperature dependence of the phonon features i.e., Frequency, FWHM may be comprehended via a anharmonic phonon-phonon interaction model. Anharmonic effect arises due to phonon-phonon interaction and is reflected in change in self-energy parameters whereas thermal expansion is a pure volume effect. To comprehend the microscopic origin of anharmonicity in case of periodic solids one can think of atoms vibrating about their equilibrium position around which potential energy can be Taylor expanded as [215,310]:

$$U(q) = U(q_o) + q \left. \frac{\partial U}{\partial q} \right|_{q=q_o} + q^2 \left. \frac{\partial^2 U}{\partial q^2} \right|_{q=q_o} + q^3 \left. \frac{\partial^3 U}{\partial q^3} \right|_{q=q_o} + \dots \qquad \text{-- (4.1)}$$

or equivalently: $\qquad U = const. + U_{harmonic} + U_{anharmonic} \qquad \text{-- (4.2)}$

$$U_{anharmonic} = gq^3 + mq^4 + \dots \equiv \Pi \ (a^+ a^+ a + a^+ aa) + \Lambda \ (a^+ a^+ a^+ a + a^+ a^+ aa + ..) + \dots \text{-- (4.3)}$$

where '$a^+$' and '$a$' are the creation and annihilation operators, $q$ is the normalized coordinate. Keeping energy and wavevector conservation in consideration, cubic term $g\,q^3\ or\ \Pi\ (a^+ a^+ a + a^+ aa)$ reflects the three-phonon process where an optical phonon mode decays into two equal energy acoustic phonon modes $(\omega_1 = \omega_2 = \omega/2; k_1 + k_2 = 0)$. While quartic term $m\,q^4\ or\ \Lambda\ (a^+ a^+ a^+ a + a^+ a^+ a a +..)$ indicates four-phonon process where optical phonon decomposes into three acoustic phonons $(\omega_1 = \omega_2 = \omega_2 = \omega/3; k_1 + k_2 + k_3 = 0)$. Coefficients $g\ or\ \Pi$ and $m\ or\ \Lambda$ and $\Pi$ phenomenologically indicate the strength or contribution of three and four-phonon processes in anharmonicity. At low temperatures, three-phonon process is dominating with a minor contribution from four-phonon process. Temperature dependence of the phonon frequencies having the contribution of both three and four-phonon processes can be described using the following functional form [214]:

$$\delta\omega_{an}(T) = \omega(T) - \omega_o = A\left(1 + \frac{2}{e^x - 1}\right) + B\left(1 + \frac{3}{e^y - 1} + \frac{3}{(e^y - 1)^2}\right) ; \qquad \text{-- (4.4)}$$

Whereas the FWHM, $\Gamma_{an}(T)$, can be expressed as follows [311]:

$$\Gamma_{an}(T) = \Gamma_{an}(0) + C\left(1 + \frac{2}{e^x - 1}\right) + D\left(1 + \frac{3}{e^y - 1} + \frac{3}{(e^y - 1)^2}\right) \qquad \text{-- (4.5)}$$

respectively, here $\omega_0$ and $\Gamma_{an}(0)$ is mode frequency and line width at absolute zero temperature; $x = \dfrac{\hbar\omega_o}{2k_B T}$, $y = \dfrac{\hbar\omega_o}{3k_B T}$ and $k_B$ is the Boltzmann constant. Coefficient A/B and

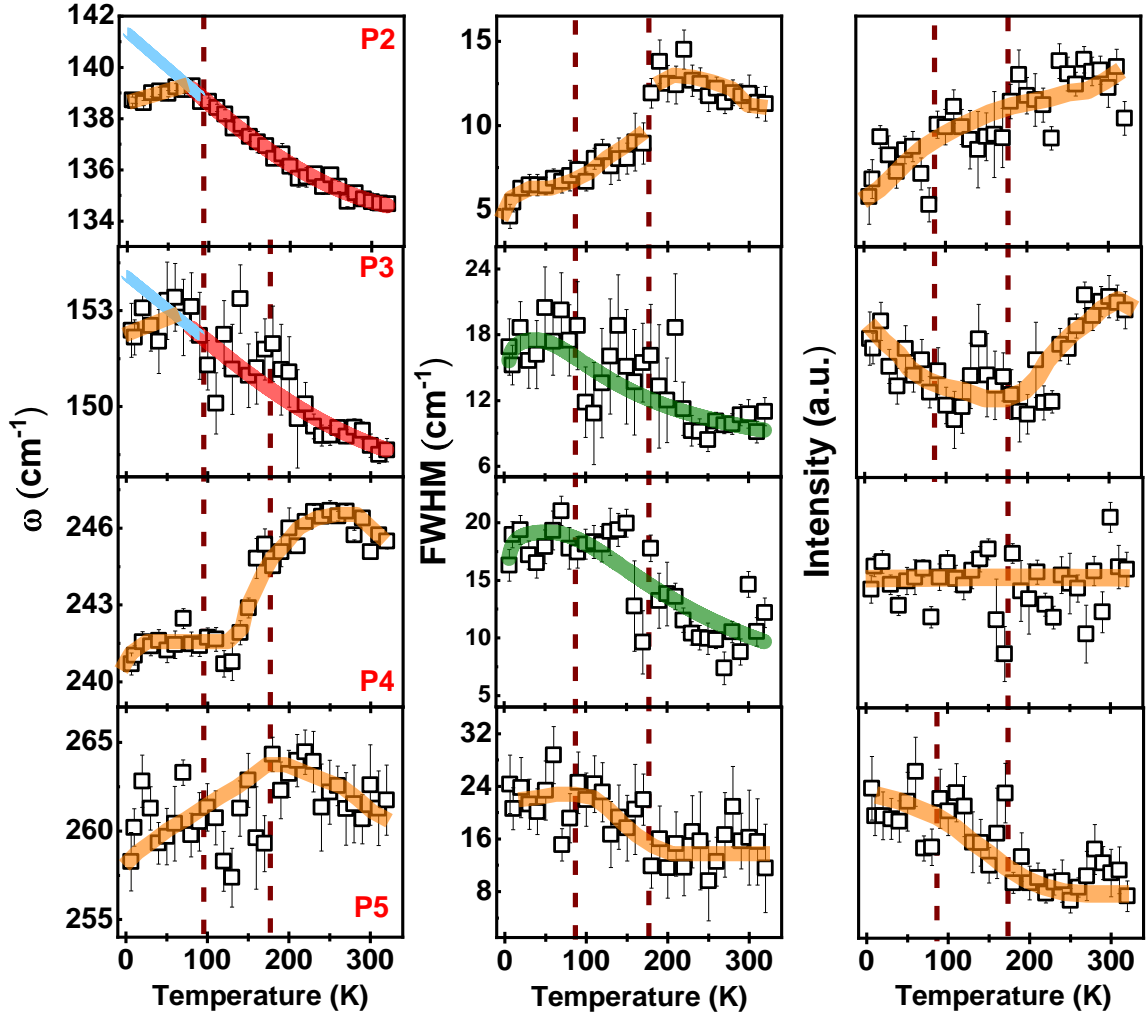

**Figure 4.8:** Temperature-dependent evolution (frequency, FWHM, and intensity) of the P2-P5 phonon modes. The red and light blue solid line shows the anharmonic fitting, and linear extrapolation, respectively for frequency. The orange solid line is a guide to the eye. Solid Olive line is a fit for FWHM of P3 and P4, with a model mentioned in the text. The dashed line shows the position of discontinuities.

C/D represent the strength of phonon-phonon interaction involving three/four phonon processes respectively. Higher-order terms can be neglected as they start contributing at much higher temperatures. Temperature evolution of the frequency, FWHM, and Intensity of some of the prominent phonon modes P2, P3, P4, and P5 are shown in **figure 4.8** (the solid orange line is a guide to the eye, (the dashed line shows the position of discontinuity). We note that in the case of frequency, only P2 and P3 show a significant anharmonic effect so we have fitted these with the equation. 4.4 in temperature range 100K-320K and fitted parameters are mentioned in **table 4.3**. The fitted curve is shown in a thick red line and the solid light blue line is a linear extrapolation which reflects the deviation from the anharmonic prediction. On decreasing the temperature both modes i.e., P2 and P3, frequency blueshifts in temperature range from 320K to ~100K and show significant softening below 100K till 6K. P4 mode frequency shows normal behavior from 320K to ~220K i.e., mode frequency increases with a decrease in temperature. From ~ 220K to ~120K it shows a sharp decrease and thereafter it remains nearly constant till 6K. Similar anomalous behavior is seen in P5 where it blueshifts almost linearly till ~200K and on further decreasing the temperature it monotonously redshifts till 6K. On decreasing the temperature, the linewidth of the P2 mode increases linearly from 320K to ~200K and then decreases continuously till 6K.

| Modes | $\omega_{EXP.}$ | $\omega_o$ | A/B |
|---|---|---|---|
| **M1** (100K) | 60.2 ± 1.4 | | |
| **M2** (100K) | 77.4 ± 0.9 | | |
| **P1** (6K) | 133.4 ± 2.5 | | |
| **P2** (6K) | 138.7 ± 0.1 | 142.5 ± 0.3 | -26.4 |
| **P3** (6K) | 152.4 ± 0.8 | 155.0 ± 1.3 | -28.3 |
| **M3** (100K) | 164.9 ± 0.8 | | |
| **M4** (180K) | 182.5 ± 1.6 | | |
| **M5** (160K) | 235.2 ± 1.7 | | |
| **P4** (6K) | 240.7 ± 0.4 | | |
| **P5** (6K) | 258.3 ± 1.6 | | |

**Table 4.3:** Experimentally observed modes and anharmonic fit parameter. Units are in cm$^{-1}$.

In the case of linewidth of P3 and P4 mode an intriguing behavior is observed that, other than the discontinuity at ~100K and 150K respectively, it monotonously increases with decreasing temperatures. Such an anomalous nature indicates a possible presence of electron-phonon coupling which is discussed in the next section i.e., 4.3.3. For mode P5, linewidth shows anomalous behavior i.e., increases with decreasing temperature. Now we will discuss the temperature evolution of the intensity of these modes (P2-P5). We know that generally bosonic (phonons) population decreases with a decrease in temperature. P2 shows an expected behavior as its intensity decreases with a decrease in the temperature. An interesting behavior is observed in the case of P3 where the intensity decreases from 320K to ~150K and increases on further decreasing the temperature till 6K. P4 shows anomalous behavior where it remains almost constant throughout the temperature range. For P5 intensity remains constant from 320K to ~150 K and then increases monotonously on further decreasing the temperature which is anomalous as well.

Interestingly we observed a very strange temperature-dependent nature of the modes labelled as M1 (~ 60 cm$^{-1}$), M2 (~ 77 cm$^{-1}$), M3 (~ 165 cm$^{-1}$), M4 (~ 182 cm$^{-1}$), and M5 (~ 235 cm$^{-1}$). M1, M2, and M3 disappear below ~100K whereas M4 and M5 disappear below ~ 150K respectively. To further investigate the temperature-dependent behavior of these modes i.e., frequency, FWHM, and intensity is shown in **figure 4.9** where a solid blue line is a guide to the eye. Interestingly most of these emergent modes (M1-M4) show an asymmetry in the line shape i.e., Fano line shape, which suggests the presence of interaction of discrete states (phonons) with an underlying continuum possibly electronic in nature [232]. We have qualitatively analyzed and discussed the Fano asymmetry of these phonon modes using a slope method as mentioned in Chapter 3A.3.7 and the temperature evolution of asymmetry is plotted in **figure 4.10**.

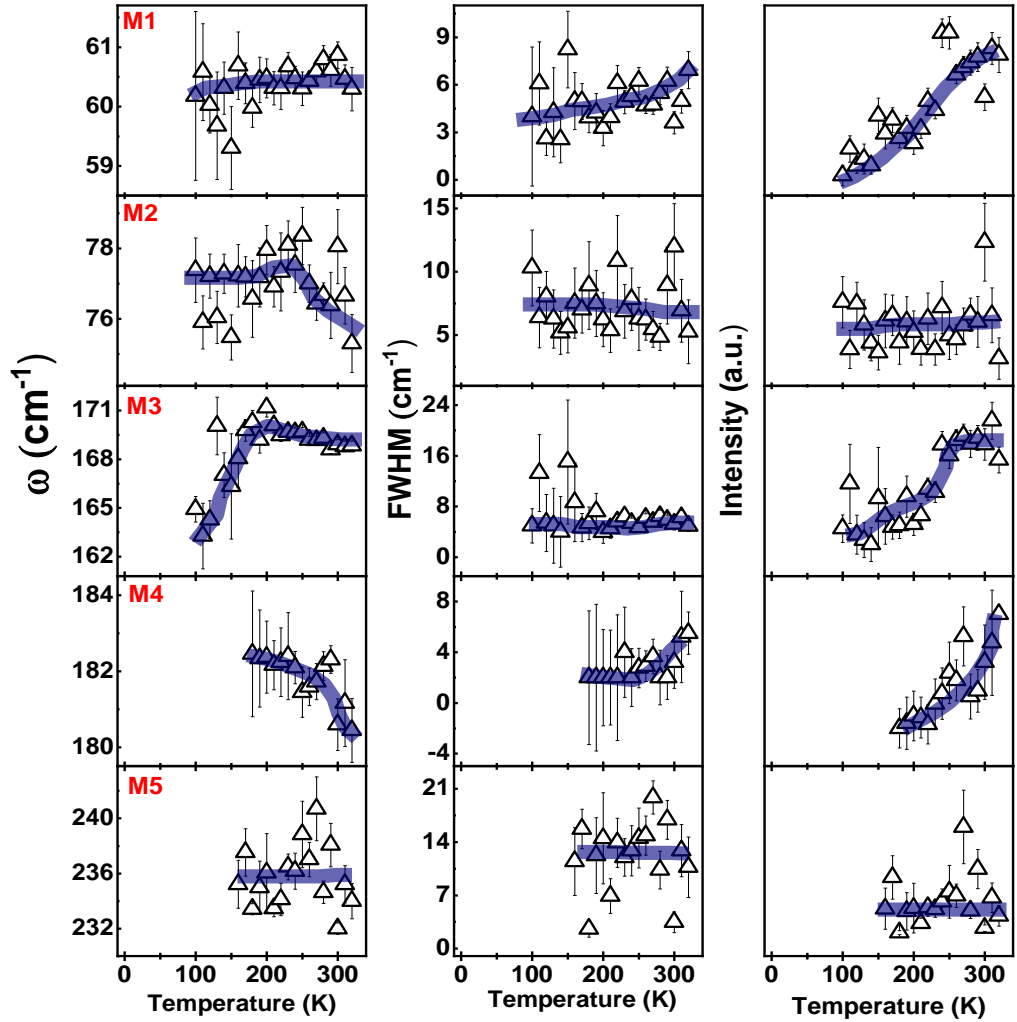

**Figure 4.9:** Temperature-dependent evolution (frequency, FWHM, and intensity) of the M1-M5 phonon modes. The blue solid line is a guide to the eye.

The frequency of M1 and M5 remains nearly constant throughout. M2 mode shows a linear increase with a decrease in temperature till ~200K and remains nearly independent of temperature on further decreasing the temperature till 100K. M3 frequency increases marginally from 320K to ~150K and then redshifts continuously till 100K, whereas M4 continuously blueshifts till 150K. FWHM of M2, M3, and M5 demonstrates nearly temperature-independent behavior throughout. M1 shows a continuous decrease in linewidth with a decrease in temperature till 100K, whereas for M4 it decreases continuously ~220K and remains constant below till ~150K. As far as temperature-dependent variation of intensity is concerned. M2 and M5 mode exhibits nearly constant behavior. M1, M3, and M4 show expected behavior i.e., decreasing with a decrease in temperature.

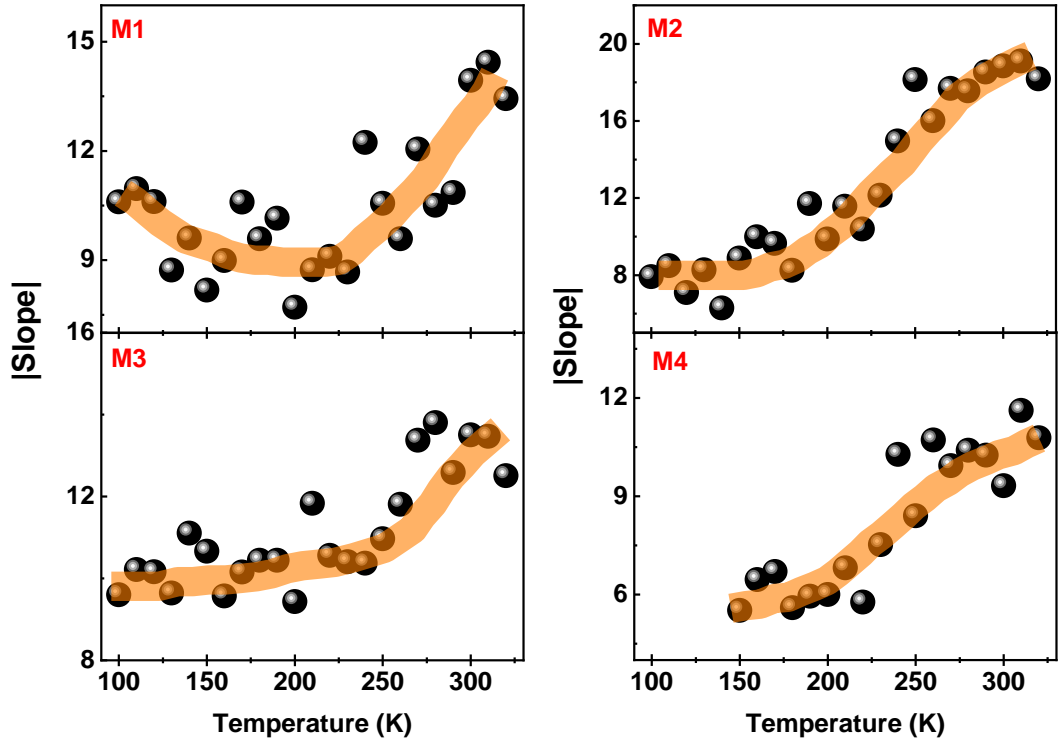

**Figure 4.10:** Temperature dependence of mode of slope for M1-M4 phonon modes, indicating the Fano -asymmetry. The solid orange line is a guide to the eye.

### 4.3.3 Electron-Phonon Coupling (EPC)

As electrons move through the crystal lattice, they can disperse off lattice vibrations, leading to a change in momentum and energy of both the electron and the phonon. In metals for example, low-energy excitations are significantly modified due to interaction with the lattice vibrations which consequently impact the electric and thermodynamic behavior. In the context of superconductivity, strong electron-phonon coupling leads to the formation of Cooper pairs. Electron-phonon coupling dictates significantly both phonon dispersions and linewidths [312]. In ideal crystals, the linewidth ($\Gamma$) of a phonon is dictated by its various decay channels due to interaction with elementary excitations and can be written as: $\Gamma = \Gamma^{an} + \Gamma^{el-ph}$, here $\Gamma^{an}$ is due to phonon-phonon interaction and is governed by anharmonic terms in the interatomic potential. $\Gamma^{el-ph}$ comes into the picture due to electron-hole pairs and governed by the magnitude of electron-phonon coupling. $\Gamma^{el-ph}$ can be evaluated if the anharmonic contribution is known or negligible which is the case at lower temperatures. Generally, $\Gamma^{el-ph}$ is determined by EPC and present in cases where the electronic bandgap is zero which means the electrons in the vicinity of the Fermi-level contribute. PbTaSe$_2$ is a topological semi-metal but bears a definite possibility of finite however weak electron-phonon coupling as also evident from the temperature evolution of phonon linewidth in **figure 4.8**. The presence of weak electron-phonon coupling has also been proposed for topological insulators Bi$_2$Se$_3$ and Bi$_2$Te$_3$ [313]. In the case of EPC, the scattering probability is dictated by Fermi's golden rule, where a phonon excites an electron into a higher momentum state. Hence within mean-field approximation, the contribution of EPC to $\Gamma_{\vec{q}n}^{el-ph}$ for a phonon of wave vector $\vec{q}$, branch index $n$ and frequency $\omega_{qn}$ can be given as [314,315]: $\Gamma_{qn}^{el-ph} = \dfrac{4\pi}{N_k} \sum_{k,i,j} \left| g_{(\vec{k}+\vec{q})j,\vec{k}i} \right|^2 \left[ f_{\vec{k}i} - f_{(\vec{k}+\vec{q})j} \right] * \delta\left[ \varepsilon_{\vec{k}i} - \varepsilon_{(\vec{k}+\vec{q})j} + \hbar\omega_{qn} \right]$ here the summation is over the electron vectors $\vec{k}$ and bands $i$ and $j$. $N_k$ is the number of $\vec{k}$ vectors, $f_{\vec{k}i}$ is the occupation of the electron state $\left| \vec{k}, i \right\rangle$, with energy $\varepsilon_{\vec{k}i}$; $\delta$ is the Dirac delta function. And, $g_{(\vec{k}+\vec{q})j,\vec{k}i} = D_{(\vec{k}+\vec{q})j,\vec{k}i} \sqrt{\hbar/(2M\omega_{qn})}$, here $M$ is the atomic mass. $D_{(\vec{k}+\vec{q})j,\vec{k}i} = \left\langle \vec{k}+\vec{q}, j \left| \Delta V_{qn} \right| \vec{k}, i \right\rangle$ is the EPC and $\Delta V_{qn}$ is the derivative of the interatomic potential respective to phonon displacement. Hence, for a phonon to contribute to EPC must have a finite value of $D_{(\vec{k}+\vec{q})j,\vec{k}i}$. This term also dictates the extent of EPC for different energy phonons with particular symmetry, which is also reflected in our DFT-based phonon calculation where the value of EPC is found to be different for different modes as tabulated in supplementary **table 4.1**. In the DFT calculation, the modes at 150.3 cm$^{-1}$ ( $E''$) and 221.1 cm$^{-1}$ ($E'$) and 228.3 cm$^{-1}$ ( $A'$) show a significantly large value of electron-phonon coupling as compared to other modes. Interestingly, these modes are in close proximity to P3 (152.4 cm$^{-1}$), P4 (240.7 cm$^{-1}$), and P5 (258.3 cm$^{-1}$) as observed in Raman spectra at 6K, which also shows considerable evidence of electron-phonon coupling effect reflected via the anomalous line-width of these modes as shown in **figure 4.8**.

Temperature-dependent Raman scattering is a sensitive technique to detect the presence of electron-phonon scattering. Generally, phonon mode frequency blueshifts with a decrease in temperature due anharmonic effect [215,316]. The observation of anomalous redshift in frequency for P2 and P3 mode below ~ 100K can be construed as evidence of the presence of electron-phonon coupling due to intra-band fluctuations in this semimetal [317]. The broadening in the phonon linewidth is related to the phonon lifetime. Normally it arises from the anharmonic decay of optical phonons to acoustic phonons and the linewidth is expected to scale with the Bose factor $n_B(\omega,T)$ i.e., decreases with decreasing temperatures. Its value depends on the interaction of phonon with various quasi-particles for e.g., phonon-phonon, electron-phonon interaction, etc. Non-monotonic behavior of FWHM of P3, P4, and P5 modes also suggests a strong electron-phonon coupling [318]. Such an anomalous nature of FWHM can be captured by a model based on phonon decaying into electron-hole pairs. In this case, linewidth behavior is dictated by the Fermi-Dirac occupation factor, $n_F(\omega,T)$. Such a phenomenological expression for linewidth is given as follows [296,319]:

$$\Gamma(T) = \Gamma_o + \eta \ [n_F(\hbar\omega_a, T) - n_F\{\hbar(\omega_a + \omega_p), T\}] \qquad \text{-- (4.6)}$$

here $\Gamma_o$ is a temperature-independent term that incorporates any anharmonic or electron-phonon contribution to linewidth at absolute zero. $\eta$ is an indicator of coupling strength and the available density of states for decay [314]. $\omega_p$ is the optical phonon frequency whereas $\omega_a$ is the energy difference between the electron's initial state and the Fermi energy in the phonon-mediated inter-band scattering. We have fitted the FWHM of P3 and P4 modes using equation 4.6 as shown in **figure 4.8** (Olive solid line). Phenomenologically it can be understood as, when $\omega_a \approx \omega_p$ it corresponds to the case where phonon decays into an electron-hole pair. As temperature increases phonon no longer decomposes into an electron-hole pair as the final state (hole) is already occupied hence the linewidth decreases with an increase in temperature. If $\omega_a \approx 0$, it indicates that the electronic states are already occupied at $T = 0K$, hence linewidth decreases monotonously with increasing temperature [319]. That is what is observed in the case of P3, P4, and P5 mode and the obtained value of $\omega_a$ is very close to zero. So, we find that equation 4.6 dictates well the behavior of linewidth throughout the temperature range.

### 4.3.4 Polarization-Dependent Phonons Analysis

To unveil the symmetry of the phonon modes we performed an angle-resolved polarized Raman scattering experiment in a configuration where we controlled the direction of polarization of the incident light ranging from $0^o$ to $360^o$ using a half wave retarder ($\lambda/2$ plate), whereas the analyzer principal axis has been kept fixed. The experiment is performed at two different temperatures i.e., room temperature (~300K) and 14K, and angular variation of intensities of modes are shown in **figure 4.11**. Within the semiclassical approach, the Raman scattering intensity from first-order phonon modes is related to the Raman tensor, polarization configuration of incident and scattered light which is given by equation 3.4 of chapter 3A.3.5. A schematic diagram of polarization vectors of incident and scattered light projected on a plane is shown in **figure 4.1(c)**. In the matrix form, incident and scattered light polarization direction vectors can be decomposed as: $\hat{e}_i = \begin{bmatrix} \cos(\alpha + \beta) & \sin(\alpha + \beta) & 0 \end{bmatrix}$ ; $\hat{e}_s = \begin{bmatrix} \cos(\alpha) & \sin(\alpha) & 0 \end{bmatrix}$, where '$\beta$' is the relative angle between '$\hat{e}_i$' and '$\hat{e}_s$' and '$\alpha$' is an angle of scattered light from the x-axis, when polarization unit vectors are projected in the x ($a$-axis) - y ($b$-axis) plane. The Raman tensors for various modes are summarized in **table 4.2**. The angular dependency of intensities of the Raman active modes can be written as:

$$I_{E'(x)} = \left| d \, \cos(2\alpha + \beta) \right|^2, \; I_{E'(y)} = \left| d \, \sin(2\alpha + \beta) \right|^2, \; I_{E''} = 0 \text{ and } I_{A_1'} = \left| a \, \cos(\beta) \right|^2 \quad \text{-- (4.7)}$$

Here $\alpha$ is an arbitrary angle from the *a*-axis and is kept constant. Therefore, without any loss of generality, it can be taken as zero, giving rise to the expression for the Raman intensity as $I_{E'(x)} = |\mathrm{d}\cos(\beta)|^2$ and $I_{E'(y)} = |\mathrm{d}\sin(\beta)|^2$.

The predicted angular dependence of the intensity of phonon modes for α-phase and experimentally observed P2-P5 and M1-M5 modes is shown in **figure 4.11** at 14K and room temperature (∼ 300K). Fitted spectra at different angles for room temperature is shown in **figure 4.12**. First, we will discuss the prominent modes i.e., P2-P5. We can see that at room temperature P2, P3, P4 shows $E'(x)$ or $A_1'$ kind of behavior whereas P5 shows elliptical type behavior. While at 14K, P2 mode exhibits nearly similar behavior as compared to that at room temperature. But one can notice a rotation in the major axis by ∼10°, and the area in the 3rd and 4th quadrants has decreased. At room temperature maxima of P3 is observed at ∼10° but at 14K it has shifted to 30°/160°. Whereas P4 and P5 show very distinct variations of intensity as compared to the room temperature. Change in the symmetry of the modes at different temperatures may be owing to a possible structural phase transition. Further, interestingly at room temperature mode M1- M5 appears only at certain angles, discussed in detail in the next section.

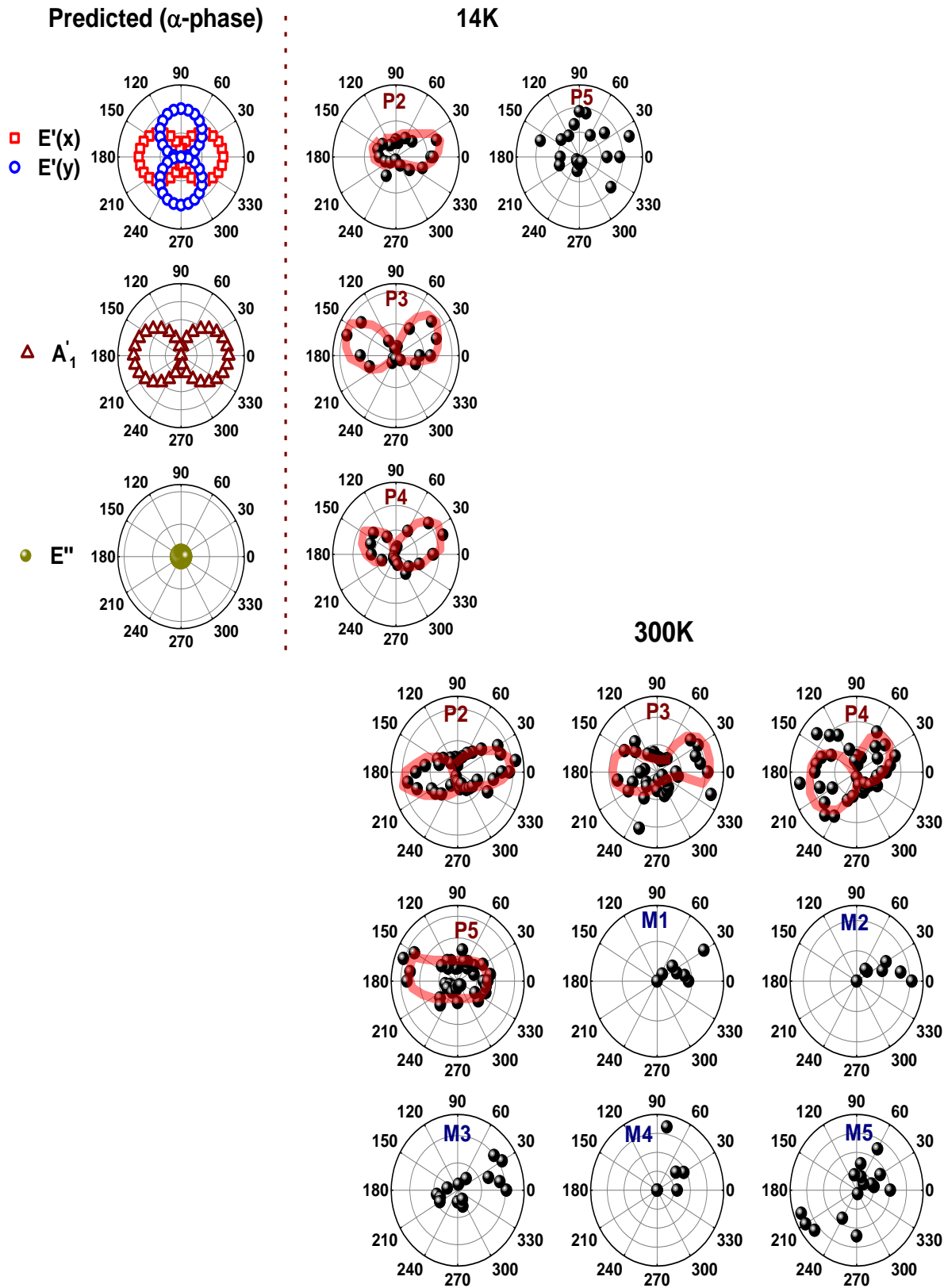

**Figure 4.11:** Polarization-dependent intensity variation of the phonon modes as per theoretical prediction for α-phase; experimentally observed phonons at 14K and 300K. The red solid line is a guide to the eye.

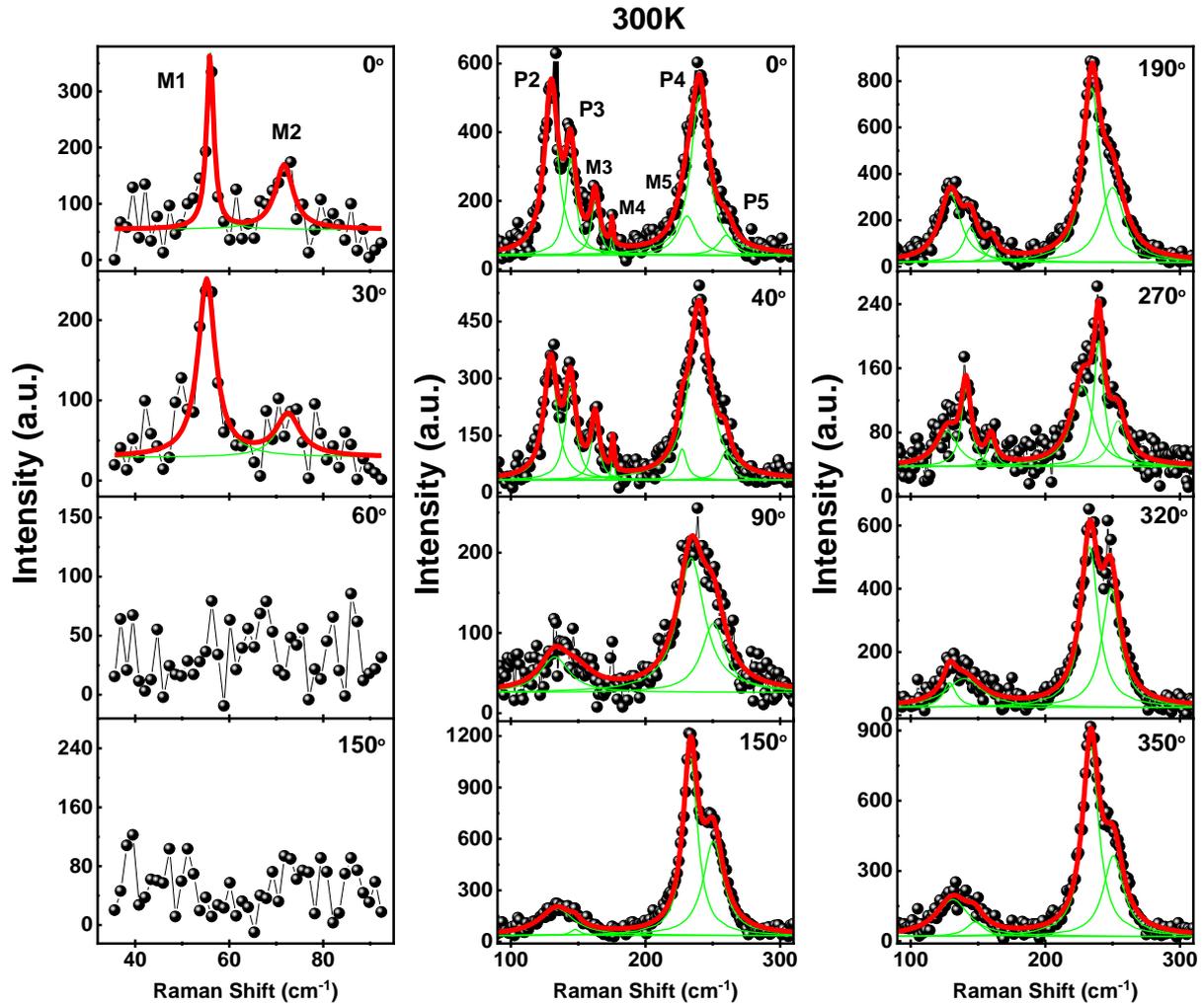

**Figure 4.12:** Lorentzian fitted raw spectra at different polarization angles at 300K.

### 4.3.5 Discussion

We will be focusing here only on the possible origin of weak modes M1-M5, which appears in the vicinity of 100K- 150K. Kung *et al.*, in their polarization-resolved Raman scattering and DFT-based study on a topological insulator $Bi_2Se_3$ reported the presence of four additional modes with lower intensity and energy than the bulk counterparts [320]. These modes appear due to c-axis lattice distortion and surface van der Waals gap expansion where these modes become Raman active due to the reduction of crystalline symmetry from $D_{3d}$ in bulk to $C_{3v}$ on the surface. Additionally, these modes were observed to have an asymmetry in the line shape i.e., Fano asymmetry, suggesting interference of electron-phonon coupling with the surface excitations [320]. Here a finite phonon density of states exists across the entire BZ and because of that surface modes decay into bulk phonons which reflects that the surface is not entirely separated from the bulk. A surface resonance can be expected with a lower energy profile than the bulk [320,321]. The concept of surface phonons was initially developed by Lifshitz and Rosenzweig [322-324] where the symmetry of the crystal breaks when we move from bulk to the surface. These modes are localized at the surface where dispersion can be very different than the bulk and phonon frequencies at the surface get consequently modified for $\Gamma$-point [324,325]. It can be considered as a perturbation of an infinite lattice to derive the surface modes from the bulk mode spectrum. Generally, these modes are only modified to a minimal value as compared to the bulk. However, the presence of a gap in the phonon density of states with large distortion can separate these modes from the bulk [326]. In the case of topological insulator $Bi_2Se_3$ surface effects are also reflected in both bulk and surface electronic band structure [321,328,329].

| Phase | Space group | Point group | Raman active modes | |
|-------|-------------|-------------|--------------------|--|
| α | P$\bar{6}$m2 | $D_{3h}$ | $A_1' + 3E' + E''$ | [306,327] |
| β | P6$_3$mc | $C_{6v}$ | $3A_1 + 4E_2 + 3E_1$ | [305] |

**Table 4.4:** Space group, point group and Raman active modes in different phases.

Recently, Murtaza *et al.*, reported a series of pressure-driven phase transitions in $PbTaSe_2$ up to 56 GPa, with space group P$\bar{6}$m2 (α-phase) at ~ 0 GPa, P6$_3$mc (β-phase) at ~0.5 GPa [279]. The P6$_3$mc structure can be obtained by doubling the unit cell of the P$\bar{6}$m2 structure along the c-axis and relocating the upper half of the unit cell by 1/3 along the diagonal of the basal plane [1/3(b-a)] [277]. Corresponding group theoretical prediction of Raman active modes for different phases are mentioned in **table 4.4**. In the α-phase, there are two bands crossing each near the Fermi level where it forms two nodal lines on the plane of $k_z = \pi/c_\alpha$ (where $c_\alpha$ is the c lattice constant for α-phase, etc.). Whereas for β-phase multiple types of band crossing has been observed, including type-II Weyl nodal lines, type-II Dirac points, and a twofold nodal surface, near the Fermi level [279]. There is a twofold nodal surface located on the plane of $k_z = \pi/c_\beta$ as a protection of $S_{2z}$ and T [330,331].

Generally, on decreasing the temperature a CDW phase is accompanied by the appearance of new Raman active modes which are denoted as amplitude or phase mode and are distinguished based on their dispersion characteristics [286,298,332,333]. The study of such modes in Raman spectra helps to comprehend the stability of CDW phases [334-336]. The disappearance of modes M1-M5 on decreasing the temperature is unlike a CDW-like transition. In addition to that the presence of Fano asymmetry in modes M1-M4 is a signature of electron-phonon coupling with an underlying electronic continuum of the same symmetry which is crucial to understanding the relaxation and scattering of surface state excitations. We observed an increase in asymmetry of modes M1-M4 with increasing temperature as in **figure 4.13**. An equally effective 'Slope method' can be utilized to calculate the qualitative Fano-asymmetric nature of the phonon modes as mentioned in Chapter 3A.3.7**.** As observed M1-M3 modes appear at ~100K and M4 appears at ~150K. In **figure 4.13** we have plotted the mode of the slope (solid orange line is a guide to the eye) in the temperature range of 100K-320K for modes M1-M4 for which we observed Fano asymmetry as shown in **figure 4.5**. On decreasing temperatures slope of mode M1 shows a linear decrease till ~ 200K before increasing slightly with a further decrease in the temperature till 100K. For M2 it decreases monotonously till ~200K and remains nearly constant till 100K. The slope of M3 decreases till ~200K and then remains nearly constant on further decreasing temperatures. Mode M4 slope decreases monotonously till ~150K. We observed an overall increase in asymmetry with increasing temperatures for M1-M4 modes. Such a behavior is expected as the topological nature is active in this phase. Increasing asymmetry corresponds to the increasing electron-phonon coupling with the topological surface excitations with increasing temperatures and these modes gain more intensity consequently. Such a behavior is expected as the topological nature is active in this phase.

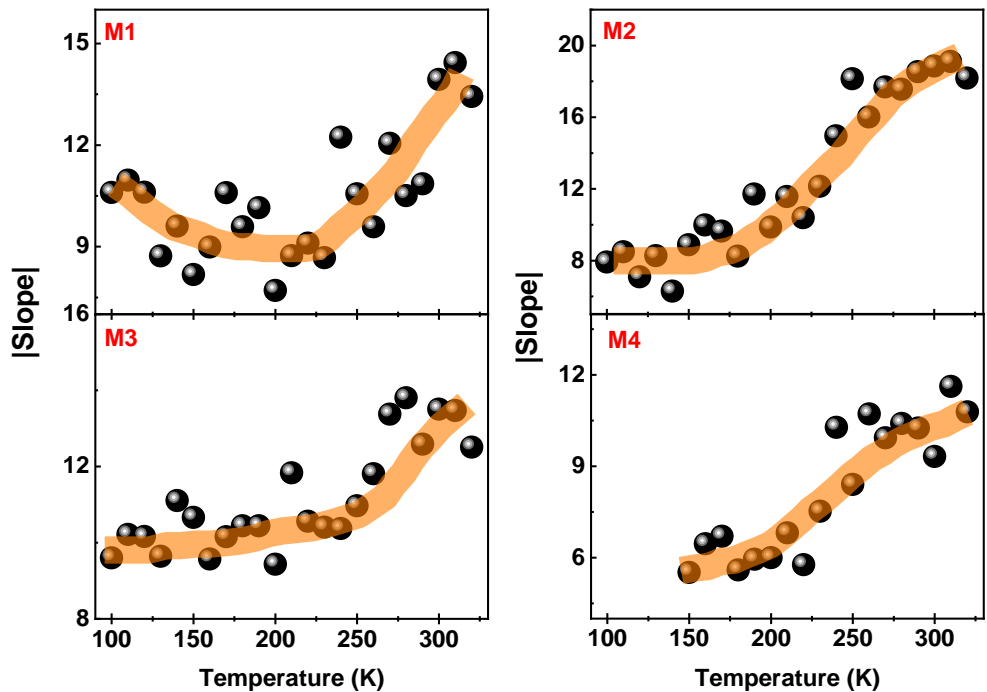

**Figure 4.13:** Temperature dependence of mode of slope for M1-M4 phonon modes, indicating the Fano -asymmetry. The solid orange line is a guide to the eye.

The fact that caught our attention is the appearance of modes i.e., M1-M5, and the discontinuities present in P2-P5 modes i.e., in frequencies, FWHM, and Intensities, occurs around the similar temperature range. We admit that further studies, both theoretical as well as experimental, especially thickness-dependent may be useful to probe the effect of local strain, lattice distortion, etc. on the surface phonons and unraveling their origin.

Based on the above different possible scenarios, discussed above and our Raman spectroscopic analysis, we assign these weak modes as surface phonon modes. Generally increasing pressure [337,338] or decreasing the temperature [172,215] causes a similar effect as in both cases a decrease in the bond length is observed which causes a blueshift in the phonon mode frequencies [339,340]. So, there might be a possibility that as we decrease the temperature it undergoes from α to β phase at $T_{\alpha \rightarrow \beta} \sim 150$ K. Due to this transition, the surface phonon spectrum is being modified because in the β-phase, the surface states are affected considerably as discussed above and consequently, we observe the disappearance of these weak modes M4-M5. Further, we observe another transition $T_{CDW+\beta} \sim 100K$, below this temperature there is a CCDW phase reported for 1H-TaSe$_2$. Such a transition in PbTaSe$_2$ could be due to the interplay of the underlying CCDW phase and surface topology of the material because of which M1-M3 also disappear. The possibility of structural transition is also supported by our SC-XRD measurements, where we find the appearance of new XRD peaks on decreasing temperatures. Interestingly mode M1-M5 appeared only at certain angles in our polarization-dependent analysis at room temperature. There might be a possibility that these modes are getting activated only at certain angles depending on the crystal symmetry at the surface and the topological nature of the states in a particular phase which couples with the surface lattice dynamics. We note that the structure is quite similar to the van der Waals heterostructures [341] and sensitive to the surface topology. Hence, it would be very interesting to explore this material in a low thickness regime especially down to a few layers.

## 4.4 Conclusion

In conclusion, we conducted a temperature-dependent SC-XRD to unveil the structural transition and inelastic light scattering Raman measurements (temperature and polarization dependent) on single crystals of PbTaSe$_2$. We observed discontinuities in temperature evolution of frequency, FWHM, and Intensity for prominent phonon modes which are consistent throughout the temperature range i.e., 6K – 320K (P2-P5). The polarization-dependent results for these modes show a change in symmetry at different temperatures (14K and 300K), which suggests a structural change. We observed the disappearance of some weak modes on decreasing temperature i.e., M1-M5. These modes vanish at around the temperatures where the discontinuities in prominent phonon modes arise i.e., ~100K (M1-M3) and ~150K (M4-M5). Most of these modes (M1-M4) exhibit Fano asymmetry in the line shape which is an indication of the presence of electron-phonon coupling. Further at room temperature the polarization dependence analysis of these modes reveals that M1- M5 modes appear at certain angles only. Hence based on our SC-XRD and Raman analysis we suggest that there is a thermally induced structural phase transition happening which is possibly causing it to undergo from α to β phase ~ 150K. The transition ~100K could be owing to the interplay of the underlying CCDW phase of $1H\text{-}TaSe_2$ and β phase. We propose the modes (M1-M5) that are disappearing on decreasing temperature are surface phonon modes. Instead of speculations, however, the precise understanding of the proposed surface phonon modes still needs to be understood. So, we would like to keep this an open problem for the condensed matter community for now and hope that our result will motivate more research groups for further theoretical and experimental investigation.

# Chapter 5

# Hidden Quantum State and Signature of Mott Transition in 2D 1T-TaS$_2$

## 5.1 Introduction

Quasi-two-dimensional (2D) transition metal dichalcogenides (TMDs) have been a center of interest in the scientific community for the past several decades due to their rich properties. Group V TMDs (V, Nb, Ta, and Tb), specifically have been known to feature charge density waves (CDWs) and superconductivity. CDWs and superconductivity are two such electronic ground states that often compete and/or coexist in materials such as Cuprates and TMDs. Such an intriguing interplay has been vastly studied theoretically as well as experimentally, but still, it has perplexed scientists in the complete understanding of such a complex phenomenon [342-344]. The CDW transitions in these materials occur at relatively higher temperatures which is a favorable feature for investigating hysteresis effects and is a defining aspect for potential applications such as data storage [345], information processing [346], or voltage-controlled oscillators [347].

Generally, the factors that dictate CDW phase transitions in two-dimensional and three-dimensional systems are Fermi surface nesting [348,349], electron-phonon coupling [350,351], and excitonic insulators [352,353]. In 1930, Rudolf Peierls first proposed a new kind of phase transition wherein the one-dimensional metals turn insulators resulting in a gap opening at the Fermi-level ($\pm K_F$) owing to the inbuilt lattice instability at low temperatures [134]. Later in 1954, Herbert Fröhlich proposed a detailed microscopic theory of 1-D superconductivity which predicted CDW formation and resulted in collective charge transport below transition temperatures [135] . In the CDW state the periodic lattice instability is accompanied by fluctuations in the electron density which can be described phenomenologically as $\rho(\vec{r}) = \rho_o + \rho_A \cos(2\vec{K}_F.\vec{r} + \theta)$ below the transition; here $\rho_o$ is the undistorted lattice electron density, $\rho_A$ is the amplitude of the modulations and $\theta$ is the phase of the CDW that determines its position relative to the underlying lattice. It can be characterized by an order parameter $\xi = \Delta e^{i\theta}$, the magnitude '$\Delta$' dictates the electronic energy gap and atomic displacement. The modulation in $\theta$ and $\Delta$ give rise to collective excitations as phasons and amplitudons, respectively; and have been observed in Raman spectroscopic measurements [354]. In the commensurate phase (C-CDW) the altered periodic phase lattice constant is an integer multiple of the undistorted one, where $\cos\theta = \pm 1$ and also known as the pinned phase. Brillouin zone folding effects can be seen in this phase as the emergence of new phonon modes at $\Gamma$ - point. On the other hand, if it is not an integer multiple then it is known as incommensurate phase (IC-CDW). Here, the charge density wave is independent of the choice of phase or position and thus could move hence the Peierls gap in the k-space which induces an electric current that is proportional to $\partial\theta/\partial t$. This phase lacks translational symmetry and the vibrational modes are no longer plane-waves, hence the corresponding relaxation of the phonon momentum conservation mimics that of amorphous materials with broad Raman peaks [355,356]. In a nearly or partially commensurate phase (NC-CDW), the IC-CDW and C-CDW coexist [357,358].

It has been a matter of debate among the condensed matter community regarding the nature of the low-temperature insulating state. Some consider that, 1T-TaS$_2$ is a correlation-driven Mott insulator, owing to an imperfect nesting in 2D the CDW gaps out only a certain part of the Fermi surface which leaves behind a metallic state that often becomes superconducting [359]. In a standard description of a Mott insulator where an anti-ferromagnetic (AFM) ground state is formed due to the exchange coupling. However, there are several studies that have advocated it to be a band insulator. Wang *et al.* [360] in their angle-resolved photo emission spectroscopic and X-ray diffraction investigation found indication of band insulating phase at low temperature upon cooling. The Mott-insulating state exists only in a narrow temperature window upon heating. Such a behavior is dictated by the relative strength of onsite Coulomb repulsion and interlayer hopping of electrons, which plays a crucial role in dictating the material's electronic properties. Another recent study has reported a dualistic nature of the insulating state due to the effect of dimensionality and interlayer coupling on varying thickness [361]. Monolayer 1T-TaS$_2$ is reported as a Mott-insulator; however, a band-insulator in a bulk crystal. The band-insulating behavior of bulk has been studied both experimentally and theoretically [362-364], which is attributed to the on-top David-stars stacking configurations.

As far as the magnetic properties of 1T-TaS$_2$ are concerned there is no such ordering of spins has been reported. Magnetic susceptibility drops at CDW transitions with decreasing temperatures but remains nearly constant below 200K which is unlike Curie Weiss behavior [241,359,365]. It has been advocated that this material shows a quantum spin liquid state, either a fully gapped Z$_2$ spin liquid or a Dirac spin liquid [241]. Upon cooling 1T-TaS$_2$ undergoes a series of different CDW transitions and at each transition, the lattice goes under reconstruction and is reflected in an abrupt increase in its electrical responses. The transition from a normal undistorted metallic phase to a metallic incommensurate (IC-CDW) phase occurs at T$_{IC}$ ~ 550K, followed by nearly commensurate (NC-CDW) phase at T$_{NC}$ ~ 350K and finally, to an insulating completely commensurate state (C-CDW) which triggers in a temperature range of 180K $\leq T_C \leq$ 230K [363] and occurs at $T$ ~ 230K on heating [167]. C-CDW phase is well known for the formation of a $\sqrt{13}$ $X$ $\sqrt{13}$ structure, known as 'Stars of David' where the sites of stars move inwards towards the middle site. It forms a triangular lattice and the unit cell contains 13 Ta sites. As the valency of Ta is 4+ ($d^1$), and each Ta site has a single 5d electron. Band theory predicts the metallic nature of the ground state, but in contrast, it is insulating which indicates towards correlation-driven Mott insulator [366-368]. Recent reports have shown the presence of another hidden quantum CDW state below $T_H \sim 80K$ which has revived the interest in further investigation of this material [369,370]. In particular, a step like rise in the resistivity is reported between $\sim 50-100K$, which is much lower than the transition temperature between C-CDW and NC-CDW states [32]. It is tentatively attributed to the different possible scenarios such as Mott transitions or interlayer re-ordering of the stacking structure.

The study of the dimensionality effects and interlayer coupling in different CDW phases has been an area of deep interest [371,372]. The electronic structures of bulk CDW phases strongly depend on interlayer coupling [364,373,374]. The occupied states of $TaS_2$ are delocalized by the interlayer interactions and form a quasi-1D metallic state in bulk form [375]. It has been reported that in ultrathin flakes the C-CDW phase becomes more conducting whereas the NC-CDW phase becomes more metastable [376]. Further, it has been advocated that bulk and surface CDW transitions are similar for thick crystals, which means the interlayer coupling strength decreases in ultrathin limits and decouples the surface layers [373,377,378].

Clearly, 1T-$TaS_2$ provides a rich theoretical playground where still a lot to unveil. Raman spectroscopy has been proven to be an excellent probe for studying CDWs, superconductivity, and quantum spin liquids [58,69,166,379]. The presence of soft, amplitude, phase modes, and discontinuities in the phonon dynamics in Raman spectra, provides a signature of CDW transitions. The quantum phenomenon becomes strong as we decrease the dimensionality of the material and consequently becomes an important tuning factor for manipulating the properties of such materials. Hence, we decided to carry out an in-depth thickness-dependent Raman investigation on 1T-$TaS_2$.

## 5.2 Results and Discussion

## 5.2.1 Crystal Structure Details and Group Theory

1T-TaS$_2$ contains the planes of tantalum (Ta) atoms, which are surrounded by Sulphur (S) atoms in an octahedral arrangement as shown in **figure 5.1 (a, b)**. At ambient pressure and high temperatures, it possesses a trigonal symmetry [ P$\bar{3}$m1 (#164); D$_{3d}$ (-3m)]. The group theoretical prediction of the symmetry of modes is given as $\Gamma = A_g + E_g + 2E_u + 2A_{2u}$ where $\Gamma_{Raman} = A_g + E_g$ are Raman active while $\Gamma_{infrared} = E_u + A_{2u}$ are infrared active modes. Here, E-kind modes are doubly degenerate and correspond to in-plane atomic vibrations whereas A-kind of modes correspond to out-of-plane vibrations [380]. Raman tensor, Wyckoff positions, and corresponding phonon modes are summarized in **table 5.1**.

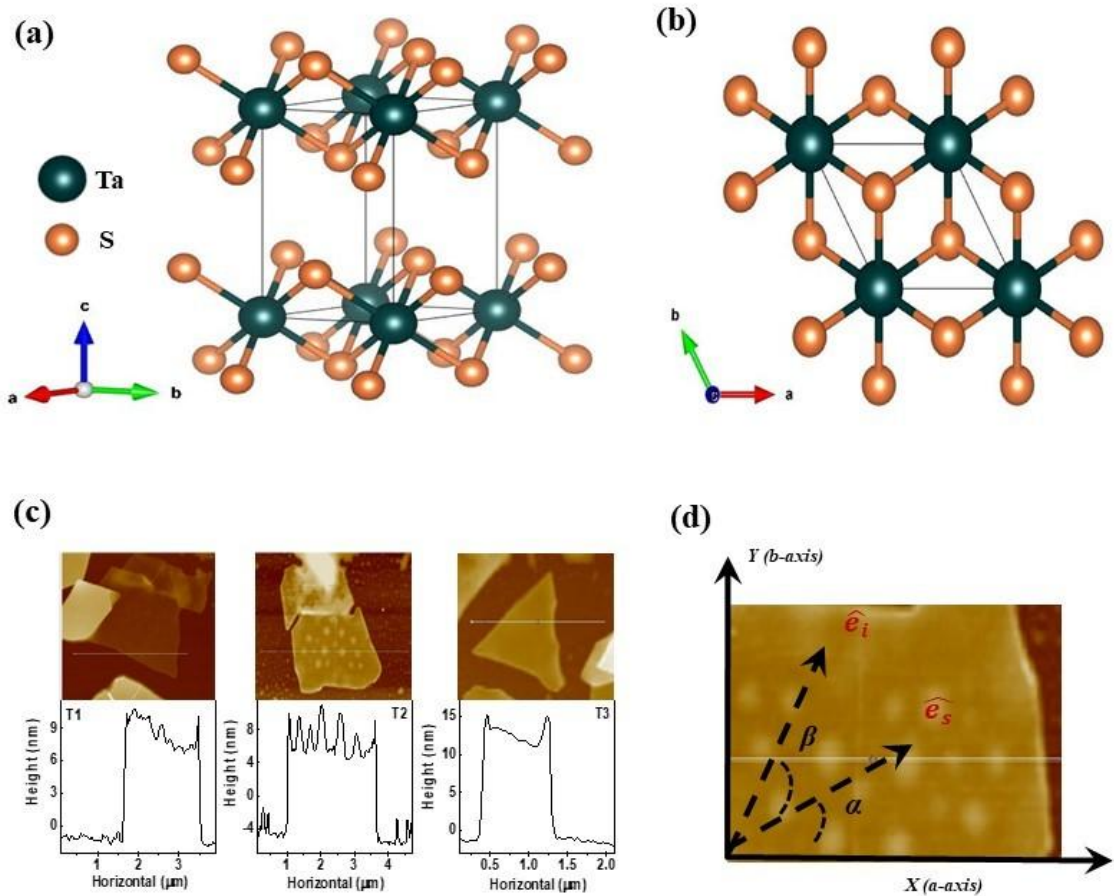

**Figure 5.1:** Shows structure of the undistorted phase of 1T-TaS$_2$ in **(a)** standard orientation **(b)** along c-axis **(c)** AFM Image of T1, T2 and T3 and **(d)** Plane projection of polarization direction of incident and scattered light.

| Atoms | Wyckoff site | $\Gamma$-point mode decomposition | Raman Tensors |
|---|---|---|---|
| **Ta** | **1a** | $A_{2u}\,(I.R) + E_u\,(I.R)$ | $A_g = \begin{pmatrix} a & 0 & 0 \\ 0 & a & 0 \\ 0 & 0 & b \end{pmatrix}$ |
| **S** | **2d** | $A_{2u}\,(I.R) + E_u\,(I.R) + A_g(R) + E_g(R)$ | $E_{g,1} = \begin{pmatrix} c & 0 & 0 \\ 0 & -c & d \\ 0 & d & 0 \end{pmatrix}; E_{g,2} = \begin{pmatrix} 0 & -c & -d \\ -c & 0 & 0 \\ -d & 0 & 0 \end{pmatrix}$ |
| $\Gamma = A_g + 2A_{2u} + 2E_u + E_g$ | | $\Gamma_{\text{Raman}} = A_g + E_g$ | $\Gamma_{\text{Infrared}} = 2A_{2u} + 2E_u$ |

**Table 5.1:** Wyckoff positions, irreducible representations of the phonon modes of trigonal [#164; P$\bar{3}$m1; $D_{3d}$ (-3m)] 1T-TaS$_2$ at the gamma point and Raman Tensors of Raman active phonon modes.

In the completely commensurate phase where Ta atoms form star-of-David clusters. Uchida *et al.*, [381] reported that in this phase the new unit-cell is triclinic having $C_i^1$ symmetry and contains 39 atoms which means a total of 114 vibrational modes (57 Raman active + 57 IR-active). Polarization analysis reveals $A_g$, $E_g$ symmetry of the modes in this phase, which indicates a possibility of triagonal or hexagonal symmetry [285]. Further, it was advocated that due to weak coupling between layers in the C-CDW phase, it can even be assigned the symmetry of a single layer i.e., $C_{3i}$, which means a total of 117 modes given as: $\Gamma_{\text{C-CDW}} = 19\,A_g + 19\,E_g + 20\,E_u + 20\,A_{2u}$ [381].

## 5.2.2 Experimental and Computational Details

Raman spectroscopic technique with configuration as mentioned in Chapter 2 is utilized to probe 1T-TaS$_2$ with a 532nm (2.33 eV) laser, on varying temperatures over a range of 4K-330K. The experiment is performed on three flakes of thickness $\sim$ 8.6 nm (T1), $\sim$ 10.4 nm (T2) and $\sim$ 12.5 nm (T3). These flakes are mechanically exfoliated using the scotch-tape method and placed on a SiO$_2$/Si substrate. The AFM image of these flakes is shown in **figure 5.1(c)**. Further polarization-dependent study is done in a configuration where the incident light polarization direction is rotated keeping the analyzer fixed (refer to Chapter 2 Figure 2.8 and 2.10) at 4K and 230K for flake T2. A pictorial representation of plane projection of this configuration is shown in **figure 5.1 (d)**.

Structural optimization and Zone-centered phonon frequencies were calculated utilizing a plane-wave approach as implemented in the QUANTUM ESPRESSO package [174]. Calculations are done for 1T undistorted phase. The linear response method within the Density Functional Perturbation Theory (DFPT) is used to get a dynamical matrix. Ultrasoft pseudopotentials (PBESOL) are used as an exchange-correlation functional with non-linear core correction. The Kinetic-energy cutoff of 50 Ry and charge-density cutoff of 400 Ry were used. The Monkhorst-pack scheme with 15 x 15 x 15 k-point dense mesh is used for the numerical integration of the Brillouin zone.

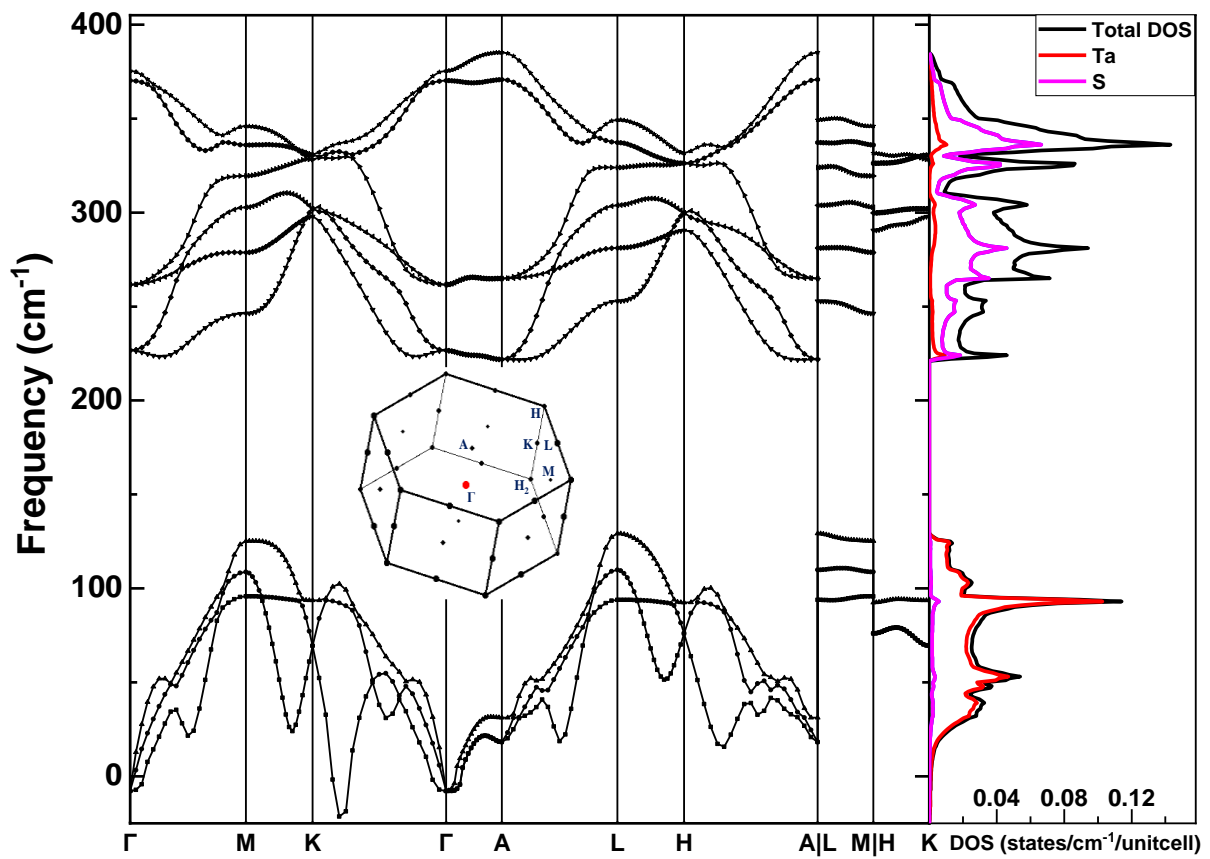

**Figure 5.2:** Phonon dispersion and DOS for 1T undistorted phase.

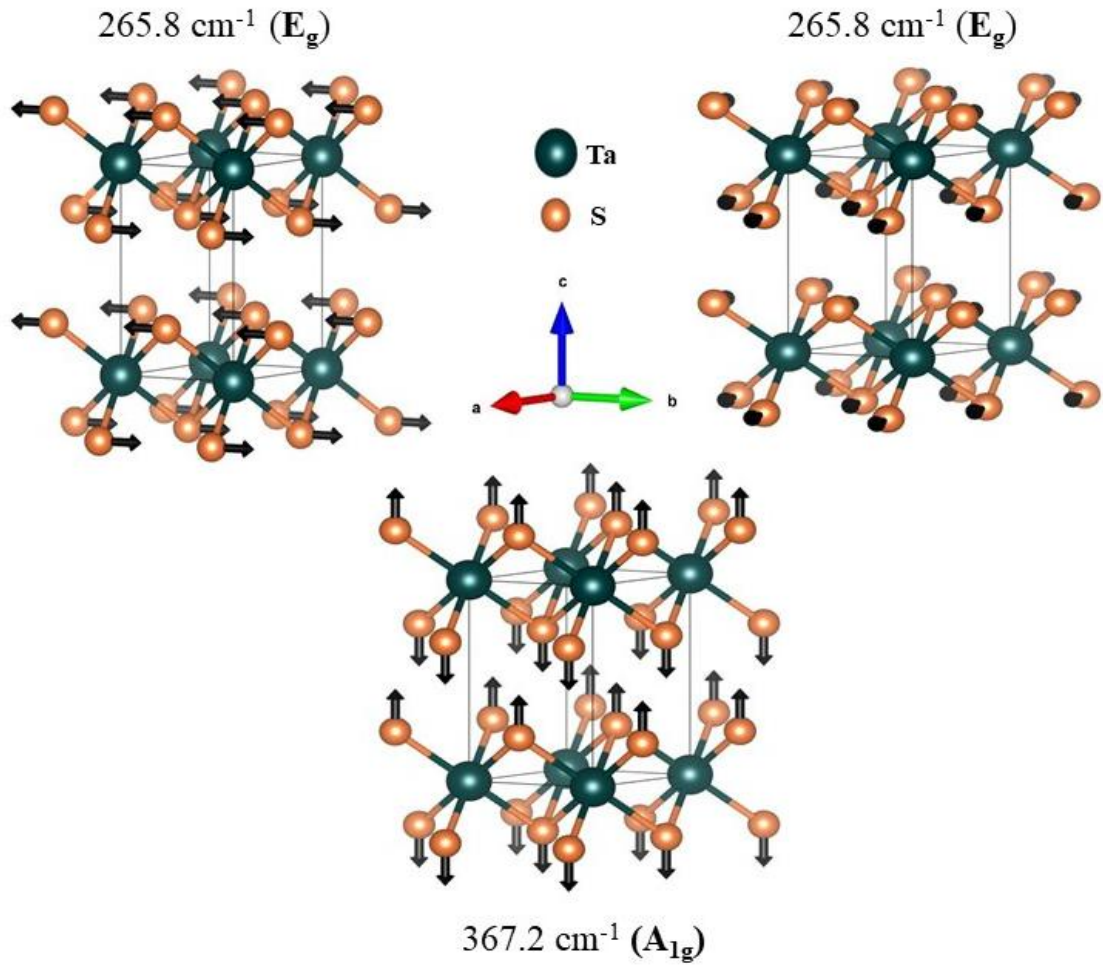

**Figure 5.3:** Pictorial representation of Raman active modes atomic displacement at the $\Gamma$-point in 1T undistorted phase.

The calculated phonon dispersion curve along the high symmetry points ($\Gamma - M - K - \Gamma - A - L - H - A \mid L\,M \mid H$ K) and phonon density of states (Partial and total) are also plotted for this phase as shown in **figure 5.2**. A pictorial representation of atomic displacement of Raman active phonon modes at $\Gamma$-point is shown in **figure 5.3** and obtained phonon frequencies, and optical activity at $\Gamma$-point are tabulated in Table 5.2 which are also consistent with the previous calculation done by Oliver *et al.* [382].

| Modes # (Symmetry) | $\omega_{DFT}$ (cm$^{-1}$) | Optical Activity |
|---|---|---|
| 1 (A$_{2u}$) | -7.8 | I |
| 2 ( E$_u$ ) | -7.6 | I |
| 3 ( E$_u$ ) | -7.6 | I |
| 4 ( E$_u$ ) | 229.4 | I |
| 5 ( E$_u$ ) | 229.4 | I |
| 6 ( E$_g$ ) | 265.8 | R |
| 7 ( E$_g$ ) | 265.8 | R |
| 8 (A$_{1g}$) | 367.2 | R |
| 9 (A$_{2u}$) | 375.5 | I |

**Table 5.2:** DFT calculation for D$_{3d}$ point group (Bulk).

## 5.2.3 Temperature Evolution of the Phonon Modes

We observed a total of 25 modes and a broad two phonon mode (S$_{2\text{-ph}}$) at lowest recorded temperature i.e., 4K, for all the flakes (T1, T2 and T3). **figure 5.4** shows comparison of raw spectra for T1, T2 and T3 at (a) 4K along with mode labels S1-S25, S$_{2\text{-ph}}$, (b) 200K and (c) 330K. Temperature evolution of raw spectra for T1, T2 and T3 flakes is shown in **figure 5.5**. We have fitted the Raman spectra at different temperatures with Lorentzian function while S$_{2\text{-ph}}$ is fitted with a Gaussian and extracted corresponding evolution of phonon self-energy parameters (frequency, linewidth) and intensity with temperature. Interestingly, we also observed a peak at ∼ 5 cm$^{-1}$ as shown in the shaded orange region in **figure 5.5**. It emerges and gains intensity with an increase in temperature for all thicknesses. Its low energy suggests that it may have its origin due to interlayer interactions as sheer or breathing modes [383].

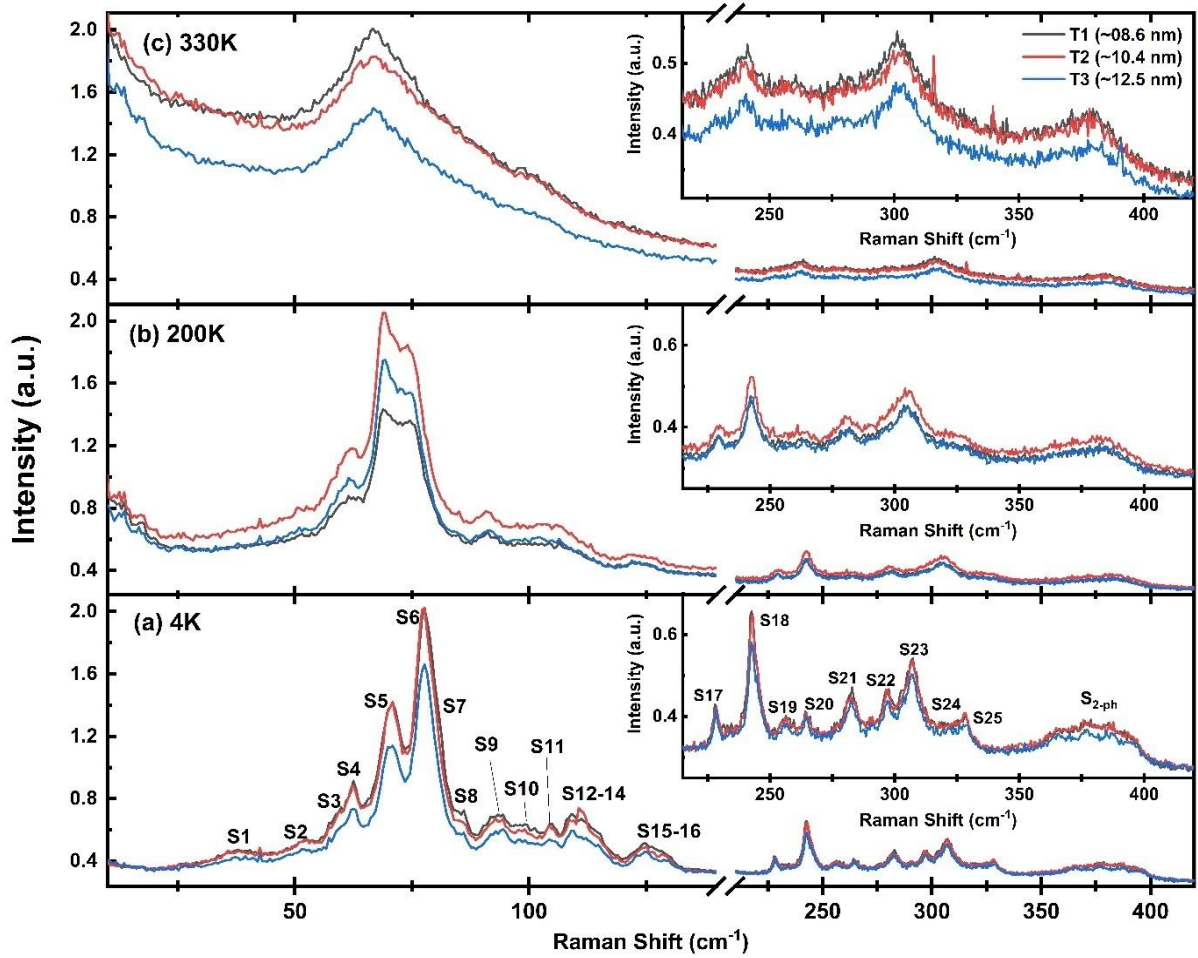

**Figure 5.4:** Raman Spectra for T1, T2 and T3 at 4K, 200K and 330K along with mode labels. Surprisingly, it disappears with lowering the temperature, suggesting other possible origins of this ultra-low frequency mode. We suspect it may be arising due to CDW effects in different phases and further detailed study is required to unveil its origin. Experimentally observed Raman active phonon modes for different thicknesses at 4K are mentioned in **Table 5.3**.

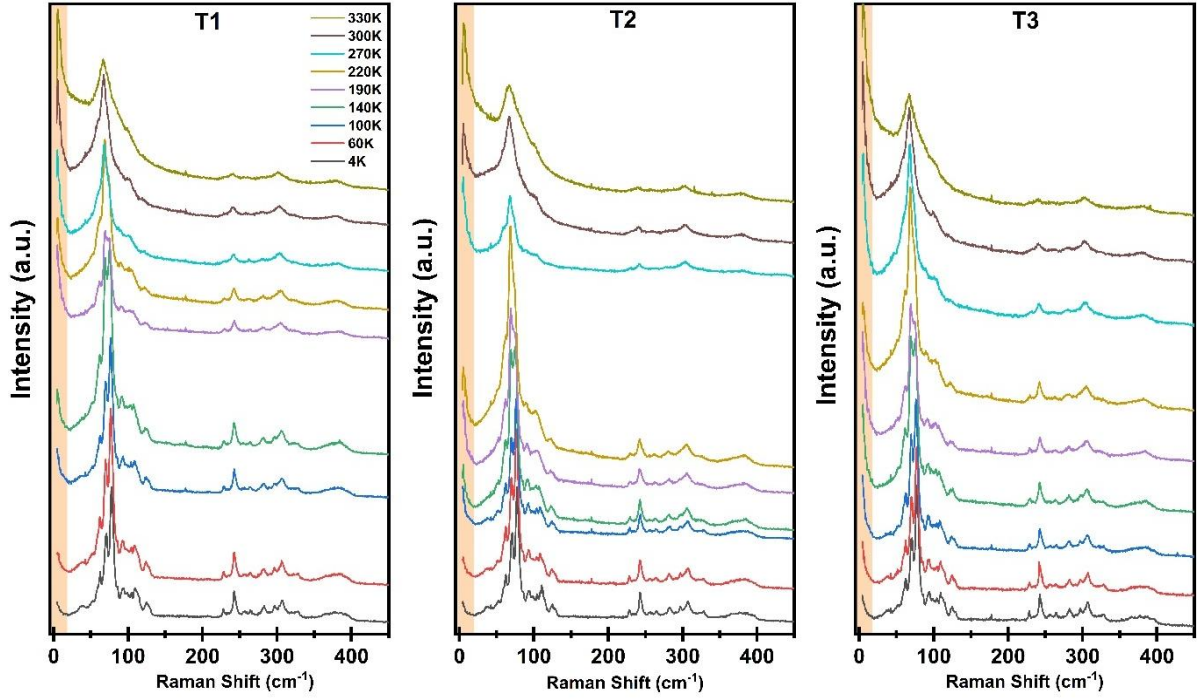

**Figure 5.5:** Temperature evolution of the Raman Spectra for T1, T2, and T3. The shaded orange area shows a low-frequency peak.

| Modes # | $\omega_{EXP.}$ T1 | $\omega_{EXP.}$ T2 | $\omega_{EXP.}$ T3 | Modes # | $\omega_{EXP.}$ T1 | $\omega_{EXP.}$ T2 | $\omega_{EXP.}$ T3 |
|---|---|---|---|---|---|---|---|
| S1 | 38.3 ± 0.3 | 38.0 ± 0.3 | 38.3 ± 0.3 | S15 ($A_g$) | 124.7 ± 0.2 | 124.6 ± 0.1 | 124.8 ± 0.1 |
| S2 ($E_g$) | 51.5 ± 0.3 | 51.7 ± 0.3 | 51.3 ± 0.3 | S16 | 128.6 ± 0.2 | 128.9 ± 0.2 | 129.4 ± 0.2 |
| S3 ($E_g$) | 59.1 ± 0.2 | 58.88 ± 0.2 | 59.5 ± 0.3 | S17 ($A_g$) | 228.4 ± 0.1 | 228.6 ± 0.1 | 228.4 ± 0.1 |
| S4 ($E_g$) | 62.5 ± 0.1 | 62.3 ± 0.1 | 62.5 ± 0.1 | S18 ($E_g$) | 242.9 ± 0.0 | 243.0 ± 0.0 | 243.1 ± 0.0 |
| S5 ($A_g$) | 70.4 ± 0.0 | 70.4 ± 0.0 | 70.3 ± 0.0 | S19 ($E_g$) | 256.3 ± 0.3 | 256.2 ± 0.3 | 256.2 ± 0.3 |
| S6 ($A_g$) | 77.4 ± 0.0 | 77.2 ± 0.0 | 77.2 ± 0.1 | S20 | 264.9 ± 0.3 | 264.9 ± 0.2 | 264.8 ± 0.2 |
| S7 ($A_g$) | 79.8 ± 0.1 | 79.4 ± 0.1 | 78.9 ± 0.4 | S21 ($E_g$) | 282.4 ± 0.1 | 282.5 ± 0.1 | 282.4 ± 0.1 |
| S8 | 85.9 ± 0.1 | 85.9 ± 0.1 | 85.9 ± 0.1 | S22 | 296.8 ± 0.1 | 296.9 ± 0.1 | 296.9 ± 0.2 |
| S9 ($E_g$) | 93.0 ± 0.1 | 93.1 ± 0.1 | 93.5 ± 0.2 | S23 ($A_g$) | 306.8 ± 0.1 | 306.9 ± 0.1 | 306.8 ± 0.1 |
| S10 ($A_g$) | 99.5 ± 0.2 | 99.2 ± 0.2 | 99.6 ± 0.4 | S24 ($A_g$) | 322.2 ± 0.9 | 322.9 ± 0.8 | 322.8 ± 0.9 |
| S11 ($A_g$) | 104.7 ± 0.1 | 104.6 ± 0.1 | 104.7 ± 0.2 | S25 ($A_g$) | 328.3 ± 0.2 | 328.2 ± 0.1 | 328.5 ± 0.2 |
| S12 ($A_g$) | 108.8 ± 0.2 | 108.0 ± 0.1 | 109.1 ± 0.1 | S2-ph($A_g$) | 379.9 ± 0.2 | 380.1 ± 0.2 | 379.6 ± 0.3 |
| S13 ($A_g$) | 11.6 ± 0.3 | 111.2 ± 0.1 | 112.05 ± 0.3 | | | | |
| S14 ($A_g$) | 114.7 ± 0.4 | 114.7 ± 0.2 | 115.1 ± 0.3 | | | | |

**Table 5.3:** Experimentally observed modes frequency ($\omega_{EXP}$) in cm$^{-1}$ at 4K for different thickness i.e., T1, T2 and T3 and their tentative symmetry assignment.

A thickness-dependent comparison of temperature evolution of phonon frequency and line width for some modes (S5, S6, S9, S18, S21, and S23) is shown in **figure 5.6** and for other modes (S11, S12, S14, S15, S17, and S22) is shown in **figure 5.7**. In **figure 5.6** we clearly observed CDW transition in the temperature evolution of the phonon modes shown by the dashed orange line for both frequency and FWHM at ~ 210K. Solid yellow and cyan line shows anharmonic and amplitude mode fit which is discussed ahead. In addition to that, we observed the signature of some other hidden quantum state transition at ~ 80K as clear from **figure 5.6** (inset) shown in the dashed red line and **figure 5.7**. A clear change in slope is observed in the mode frequencies and line widths for the most prominent modes S5 and S6.

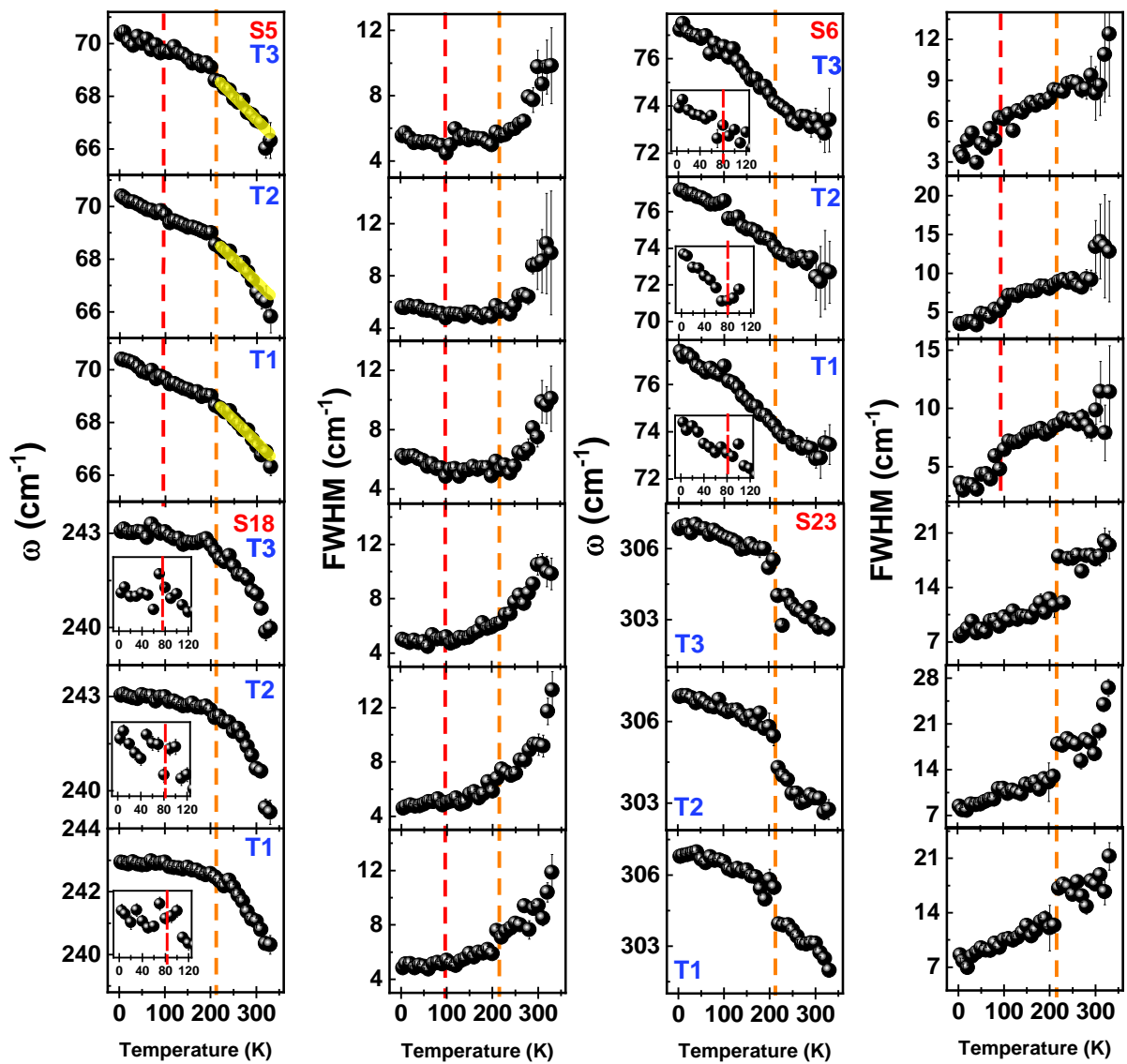

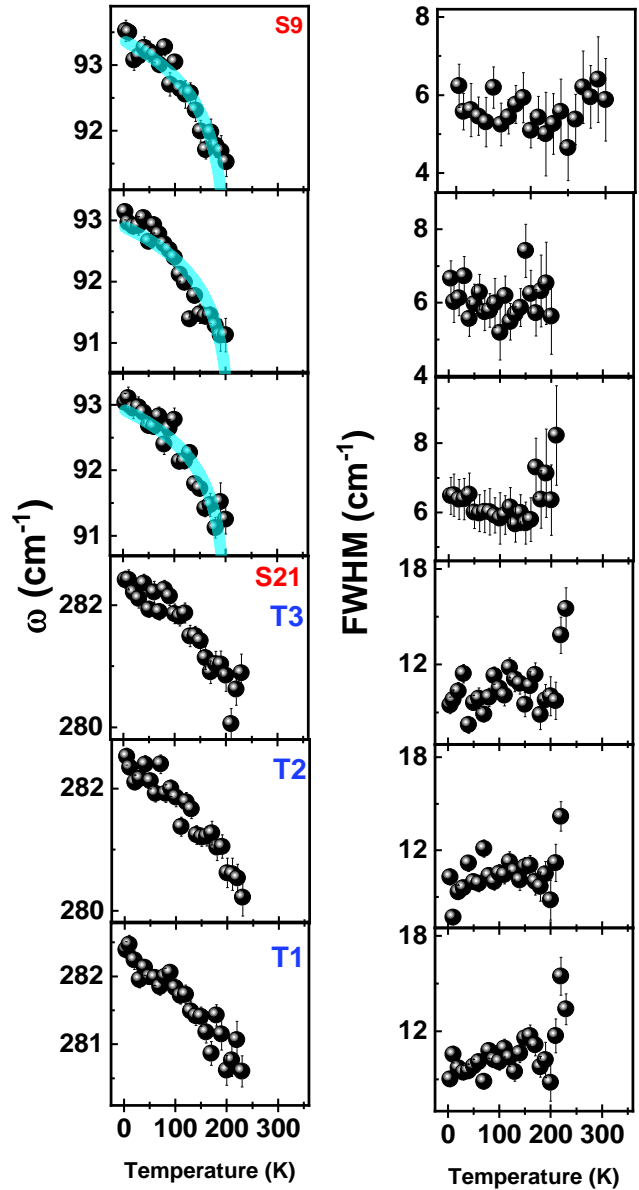

**Figure 5.6.** (Top and bottom panels) Temperature dependent frequency and linewidth for T1, T2 and T3 of some of the modes i.e., S5, S6, S9, S18, S21 and S23. The yellow and cyan solid line shows fit as mentioned in the text. The red and orange dashed line shows $T_H$ and $T_{CDW}$ transition temperatures.

The possible reason for the hidden phase will be discussed in later section. The Hamiltonian for a typical CDW systems at finite temperatures can be formulated as follows:

$$H = H_{elc.} + H_{lat.} + H_{elc.+lat.} + H_{anh.}$$ here $H_{elc.} = \sum_k E(k)\, c^\dagger(k)\, c(k)$ represents the electronic contribution, whereas $H_{lat.} = \sum_k \hbar\omega(k)\, b^\dagger(k)\, b(k)$ is a contribution from the lattice, the term,

$$H_{elc.+lat.} = \sum_{k+q} g(q)\, c^\dagger(k+q)\, c(k)[b^\dagger(q) + b^\dagger(-q)]$$ represents electron-lattice interaction term and

$H_{anh.}$ comes into the picture due to cubic and quartic anharmonic terms present in the lattice potential. $c\,(k)$ and $b\,(k)$ are electron and phonon operators;

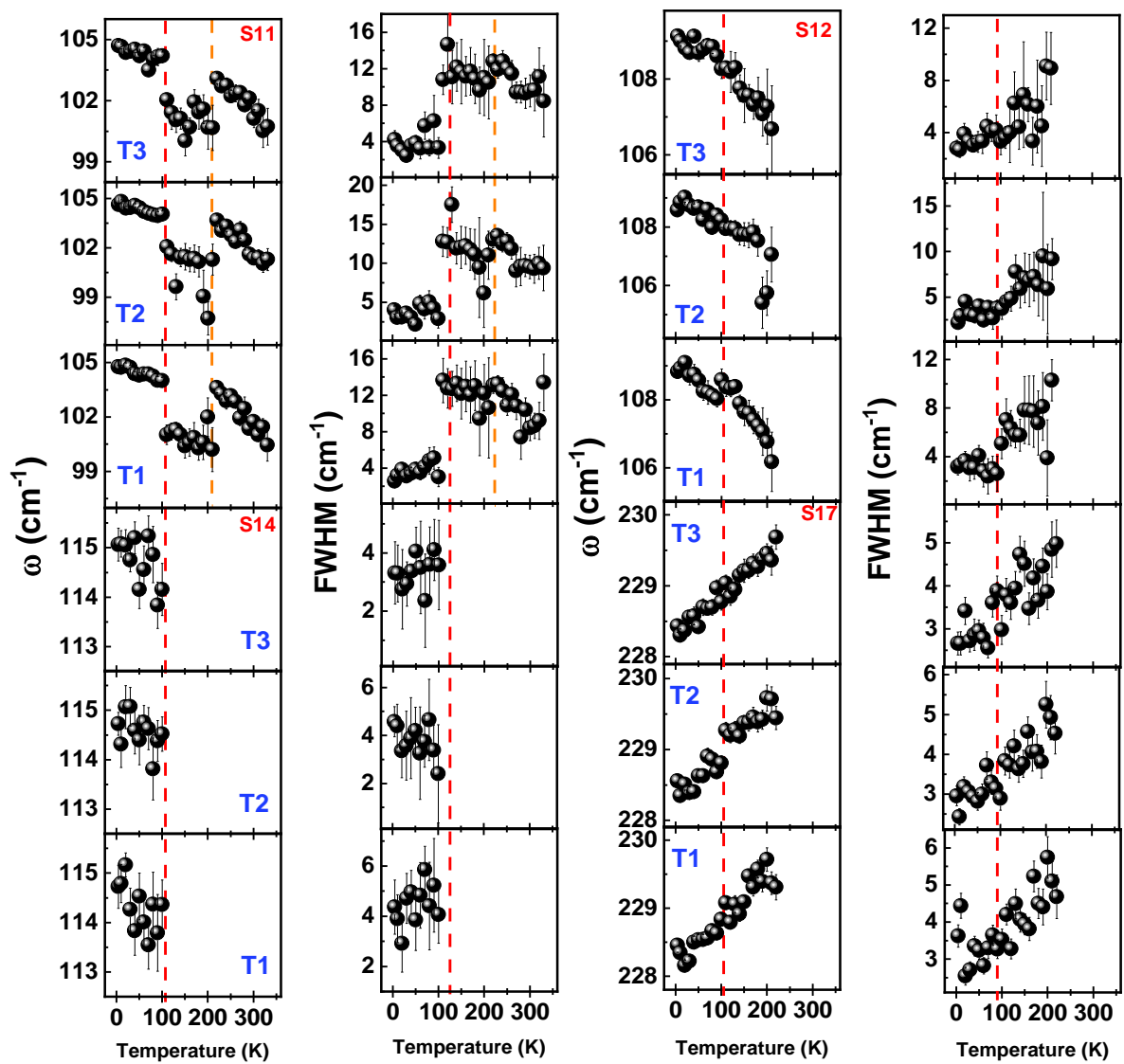

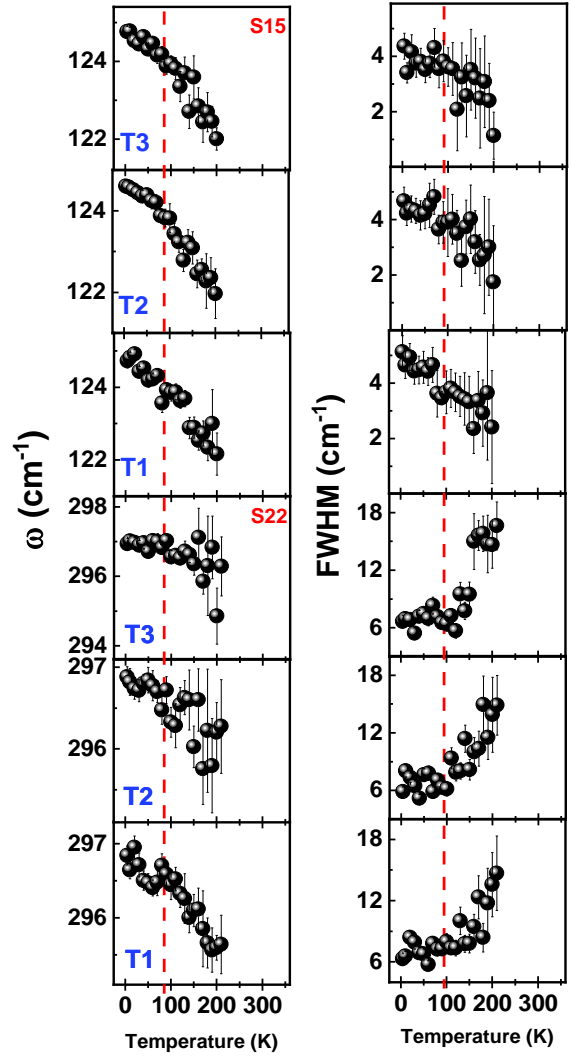

**Figure 5.7:** (Top and bottom panels) Temperature-dependent frequency and linewidth variation for T1, T2, and T3 of some of the modes S11, S12, S14, S15, S17, and S22. The red and orange dashed line shows $T_H$ and $T_{CDW}$ transition.

$E(k)$ and $\omega(k)$ are their respective energies; $g(q)$ is a measure of electron-phonon coupling [384]. First, we will discuss the effect of anharmonicity here. The effect of anharmonicity on frequency and FWHM comes into the picture due to the presence of anharmonic terms in the lattice potential as also discussed in chapter 3A.3.3 and chapter 4.3.2.

We observed a significant anharmonic effect in the temperature range of ~ 230K-330K. According to the anharmonic model, generally frequency of a phonon mode increases with lowering the temperature. We fitted frequency for modes S5, S6, and S18 for T1, T2, and T3 using a three-phonon model and equation 3.1 (shown for only S5 in **figure 5.6** as a yellow solid line). The extracted parameters are mentioned in the **table 5.4**.

| Modes | T1 | | T2 | | T3 | |
|---|---|---|---|---|---|---|
| # | $\omega_o$ | A | $\omega_o$ | A | $\omega_o$ | A |
| S5 | 72.4 ± 0.3 | -0.4 ± 0.0 | 72.3 ± 0.4 | -0.4 ± 0.1 | 72.5 ± 0.3 | -0.5 ± 0.0 |
| S6 | 76.3 ± 0.3 | -0.3 ± 0.0 | 75.5 ± 0.4 | -0.2 ± 0.0 | 76.6 ± 0.5 | -0.3 ± 0.1 |
| S18 | 246.8 ± 0.5 | -1.6 ± 0.2 | 247.2 ± 0.6 | -1.8 ± 0.2 | 246.3 ± 0.6 | -1.5 ± 0.2 |

**Table 5.4:** Value of anharmonic fit parameters for different thickness.

One of the important features of CDW transitions is the presence of the amplitude modes which softens and broadens on warming. These CDW amplitude modes are related to the complex order parameter and follow a mean-field like temperature dependence which is given as:

$$\omega^i_{CDW}(T) = D \; \omega^i_o \left(1 - \frac{T}{T_{CDW}}\right)^{\gamma} \qquad\qquad \text{-- (5.1)}$$

Here $\omega^i_o$ is the unscreened, or high-temperature phonon frequency and $D \propto N \, g^2(q)$ represents the electron-phonon coupling constant. $N$ is the joint density of states of electrons or holes involved in CDW transition, $g(q)$ is $q$ dependent electron-phonon coupling matrix element [385-387]. The solid cyan line for S9 mode in **figure 5.6** represents the fit and is quite a good fit, where D and $\gamma$ are varied to get a reasonable value of $T_{CDW}$. We have fitted other modes which emerged below $T_{CDW}$ ~230K i.e., S2, S12, S15, S16, S20, S21 and S22. The obtained values of $\gamma$ is mentioned in the **table 5.5**. Theoretically expected value for $\gamma \sim 0.5$ , but we found that this expected nature failed for the modes in our case, owing to an incomplete softening as the mode acquires a finite value of frequency instead of zero just before the phase transition as predicted. This behavior has been observed in different CDW materials and is attributed to short-range fluctuations that are out of phase with local CDW order as found for 2H-NbSe$_2$ [388] and TiSe$_2$ [387,389]. **Figure 5.8** shows a bird's eye view of the emergence of new modes for T1 i.e., S1, S2, S3, S8, S9, S12, S13, S15, S16, S17, S19, S20, S21, S22 and S25 below ~230K. The emergence of new modes for T2 and T3 is also shown in **figure 5.9**. In addition to that on further lowering the temperature we observed the emergence of some weak modes i.e., S7, S10, S14, and S24 below T$_H$~80K. This temperature is marked by the presence of a hidden quantum CDW state.

| Modes # | $\gamma$ $10^{-3}$ T1 | $\gamma$ $10^{-3}$ T2 | $\gamma$ $10^{-3}$ T3 |
|---------|------------------------|------------------------|------------------------|
| **S9**  | **7.8 ± 2.1** | **8.8 ± 2.2** | **8.7 ± 1.8** |
| **S12** | **8.1 ± 1.2** | **6.9 ± 1.4** | **9.6 ± 1.2** |
| **S15** | **2.3 ± 1.5** | **8.5 ± 1.2** | **5.8 ± 1.4** |
| **S16** | **9.9 ± 2.7** | **9.3 ± 2.0** | **9.3 ± 2.6** |
| **S20** | **4.3 ± 0.8** | **3.8 ± 1.3** | **3.7 ± 0.6** |
| **S21** | **2.5 ± 0.7** | **2.8 ± 0.6** | **2.6 ± 0.7** |
| **S22** | **1.3 ± 0.2** | **1.0 ± 0.3** | **0.9 ± 0.4** |

**Table 5.5:** Value of $\gamma$ for modes corresponding to different thicknesses as mentioned in the text.

The presence of this hidden quantum CDW state is also observed in the modes S5, S6, S18, and S23 as shown in the inset (S6 and S18) of **figure 5.6**. The effect is also visible in the modes S11, S12, S14, S17, and S22 as shown in **figure 5.7** (dashed red line). Stojchevska *et al.* [370], also showed the presence of this state by probing its insulating ground state using a single intense femto-second laser pulse. As compared to other states in this system the hidden state shows larger changes in the resistance, strong modification in single-particle and collective-mode spectra, and is also reflected in the optical reflectivity. Stahl *et al.* [390], proposed that this state involves a rearrangement of the charge and orbital order in the direction of perpendicular to the $TaS_2$ layer which results in the collapse of inter-layer dimerization. Further, we observed a broad spectral feature at ∼ 380 cm⁻¹ **(figures 5.4 and 5.5)**. Such broad Raman signatures are commonly found in CDW materials. It is associated with the strong momentum-dependent electron-phonon coupling near the CDW wave vector. It is a second-order process where the phonon-assisted scattering of the electrons creates two phonons with equal and opposite wave-vectors; $\vec{q_1} + \vec{q_2} = \vec{0}$. All phonons which satisfy this condition contribute to the two-phonon scattering which leads to large FWHM. Generally multi-phonon processes are weaker than one-phonon processes but in the case of CDW materials, it is found to be much stronger which is attributed to the electron-phonon coupling [391]. The temperature evolution of this mode gives an indirect measure of phonon branch renormalization during the CDW transition [392]. We label this broad two-phonon mode as $S_{2\text{-ph}}$ and the temperature evolution of its peak frequency, linewidth width, and intensity are shown in **figure 5.10**. The frequency of this mode increases linearly on decreasing temperatures till ~210K ($T_{CDW}$) and shows a discontinuity then remains nearly constant (for T3 and T2) till ~100K which is the hidden quantum phase transition temperature and finally softens till 4K for all flakes. Surprisingly, the FWHM of this mode increases with decreasing temperatures which is opposite to the normal phonon mode behavior suggesting electron-phonon coupling. The intensity variation is also anomalous as it increases with a decrease in temperature.

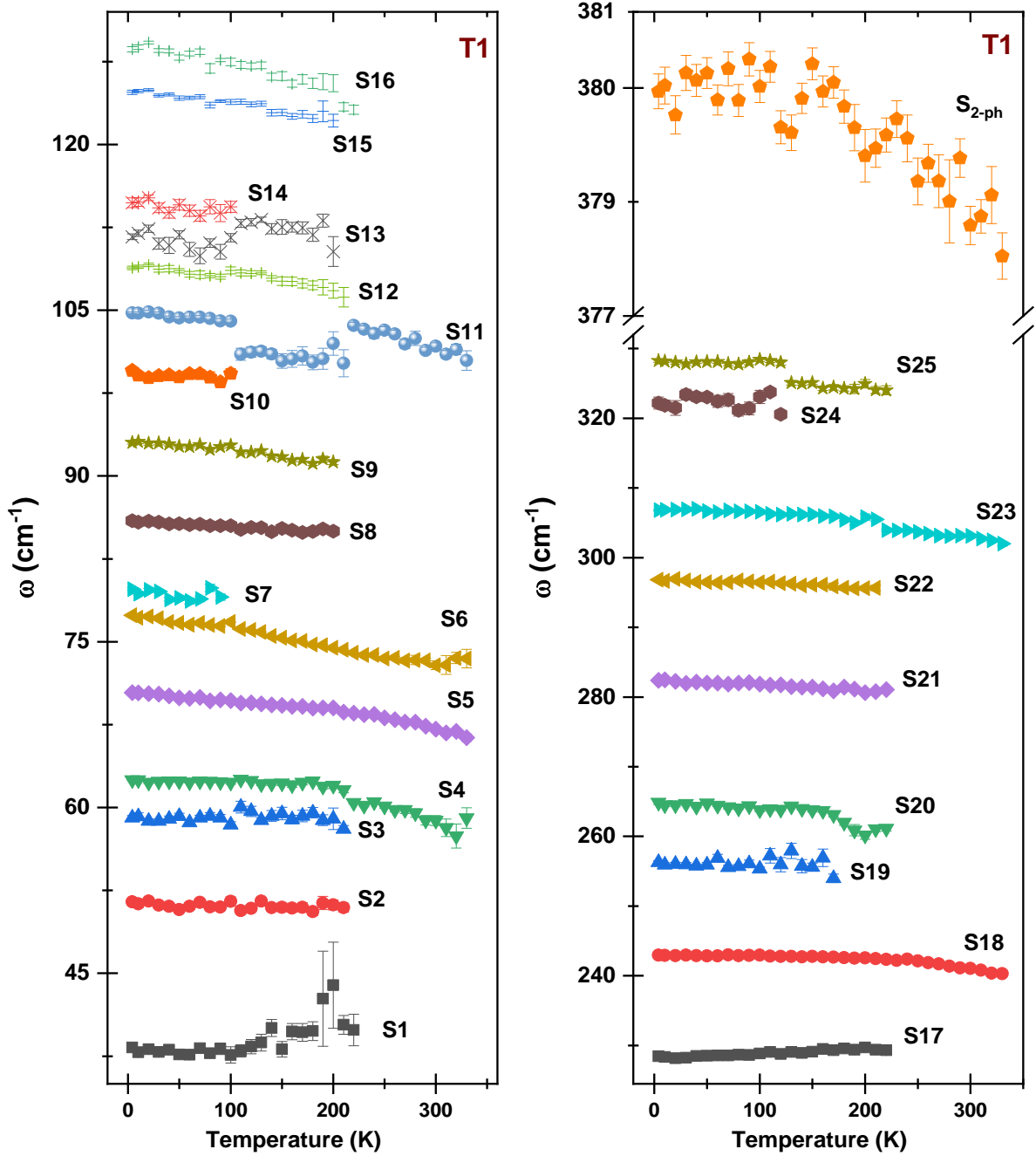

**Figure 5.8:** Temperature evolution of the observed phonon modes frequency for T1.

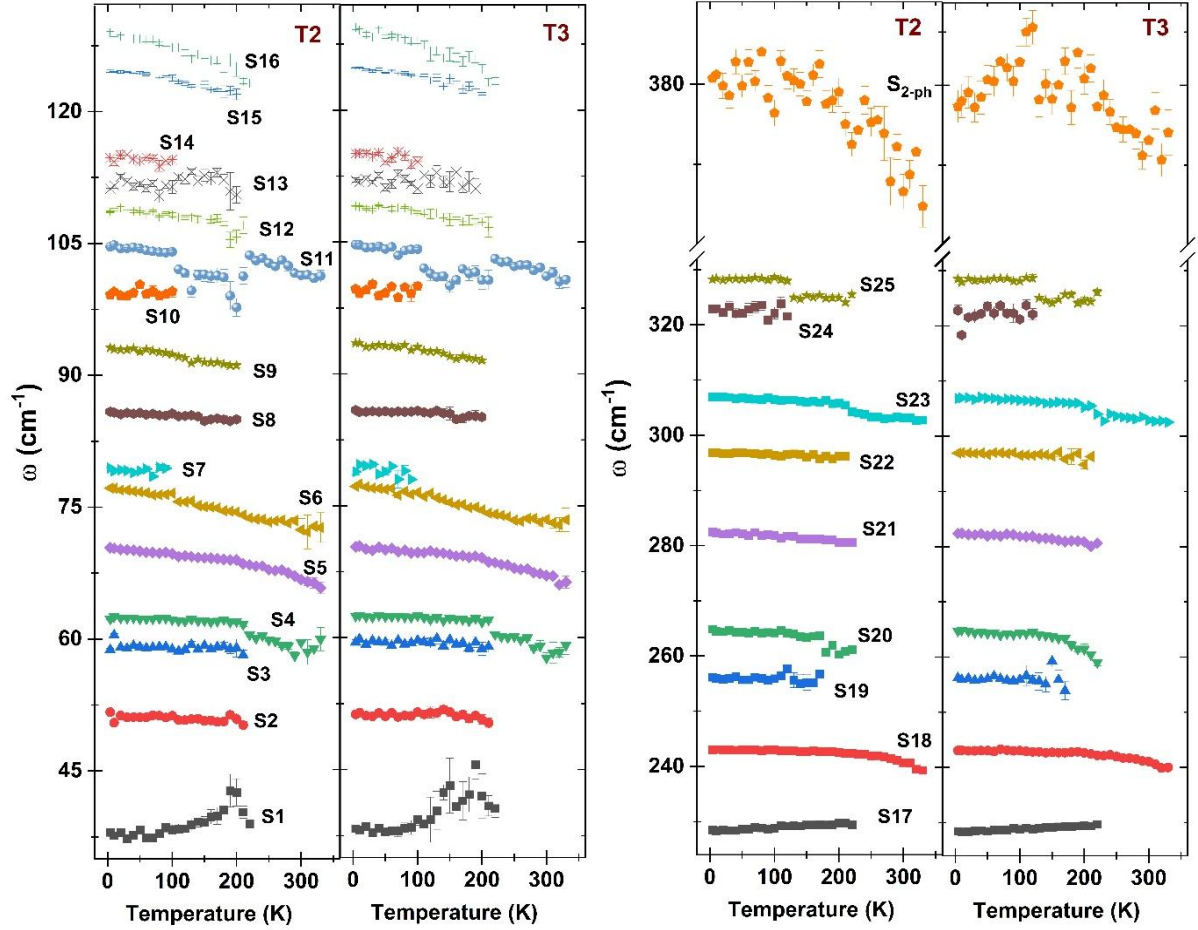

**Figure 5.9:** Temperature variation of frequency for T2 and T3 of all modes.

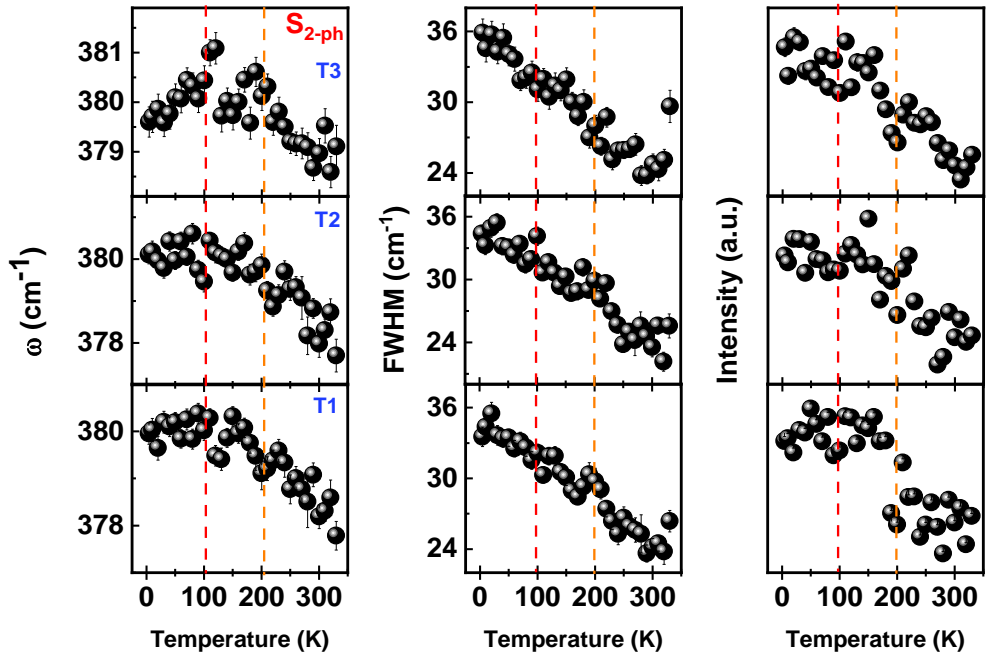

**Figure 5.10:** Temperature-dependent frequency and linewidth variation for T1, T2, and T3 of phonon modes $S_{2\text{-ph}}$. The red and orange dashed line shows $T_H$ and $T_{CDW}$ transition temperatures.

## 5.2.4 Polarization-Dependent Analysis

The Raman scattering cross-section is affected by the symmetry of the crystal through second-order susceptibility/electron-phonon interaction in macroscopic/microscopic descriptions, respectively; and the intensities depend on the polarization direction of the incident light [218,229]. Polarization measurements can unveil the symmetry of the phonon modes, hence we performed an angle-resolved polarized Raman scattering (ARPRS) experiment in a configuration where the direction of polarization of the incident light is tuned ranging from 0$^{o}$ to 360$^{o}$ using a half wave retarder ($\lambda$/2 plate), whereas the analyzer's principal axis has been kept fixed. The experiment is performed at two different temperatures 5K and 230K for T2 flake. **Figure 5.11** shows angular variation in intensity and the red solid line shows the fitting which is discussed below. Within the semiclassical approach, the Raman scattering intensity from first-order phonon modes is related to the Raman tensor, polarization configuration of the incident, and scattered light which can be defined as mentioned in equation 3.4 of Chapter 3.

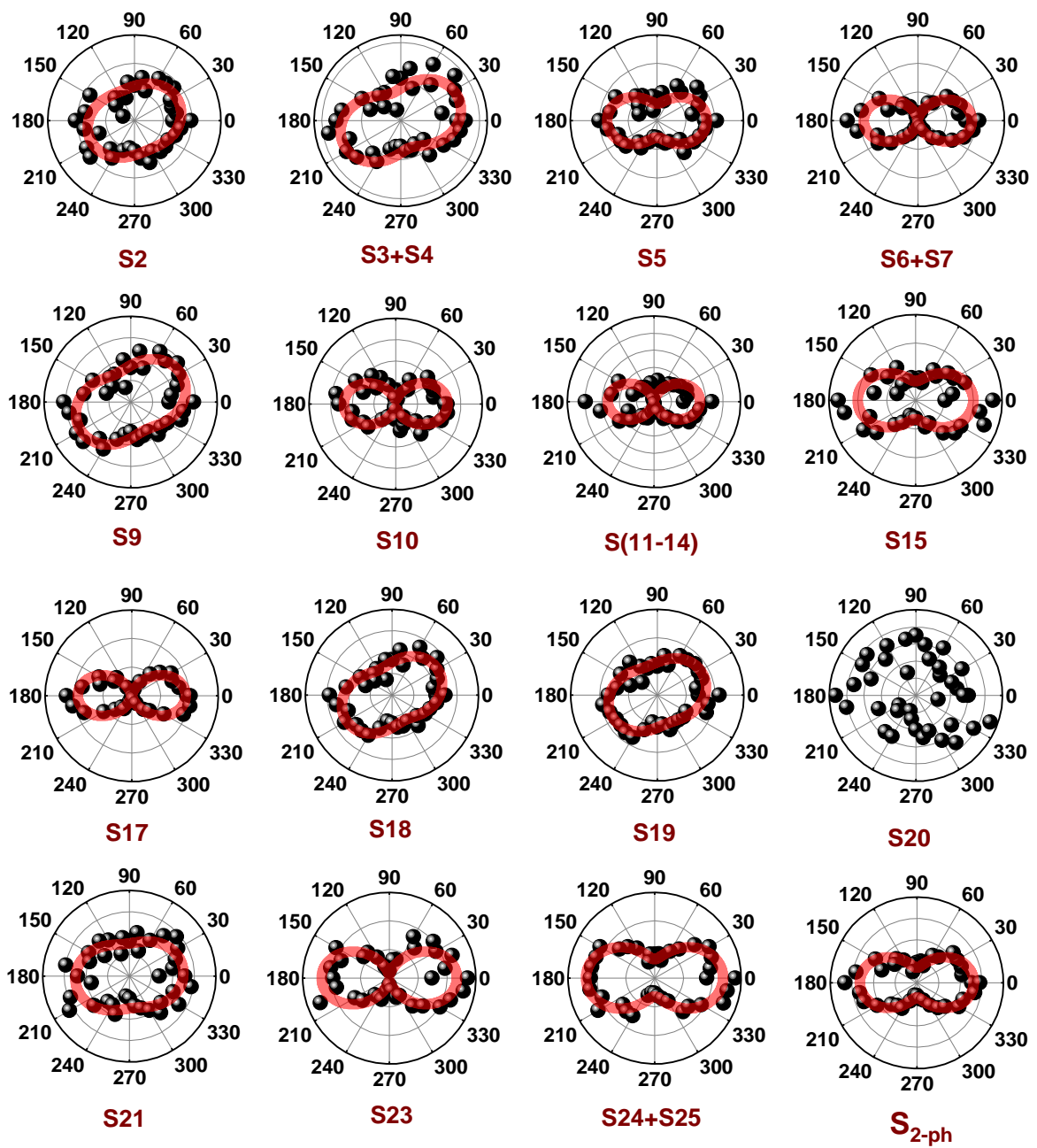

**230K**

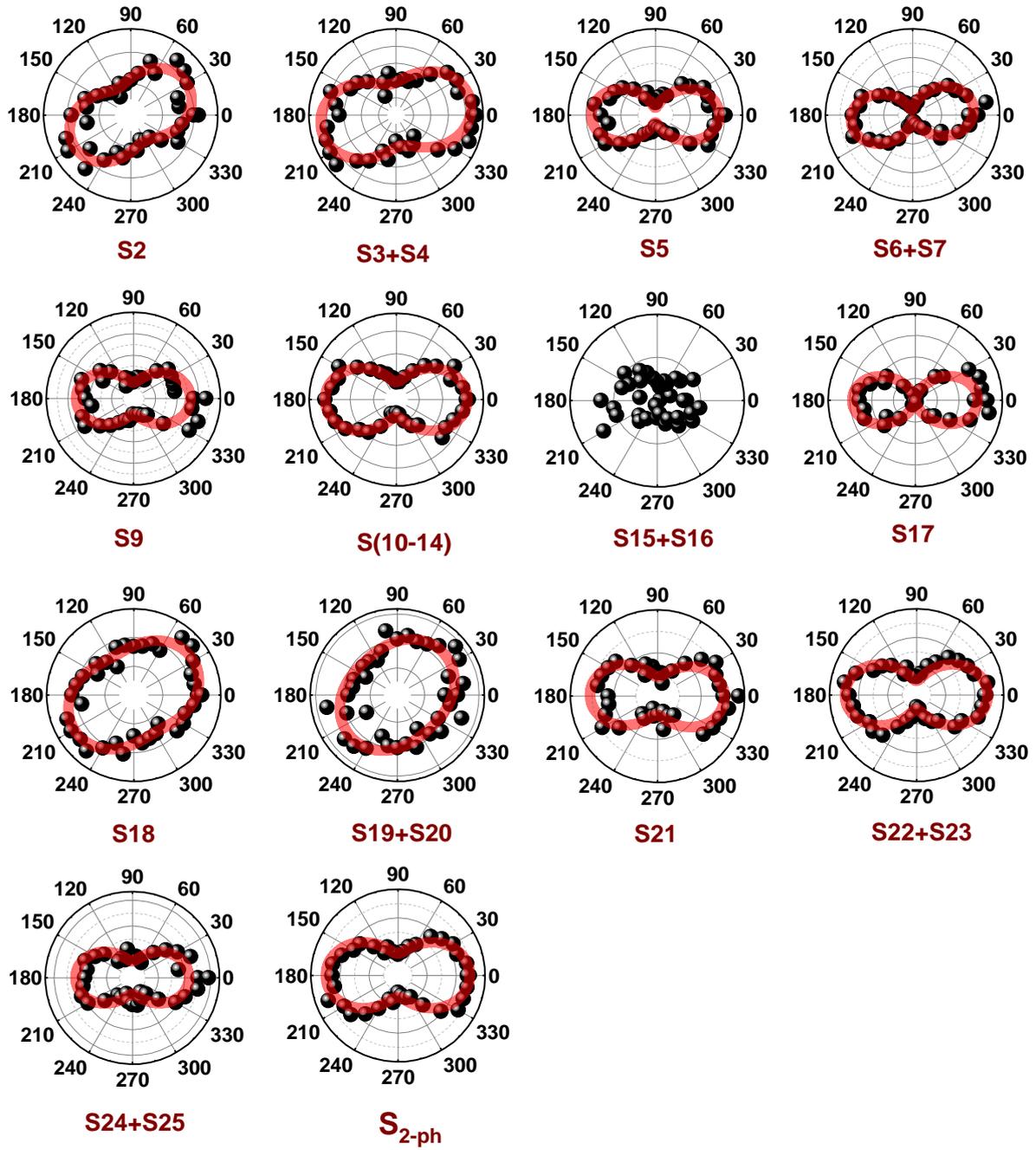

**Figure 5.11**: ( Top and bottom panel ) Polarization-dependent intensity variation of modes at 5K and 230K for T2. A red solid line is a fit as mentioned in the text.

A schematic diagram of polarization vectors of incident and scattered light projected on an x (*a*-axis) - y (*b*-axis) plane is shown in **figure 5.1(d)**. In the matrix form, incident and scattered light polarization direction vectors are written as: $\hat{e}_i = \begin{bmatrix} \cos(\alpha + \beta) & \sin(\alpha + \beta) & 0 \end{bmatrix}$ ;

$\hat{e}_s = [\cos(\alpha)\ \sin(\alpha)\ 0]$, where '$\beta$' is the relative angle between '$\hat{e}_i$' and '$\hat{e}_s$' and '$\alpha$' is an angle of scattered light from x-axis. As discussed in section 5.2.1 the symmetry of the modes in the C-CDW and NC-CDW phase is $A_g / E_g$. Hence, we have used the Raman tensors of the metallic phase with space group #164; P$\overline{3}$m1 and point group $D_{3d}$ (-3m) for fitting purpose for different modes and are summarized in Table 5.1. The angular dependency of intensities of the Raman active modes can be written as $I_{A_g} = |a\cos(\beta)|^2$ and $I_{E_g} = I_{E_{g,1}} + I_{E_{g,2}} = |c\cos(2\alpha+\beta)|^2 + |c\sin(2\alpha+\beta)|^2 = |c|^2$. Here, $\alpha$ is an arbitrary angle from the $a$-axis and is kept constant. Therefore, without any loss of generality, it can be taken as zero.

At 5K we observed that modes S2, S3+S4, S9, S18, S19, and S21 show an $E_g$ kind of symmetry. While modes S5, S6+S7, S10, S (11-14), S15, S17, S23, S24+S25 and $S_{2\text{-ph}}$ show $A_g$ kind of symmetry. Here intensity at 90$^o$ does not go to exact zero as theoretically predicted due to experimental limitations but obtains a relative minimum value. At 230K mode S2, S3+S4, S18 and S19+S20 show similar $E_g$ kind of symmetry, while modes S5, S6+S7, S(10-14), S17, S22+S23, S24+S25 and $S_{2\text{-ph}}$ shows $A_g$ symmetry. Interestingly modes S9 and S21 showed a change in symmetry from nearly $E_g$ to $A_g$ with changing temperature. The polarization study reveals that in different CDW phases, the symmetry of the modes is almost unchanged as predicted in previous reports [381].

## 5.2.5 Low-Frequency Raman Response

In CDW systems, the gap in the electronic excitation is found to be momentum-dependent [380]. The Mott transition leads to an additional opening in bands around the $\Gamma$-point [368]. The opening of a gap at the Fermi level reduces $N_F$, which consequently increases resistivity and this reduction of $N_F$ may manifest itself in the Raman spectra resulting in the change of the initial slope of the Raman response [156,393]. Raman response; $\chi^{''}(\omega,T)$ is calculated by dividing raw Raman intensity with the Bose factor, $I(\omega,T) \propto \left[1 + n(\omega,T)\right] \chi^{''}(\omega,T)$. The

Raman response, $\chi^{''}(\omega,T)$, is the imaginary part of the susceptibility and is proportional to

Stokes Raman intensity given as: $I(\omega,T) = \int_0^\infty dt\, \mathrm{e}^{i\omega t} \left\langle R(t)\, R(0) \right\rangle \propto \left[1 + n(\omega,T)\right] \chi^{''}(\omega,T)$;

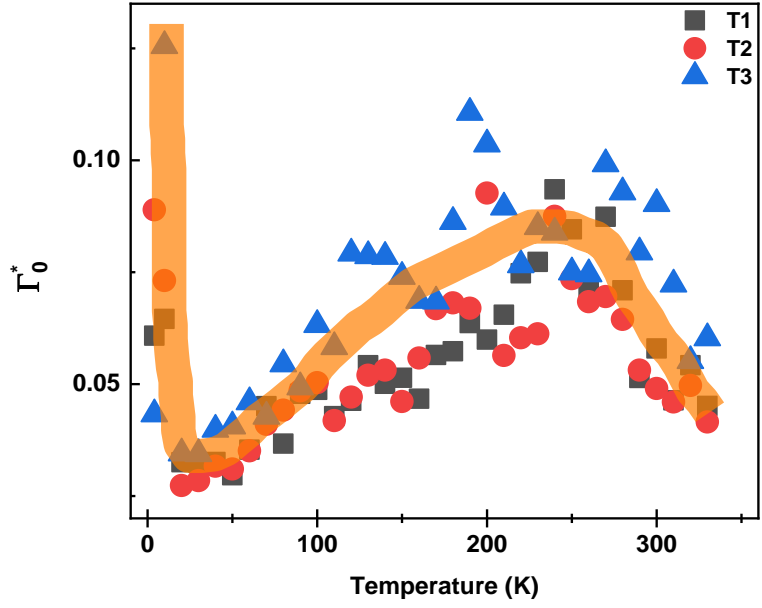

**Figure 5.12**: Raman response slope magnitude variation with temperature for T1, T2, and T3. The orange solid line is a guide to the eye.

where *R(t)* is the Raman operator and $\left[1 + n(\omega,T)\right] = 1/\left[1 - e^{-h\omega_b/k_B T}\right]$ is the Bose factor. The initial slope of the Raman response is $R \lim_{\omega \to 0} \dfrac{\partial \chi^{''}}{\partial \omega} \propto N_F \tau_o \propto \dfrac{1}{\Gamma_0^*}$, here *R* incorporates experimental factors and $\Gamma_0^*$ represent electronic scattering rate. The scattering rate of the conduction electrons is also referred to as Raman resistivity [394]. It is anticipated that with increasing correlation at low temperatures, as the system under study goes from metal to a correlated one and finally to a Mott insulating state inside the CDW phase, the lifetime of putative quasiparticles decreases and as a result inverse of slope $\left(\Gamma_0^*\right)$ should increase. **Figure**

**5.12** shows the inverse of the slope for all thicknesses as a function of temperature. To extract the slope, we linearly fitted the low-frequency region in the range 8 cm$^{-1}$ – 21 cm$^{-1}$. We observe that with decreasing temperature it increases till ~ 210K and then starts decreasing till ~ 100K and with further lowering the temperature it increases below ~ 50-60K. With decreasing temperature below $T_{CDW}$, there is a small but finite $U$ (onsite correlations) contribution and the correlated metal displays an inverse slope $\propto T^2$ reflecting a canonical Fermi liquid behavior i.e. correlated metallic phase. Interestingly, on further decreasing the temperature below ~ 50-60K we saw an upturn in the inverse of the slope suggesting that the system becomes more insulating with decreasing the temperature.

## 5.2.5.1 Quantum Spin Liquids

Quantum Spin Liquid (QSL) is a state in which the quantum fluctuations of the spins are strong enough to preclude the ordering even at absolute zero. The key feature of spin liquids is the presence of fractionalized excitations. Such quantum particles carry spin ½ but lack charge and are known as spinons [16]. 1T-TaS$_2$ in its C-CDW state forms a star of David structure with 12 Ta-atoms pair and forms 6 occupied bands, which leaves the central Ta atom localized and unpaired. The band structure calculations in the commensurate state have pointed out that mostly Ta 5d orbitals contribute to the conduction and valence bands [372,395], while the atomic spin-orbit coupling (SOC) from $d_{x^2-y^2}$ and $d_{xy}$ orbitals go under reconstruction and modify the band structure. This combined effect of structural deformation and atomic SOC give rise to a narrow band gap at the Fermi level. Thus, there is a weak repulsive interaction and the resulting ground state has been proposed to be a Mott insulator [395-397]. Mott insulators generally show an antiferromagnetic ordering in the insulating phase which is not in the case of 1T-TaS$_2$. It has been shown that due to the large paramagnetic contribution to the magnetic susceptibility, it does not show any magnetic ordering even down to 20mK [40]. The localized unpaired electrons carry spin-1/2 and form a triangular lattice which makes it a putative candidate for the realization of quantum spin liquid [40,46,397]. The experimental realization of QSL has been an active area of research and Raman spectroscopy has proved to be an excellent probe for that [58,205]. Magnetic Raman scattering give rise to a broad continuum originating from underlying dynamical spin fluctuations and may be used to gauge the fractionalization of quantum spins expected for proximate spin liquid candidates [57,58,205,221,223-225,240].

Hence, we have analyzed the possible existence of a very much anticipated QSL state and the effect of dimensionality for 1T-TaS$_2$ via Raman Spectroscopy. Signatures of a QSL can be uncovered in the presence of underlying dynamic quantum spin and/or orbital fluctuations and the Raman technique is an excellent probe for probing these underlying dynamical fluctuations and is reflected in the emergence of quasi-elastic peaks at low frequency and $\chi^{dyn}(\omega,T)$ [57,58,240]. We observed a profound temperature variation in the low-frequency region which we consider the quasi-elastic part. It arises from diffusive fluctuations of a four-spin time correlation function or fluctuations of the magnetic energy density [257].The temperature dependence of quasi-elastic scattering intensity can provide information about the evolution of low-energy excitations. The analysis of QES is done by analyzing the temperature dependence of intensity under the curve for the low-frequency range of ~ 6 cm$^{-1}$ - 21 cm$^{-1}$. **Figure 5.13 (a)** shows the temperature-dependent raw spectra and Raman conductivity for T1. The variation of raw intensity spectral weight with temperature for different thicknesses is shown in **figure 5.13 (b)**.

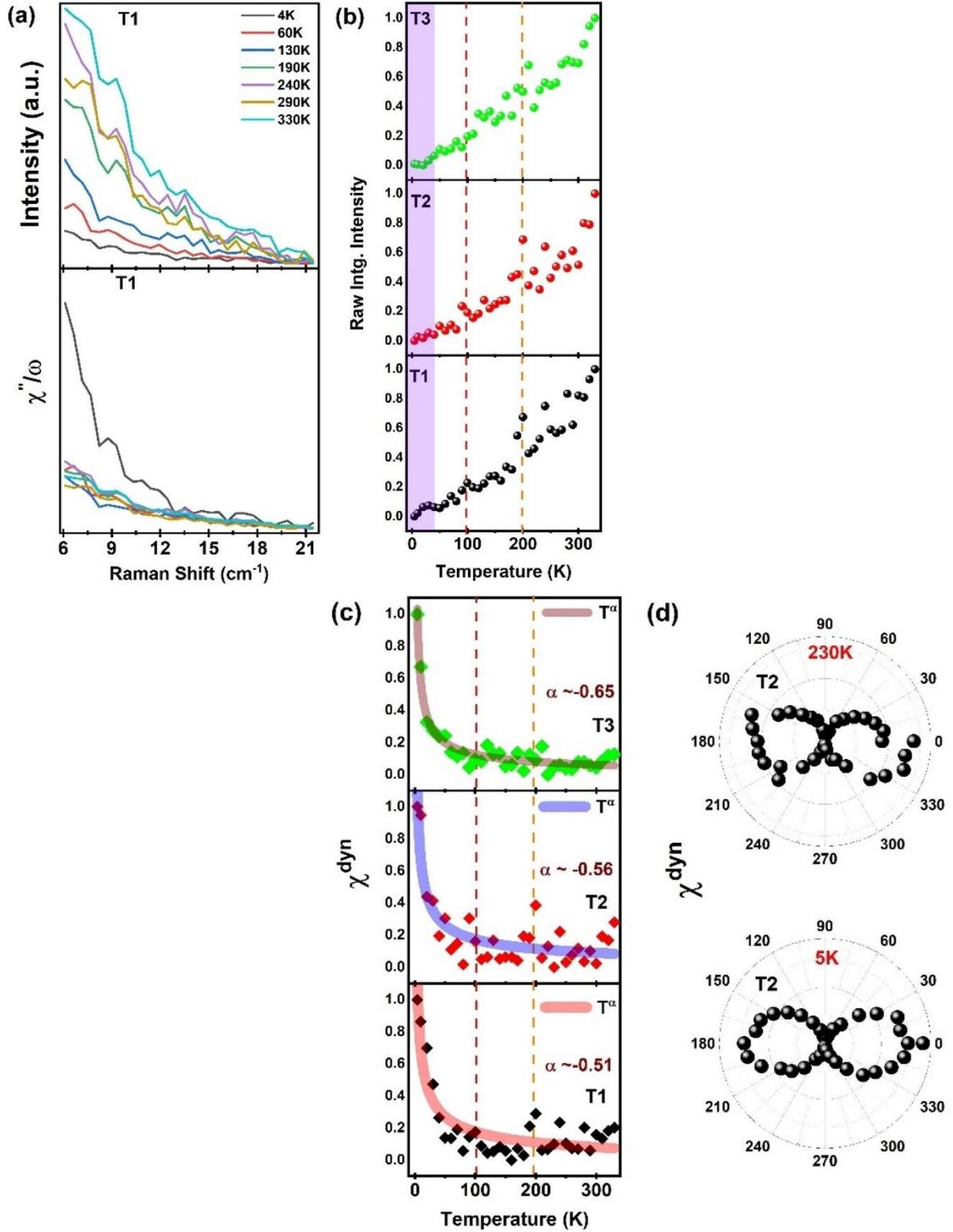

**Figure 5.13:** (Top and bottom panels) **(a)** Raw spectra and Raman conductivity of the low-frequency region; **(b)** integrated raw intensity and **(c)** $\chi^{dyn}(\omega,T)$ thick solid lines are the power law fit as mentioned in the text and **(d)** polarization-dependent $\chi^{dyn}(\omega,T)$ at 5K and 230K for low-frequency region for flake T2.

In addition to that, we also observed the signature of the proposed hidden phase as discontinuity at $T_H \sim 80K$, which has also been reflected in the phonon dynamics. We found another possible signature of a quantum transition below 50K as shown in the shaded violet region, which may be related to the quantum spin liquid and needs further investigation. For further probing the evolution of the underlying quantum spin fluctuations we quantitatively evaluated $\chi^{dyn}(\omega, T)$, shown in **figure 5.13 (c)**. $\chi^{dyn}(\omega, T)$ is evaluated using Kramers - Kronig relation as mentioned in equation 3.7 of chapter 3A.3.6 which relates the real and imaginary part of the susceptibility. $\chi^{"}(\omega)/\omega$ is Raman conductivity, $\Omega$ is the upper cutoff value of integrated frequency and is taken as 21 cm$^{-1}$ for the Quasi-elastic scattering (QES) region. For all thickness with lowering temperatures other than small discontinuities at $T_{CDW} \sim 200K$ which is CDW transition temperature, $\chi^{dyn}(\omega, T)$, remains almost independent of temperature till ~50K which is expected in a paramagnetic regime where spins remain uncorrelated. Interestingly it starts to increase drastically on further lowering the temperature below, ~ 50K for T1, T2, and T3. We consider these temperatures as the onset point of dynamical fluctuation as cross-over temperature. For typical antiferromagnets is expected to quench below Néel temperature ($T_N$). Such a behavior is a typical characteristic scattering from underlying quantum spin fluctuations [58,225].

The variation of $\chi^{dyn}(\omega, T)$ is well described by a power law $\chi^{dyn}(\omega, T) \propto T^{\alpha}$. The power law is associated with the inherent long-range entangled spin-liquid and is very different from conventional magnets. Hence, we have fitted it with power law as shown in **figure 5.13 (c)**, $\alpha$ reflects on the underlying valence bond randomness due to spins [398]. Further, we inquired about the polarization dependence of the quasi-elastic $\chi^{dyn}(\omega, T)$ response for T2 flake at 5K and 230K which is shown in **figure 5.13 (d)**. At both temperatures, it shows an $A_g$ kind of symmetry i.e., two-fold rotational anisotropy similar to the $A_g$ phonon modes.

## 5.3 Conclusion

In conclusion, we performed detailed temperature (4K-330K) and polarization-dependent Raman measurements on three flakes with varying thicknesses of 1T-TaS$_2$. A clear signature of CDW transition is seen as we observed the emergence of new modes on going from the high-temperature NC-CDW towards the low-temperature C-CDW phase. We observed discontinuities in the phonon dynamics around T$_{CDW}$. In addition to the CDW transition, a strong signature of a hidden quantum CDW state is also observed at $T_H \sim$ 80K marked by the emergence of new modes and discontinuities in the most prominent modes. Our polarization-dependent analysis shows that the symmetry of most of the modes remains the same even in the CDW phases which is consistent with previous reports. Our analysis of the low-frequency quasi-elastic region also shows that there are some additional competing interactions at low temperatures, which is reflected in the slope of the Raman response. The analysis of dynamic fluctuations qualitatively showed by $\chi^{dyn}(\omega, T)$ and the power law fit reflects an increase in quantum fluctuation effects on decreasing thickness. Our results show a possible Mott insulating state below $\sim$ 50-60 K. Our results reflect on the interesting interplay of dimensionality with the underlying hidden quantum state as well as normal CDW states in 1T-TaS$_2$.

# Chapter 6: Summary and Future Scope

This chapter summarizes the research work done in the thesis which is described in the earlier chapters along with the significance of the finding. The thesis research work focuses on probing the underlying rich physics of the quantum materials via phonon dynamics using both experimental i.e., Raman Spectroscopy, and computational i.e., DFPT. The findings of the research work set a paradigm in the search of future quantum materials for advanced technologies such as Quantum computing and communication. A summary of the research work findings as mentioned in detail in the previous chapters and future scope is provided in the following sections.

## 6.1 Summary

In **part A** of **Chapter 3** an experimental signature of the QSL state in a single crystal of high spin ($S =1$ and $3/2$) quasi-two-dimensional magnetic $V_{1-x}PS_3$ via the inelastic light (Raman) scattering and transport measurements studies is presented. Our analysis of dynamic Raman susceptibility, which reflects the underlying dynamics of collective excitations, evidences the signature of the spin fractionalization into Majorana fermions marked by a crossover temperature $T^*$ ~ 200 K from a pure paramagnetic state to the fractionalized spins regime reflected as a quasi-elastic response at low energies and temperature along with a broad magnetic continuum, qualitatively identified in dynamic Raman susceptibility. We further evidenced anomalies in the phonons' self-energy parameters, in particular phonon line broadening and Fano line asymmetry evolution starting from this crossover temperature ($T^*$), attributed to the decaying of phonons into itinerant Majorana fermions signaling a proximate QSL phase. Our results provide the first experimental realization of fractionalized quasiparticle excitation in a high spin-based quasi-2D magnetic system.

In **part B** of **Chapter 3,** we presented a comprehensive thickness, down to ~ 8- layers, dependent on inelastic Raman spectroscopic measurement on a single crystal magnetic insulator $V_{0.85}PS_3$. This system in the high spin configuration remains mostly unexplored in particularly as a function of thickness, especially down to a few layers. In the temperature evolution of the Raman spectra, we observed signatures of a broad background magnetic continuum, quasi-elastic scattering, phonon anomalies, and Fano asymmetry. The presence of anomalies in phonon dynamics (energy and linewidth) such as discontinuities at cross-over temperature $T^*$ ~200K and an increase in linewidth of low-energy phonon modes on decreasing temperature, indicates the presence of an additional decay channel. Interestingly, we observed suppression of the spin-phonon coupling effect in the spin-solid phase below 60K, which suggests the presence of competing magnetic interactions in low dimensions and is attributed to enhanced quantum spin fluctuations. The low-energy modes show Fano asymmetry, which is evidence that the magnetic continuum strongly couples with the lattice vibrations. The qualitative analysis of dynamic Raman susceptibility for the quasi-elastic scattering and the broad magnetic continuum reveals the presence of non-bosonic excitations which increase with decreasing temperature and thickness and are attributed to the fractionalization of the spins deep into the paramagnetic regime. The investigation is also supported by DFT-based zone-center phonon calculations.

In **Chapter 4** we presented an in-depth temperature and polarization-resolved Raman spectroscopic investigation for a topological nodal line non-centrosymmetric semimetal $PbTaSe_2$ where we found evidence of surface phonons and a possible structural transition from α to β phase on decreasing temperatures. We observed the disappearance of some phonon modes (M1-M5) on decreasing temperature, which are found to be weak in intensity and lower in energy as compared to other (Bulk modes, P1-P5) that are consistent throughout the temperature range. Interestingly, only these weak modes (M1-M4) exhibit Fano asymmetry in their line shape which is an indication of electron-phonon coupling with the surface excitation. We assign these modes as surface phonons. Also, these modes are disappearing in the temperature range where discontinuities in bulk phonon dynamics are observed, which is an indication of the potential structural transition associated with it. Further, this argument is also corroborated by zone-centered phonon calculations and temperature-dependent single crystal XRD measurements where we observe new peaks at lower temperatures, and polarization-dependent study where we observed change in symmetry of the bulk modes at different temperatures i.e., 6K and 300K. Based on our results, at ambient pressures and on decreasing temperature it possibly undergoes from $\alpha$ to $\beta$ phase at $T_{\alpha \rightarrow \beta} \sim 150$ K or is in close proximity to the $\beta$ phase and another phase at $T_{CDW+\beta} \sim 100$K which could possibly be due to the interplay of remanent CCDW phase of $1H$-$TaSe_2$ and surface topology of the material. Our results provide the first experimental evidence of surface phonons in this material, which demonstrate an intriguing interplay of topology and thermally driven structural transition in $PbTaSe_2$.

In **Chapter 5** we present a temperature (4K-330K) and polarization-resolved Raman spectroscopic investigation on mechanically exfoliated flakes of $1T$-$TaS_2$ of thickness $\sim 8.6$ nm, $\sim 10.4$ nm, and $\sim 12.5$ nm. A DFT-based theoretical calculation of phonon dispersion and zone-centre phonon frequencies is also computed. In addition to the completely commensurate CDW state, we have found strong evidence of a 'Hidden CDW' state below $\sim 80$K reflected in phonon anomalies and, the emergence of new phonon modes. The nature of the low-frequency quasi-elastic region shows that there are some additional competing interactions present at low temperatures, which is reflected in the slope of the Raman response. The analysis of dynamic fluctuations qualitatively by dynamic Raman susceptibility ($\chi^{dyn}$) and the power law fit reflects an increase in quantum fluctuation effects on decreasing thickness. The results indicate a possible Mott insulating state below ~ 50-60 K. Our results reflect on the interesting interplay of dimensionality with the underlying hidden quantum state as well as normal CDW states in quasi-two-dimensional 1T-TaS$_2$.

## 6.2 Future Scope

In this thesis, we have mostly focused on the temperature and polarization-dependent inelastic light scattering investigation of quantum materials having properties like quantum spin liquids, charge density waves, topological surface phonons, structural transitions and Mott-insulators. Raman scattering has been very crucial in unveiling the underlying physics of exotic quasi-particles and the ground state of these materials. However, we believe that furthermore, light can be shed on the true nature of these systems in a Raman experiment by studying them under external perturbations such as magnetic and electric fields and pressure. Also, it will be interesting to explore the physics of these correlated systems in ultra-low dimensions, such as mono and bi-layers. We have used van der Waals single crystals in bulk and mechanically exfoliated low-dimensional counterparts up to a few layers. We believe that a heterostructure configuration of these systems will provide an interesting playground for exciting and novel phenomena and will provide a new dimension and paradigm for the future of quantum materials.